\documentclass[a4paper,10pt,titlepage,twoside,openright]{book}
\usepackage[english]{babel}
\usepackage[T1]{fontenc}
\usepackage[utf8,latin1]{inputenc}
\usepackage[Lenny]{fncychap}
\usepackage{gensymb}
\usepackage{xcolor}
\usepackage{ragged2e}
\usepackage{booktabs}
\usepackage{multirow}
\usepackage{footnote}
\usepackage{setspace}
\usepackage{enumitem}
\usepackage{fancyhdr}
\usepackage{amsmath}
\usepackage{wasysym}
\usepackage{amssymb}
\usepackage{amsthm}
\usepackage{mathrsfs}
\usepackage{setspace}
\usepackage{amscd}
\usepackage{graphicx}
\usepackage{color}
\usepackage{multicol}
\usepackage{caption}
\usepackage{afterpage}
\usepackage{lmodern}
\usepackage{datetime}
\usepackage{lettrine}
\usepackage{epigraph}
\usepackage{hyperref}
\hypersetup{colorlinks, linkcolor={blue},citecolor={blue},urlcolor={blue}}  
\renewcommand{\chaptermark}[1]{%
\markboth{#1}{}}
\fancypagestyle{plain}{%
\fancyhf{} 

}
\makeatletter
\renewcommand{\chaptermark}[1]{\markboth{\@chapapp \space \thechapter.\ \textit{#1}}{}}
\makeatother

\newcommand\ChangeRT[1]{\noalign{\hrule height #1}}

\newcommand{\qm}[1]{``#1''}
\theoremstyle{definition}

\theoremstyle{remark}

\DeclareMathOperator{\sign}{sign}
\def\compps{{\sc compps}} 
 
\def\bbodyrad{{\sc bbodyrad}}

\def\gr{$\gamma$}
\def\Integ{{\em INTEGRAL}} 
\def\maxi{{\em MAXI}} 
\def\rxte{{\em RXTE}} 
\def\swift{{\em Swift}} 
\def\Fermi{{\em Fermi}} 
\def\chandra{{\em Chandra}} 
\def\xmm{{\em XMM-Newton}} 
\def\igr{IGR~J00291+5934}
\def\igrj{IGR~J18245--2452} 
\def\sax{SAX~J1748.9--2021}  
\def\be{\begin{equation}} 
\def\ee{\end{equation}} 
\begin{document}
%
%
%
%
%
%
%
%
%
%
%
%
%
%
%
%
%
%
%
\frontmatter

\thispagestyle{empty}

\begin{center}
\vspace{0.2cm}
\textsf{{ \LARGE \bfseries Coupling Poynting-Robertson Effect\\ \vspace{0.3cm} in Mass Accretion Flow Physics}}
\end{center}                                  
\vspace{2cm}

\begin{center}
\textbf{Inauguraldissertation}\\
zur\\
Erlangung der W{\"u}rde eines Doktors der Philosophie\\
vorgelegt der\\
Philosophisch-Naturwissenschaftlichen Fakult{\"a}t\\
der Universit{\"a}t Basel\\
\vspace{0.2cm}
von
\end{center}                                  
\vspace{1cm}

\begin{center}
\textsf{{\large Vittorio De Falco}}\\
aus Italien
\end{center} 
\vspace{1cm}

\begin{center}
2019\\
\vspace{1cm}
Originaldokument gespeichert auf dem Dokumentenserver der Universität Basel.\\
Namensnennung 4.0 International (CC BY 4.0)\\
edoc.unibas.ch\\
\href{https://edoc.unibas.ch/69488/}{doi: 10.5451/unibas-007073960}
\end{center}                                  

\begin{figure}[h!]
\centering
\includegraphics[scale=0.5]{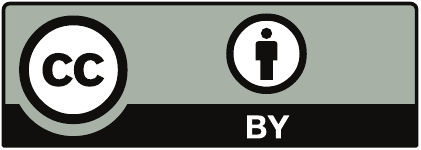}
\end{figure}

\clearpage
\thispagestyle{empty}

\noindent
Genehmigt von der Philosophisch-Naturwissenschaftlichen Fakultät\\
auf Antrag von
\vspace{1cm}\\
{\large Prof. Dr. \ Friedrich-K. Thielemann\\
\\
PD Dr. \ \ \ Maurizio Falanga\\
\\
Prof. Dr.\ \  Luigi Stella}
\vspace{5cm}\\
Basel,  23 May 2017\\
\vspace{2cm} 

\begin{flushright}
Prof. Dr. Martin Spiess\\
The Dean of Faculty
\end{flushright}
\thispagestyle{empty}
\cleardoublepage
\clearpage

\begin{center}
\emph{Few people are able to express opinions that \\
dissent from the prejudices of their social group. \\
The majority are even incapable of forming \\ 
such opinions at all. }
\end{center}
\vspace{0.5cm}
\epigraph{Albert Einstein} 

\thispagestyle{empty}
\begin{figure}[h]
\centering
\includegraphics[scale=0.9]{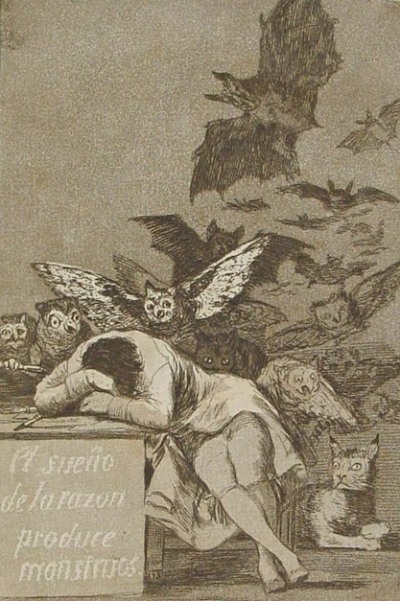}
\\
\justifying
{\bf The Sleep of Reason Produces Monsters} is an etching by the Spanish painter and printmaker Francisco Goya. Created between 1797 and 1799. The work is held at the Metropolitan Museum of Art in New York.
\label{fig:Dedica}
\end{figure}
\thispagestyle{empty}
\clearpage
\thispagestyle{empty}
{\Huge \bf Abstract}\\
\\
The physics of accretion onto compact objects has been experiencing for several decades by now a golden age in terms of theoretical knowledges and observational discoveries. Compact objects release the gravitational energy of the accreted matter in the form of persistent emission or thermonuclear type-I X-ray burst. This radiation field carries out energy and momentum that is transferred back to the interacting plasma inside the accretion disk. The radiation field entails a radiation pressure and a radiation drag force, which both can drastically change or even halt the whole mass transfer (especially when their intensity reaches the Eddington limit). The radiation drag force, known as Poynting-Robertson effect, acts as a dissipative force against the matter's orbital motion, removing very efficiently angular momentum and energy from it.\\
To describe suitably the radiation processes around static compact objects, the Schwarzschild metric is usually employed. To this aim, I have developed a mathematical method for deriving a set of high-accurate approximate polynomial formulae to easily integrate photon geodesics in a Schwarzschild spacetime.\\ 
Starting from the general relativistic treatment of the Poynting-Robertson effect led by Bini \emph{et al.}, I gave two fundamental contributions in such research field. In a first work, I proved through the introduction of an integrating factor that such effect admits a Lagrangian formulation, very peculiar propriety for a dissipative system in General Relativity. In the other work, I have extended the two dimensional general relativistic PR model in three dimensions. \\
Once the theoretical apparatus has been developed, it is important to learn the state of art about the observational high-energy astrophysics. For such reasons, I focussed my energy on the data analysis of three accreting millisecond X-ray pulsars: \igr, \igrj, and \sax.\\
This thesis offers innovative ideas in the field of radiation processes involving the Poynting-Robertson effect in high-energy astrophysics, opening thus up future interesting perspectives both in theoretical and observational physics. 
As conclusion, we propose possible further developments and applications.
\thispagestyle{empty}
\tableofcontents 

\mainmatter
\chapter{Introduction}
\epigraph{The scientist is not a person who gives the right answers, he's one who asks the right questions.}{Claude L\'evi-Strauss}

\begin{figure}[h]
\centering
\includegraphics[scale=0.265]{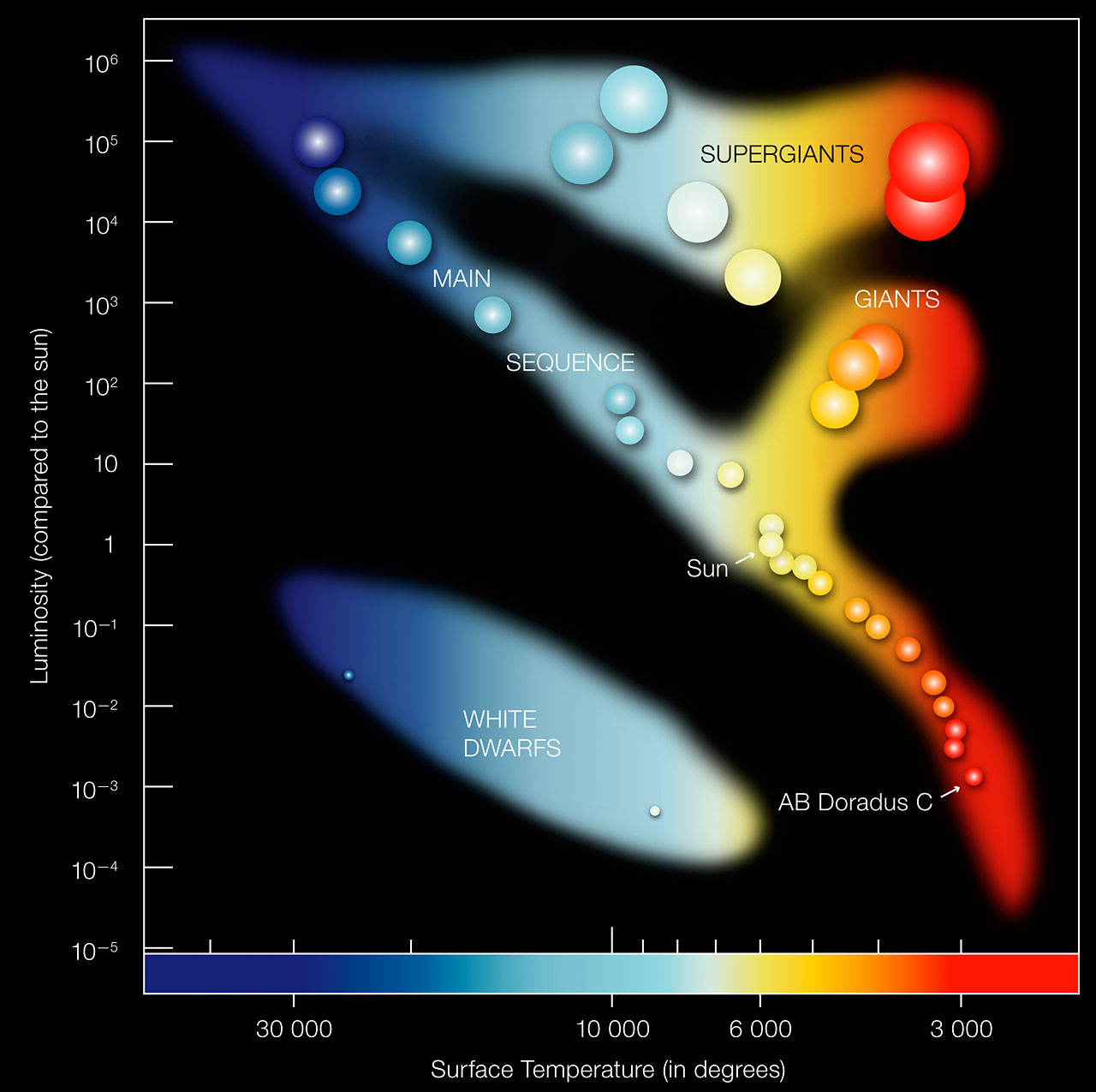}
\caption{Hertzsprung-Russell diagram is a plot reporting the temperature in terms of the luminosity. Credit: European Southern Observatory (image taken from \url{https://www.eso.org/public/images/eso0728c/}).} 
\label{fig:Fig0_1}
\end{figure}

\lettrine{T}{he} stellar evolution is the process by which a star changes over the course of time, where its life cycle is closely related on its mass \cite{Kippenhahn12}. Since stellar changes occur over many centuries, the stellar evolution is studied observing a star population, containing stars in different phases of their lives, and astrophysicists come to understand how stars evolve by simulating the stellar structure through computer models. For understanding the evolutionary mechanisms the \emph{Hertzsprung-Russell diagram} is a valid instrument to know in which phase of the life an observed star is (see Fig. \ref{fig:Fig0_1} and Ref. \cite{Longair11}).  

A star can be seen as a luminous sphere of plasma held together by its own gravity \cite{Shapiro86}. For most of its life, a star shines due to thermonuclear fusion of hydrogen into helium in its core, releasing energy that traverses the star's interior and then radiates into outer space. During the star's lifetime, elements heavier than helium are created by stellar nucleosynthesis and fused in the star's core \cite{Shapiro86,Misner73}. The internal pressure prevents the star from collapsing further under its own gravity. Therefore, it can be said that stars are objects where the gravity force is balanced by the pressure gradient of the hot gas contained inside them \cite{Shapiro86}. The stellar evolution expects three possible endpoints: \emph{(i) white dwarfs}, where the inward pull of gravity is balanced by the electron degeneracy pressure; \emph{(ii) neutron stars} (NSs), where the internal pressure support is provided by the neutron degeneracy pressure;\emph{(iii) black holes} (BHs), where the neutron degeneracy pressure is insufficient to prevent collapse. These three kinds of objects are often referred to as stellar remnants or compact stars, because they are very massive and small in volume, conferring them a very high density. In this thesis, I focus my attention only on the latter two, i.e., NSs and BHs.  

The exotic idea of a BH, as an astrophysical body endowed with sufficiently large mass and small radius such that its gravitational pull is so strong that even the light cannot escape, was ideally introduced for the first time by the astronomical pioneers Michell and Laplace in 1783--1795. Later on in 1915, Einstein proposed his theory of General Relativity (GR), and one year later, in 1916, Schwarzschild proposed the first exact solution of Einstein's field equations, describing the gravitational field surrounding a static spherical mass. In 1924, Eddington fiercely opposed against the possibility to detect a massive star compressed in its Schwarzschild radius, because: ($i$) the gravitational force is so strong that light would be unable to escape from it; ($ii$) the redshift of the spectral lines would be shifted out of existence; ($iii$) the mass would curve so much the space around, leaving us outside. 

In 1931, Chandrasekhar discovered the existence of an upper limit for the mass of a completely degenerate configuration (now called \emph{Chandrasekhar limit}). However in 1932--1935, Eddington and Landau did not accept Chandrasekhar's result, because it would have implied that the inevitable fate of massive star evolutions are the formation of BHs. In 1939, Oppenheimer and Snyder predicted that NSs with mass $M\gtrsim3M_\odot$ (also known as \emph{Tolman Oppenheimer Volkoff limit}) collapse into BHs. They calculated rigorously that for a homogeneous sphere of pressureless gas in GR there is no physical law that can halt the collapse, demonstrating thus the formation of a BH. 

In the late 1950, Wheeler and his collaborators began a serious investigation of the collapse problem and they coined the name "BH". In 1963--1965, other two important exact solutions of Einstein's field equations were presented: Kerr found it for rotating BHs and Newman for both rotating and electrically charged BHs. From these results the \emph{no hair theorem} emerged, stating that a stationary BH solution is completely described by only three fundamental parameters: mass, angular momentum, and electric charge. Until that time, NSs and BHs were regarded as just theoretical curiosities, since they have never been observed.

However, with the discovery of quasars in 1963, pulsars in 1968, and compact X-ray sources in 1962, the theoretical studies on compact objects forming by gravitational collapse were intensively stimulated. NSs were detected in two possible states: \emph{(i)} as \emph{radio pulsars}, which are rotating, magnetized NSs; \emph{(ii)} as compact invisible stars of \emph{binary X-ray sources}, where the X-ray luminosity emission is produced by the accreting matter falling from the companion star onto the NS polar caps. In addition in 1979, the detection of the binary X-ray source Cygnus X-1 represented a great success, since they proved the first evidence of BHs existence in the Universe.

A BH is actually defined as a region of spacetime that cannot comunicate with the external universe, where its boundary is called \emph{event horizon} \cite{Misner73}. The Einstein equations inside a BH break down, showing a \emph{singularity} due most likely to the fact that there is not yet a complete quantum theory of gravitation, able to explain what is happenning in that region. A BH behaves like an ideal black body because it reflects no light and, for the quantum field theory in curved spacetimes, it emits Hawking radiation with the same spectrum as a black body of a temperature inversely proportional to its mass \cite{Misner73,Shapiro86}.

Since BHs, for their nature, do not directly emit any electromagnetic radiation (other than Hawking radiation, that is very faint and almost undetectable with the actual technologies), the astrophysicist hunting for them must generally rely on indirect observations. The presence of BHs (and in general also of NSs) can be inferred through their gravitational interactions with their surroundings. One of possible and most studied interactions is with the accreting matter from a companion star, that falls on the compact object forming an external accretion disk. The accreting matter is heated by the internal viscosity (whose nature still remains a matter of discussion), emitting thus thermal X-ray energy, making thus them as the brightest objects in the Universe. For several decades, the physics of accretion onto compact objects experienced a golden age in terms of all the performed discoveries. Space satellites like XMM-Newton, INTEGRAL (ESA missions), RXTE, Swift, and Chandra (NASA missions) have collected over the years a wealth of information, which have in turn provided astronomers with new insights into the physics of X-ray sources. 

For all compact objects, the emitted X-ray spectrum offers relevant information about the processes occurring in the innermost regions of an accretion disk. In particular for BHs, the motion of matter in the vicinity of the event horizon leads to investigate the spacetime distortion generated by the central object, allowing to infer important features, such as mass and spin. In addition, this represents a powerful diagnostic both to study gravity in the strong-field regime and to validate the predictions of GR in strong field regimes.

Compact objects can release the gravitational potential energy of the accreted matter in different forms, e.g.,  
persistent radiation (accretion disk), thermonuclear burning radiation (type-I X-ray bursts in the case of NSs). The radiation field carries energy and momentum that interacts with the plasma inside an accretion disk structure through a radiation pressure, which can damp the accretion rate or even halt the whole mass transfer when it is near the Eddington limit (maximum allowed luminosity). It has been observed that during such phenomena there is an enhancement of the accretion rate, because the radiative drag force removes angular momentum from the accreting gas, forcing it to spiral inward or outward according to the strength of the radiation field, the so called Poynting-Robertson (PR) effect. In the last few years, many efforts have been made to derive a fully general relativistic treatment of the PR effect (see Refs. \cite{Bini09,Bini11}), with the intention of understanding how and to what extent the emitted radiation can influence the motion of matter in highly-warped spacetimes and how that would impact on the observational features.

This thesis focuses on the general relativistic PR effect and its connection with the observations, mainly related to accretion physics phenomena around compact objects (as BHs and NSs). During my PhD years, this project has been developed along three directions: $(i)$ theoretical works on static compact objects in GR in order to derive a simpler mathematical formalism to describe photon ray-tracing; $(ii)$ numerical-modeling attempts to describe complex phenomena that cannot be approached analytically; $(iii)$  data analysis in high-energy astrophysics to acquire the state of art on the actual observational knowledge. Following this line of thought, the thesis is organized in three chapters, where below an outline of their contents is reported. 

\begin{itemize}
\item {\bf Chapter one.} In GR, static compact objects are well described by the Schwarzschild metric, that possesses several advantageous proprieties thanks to its spherical symmetry, like: conservation of energy and angular momentum, the space outside the compact object is static and asymptotically flat (\emph{Birkoff theorem}), all kinds of geodesics lie in an invariant plane. The investigation of orbits in the Schwarzschild metric is very interesting for exploring and deeply understanding the geometrical proprieties of this spacetime. Comparing such orbits with the respective ones in the Newtonian framework permits to note the role played by the general relativistic effects. The first astrophysical researches are primary set up in such spacetime, sine in first approximation  objects can be considered static or very slowly rotating. A huge variety of phenomena involving the emission of radiation in GR can be described in terms of three main effects: light bending, travel time delay, and gravitational lensing. Mathematically, they are expressed through elliptic integrals (not solvable analytically in terms of elementary functions). For this reason, there is a common attitude to exploit numerical codes to calculate them. However in 2002--2006, a high accurate polynomial approximation of the light bending \cite{Beloborodov02} and time delay \cite{Poutanen06} have been empirically found by Beloborodov and Poutanen. Since a mathematical systematical procedure to deal with those issues was missing, the gravitational lensing was not yet accurately approximated (due to its more complicate functional form with respect to the previous cases). In 2016, I was able to introduce a mathematical method to approximate elliptical integrals for photon geodesics in the Schwarzschild spacetime, explaining formally how to derive the light bending and travel time delay effects, and for the first time to approximate the solid angle formula.  These approximations permit to reduce substantially the computational integration times and speed up the calculations in a wide range of astrophysical contexts. I discussed the accuracy and range of applicability of the new equations and presented a few applications of them to known astrophysical problems. This topic is the subject of the published paper in the peer-reviewed Astronomy \& Astrophysics Journal \cite{Defalco16}.
\item {\bf Chapter two.} This chapter is focused on the most fundamental part of my PhD program, that is the PR effect in accretion physics phenomena. In the first sections it is presented the PR effect, starting from Poynting, who in 1904 introduced for the first time this effect in the Newtonian frame until Robertson, who in 1937 extended it to the special relativistic case. Only several years later in 2009 -- 2011, it has been extended to the general relativistic case by Bini, Jantzen and Stella \cite{Bini09,Bini11}. I analysed and compared the orbits in the case of a flat spacetime with the curved spacetimes of Schwarzschild and Kerr. It is interesting to analyse the influences of the PR effect in combination with the general relativistic contributions. Before to study the possible applications in high energy astrophysics, I have developed two important works aimed at better comprehension of this effect and profound connection between theory and observations. First, I have proved the possibility to describe the PR effect through a Lagrangian formalism, introducing a new method based on the introduction of an integrating factor, which permits to integrate more physical systems involving dissipation. In the other work, I have extended the previous two-dimensional general relativistic models in three dimensions. The method used to derive the test particle equations of motion are based on the relativity of observer splitting formalism (that is a powerful method in GR to distinguish between the fictitious forces arising from the relative non-inertial motion of two observers and the gravitational effects). These contributions constitute fundamental works on such topic. Both works are published on Physical Review D Journal \cite{Defalco2018,Defalco20183D}.
\item {\bf Chapter three.} The theoretical apparatus, developed in the previous two chapters, permits to encompass a wide class of phenomena. It is equally important to learn more also about observational high-energy astrophysics. In this chapter, I begin illustrating the main proprieties of the pulsars, focusing then the attention on those coupled with a companion star of low mass ($<1M_\odot$), the so called low mass X-ray binaries. They are characterized by the formation of an accretion disk, via Roche lobe overflow from the companion star, around the compact object. A peculiar subclass of pulsars are represented by \emph{the accreting millisecond X-ray pulsars} (AMXPs), hosted in low mass X-ray binaries and believed to be their progenitors. They are old (order of Gyr) NSs endowed with a relatively low magnetic field ($B\approx 10^{8-9}$ G), spin frequencies typically between 180 -- 600 Hz, and an orbital period ranging from 40 min to 5 hr. All AMXPs are \emph{X-ray transients}, spending most of their time in a quiescent state (X-ray luminosities of the order of $10^{31-32}$ erg s$^{-1}$) and sporadically undergoing outbursts that can last for a few weeks reaching X-ray luminosities of $10^{38}$~erg~s$^{-1}$ (see, e.g., Ref. \cite{wijnands06,poutanen06b,patruno12} for reviews). These sources have the peculiar proprieties that once they finish to emit in radio, they come back again to life as X-ray sources because they couple with a low mass companion star (the so called \emph{recycling scenario}). 

I reduced and analysed INTEGRAL data of three interesting AMXPs (PI M. Falanga): IGR J00291+5934, IGR J18245--2452, and SAX J1748.9--2021. In this thesis I report only the analysis of the first two sources, while the latter can be found in the paper \cite{Li2018}. I concentrated to analyse into details spectra, timing, and type-I X-ray bursts analysis, using the INTEGRAL data, together also with XMM-Newton and Swift data, while these sources were in outbursts. Type-I X-ray bursts are thermonuclear explosions that occurs on the surface of a NS when the matter reaches a critical temperature and density. Such intense radiation fields are of high relevance for the aims of my thesis, because, together with the PR effect, they can be very determining to alter the dynamics of an affected accretion disk. The contents of these topics are both contained in papers published on Astronomy \& Astrophysics Journal \cite{Defalco2017a,Defalco2017b}
\end{itemize}

\chapter{Approximation of photon geodesics in Schwarzschild metric}
\label{chap:1}
\epigraph{Imagination is the only weapon in the war against reality.}{Alice in Wonderland, Lewis Carroll}

\lettrine{I}{n this chapter}, I present a mathematical method for approximating through polynomial functions photon geodesics in the Schwarzschild spacetime. Based on this, I derive the approximate equations for light bending and propagation delay (already introduced empirically in the literature). Then, I derive for the first time an approximate for the solid angle. I discuss the accuracy and range of applicability of the new equations and present a few simple applications of them to known astrophysical problems. This topic is the subject of the published paper in the peer-reviewed Astronomy \& Astrophysics Journal \cite{Defalco16}. 

\section{Astrophysical motivations}
\label{sec:intro}
\begin{figure}[h]
\centering
\includegraphics[scale=0.213]{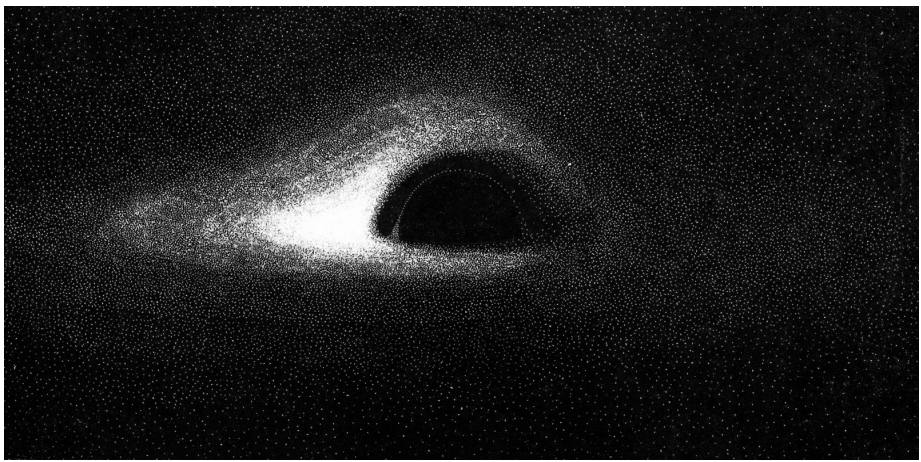}
\caption{First simulated photograph of a spherical BH with a thin accretion disk employing the ray-tracing technique (image taken from \cite{Luminet79}).} 
\label{fig:Fig1_1}
\end{figure}

In the 80s with the discovery of X-ray emission coming from the accretion disks around BHs, studies \cite{Luminet79, Pechenick83} began a great interest in photons emitted by matter in a strong gravitational field. They stimulated intensively theoretical researches to understand how the matter around a BH appears to an observer located at infinity. The relevant computations are carried out with \emph{ray-tracing techniques} that are based on following the photon geodesics until the observer frame in general relativistic spacetimes. In 1979 Luminet proposed the first numerical simulation reproducing the simulated photograph of a spherical BH with a thin accretion disk (see Fig. \ref{fig:Fig1_1}) \cite{Luminet79}.
Effects to be considered are: ({\it i}) light bending, ({\it ii}) travel time delay, and ({\it iii}) gravitational lensing (known also as solid angle) \cite{Misner73}. The basic equations for the Schwarzschild metric are expressed through \emph{elliptic integrals} that can be solved numerically. 

An elliptic integral is an integral of the following form:
\begin{equation} \label{ei}
\int \frac{A(x)+B(x)\sqrt{S(x)}}{C(x)+D(x)\sqrt{S(x)}}dx,
\end{equation}
where $A(x),B(x),C(x),$ and $D(x)$ are polynomials in $x$, while $S(x)$ is a polynomial of degree 3 or 4 with no repeated roots \cite{c18}. These integrals cannot be expressed in terms of elementary functions. However, with appropriate transformations, every elliptic integral can be expressed as a combination of the three Legendre canonical forms, i.e. incomplete elliptic integrals of the first $F(x,k)$, second $E(x,k)$, and third kind $\Pi(n;x,k)$ \cite{c18}:
\begin{equation} \label{cei}
\begin{aligned}
F(x,k)&=\int_0^x\frac{d\theta}{\sqrt{1-k^2\sin^2\theta}},\\
E(x,k)&=\int_0^x \sqrt{1-k^2\sin^2\theta} d\theta,\\
\Pi(n;x,k)&=\int_0^x\frac{d\theta}{(1-n\sin^2\theta)\sqrt{1-k^2\sin^2\theta}},
\end{aligned}
\end{equation}
where the constant $n$ is the \emph{elliptic characteristic} and $k$ is the \emph{elliptic modulus}. Eqs. (\ref{cei}) are called complete elliptic integrals, when $x=\pi/2$. The elliptic integrals, which I will deal with, are of the following form:
\begin{equation} \label{oei}
\int \frac{dx}{\sqrt{1-b^2P(x)}},\quad \int \frac{dx}{P(x)\sqrt{1-b^2P(x)}},
\end{equation} 
where $P(x)=x^2(1-2Mx)$.

A powerful analytical approximation was introduced by Beloborodov \cite{Beloborodov02}, who derived an approximate linear equation to describe the gravitational light bending of photons emitted at radius $r \geqslant r_s$ ($r_s = 2GM/c^2$). In the same vein, Poutanen \& Beloborodov \cite{Poutanen06} derived an approximate polynomial equation for photon travel time delays. These two analytical approximations were obtained by introducing an {\it {ad hoc}} parametrization of the photon emission angle. Nevertheless, the equation for gravitational lensing, also known as solid angle equation, was still solved numerically by these authors. 

In the next sections, I present a mathematical method through which the approximate polynomial equations for light bending and travel time delay in a Schwarzschild spacetime are derived without any ad hoc assumption. I then apply the same method to derive for the first time an approximate polynomial equation for gravitational lensing. High-accuracy approximate equations for photon geodesics translate into high-speed ray-tracing codes for different astrophysical applications in the strong gravitational field of Schwarzschild BHs. As examples I apply my approximate equations to calculate the light curve from a hot spot on the surface of a rotating NS and a clump in a circular orbit around  BH. Moreover, I calculate the fluorescent iron $K_{\alpha}$ line profile from an accretion disk around a BH (see e.g., Ref. \cite{Fabian89}).

\section{The Schwarzschild spacetime}
\label{subsec:Schwarzschild}
For static and spherically symmetric BHs of mass, $M$, the Schwarzschild metric in spherical coordinates ($t, r, \varphi, \psi$) reads as \cite{Misner73}
\begin{equation} \label{msch} 
ds^2=dt^2\left(1-\frac{2M}{r} \right)-dr^2\left(1-\frac{2M}{r} \right)^{-1}-r^2d\Omega^2,
\end{equation}
where $G=c=1$ and $d\Omega^2=d\varphi^2+\sin^2\varphi\ d\psi^2$. In this standard system, the coordinate variables are time $t$, radius $r$, polar angle $\varphi$, and azimuthal angle $\psi$. 

\subsection{Physical interpretation of Schwarzschild coordinates}
It is important to clarify the geometric meaning of the Schwarzschild coordinates and how to measure them \cite{Misner73}. Normally, the names given to the coordinates have no intrinsic significance, because they are simply mathematical parameters. Nevertheless, even if another coordinate transformation is perfectly admissible, it is important that exists an easy communication between the investigator, who adopts it, and his colleagues. The Schwarzschild coordinates $(t,r,\varphi,\psi)$ provide an immediate link with their geometric-physical contents. 
\begin{itemize}
\item The coordinates $(\varphi,\psi)$ are respectively the polar and azimuthal angle on a two dimensional surface of constant $r$ and $t$. 
\item The distance between two nearby events is given by $ds^2=r^2d\Omega^2$, that is a two dimensional sphere of area $A=\int(rd\varphi)(\sin\varphi d\psi)=4\pi r^2$. Therefore, the radial coordinate $r$ in a point $\mathcal{P}$ is measured as $r=\sqrt{A_{\mathcal{P}}}/4\pi$, where $A_{\mathcal{P}}$ is the sphere through the point $\mathcal{P}$. 
\item The parameter $t$ is the time coordinate and is connected with the proper time $\tau$ through the formula $d\tau=\sqrt{(1-2M/r)}\ dt$. In this case it is difficult to construct a device, that measures the time, because it should be a radar-clock that takes into account the geometrical structure of the spacetime in the measurement process. However, the measurement of $t$ is coincident with $\tau$ for a stationary clock located at infinity. 
\end{itemize}

\subsection{Birkhoff theorem}
The metric (\ref{msch}) is written in an advantageous form, because for $r\rightarrow \infty$ the Schwarzschild metric easily reduces to the Minkowski metric written in spherical coordinates, i.e., $ds^2=dt^2-dr^2-r^2d\Omega^2$. Since the metric coefficients are explicitly independent of time and there is no frame-dragging effect, the spacetime is \emph{static}, as experienced by an observer external to the horizon. These notions are the content of the \emph{Birkhoff's theorem}, stating that any spherically symmetric solution of the vacuum Einstein's field equations must be static and asymptotically flat. This proves that the exterior Schwarzschild metric solution is unique and in the same time is the most general metric describing a spherically symmetric spacetime \cite{Misner73}.

\subsection{Analysis of the singularities}
The metric (\ref{msch}) presents two singularities at $r=2M$ and $r=0$. The first one is called a \emph{coordinate singularity}, because it can be removed with an appropriate change of coordinates. This can be achieved choosing the Kruskal coordinates, that retain the usual angular coordinates $(\psi,\varphi)$, but $(r,t)$ coordinates are replaced by new coordinates $(u,v)$, defined as follows \cite{Synge64,Misner73}
\begin{eqnarray}  
u&=&\left(\frac{r}{2M}-1\right)^{1/2}e^{r/4M}\ \cosh\left(\frac{t}{4M}\right),\label{CK1}\\ 
v&=&\left(\frac{r}{2M}-1\right)^{1/2}e^{r/4M}\ \sinh\left(\frac{t}{4M}\right). \label{CK2} 
\end{eqnarray}
These equations can be inverted to give
\begin{eqnarray}  
&u^2-v^2=\left(\frac{r}{2M}-1\right)\ e^{r/2M},\label{ICK1}\\ 
&2M\log\left|\frac{u+v}{u-v} \right|=t. \label{ICK2} 
\end{eqnarray}
From Eqs. (\ref{ICK1}) and (\ref{ICK2}), the differentials $dr$ and $dt$ can be derived as
\begin{eqnarray} 
dr&=\frac{8M^2}{r}e^{-r/2M}(udu-vdv), \label{diff1}\\ 
dt&=\frac{8M^2}{r-2M}e^{-r/2M}(du^2-dv^2).  \label{diff2} 
\end{eqnarray}
Substituting Eqs. (\ref{diff1}) and (\ref{diff2}) in the metric (\ref{msch}), I obtain the Schwarzschild metric expressed in Kruskal coordinates, i.e.,
\begin{equation} \label{KM} 
ds^2=-\frac{32M^3}{r}e^{-r/2M}(du^2-dv^2).
\end{equation}
In this metric the singularity for $r=2M$ disappears, as I have claimed. 

Instead, $r=0$ is an \emph{essential singularity}, meaning that the gravitational field in that point becomes infinity. To prove this is a true singularity, one must look at quantities that are independent of the coordinate systems. For example, the Kretschmann invariant, given by
\begin{equation} \label{Kinv} 
K\equiv R^{\alpha\beta\gamma\delta}R_{\alpha\beta\gamma\delta}=\frac{48M^2}{r^6},
\end{equation}
at $r=0$ blows up, meaning that the spacetime curvature becomes infinite and in this point the spacetime itself is no longer well-defined \cite{Synge64}.

\subsection{Orbits in the Schwarzschild metric}
The geodesic equations in a spacetime with a metric $ds^2=g_{ij}dx^idx^j$ can be derived using the Euler-Lagrange equations $\frac{d}{d\tau}\left(\frac{\partial\mathcal{L}}{\partial\dot{x}^\alpha}\right)-\frac{\partial\mathcal{L}}{\partial x^\alpha}=0$, where $x^\alpha$ are the coordinates, $2\mathcal{L}=g_{ij}\frac{dx^i}{d\tau}\frac{dx^j}{d\tau}$ is the Lagrangian, and $\tau$ the affine parameter along the geodesic \cite{Shapiro86,Chandrasekhar92}. For the Schwarzschild spacetime using $x^\alpha=(t,r,\varphi,\psi)$, the Lagrangian assumes the following form
\begin{equation} \label{LagrSch} 
2\mathcal{L}=\left(1-\frac{2M}{r}\right)\dot{t}^2-\frac{\dot{r}^2}{1-2M/r} -r^2\dot{\varphi}^2-(r^2\sin^2\varphi)\dot{\psi}^2,
\end{equation}
where the dot stands for $d/d\tau$. The Euler-Lagrange equation for $\varphi$ is
\begin{equation} \label{ELtheta} 
\frac{d}{d\tau}(r^2\dot{\varphi})=r^2\sin\varphi\cos\varphi\dot{\psi}^2.
\end{equation}
This equation shows that if one orients the coordinate system such that the test particle initially is moving in the equatorial plane (i.e., $\varphi=\pi/2\Rightarrow\dot{\varphi}=0$), then it remains there forever, since $\varphi=\pi/2$ satisfies Eq. (\ref{ELtheta}) for all $\tau$. This result is based on the uniqueness theorem for differential equations. Physically, it means that any geodesic lies always in the same plane, called \emph{invariant plane}. From now on, I consider always $\varphi=\pi/2$. 

For $t$ and $\psi$, I derive the following two integrals of motion:
\begin{eqnarray}
\left(1-\frac{2M}{r} \right)\frac{dt}{d\tau}=E={\rm constant},  \label{IntMot1} \\ 
r^2\frac{d\psi}{d\tau}=L={\rm constant},  \label{IntMot2} 
\end{eqnarray}
where $E$ is the energy and $L$ the angular momentum orthogonal to the invariant plane where the geodesic lies. Substituting Eqs. (\ref{IntMot1}) and (\ref{IntMot2}) in Eq. (\ref{LagrSch}), the Lagrangian (\ref{LagrSch}) becomes:
\begin{equation} \label{LagrSchMod} 
\frac{E^2}{1-2M/r}-\frac{\dot{r}^2}{1-2M/r}-\frac{L^2}{r^2}=2\mathcal{L}.
\end{equation}
The corresponding Hamiltonian is $\mathcal{H}=\sum_{\alpha=t,r,\varphi,\psi}p_{\alpha}\dot{\alpha}-\mathcal {L}=\mathcal {L}$, where $p_{\alpha}=-(-1)^{\delta^t_{x^\alpha}}\frac{\partial \mathcal{L}}{\partial x^\alpha}$ are the canonical momenta. The equality of Hamiltonian and Lagrangian means that there is no potential energy, and the total mechanic energy, represented by $\mathcal{H}$, derives only from the kinetic energy, represented by $\mathcal{L}$. I have thus $\mathcal{H}=\mathcal{L}={\rm constant}$, therefore it is possible to rescale $\tau$ in a way that $2\mathcal{L}$ assumes one of the following values
\begin{equation}
2\mathcal{L}=\begin{cases}
+1 & \mbox{for time like geodesics},\\
\ \ 0 & \mbox{for null geodesics},\\
-1 & \mbox{for space like geodesics}.\\
\end{cases}
\end{equation}
I will not be concerned with space like geodesics.

\begin{figure}[p]
\hbox{
\includegraphics[trim=-3.0cm 0cm 4.6cm 6cm, scale=0.32]{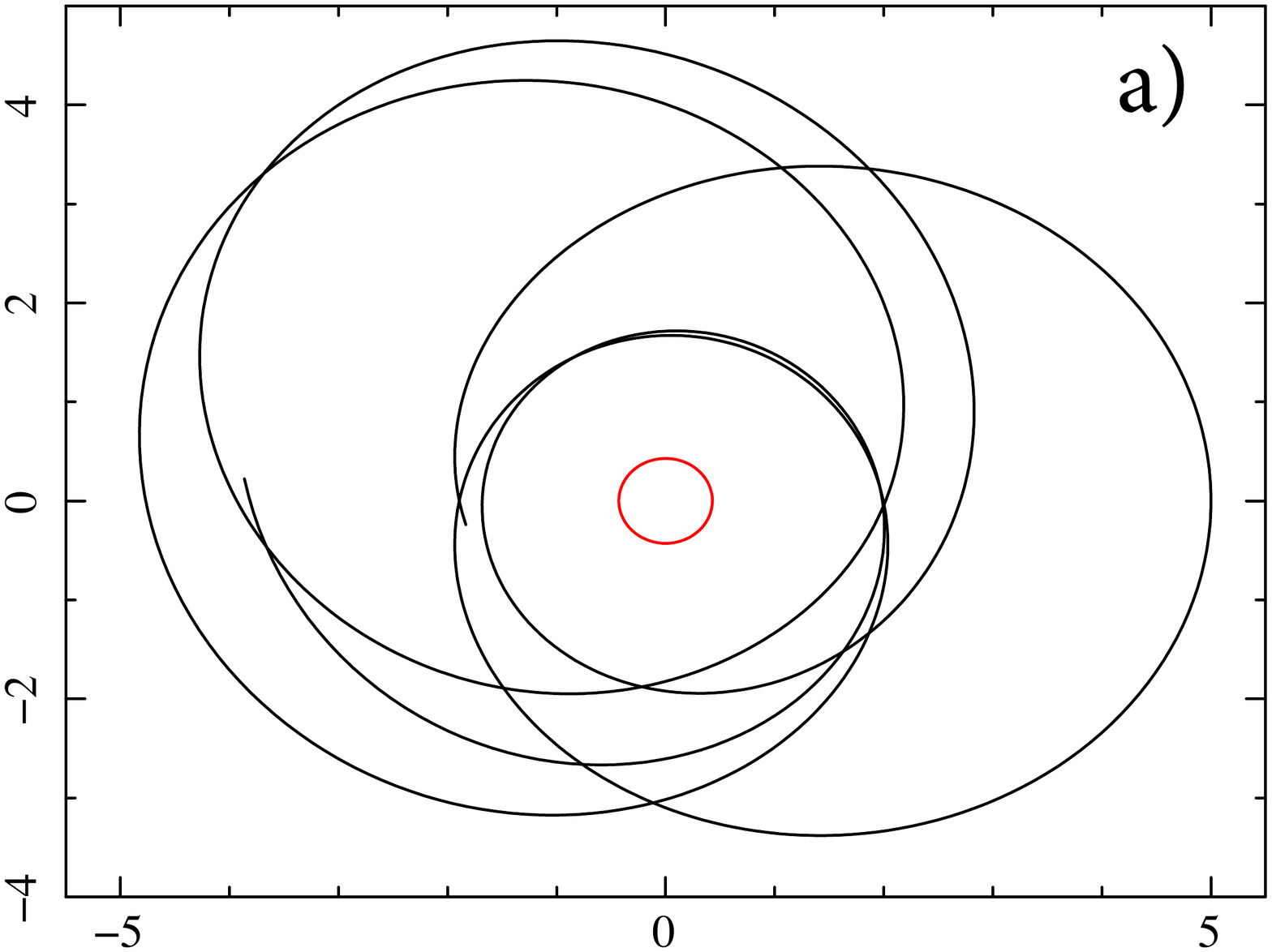}
}
\vspace{0.1cm}
\hbox{
\includegraphics[trim=-3.0cm 0cm 4.6cm 3cm, scale=0.32]{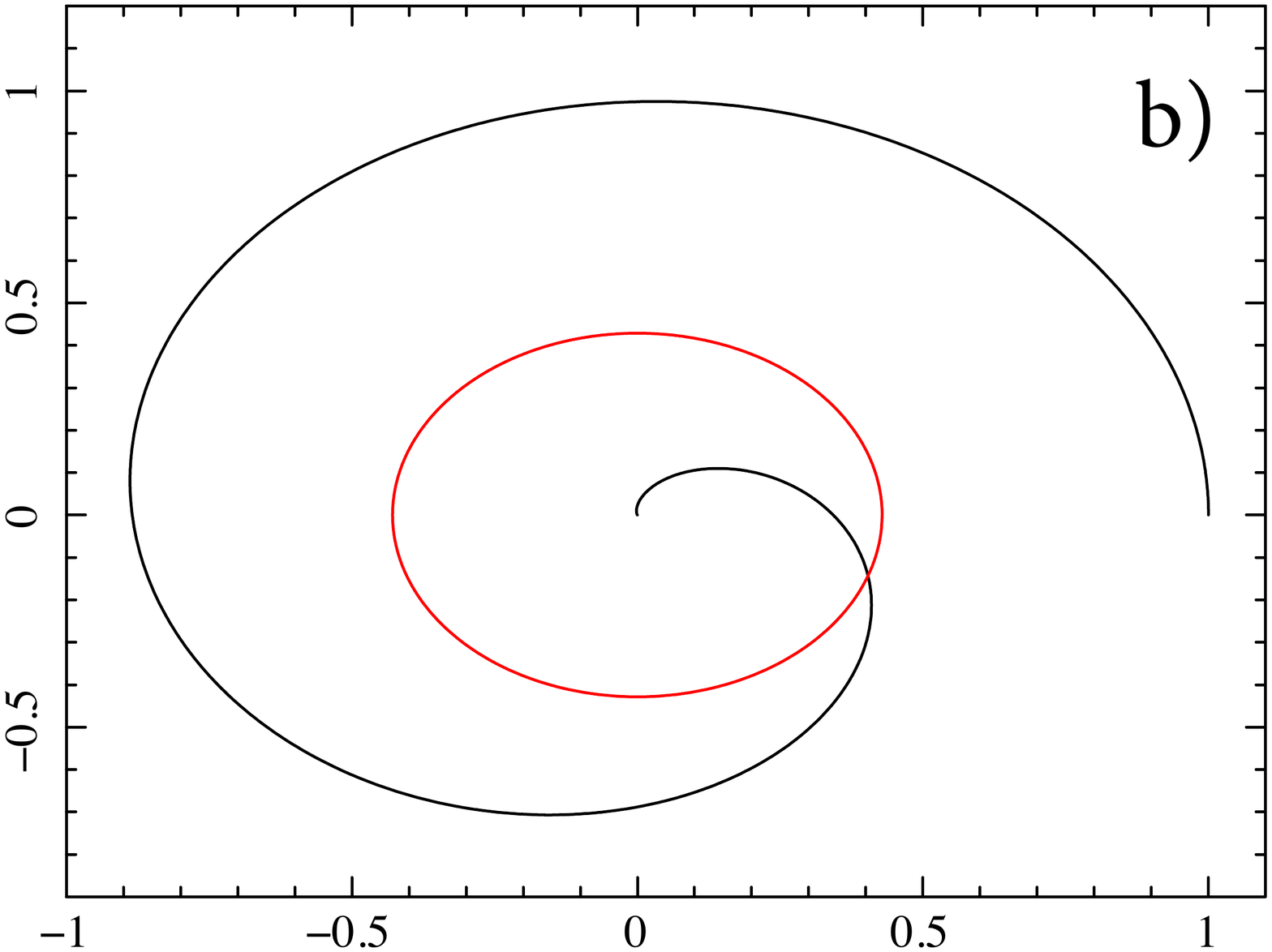}
}
\vspace{0.1cm}
\hbox{
\includegraphics[trim=-3.0cm 0cm 4.6cm 3cm, scale=0.32]{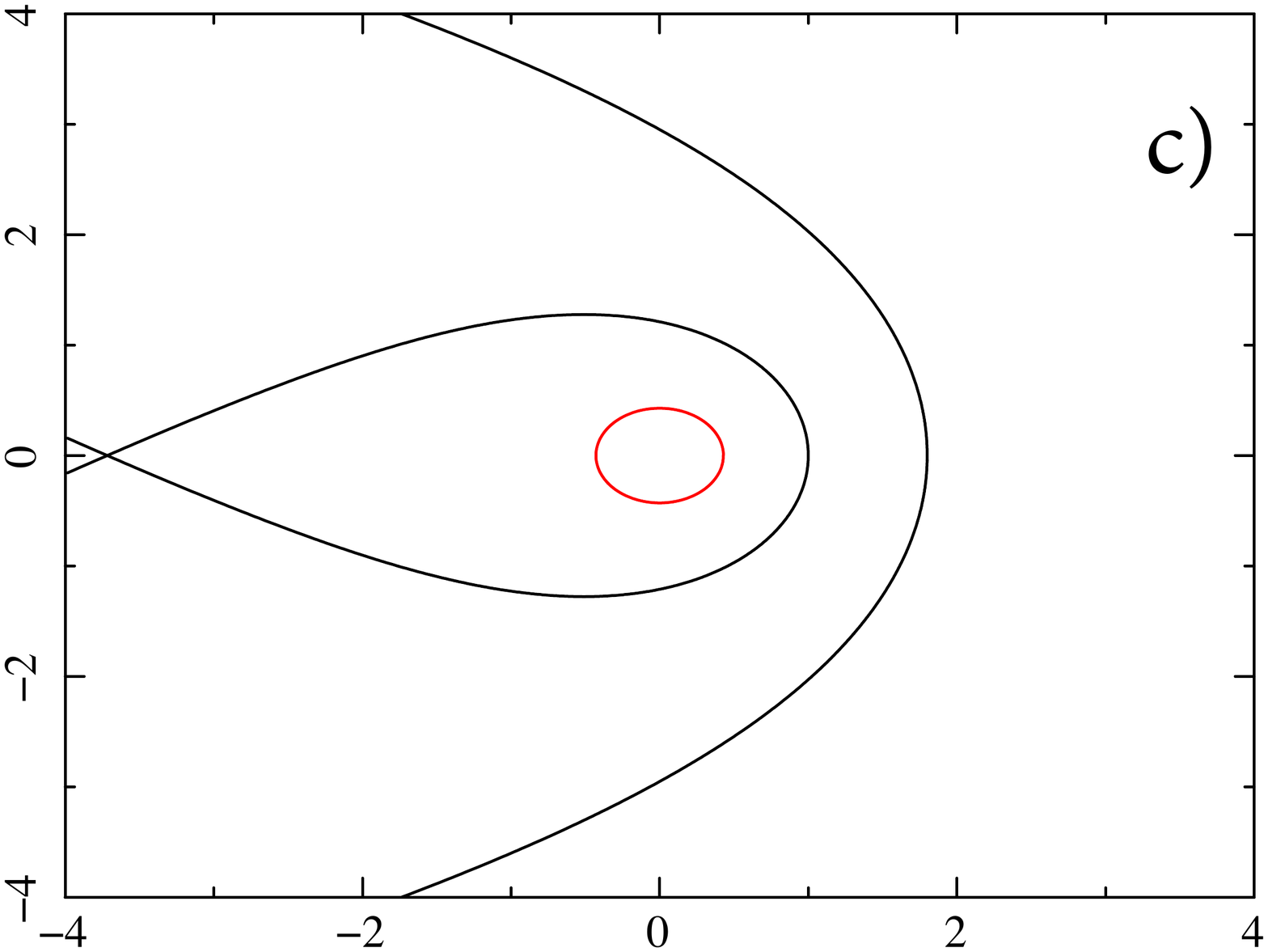}
}
\caption{Three kinds of time like geodesics in Schwarzschild metric: (a) bounded, (b) captured, and (c) unbounded. The red line delimits the BH surface. I used $M=3/14$, (a) $e=0.5,l=2.5$, (b) $e=0.2,l=1.5$, and (c) $e=1.5,l=2.5$ and $e=1.5,l=4.5$ (see Ref. \cite{Chandrasekhar92}, for further details).} 
\label{fig:OT} 
\end{figure}

For time like geodesics, Eq. (\ref{LagrSchMod}) reads as
\begin{equation} \label{LagrTemp} 
 \left(\frac{dr}{d\tau}\right)^2+\left(1-\frac{2M}{r}\right)\left(1+\frac{L^2}{r^2}\right)=E^2,
\end{equation}
Using Eq. (\ref{IntMot2}) and setting $u=r^{-1}$, I obtain  
\begin{equation} \label{LagrTemp2} 
\left(\frac{du}{d\psi}\right)^2\equiv f(u)=2Mu^3-u^2+\frac{2M}{L^2}u-\frac{(E^2-1)}{L^2},
\end{equation}
Once Eq. (\ref{LagrTemp2}) is solved for $u=u(\psi)$, the solution can be found by direct quadratures of the other equations. To study qualitatively the time like geodesics one has to solve $f(u)=0$, where the solutions are $u_1,u_2,u_3$ that can be three real or one real and two complex conjugates. I consider two separated classes: \emph{bounded} ($E^2<1$) and \emph{unbounded} ($E^2>1$) orbits \cite{Chandrasekhar92}.
\begin{itemize}
\item $E^2<1$, there is at least one positive solution. I define:
\begin{itemize}
\item \emph{orbits of the $1^\circ$ type} ($u_1\leq u\leq u_2$), which oscillate between two extreme values, where $u_1$ is called \emph{aphelion} and $u_2$ \emph{perihelion}; 
\item \emph{orbits of the $2^\circ$ type} ($u_3\leq u$), which start from the aphelion $u_3$ and plunge in the singularity $r=0$. 
\end{itemize}
\item $E^2>1$, there is at least one negative solution. As done previously, I define \emph{orbits of the $1^\circ$ type}, ($0<u\leq u_2$), and \emph{orbits of the $2^\circ$ type} ($u \geq u_3$), having both the same meanings above described.  
\end{itemize}
Summarising, there are three kinds of orbits: (1) \emph{bounded} (oscillating between two points), (2) \emph{captured} (plunging into BH), and (3) \emph{unbounded} (starting from infinity, reaching a turning point, and then going again to infinity). The orbits (1) and (3) have newtonian equivalent, instead orbits (2) are typical of the Schwarzschild metric (see Refs. \cite{Shapiro86,Chandrasekhar92}, and Fig. \ref{fig:OT}).

\begin{figure}[p]
\hbox{
\includegraphics[trim=-3.0cm 0cm 4.6cm 6cm, scale=0.32]{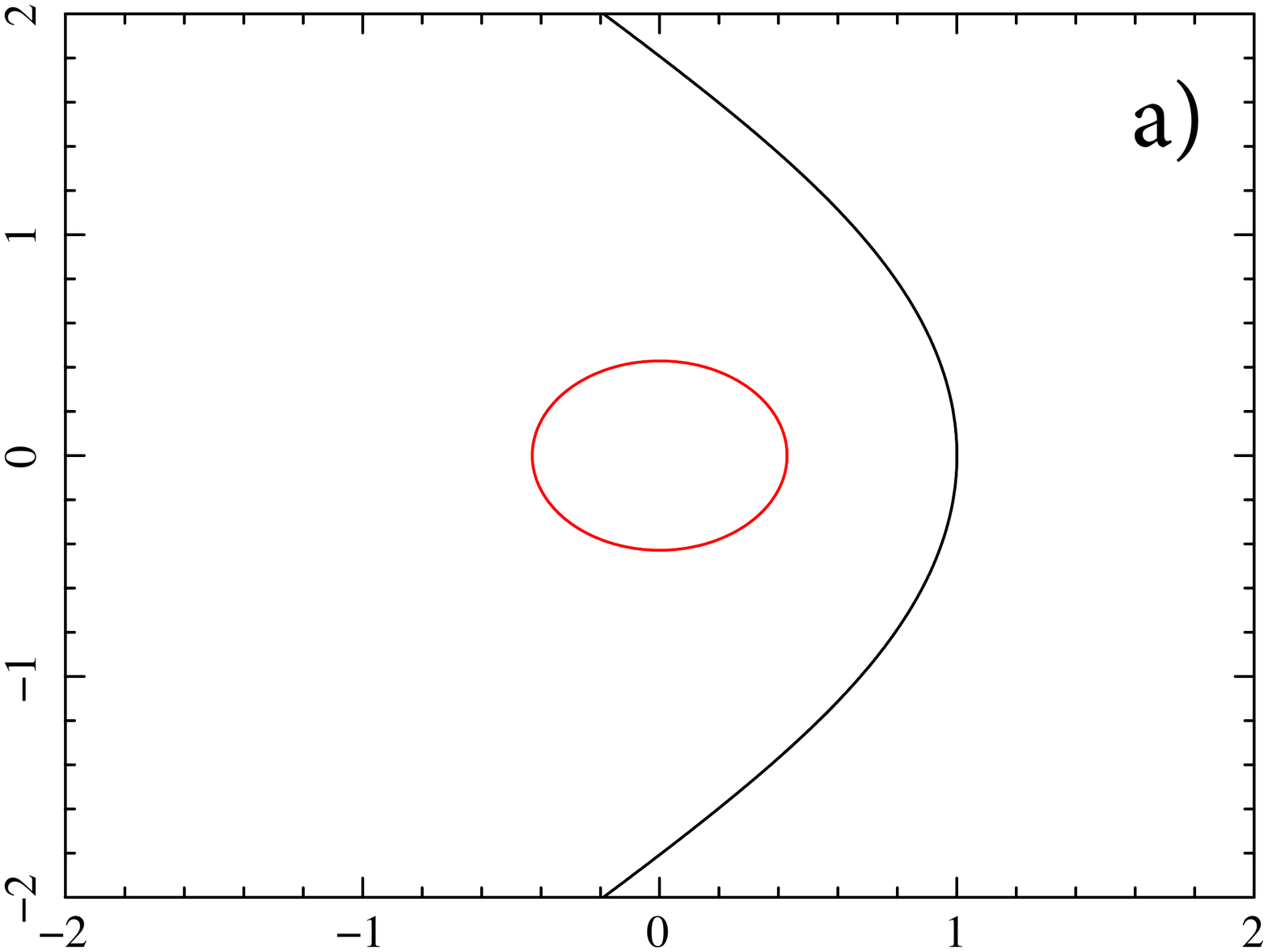}
}
\vspace{0.1cm}
\hbox{
\includegraphics[trim=-3.0cm 0cm 4.6cm 3cm, scale=0.32]{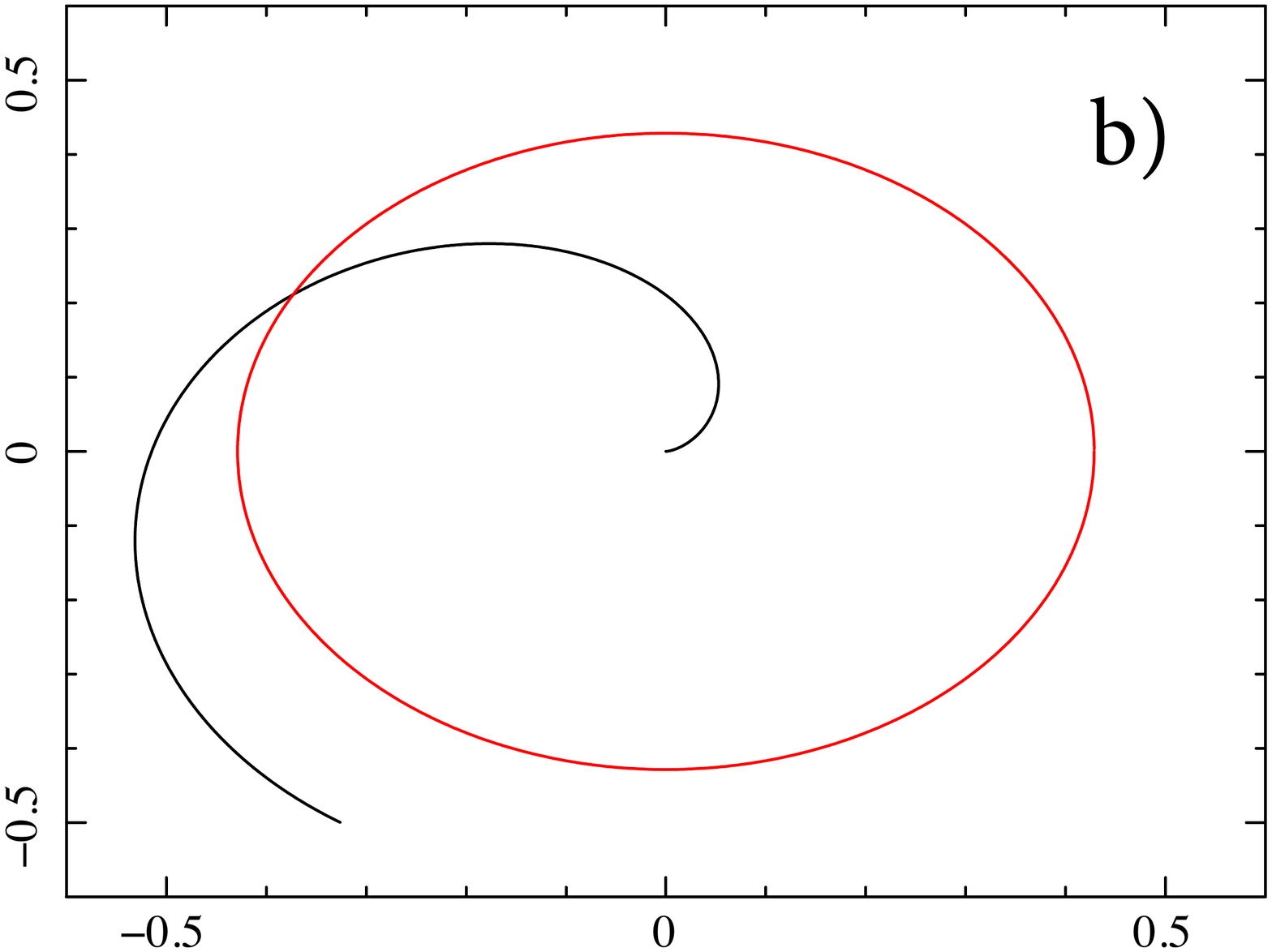}
}
\caption{Two null geodesics in Schwarzschild metric: (a) unbounded, and (b) captured. The red line delimits the BH surface. I used $M=3/14$, (a) $P=1$, and (b) $P=0.7$ (see Ref. \cite{Chandrasekhar92}, for further details).} 
\label{fig:ON} 
\end{figure}

For null geodesics, Eq. (\ref{LagrSchMod}) reads as
\begin{equation} \label{LagrNull} 
\left(\frac{du}{d\varphi}\right)^2\equiv f(u)=2Mu^3-u^2+\frac{1}{b^2},
\end{equation}
where $b=L/E$ is the impact parameter. I follow the same procedures and nomenclature developed for time like geodesics. In this case, $f(u)=0$ admits at least one negative solution. I distinguish thus between orbits with $b<b_{\rm c}$ and $b>b_{\rm c}$, where $b_{\rm c}=3\sqrt{3}M$ is the critical impact parameter \cite{Chandrasekhar92}.
\begin{itemize}
\item $b<b_{\rm c}$, there are only \emph{orbits of the $1^\circ$ type} ($0<u\leq u_2$). 
\item $b>b_{\rm c}$, there are only \emph{orbits of the $2^\circ$ type} ($u_3\leq u$).
\end{itemize}  
There are only \emph{unbounded} and \emph{captured} orbits (see Refs. \cite{Shapiro86,Chandrasekhar92}, and Fig. \ref{fig:ON}). 

\section{Photons in the Schwarzschild spacetime}
\label{sec:equations}
In this section, I introduce the elliptical integrals of photon trajectories, travel time delay, and gravitational lensing in the Schwarzschild metric. 

\subsection{Gravitational light bending}
\label{subsec:impa&libe}
Because of spherical symmetry, it is customary to use the equatorial plane at $\varphi=\pi/2$ to calculate geodesics in the Schwarzschild metric that are representative of all photon trajectories. Called $u^\alpha=dx^\alpha/d\tau$ the photon velocity, from energy's integral of motion (\ref{IntMot1}), I have
\begin{equation}
u_t=g_{tt}u^t=\left(1-\frac{2M}{r}\right)\frac{dt}{d\tau}=E=1.
\end{equation}
From angular momentum's integral of motion (\ref{IntMot2}), I have
\begin{equation}
u_\psi=g_{\psi\psi}u^\psi=-r^2\frac{d\psi}{d\tau}=L=\frac{L}{E}=b.
\end{equation}
It is important to note that the rule of $\psi$ and $\varphi$ are exchanged when I refer to the photon trajectory, because $\psi$ varies along the photon trajectory, instead $\varphi$ is constant since the photon trajectory lies in the invariant plane. A photon satisfies $g_{hk} u^h u^k = u^t u_t + u^\psi u_\psi + g_{rr} u^r u^r=0$, where $u^r$ can be determined by
\begin{equation} 
u^r = \sqrt{- \frac{(u^t u_t + u^\psi u_\psi)}{g_{rr}}}=\sqrt{1-\frac{b^2}{r^2}\left(1-\frac{2M}{r}\right)}.
\end{equation}
To find the photon impact parameter, $b$, I calculate \cite{Misner73}
\begin{eqnarray} 
\cos^2\alpha &=& u \cdot u_r =-\left[\left(1-\frac{2M}{r}\right)^{-1}-\frac{b^2}{r^2}\right],\\
\sin^2\alpha &=& u \cdot u_\theta=-\frac{b^2}{r^2},
\end{eqnarray}
The combination of these equations permits to obtain $b$, as
\begin{equation} \label{A2} 
\sin \alpha= \sqrt{\frac{\tan^2\alpha}{\tan^2\alpha^2 + 1}}=\sqrt{\frac{1}{\Big{(}1-\frac{2M}{r}\Big{)}^{-1}\frac{r^2}{b^2}}}
\end{equation}
The impact parameter $b$ represents the distance between the observer and the photon trajectory at infinity and is related to the photon emission angle, $\alpha$ by
\begin{equation} \label{eq:impact} 
b = \frac{R\sin\alpha}{\sqrt{1-r_{s}/R}}.
\end{equation}
The light bending equation is obtained integrating $d\psi=u^\psi d\tau=\frac{u^\psi}{u^r}dr$ \cite{Misner73,Beloborodov02}. A photon geodesic starting at radius $R$ is described by the following elliptical integral \cite{Chandrasekhar92,Misner73}
\begin{equation} \label{eq:libe} 
\psi=\int_{\infty}^{R}\frac{u^\psi}{u^r}dr=\int_R^\infty \frac{dr}{r^2}\left[\frac{1}{b^2}-\frac{1}{r^2}\left(1-\frac{r_s}{r} \right) \right]^{-\frac{1}{2}}.
\end{equation}
Equation (\ref{eq:libe}) is strictly valid up to $\alpha=\pi/2$, since the sine function is symmetric with respect to $\alpha=\pi/2$. The photon deflection angle, $\psi$, can be directly determined in terms of the emission angle $\alpha$ through Eq. (\ref{eq:impact}). 

\begin{figure}[h]
\centering
\includegraphics[trim=2cm 7cm 4.6cm 3cm, scale=0.55]{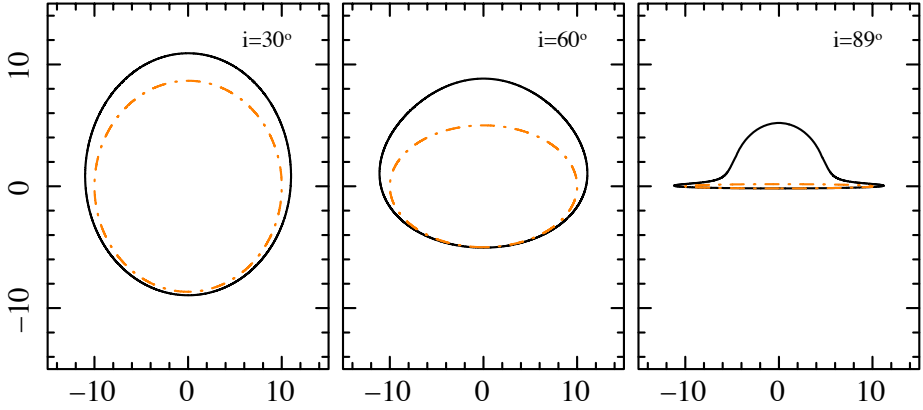}
\caption{Image of a ring at infinity located at $R=10M$ for $i=30^\circ,60^\circ,89^\circ$. The plot is in unity of $M$. The continuos black lines are given by Eqs (\ref{ibh}), instead the dashed-dotted orange lines are given by Eqs (\ref{cibh}).} 
\label{fig:LB}
\end{figure}

The effects of gravitational light bending can be visualized looking at the image of a ring around a BH seen by an observer at infinity inclined of an ingle $i$ respect to the normal at the plane containing the ring. The relative equations in the observer coordinates are \cite{Luminet79}
\begin{equation} \label{ibh}
x_{\rm obs}=-\frac{b\ \sin\varphi}{\sin\psi},\qquad y_{\rm obs}=-\frac{b\ \cos i\ \cos\varphi}{\sin\psi}.
\end{equation}
To appreciate more the general relativistic effects, it is useful to compare them with the newtonian case. In the classical limit ($M\longrightarrow0$), Eqs (\ref{ibh}) become
\begin{equation} \label{cibh}
x_{\rm obs}=-R\ \sin\varphi,\qquad y_{\rm obs}=-R\ \cos i\ \cos\varphi.
\end{equation}
In Fig. \ref{fig:LB} it is possible to note how the general relativistic effects change the shape of the ring and become more prominent increasing the inclination angle.  

\subsection{Turning points}
\label{subsec:turn_poin}
\begin{figure}[h]
\centering
\includegraphics[scale=0.7]{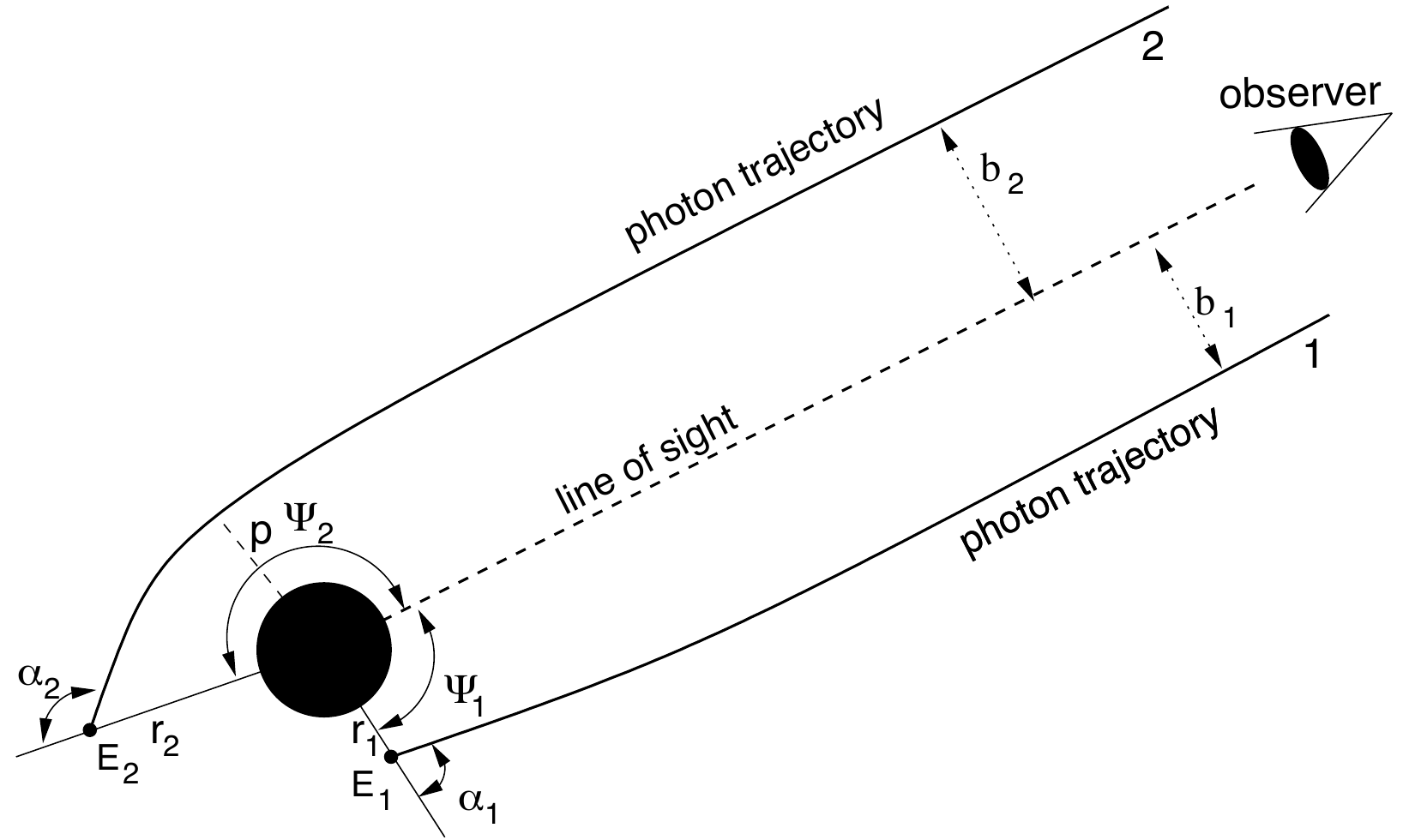} 
\caption{Two photon trajectories emitted at different radii, $r_1$ and $r_2$, and emission angles, 
$\alpha_1$ and $\alpha_2$ , with their corresponding impact parameters, $b_1$ and $b_2$. Trajectory 1 is for a direct photon, while trajectory 2 has a turning point, i.e., passes through periastron $p$, the minimum distance between the trajectory and the BH. The observer is at infinity, and photons geodesics lie in a single invariant plane.} 
\label{fig:Fig1_2} 
\end{figure}

It must be distinguished between direct photons, which have trajectories with an emission angle $0\le\alpha\le\pi/2$, and photons with a turning point, whose trajectories have an emission angle $\pi/2\le\alpha\le\alpha_{max}$ (see Fig. \ref{fig:Fig1_2}). 
Photon trajectories with a turning point can reach infinity only if their $b$ is greater than the critical impact parameter $b_c=3\sqrt{3}M$ (see, e.g., Ref. \cite{Luminet79}). Since I am interested only in photons that are not captured by the BH, the maximum possible emission angle is obtained by substituting $b_c$ into Eq. (\ref{eq:impact}) 
\begin{equation} \label{alphamax}
\alpha_{\rm max}=\pi-\arcsin\left[\frac{3}{2}\sqrt{3\left(1-\frac{r_{s}}{R}\right)}\frac{r_{s}}{R}\right].
\end{equation}   
Photons emitted between $\pi/2\le\alpha\le\alpha_{\rm max}$ follow trajectories with a turning point; therefore a periastron distance, $p$, is defined at an angle $\alpha_{\rm p}=\pi/2$, which determines the minimum distance between the compact object and the photon trajectory. The emission point of a photon at $\psi_{\rm E}$ that passes through the turning point is symmetric with respect to the periastron angle, $\psi_{\rm p}$, to the point $\psi_{\rm S}$, (with an emission angle $\alpha\le\pi/2$) along the same trajectory, as they have the same impact parameter at infinity. Based on this symmetry, I determine  $\psi_{\rm S} = 2\psi_{\rm p}-\psi_{\rm E}$, where $\alpha_{\rm S}=\pi-\alpha_{\rm E}$. 

\subsection{Travel time delay}
\label{subsec:tide}
A photon following its geodesic from an emission point, $E$, to an observer at infinity has an infinite travel time, $\Delta \tau$, value. To have a finite quantity, I calculate the relative travel time delay between a photon emitted at a distance, $R$, following its geodesic and the photon emitted radially with $b=0$, that is, $\Delta t(b)=\Delta \tau(b) - \Delta \tau(b=0)$ \cite{Pechenick83}. The travel time delay equation in the Schwarzschild metric is obtained integrating $dt=u^t d\tau=\frac{u^t}{u^r}dr$ \cite{Misner73,Poutanen06}, i.e.,
\begin{equation} \label{eq:tide} 
\Delta t =\int^{\infty}_R\frac{u^t}{u^r}dr=\int_R^\infty \frac{dr}{1-\frac{r_s}{r}} \left \{ \left [ 1-\frac{b^2}{r^2}\left(1-\frac{r_s}{r} \right) \right ]^{-\frac{1}{2}}  -1  \right \}.
\end{equation}

To calculate the time delay for photons with a turning point, I need to calculate the periastron distance, $p$. For a given $b$ I therefore consider the largest real solution of the following equation $p^3-b^2p+b^2r_s=0$. The polynomial in $p$ has three real solutions (because $b \ge b_{\rm c}$):  one is negative, one is lower than $3M,$ and I consider only the solution satisfying $p\ge r_c$, where $r_c=3M$ is the critical radius associated to $b_c$ (see, e.g., Ref. \cite{Luminet79}). The time delay is composed of the time delay $\Delta t_{\rm S}$ from point $\alpha_{\rm S}$, as determined by the Eq. (\ref{eq:tide}), plus the time delay between $[\alpha_{\rm S}, \alpha_{\rm p}]$, $\Delta t_{\rm p-S}$, and $[\alpha_{\rm p}, \alpha_{\rm E}]$, $\Delta t_{\rm E-p}$. Since the integrand is symmetric with respect to $\alpha_p$ , the latter two time delays are equal ($\Delta t_{\rm E-p} = \Delta t_{\rm p-S}$), the equation can be written (see Fig. \ref{fig:Fig1_3}) as
\begin{equation} \label{eq:tidetp} 
\Delta t =\Delta t_{\rm S}+2\Delta t_{\rm p-S} =\Delta t_{\rm S}+2\int_R^p \frac{dr}{1-\frac{r_s}{r}} \left \{ \left [ 1-\frac{b^2}{r^2}\left(1-\frac{r_s}{r} \right) \right ]^{-\frac{1}{2}}\right \}.
\end{equation}

\begin{figure}[h]
\centering
\includegraphics[trim=0.5cm 6cm 4.6cm 4.2cm, scale=0.52]{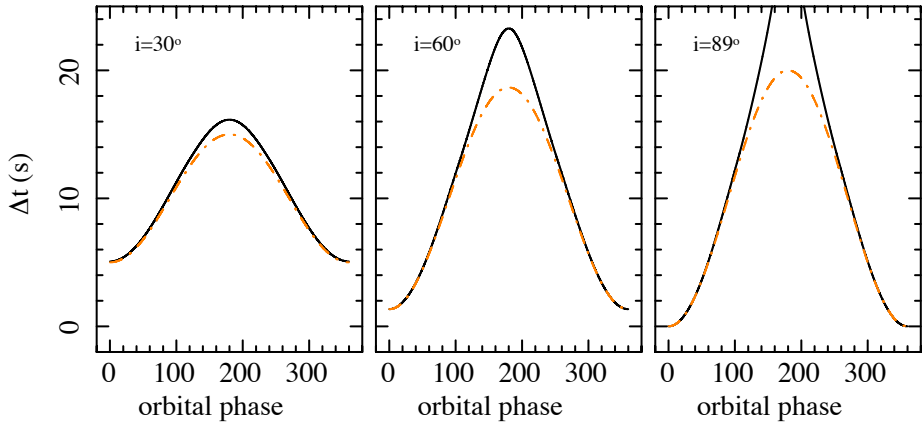}
\caption{Time delay associated to a particle moving along a circular ring at $R=10M$ for $i=30^\circ,60^\circ,89^\circ$. The continuous black lines are given by Eq. (\ref{eq:tide}), instead the dashed-dotted orange lines are given by Eq. (\ref{ctd}).} 
\label{fig:TD}
\end{figure}

\begin{figure}[ht]
\centering
\includegraphics[trim=0 0cm 0cm 1cm, scale=1.2]{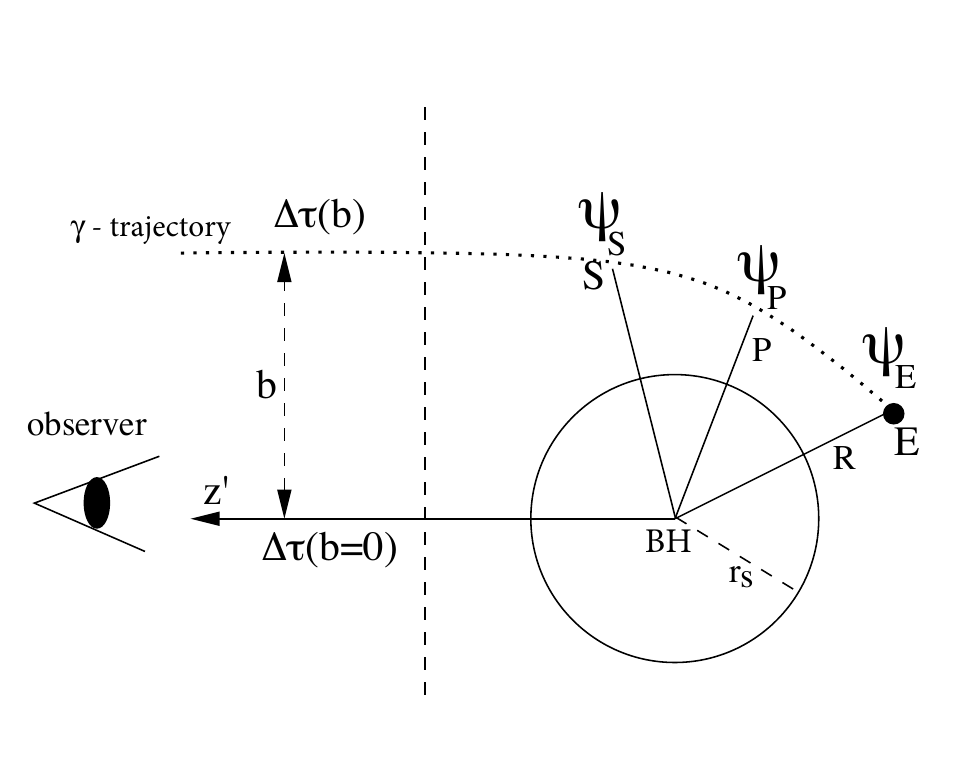} 
\caption{Calculation of travel time delay for trajectories with turning points. The photon is emitted at $E$ with radius $R$ and deflection angle $\psi_{\rm E}$. The photon trajectory passes through point, $S$, which is symmetric to $E$ with respect to periastron $p$, having the same impact parameter, $b$, and a deflection angle $\psi_{\rm S} = 2\psi_{\rm p}-\psi_{\rm E}$.} 
\label{fig:Fig1_3}
\end{figure}

The gravitational travel time delay can be better understood if it is plotted in terms of the orbital phase $\varphi$ and compared with the newtonian case. In the classical limit ($M\longrightarrow0$), Eq (\ref{eq:tide}) becomes
\begin{equation} \label{ctd}
\Delta t=R\ (1-\cos\alpha).
\end{equation}
In Fig. \ref{fig:TD}, it is possible to see how the general relativistic effects enhance the time delay increasing the inclination angle, because the photon trajectories are so bended that they have to tread a path longer compared to the newtonian straight lines.  

\subsection{Solid angle}
\label{subsec:soan}
I consider the emission reference frame of coordinates $(x,y,z)$ and the observer reference frame of coordinates $(x',y',z')$, where the two systems are rotated by an angle, $i$ around the axis $y=y'$. 
To determine the gravitational lensing equation I have to find the transformations between the BH's and the observer's reference frame. In each of such reference frames, the transformations between spherical and cartesian coordinates are given by
\begin{equation} \label{SfCor} 
\begin{minipage}[l]{0.5\textwidth}
$$
\left \{
\begin{aligned}
x &=\sin \psi \cos \varphi \\
y &=\sin \psi \sin \varphi \\
z &=\cos \psi
\end{aligned}
\right.
$$
\end{minipage}
\begin{minipage}[r]{0.5\textwidth}
$$
\left \{
\begin{aligned}
x' &=\sin \psi' \cos \varphi' \\
y' &=\sin \psi' \sin \varphi' \\
z' &=\cos \psi'.
\end{aligned}
\right.
$$
\end{minipage}
\end{equation}
Since $y\equiv y'$, I have
\begin{equation} \label{Tres1} 
\sin \psi \sin \varphi =\sin \psi' \sin \varphi',
\end{equation}
The other two coordinates are rotated of an angle $i$, so I have
\begin{eqnarray}
x' &=&x\cos i -z\sin i \Rightarrow \sin\psi' \cos\varphi'=\sin\psi \cos\varphi\cos i -\cos \psi\sin i \label{Tres2}\\   
z' &=&x\sin i +z\cos i \Rightarrow \cos\psi' =\sin\psi \cos \varphi \sin i + \cos \psi \cos i.  \label{Tres3} 
\end{eqnarray}
Since the emitting particle lies in the invariant plane, so chosing an appropriate reference frame I can set $\psi=\pi / 2$ in Eqs. (\ref{Tres1}), (\ref{Tres2}), (\ref{Tres3})  I have
\begin{equation} \label{Tres4} 
\begin{aligned}
\sin \varphi &=\sin \psi' \sin \varphi',\\
\sin \psi' \cos \varphi' &=\cos \varphi\cos i,\\
\cos \psi' &=\cos \varphi \sin i.
\end{aligned}
\end{equation}
The solid angle, $d\Omega$, in the observer reference frame reads as $d\Omega=\sin\psi \,d\psi \, d\varphi$. This equation can be expressed in terms of the impact parameter, $b$, by its first-order approximation for infinitesimally small $\psi$ as $b\approx D\cdot \psi$, where $D$ is the distance from the emission point to the observer
\begin{equation} 
\label{CASF} 
d\Omega=\frac{b\ db\ d\varphi'}{D^2}.
\end{equation}
In the emission reference frame, Eq. (\ref{CASF}) becomes
\begin{equation} \label{NCASF} 
d\Omega=\frac{b}{D^2}\frac{\partial \varphi'}{\partial \varphi}\frac{\partial b}{\partial r} dr d\varphi,
\end{equation}
where I considered the following dependencies $\varphi=\varphi(\varphi')$ and $b=b(r,\psi)$. The Jacobian of the transformation is always $\frac{\partial \varphi'}{\partial \varphi}\frac{\partial b}{\partial r}$ independent of the value of $\frac{\partial b}{\partial \psi}$, since the photon moves in an invariant plane. Therefore, Eq. (\ref{NCASF}) is valid for any emission point. 
The function $\varphi'$ is given by
\begin{equation} \label{Pasvarphi1} 
\begin{aligned}
\varphi'&=\arcsin\left [ \frac{\sin \varphi}{\sin \psi'}\right ]\\
&=\arcsin\left [ \frac{\sin \varphi}{\sqrt{1-\cos^2 \psi'}}\right ]\\
&=\arcsin \left [ \frac{\sin \varphi}{\sqrt{1-\cos^2 \varphi\sin^2i}}\right ].
\end{aligned}
\end{equation}
This permits to calculate $\frac{\partial \varphi'}{\partial \varphi}$ as
\begin{equation}  \label{Pasvarphi2} 
\begin{aligned}
\frac{\partial \varphi'}{\partial \varphi}&=\left [ \frac{1}{\sqrt{1-\frac{\sin^2\varphi}{1-\cos^2 \varphi\sin^2i}}} \right ]\left[ \frac{\cos\varphi}{\sqrt{1-\cos^2 \varphi\sin^2i}}-\frac{\sin^2\varphi\cos\varphi\sin^2 i}{\sqrt{(1-\cos^2 \varphi\sin^2i)^3}}\right]\\
&=\left [ \sqrt{\frac{1}{\cos^2 \varphi-\cos^2 \varphi\sin^2 i}} \right ]\left[ \frac{\cos\varphi-\cos\varphi\sin^2 i}{1-\cos^2 \varphi\sin^2i}\right ]\\
&=\left [ \frac{1}{\cos \varphi\cos i} \right ]\left[ \frac{\cos\varphi\cos^2 i}{1-\cos^2 \varphi\sin^2i}\right ]=\frac{\cos i}{\sin^2\psi'}.
\end{aligned}
\end{equation}
Since the photon trajectory lies in the invariant plane, it implies $d\psi'=\frac{\partial \psi'}{\partial r}dr+\frac{\partial \psi'}{\partial b}db=0$, so $\frac{\partial b}{\partial r}=-\frac{\partial b}{\partial \psi'}\frac{\partial \psi'}{\partial r}$ can be easily calculated using for $\psi'$ Eq. (\ref{eq:libe}),
\begin{equation} \label{Dervarb} 
\frac{\partial b}{\partial r}=\frac{-\frac{b}{R^2}\left[1-\frac{b^2}{R^2}\left(1-\frac{2M}{R}\right)\right]^{-1/2}}{\int_{R}^{\infty}\frac{dr}{r^2}\left[\frac{1}{b^2}-\frac{1}{r^2}\left(1-\frac{2M}{r}\right)\right ]^{-3/2}}.
\end{equation}
Using Eqs. (\ref{Pasvarphi2}), (\ref{Dervarb}), the solid angle equation in the Schwarzschild metric is thus (see, e.g., Ref. \cite{Bao1994}\footnote{Equation (\ref{EFSA}) is equivalent to the formula (A3) in Ref. \cite{Beloborodov02}.})
\begin{equation} \label{EFSA} 
d\Omega=\frac{\frac{\cos i}{D^2\ R^2\ \sin^2\psi}\frac{b^2}{\cos\alpha}} {\int_R^\infty \frac{dr}{r^2}\left[1-\frac{b^2}{r^2}\left(1-\frac{r_s}{r} \right) \right]^{-\frac{3}{2}}}\ dr\ d\varphi.
\end{equation}
This equation contains an integral with the same functional form as those of light bending Eq. (\ref{eq:libe}) and time delay Eq. (\ref{eq:tide}), except for the $-3/2$ exponent and factors depending on the impact parameter $b$ (or emission angle $\alpha$).   

\begin{figure}[h]
\centering
\includegraphics[trim=0.5cm 6cm 4.6cm 4.2cm, scale=0.52]{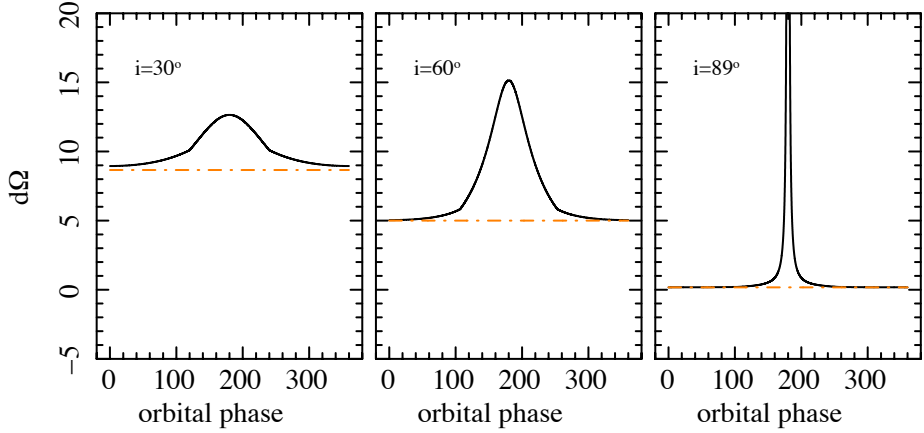}
\caption{Gravitational lensing associated to a particle moving along a circular ring at $R=10M$ for $i=30^\circ,60^\circ,89^\circ$. The continuous black lines are given by Eq. \ref{EFSA}), instead the dashed-dotted orange lines are given by Eq. (\ref{cgl}).} 
\label{fig:GL}
\end{figure}

The gravitational lensing can be fully understood if it is plotted in terms of the orbital phase $\varphi$ and compared with the newtonian case. In the classical limit ($M\longrightarrow0$), Eq. (\ref{EFSA}) becomes
\begin{equation} \label{cgl}
d\Omega=\frac{R\ \cos i}{D^2}\ dr\ d\varphi.
\end{equation}
In Fig. \ref{fig:GL} it is possible to note how the general relativistic effects enhance the observed areas increasing the inclination angle, instead in the classical case the areas remain constant. 

\section{Analytical approximations}
\label{sec:Method}
In this section I present the general mathematical method used to approximate the elliptical equations in polynomials of light bending Eq. (\ref{eq:libe}), time delay Eq. (\ref{eq:tide}), and solid angle Eq. (\ref{EFSA}). 

\subsection{Mathematical method}
\label{subsec:matmet}
Let $f$ be an integrable function of radius, $r$, mass, $M$, and sine of the emission angle, $\sin\alpha$, that is, $f=f(r,M,\sin\alpha)$ and $I$ the following elliptic integral
\begin{equation} \label{INT}
I=\int_{r_i}^{r_f} \frac{1}{\sqrt{f(r,M,\sin\alpha)}}\ dr.
\end{equation} 
I am interested in deriving a polynomial approximation of the elliptic integral $I$. I first define $\sin\alpha=g(z)$, where $g(z)$ is a generic function of $z(\alpha)$. To expand Eq. (\ref{INT}) in Taylor series I assume that $\alpha$ is very small\footnote{Therefore, $g(z)$ is small as well.}  and aim at obtaining an integrable polynomial function
\begin{equation} \label{POL} 
I=\int_{r_i}^{r_f} \frac{1}{\sqrt{f(r,M,g(z))}}\ dr\approx P(r_f,r_i,M,g(z)).
\end{equation} 
$P$ contains even powers of $g(z)$, since $f(r,M,g(z))\propto g(z)^2$. This condition is given by substituting $b = (r\sin\alpha) / (\sqrt{1-r_{s}/r})$ in the equations of the light bending Eq. (\ref{eq:libe}), time delay Eq. (\ref{eq:tide}), and solid angle Eq. (\ref{EFSA}). For an exact polynomial approximation, I therefore define $g(z)=\sqrt{Az^2+Bz}$, where $A$ and $B$ are general parameters. One of the two parameters ($A, B$) is determined by comparing Eq. (\ref{POL}) with the original integral $I$ for special values of $M=M*$, $r_f=r_f*$ and $r_i=r_i*$ that permits solving the integral $I$ easily and obtain

\begin{equation} \label{EQL} 
I(r_f*,r_i*,M*,\sin\alpha)=P(r_f*,r_i*,M*,\sqrt{Az^2+Bz}).
\end{equation}  
The other parameter can be determined through the initial condition $\sin\alpha=\sqrt{Az^2+Bz}$. I note that the polynomial approximation is valid for any emission angle $\alpha$ (not only for low values) since the parameters $A, B$ are gauged on the whole range of $I$.

\subsection{Light bending}
\label{sec:applibe}
For the light bending I Taylor-expand Eq. (\ref{eq:libe}) up to the third order and defining $u=2M/R$ and $\sin\alpha=g(z)$ I obtain
\begin{equation}\label{eq:lb}
\begin{aligned}
\psi\approx \frac{b}{R}&\left[1+\frac{g^2(z)}{6(1-u)}-\frac{g^2(z)u}{8(1-u)}+\frac{3g^4(z)}{40(1-u)^2}+\right.\\
&+\frac{3g^4(z)u^2}{56(1-u)^2}-\frac{g^4(z)}{8(1-u)^2}u+\frac{5g^6(z)}{112(1-u)^3}-\\
&\left. -\frac{g^6(z)u^3}{32(1-u)^3}-\frac{15g^6(z)}{128(1-u)^3}u+\frac{5g^6(z)}{48(1-u)^3}u^2\right].
\end{aligned}
\end{equation}
Setting $g(z)=\sqrt{Az^2+Bz}$ and neglecting all the terms up to the second order in $z$, Eq. (\ref{eq:lb})  
becomes
\begin{equation} \label{TEF} 
\psi\approx\sqrt{\frac{Az^2+Bz}{1-u}}\left[1+\left(\frac{B}{6(1-u)}-\frac{Bu}{8(1-u)}\right)z \right].
\end{equation}
To approximate this equation with a polynomial, I introduce an even trigonometric function of $\psi$ to remove the square root. The simplest choice is a cosine function expanded to the fourth order in $\psi$

\begin{equation} \label{COSESP} 
\begin{aligned}
1-\cos\psi&\approx \frac{\psi^2}{2}-\frac{\psi^4}{24}\approx\\
&\approx \frac{Bz}{2(1-u)}+\left [\frac{B^2}{6(1-u)^2}-\frac{B^2u}{8(1-u)^2}+\right .\\
&\quad\left. +\frac{A}{2(1-u)}-\frac{B^2}{24(1-u)^2} \right ]z^2,
\end{aligned}
\end{equation}
where I consider the terms to the second order in $z$. If I choose $A=-(B/2)^2$ , I obtain a simple linear approximation, $1-\cos\psi \,\approx Bz/(2(1-u))$, in which $z^2$ coefficients vanish. 

I now solve Eq. (\ref{eq:libe}) for the special values $u=0,\ R=1$ and obtain
\begin{equation} \label{TELB} 
\psi=b\int_1^\infty \frac{dr}{r^2}\left[1-\frac{\sin^2\alpha}{r^2} \right]^{-\frac{1}{2}}=\alpha.
\end{equation}
Using the same values ($u=0,\ R=1$) for the approximated polynomial equation, $1-\cos\psi \,\approx Bz/(2(1-u))$, I obtain
\begin{equation}
1-\cos\alpha=\frac{Bz}{2}.
\end{equation}
In this case, by defining $B=2$ (implying $A=-1$), I find $z=1-\cos\alpha$, which, when replaced in Eq. (\ref{COSESP}), gives the approximate light bending equation originally found by Ref. \cite{Beloborodov02}
\begin{equation} \label{AFLB} 
1-\cos\psi=\frac{(1-\cos\alpha)}{(1-u)}.
\end{equation}

\begin{figure}[h]
\centering
\includegraphics[trim=0 3cm -5cm 3cm, scale=0.48]{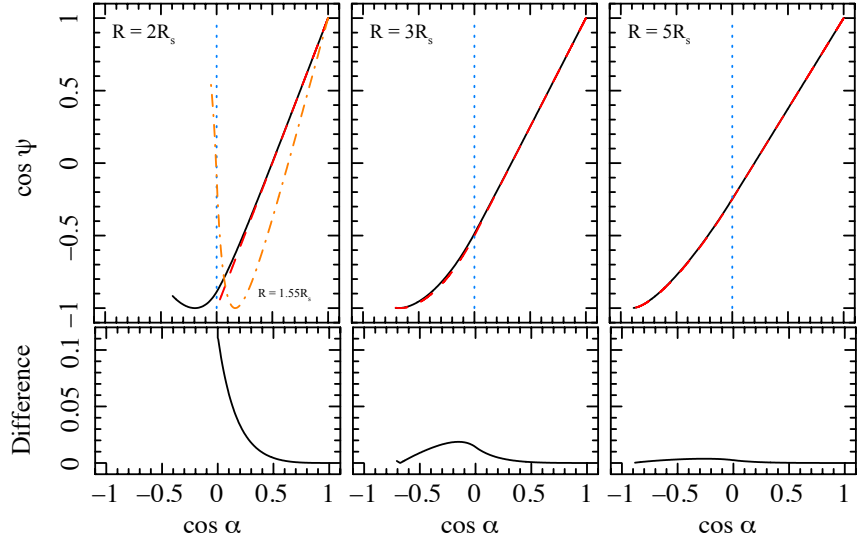} 
\caption{Light bending curves from the exact Eq. (\ref{eq:libe}) (solid lines), compared to those from the approximate Eq. (\ref{AFLB}) (dashed red lines) for $R=2r_s$, $R=3r_s$, and $R=5r_s$. The dotted blue line represents the threshold from trajectories without a turning point ($0\le\alpha\le\pi/2$) to trajectories with a turning point ($\pi/2\le\alpha\le\alpha_{max}$). The exact light bending curve for $R=1.55r_s$ is also plotted (dotted-dashed orange line) to show strong-field effects. The lower panels show the difference between the curves from the original and approximate equations.} 
\label{fig:Fig1_4}
\end{figure}

In Fig. \ref{fig:Fig1_4} I show a comparison between the exact light bending curves for different emission radii, and curves obtained from the approximate equation. The accuracy of the latter between $0\le\alpha\le\alpha_{max}$ is better than 3\% for $R=3r_s$, while for $R = 5r_s$ the error does not exceed 1\%. I note that $R = 3r_s$ corresponds to the innermost stable circular orbit (ISCO) for matter orbiting a Schwarzschild BH and is also close represent to a typical NS radius size of $\sim12$ km for mass of $1.4M_{\odot}$. For values below $R=2 r_s$ the equation is not anymore applicable after $\alpha = \pi/2$. In Fig. \ref{fig:Fig1_4} I also show the exact light bending curve for $R=1.55r_s$; after a given minimum the photons are highly bent by strong-field effects. The largest error is at $\alpha=\pi/2$ and then it tends to decrease until at $\alpha_{max}$ because of the symmetrization process around $\alpha=\pi/2$ configuring as the maximum reachable angle (see Sec. \ref{subsec:impa&libe}). For more details about the accuracy between $0\le \alpha \le \pi/2$ I refer to Ref. \cite{Beloborodov02}.

\subsection{Time delay}
\label{sec:apptide}
I now apply my method for deriving the approximate equation for the time delay. By expanding the integrand in Eq. (\ref{eq:tide}) up to the third order 
\begin{equation} \label{TDGE2}
\begin{aligned}
\Delta t = R&\left \{\frac{g^2(z)}{2(1-u)}+\frac{g^4(z)}{8(1-u)^2}-\frac{3g^4(z)}{32(1-u)^2}u+\right.\\
&\left. +\frac{g^6(z)}{16(1-u)^3}-\frac{5g^6(z)}{48(1-u)^3}u+\frac{5g^6(z)}{112(1-u)^3}u^2\right \},
\end{aligned}
\end{equation}   
I set again $g(z)=\sqrt{Az^2+Bz}$ and neglect all terms up to the third order in $z$, so that
\begin{equation} \label{TDGE3}
\begin{aligned}
\frac{\Delta t}{R} \approx &\left \{\frac{Az^2+Bz}{2(1-u)}+\frac{B^2z^2+2ABz^3}{8(1-u)^2}-\frac{3u(B^2z^2+2ABz^3)}{32(1-u)^2}+\right.\\
&\left. +\frac{B^3z^3}{16(1-u)^3}-\frac{5uB^3z^3}{48(1-u)^3}+\frac{5u^2B^3z^3}{112(1-u)^3}\right \}.
\end{aligned}
\end{equation}   
To determine $(A,B)$ I compare the original Eq. (\ref{eq:tide}) with Eq. (\ref{TDGE3}) both evaluated for $u=0$ and $R=1$; I find\footnote{For Eq. (\ref{eq:tide}) I used the following limit: $\lim_{x\to +\infty}(x^2-a)^{\frac{1}{2}}-x=0$.}  
\begin{equation} \label{AF} 
1-\cos\alpha=\frac{B}{2}z+\frac{1}{2}\left(A+\frac{B^2}{4}\right)z^2+\frac{B}{4}\left(A+\frac{B^2}{4}\right)z^3,
\end{equation}
where on the left and right hand sides are the results of Eq. (\ref{eq:tide}) and Eq. (\ref{TDGE3}), respectively. By imposing $A+B^2/4=0$ the coefficients of the second and third order in $z$ vanish. Like in the light bending case, Eq. (\ref{AF}) reduces to $1-\cos\alpha=Bz/2$; defining again $B=2$ (implying  $A=-1$) substituting in Eq. (\ref{TDGE3}), I derive the approximate travel time delay equation (see Ref. \cite{Poutanen06}, for further details)
\begin{equation} \label{FATIDE}
\frac{\Delta t}{R}=y\left [1+\frac{uy}{8}+\frac{uy^2}{24}-\frac{u^2y^2}{112} \right ],
\end{equation}
where $y=(1-\cos\psi)$.

\begin{figure}[h]
\centering
\includegraphics[trim=-1cm 3cm 3cm 3.52cm, scale=0.48]{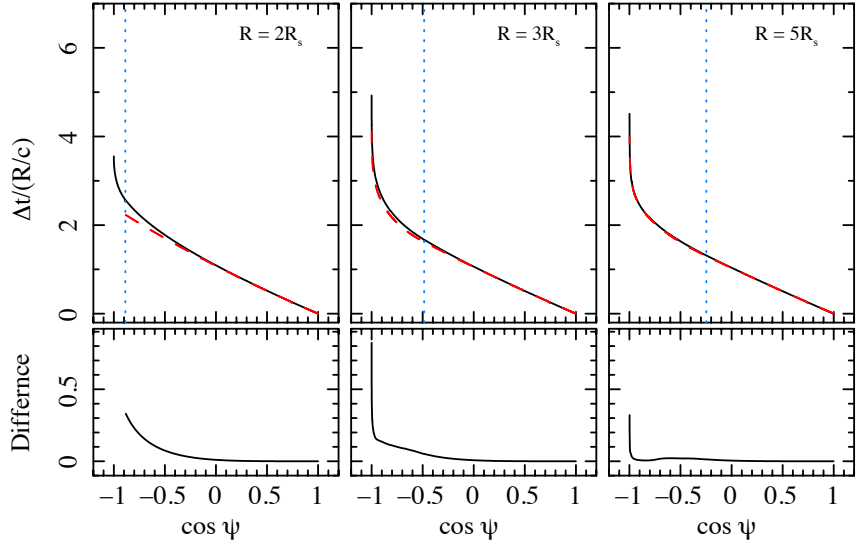} 
\caption{Continuous black curves are obtained from the original time delay Eq. (\ref{eq:tide}), while the dashed red curves are from the polynomial approximate Eq. (\ref{FATIDE}). Different panels show $R=2r_s$, $R=3r_s$ and $R=5r_s$. The dotted blue line helps distinguishing trajectories without a turning point (i.e., $0\le\cos\alpha\le1$) from those with a turning point (i.e., $\pi/2\le\alpha\le\alpha_{max}$). The lower panels show the difference between the curves from the original and approximate equations.} 
\label{fig:Fig1_5}
\end{figure}

In Fig. \ref{fig:Fig1_5} I compare for different emission radii the exact travel time delay curves with the polynomial approximated equations. I here also extend the validity of the approximation to $\alpha_{max}$-values accounting for turning points. The accuracy settles $\sim35\%$ for $R=2r_s$, while after $R=3r_s$ it is lower than 20\%, according to the same symmetry argument explained in the Sec. \ref{sec:applibe}. However, I refer to Ref. \cite{Poutanen06} for the error estimation between $0\le \alpha \le \pi/2$.

\subsection{Solid angle}
\label{sec:appsoan}
I now apply the same method to derive for the first time a polynomial approximation to the solid angle Eq. (\ref{EFSA}). I note, at variance of light bending and time delay equations, that the solid angle equation has the integral in the denominator, and moreover, the emission angle, $\alpha$, is also outside the integral. I first rewrite Eq. (\ref{EFSA}) as 
\begin{equation} \label{INI}
d\Omega=\frac{P_{\rm 1}\ P_{\rm 2}}{I}\ dr\ d\varphi,
\end{equation}
where
\begin{equation} \label{eq:soan}
\begin{aligned}
P_{\rm 1}&=\frac{\cos i}{D^2\sin^2\psi\ (1-u)},\qquad P_{\rm 2}=\frac{\sin^2\alpha}{\cos\alpha},\\
& \\
& I=\int_R^\infty \frac{dr}{r^2}\left[1-\frac{R^2\sin^2\alpha}{r^2(1-u)}\left(1-\frac{uR}{r} \right) \right]^{-\frac{3}{2}}.
\end{aligned}
\end{equation}
$P_{\rm 1}$ is a constant because $\psi$ is a function of the azimuthal angle, $\varphi$, the inclination angle, $i$, and the polar coordinate, $\theta$, (for further details see Sec. \ref{sec:applications}). As a first step, I expand the integrand of $I$ in a Taylor series up to the third order in $z$. I derive
\begin{equation}\label{AI} 
I\approx \frac{1+Cz+Dz^2}{R},
\end{equation}
with
\begin{eqnarray}
C&=&\frac{B}{2(1-u)}-\frac{3Bu}{8(1-u)},\\
D&=&\frac{A}{2(1-u)}-\frac{3Au}{8(1-u)}+\frac{3B^2}{8(1-u)^2}+\\
&+&\frac{15B^2u^2}{16(1-u)^2}-\frac{5B^2u}{8(1-u)^2}. \notag
\end{eqnarray}
The function $P_{\rm 2}/I$ is not yet a polynomial function since it contains a ratio of polynomials and square root functions in $P_{\rm 2}$. For these reasons I expand $P_{\rm 2}/I$ in a Taylor series around $z=0$ and neglect all the terms up to third order in $z$
\begin{equation}\label{AIA} 
\begin{aligned}
\frac{P_{\rm 2}}{I}&\approx \frac{Az^2+Bz}{\sqrt{1-Az^2-Bz}}\frac{R}{1+Cz+Dz^2}\approx\\
&\approx R\left [Bz+\left(\frac{B^2}{2}+A-CB \right)z^2+\right.\\
&\left. +\left(AB+\frac{3B^2}{8}-\frac{CB^2}{2}-CA+BC^2-BD \right)z^3 \right ].
\end{aligned}
\end{equation}
To determine $(A,B)$ I compare the original solid angle Eq. (\ref{eq:soan}) with the above
approximate equation, evaluating both equations for $u=0$ and $R=1$; I find
\begin{equation}
\sin^2 \alpha=Bz+Az^2.
\end{equation}
The left- and right-hand sides are the result of original Eq. (\ref{eq:soan}) and the polynomial Eq. (\ref{AIA}), respectively. I can freely define the value of $A$ and $B$ because there are no particular constraints to impose. I set, as in the previous cases, $A=-1$ and $B=2$, deriving again $z=1-\cos\alpha$. The final approximate equation for the solid angle is
\begin{equation} \label{AFSA} 
\begin{aligned}
d\Omega&\approx \frac{\cos i}{D^2\sin^2\psi\ (1-u)}\ R \left[ 2z+\left(1-2C\right)z^2+\right. \\
&\left. +\left(1-C+2C^2-2D\right)z^3\right]\ dr\ d\varphi,
\end{aligned}
\end{equation}
where
\begin{equation}
C=\frac{4-3u}{4(1-u)},\qquad D=\frac{39u^2-91u+56}{56(1-u)^2}.
\end{equation}

\begin{figure}[p]
\hbox{
\includegraphics[trim=0cm 0cm 4.6cm 8cm, scale=0.44]{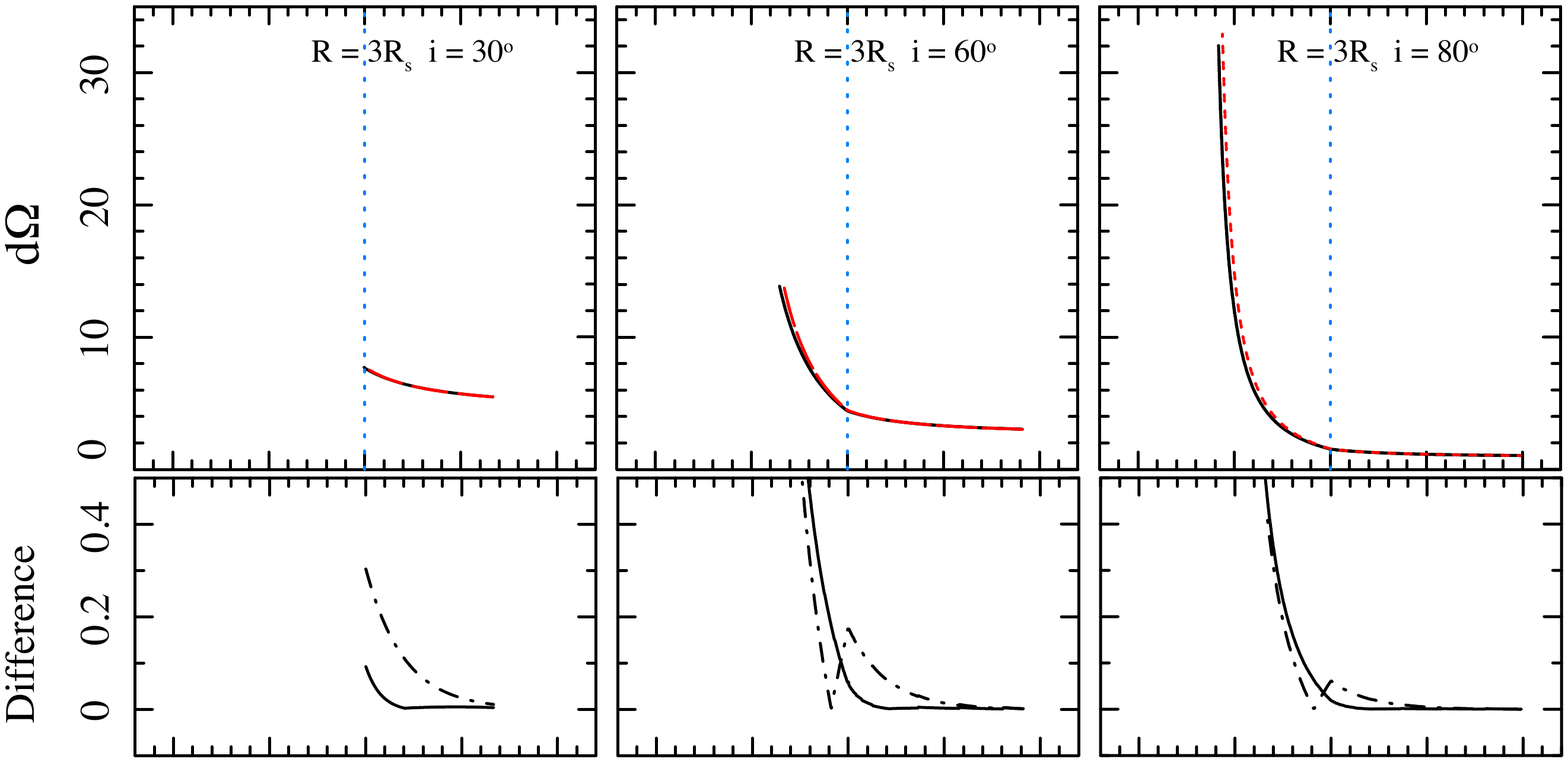}
}
\vspace{0.05cm}
\hbox{
\includegraphics[trim=0.3cm 5.5cm 4.6cm 7.97cm, scale=0.44]{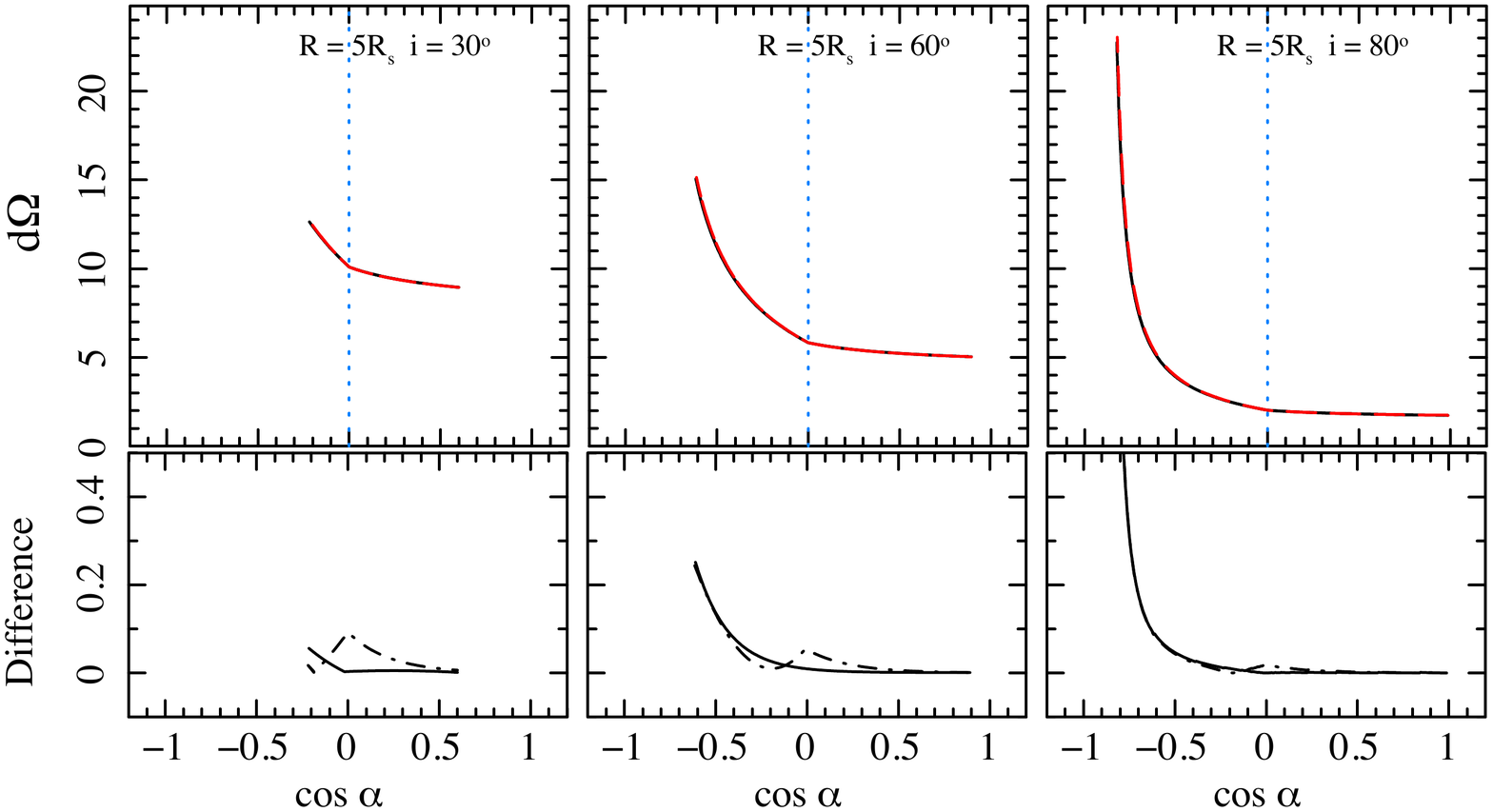}
}
\vspace{-0.1cm}
\caption{Continuous black curves are obtained from the original solid angle Eq. (\ref{EFSA}), while the dashed red curves are from the polynomial approximation in Eq. (\ref{AFSA}). Different panels are for $R=3r_s$ and $R=5r_s$ and three different inclination angles, $i=30^\circ$, $i=60^\circ$, and $i=80^\circ$. The dotted blue line helps distinguishing trajectories without a turning point (i.e., $0\le\cos\alpha\le1$) from those with a turning point (i.e., $\pi/2\le\alpha\le\alpha_{max}$). The lower panels show the difference between the curves from the original and approximate equations. The dotted-dashed lines represent the difference between the curves from the original and Ref. \cite{Beloborodov02} equations.}
\label{fig:Fig1_6}
\end{figure}

As for the previous two cases, in Fig. \ref{fig:Fig1_6} I compare the exact solid angle curves with the polynomial approximated curves for different radii and inclination angles $i$. The comparison extends to $\alpha_{max}$-values and thus accounts for trajectories with turning points in this case as well. For $R=3r_s$ the error is $\sim5\%$ and after $R=5r_s$ it is lower than $1\%$. I note that for $i=30^\circ$ the curves are fairly flat because the relativistic effects are small. Instead, passing from $i=60^\circ$ to $i=80^\circ$ , the curves become gradually steeper as general relativistic effects increase. Unlike the previous cases, I do not show here the case $R=2r_s$ because the approximate formula Eq. (\ref{AFSA}) does not give adequately accurate results.

I note that Eq. (A3) in Ref. \cite{Beloborodov02} is obtained by approximating the derivative $\frac{d\cos\psi}{d\cos\alpha}$ with the linear Eq. (\ref{AFLB}), while my Eq. (\ref{AFSA}) is a third-order polynomial that approximates the integral $I$ and all the terms depending on the emission angle $\alpha$. For example, my approximation is more accurate by a factor of $\sim$3 to 10 for $R=3r_s$ and $0\le\cos\alpha\le0.3$.
 
\section{Examples of astrophysical applications}
\label{sec:applications}
In this section I present three simple examples of astrophysical applications of the approximate equations. 
I consider the emission point at coordinates $(r,\varphi,\theta)$. The observer is located at infinity along the $z'$-axis with a viewing angle, $i$, with respect to the $z$-axis; the observer polar coordinates are ($r',\varphi',\theta'$). Photons emitted from a point are deflected by an angle, $\psi$, and reach the observer with impact parameter, $b$. The plane containing the photon trajectory rotates around the line of sight as the emission point moves around the compact object. Two unit vectors are attached to the photon emission point, $E$: ${\bf u}$ is tangential to the photon trajectory, and ${\bf n}$ points in the same direction as the radius, $R$. The photon deflection angle, $\psi$, varies as
\begin{equation}
\cos\psi=\sin i\sin\theta \cos\varphi+\cos i \cos\theta,
\end{equation}
with $\theta=\pi/2$, $\varphi = \omega_{k} t$ and $t=0$ when the emission point is closest to the observer. The photon arrival time, $T_{\rm obs}$, is the sum of the emission time, $T_{\rm orb}=\varphi/\omega_{k}$, plus the photon propagation delay, $\Delta T(b)$, from the emission point to the observer, see Eq. (\ref{eq:tide}). 

The observed flux is $F=\int_{\nu_{obs}}\int_\Omega I_{\nu_{obs}}d\Omega\ d\nu_{obs}$, where $I_{\nu_{obs}}$ is the specific intensity at the photon frequency $\nu_{obs}$. I use the Lorentz invariant ratio $I_{\nu_{obs}}/\nu^3_{obs}=I_{\nu_{em}}/\nu^3_{em}$ (see, e.g., \cite{Misner73}), where $I_{\nu_{em}}(\xi)=\frac{\epsilon_0 \xi^{q}}{4\pi}\delta(\nu_{em}-\nu_{obs})$ is the specific intensity at the emission point $E$ given by the product of the surface emissivity, varying as a power law of $\xi=R/M$ with index $q$, and the delta function peaked at $\nu_{em}$. Therefore, integrating over all the frequencies, I obtain the observed flux at frequency $\nu_{em}$, $F_{\nu_{em}}=\int_{\Omega} \frac{\epsilon_0 \xi^{q}}{4\pi}\,(1+z)^{-4}\ d\Omega$. The redshift is defined as the ratio between the observed and the emitted energy, $(1+z)^{-1}=\nu_{obs}/\nu_{em}$ \cite{Misner73} and for matter orbiting in circular orbits around a compact object or for a spot on a NS surface reads as
\begin{equation}\label{Redshift2} 
(1+z)^{-1}=\left(1-\frac{r_{\rm s}}{R}-\omega^{2}R^{2}\sin^{2}\theta\right)^{1/2}\left(1+b\omega         
\frac{\sin i\,\sin\varphi\sin\theta}{\sin\psi} \right)^{-1}. 
\end{equation}
For $\omega = \omega_k$ I consider matter orbiting with Keplerian velocity around a BH, and for $\omega = \omega_{spin}$ I consider spots rotating with the NS spin frequency. The relevant geometry is shown in Fig. \ref{fig:Fig1_7}.  

\subsection{Light curve from an emitting clump orbiting a black hole}
\label{sec:spotsBH}

\begin{figure}[ht]
\hbox{
\includegraphics[trim=-3.0cm 0cm 4.6cm 2cm, scale=0.4]{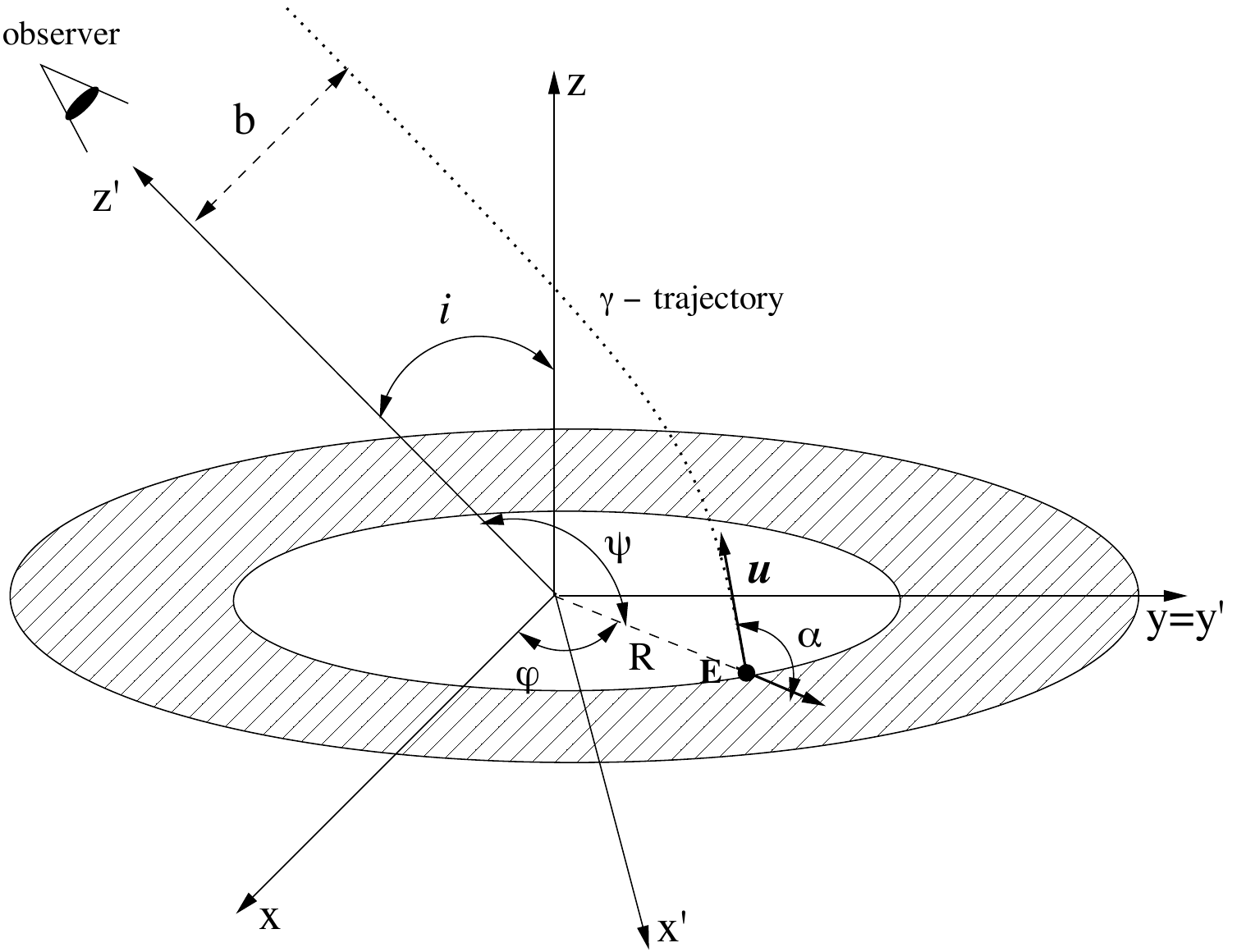}
\hspace{1.6cm}
\includegraphics[trim=-1cm -0cm 3.6cm 1cm, scale=0.4]{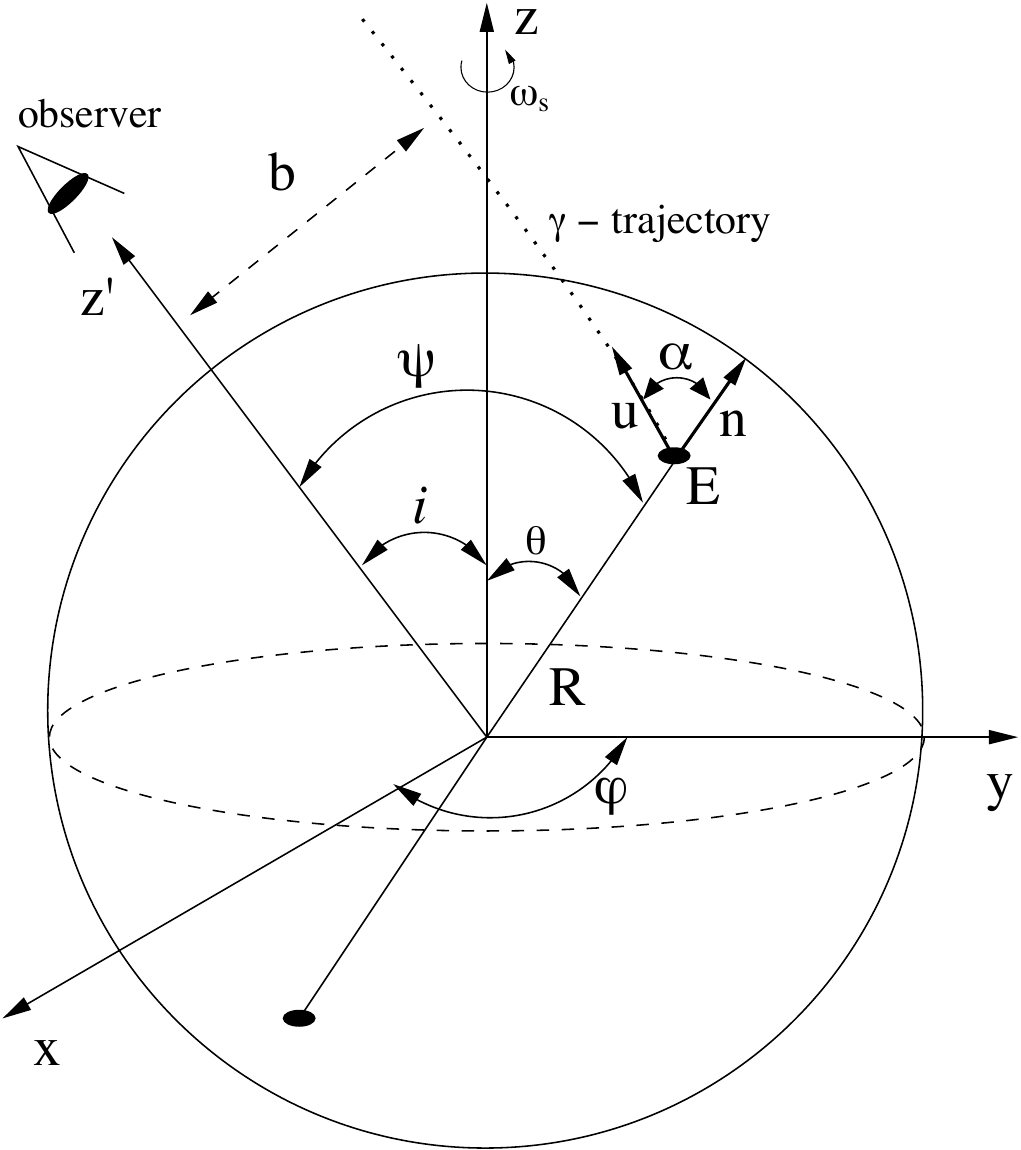}
}
\vspace{0.1cm}
\caption{Geometries adopted in the examples. \emph{Left:}  Emission from a disk or clump orbiting a Schwarzschild BH. \emph{Right:} Emission from two hot opposite spots on an NS surface.} 
\label{fig:Fig1_7} 
\end{figure}

\begin{figure}[p]
\hbox{
\includegraphics[trim=4cm 2cm 4.6cm 6cm, scale=0.35]{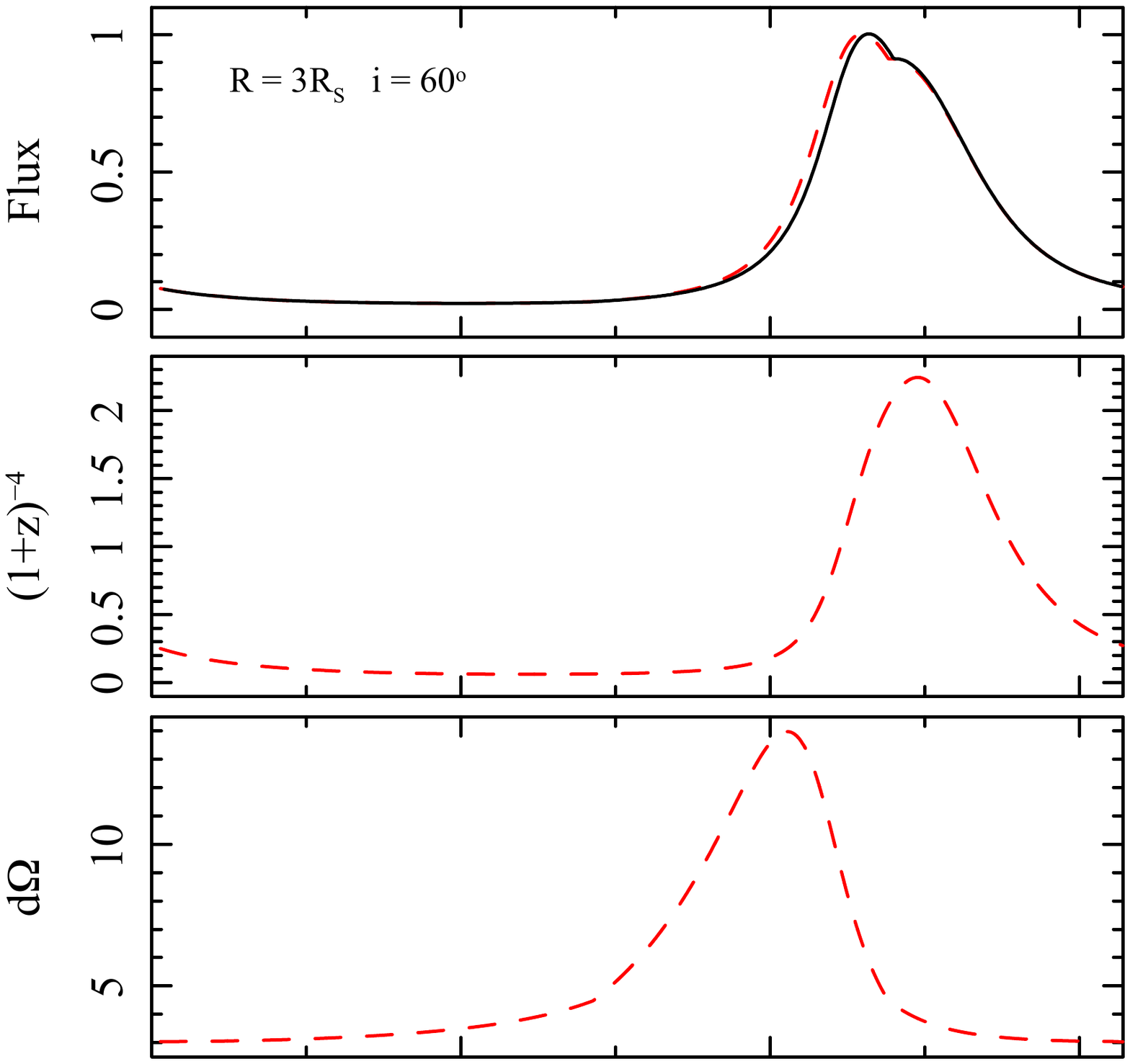}
\hspace{-1.4cm}
\includegraphics[trim=2cm 2cm 6.5cm 1cm, scale=0.35]{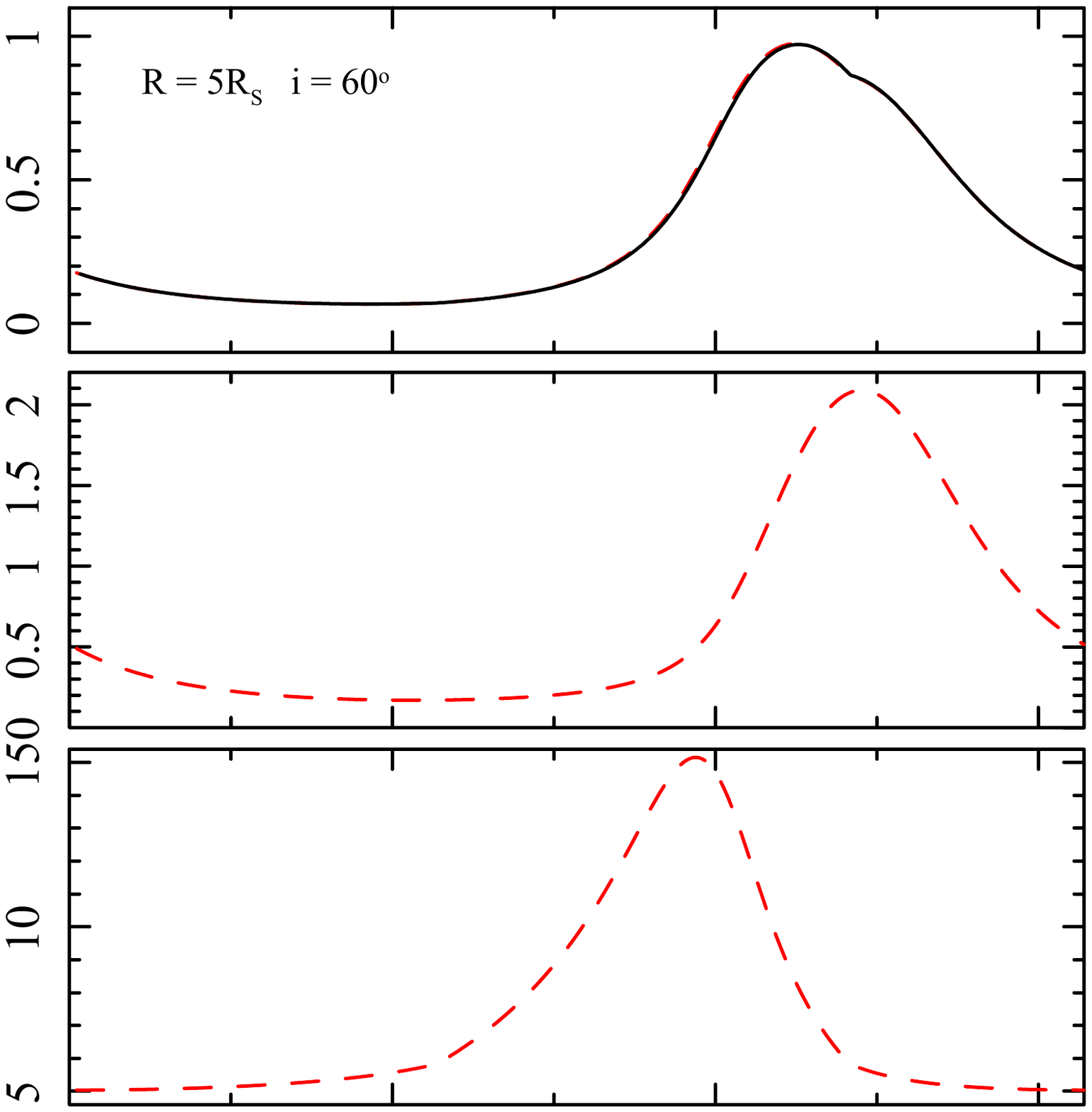}
}
\vspace{-0.1cm}
\hbox{
\includegraphics[trim=4cm 2cm 4.6cm 6cm, scale=0.35]{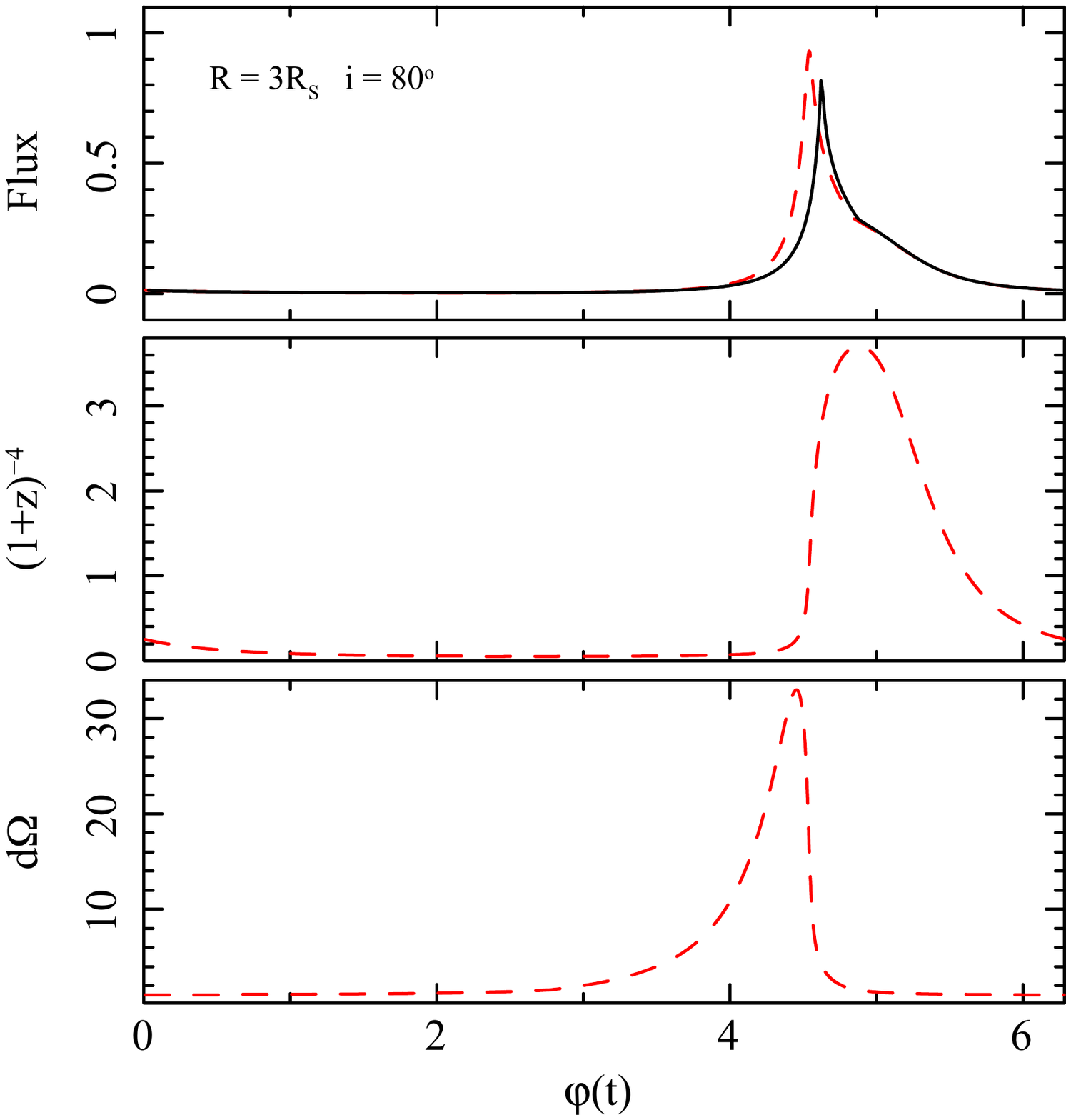}
\hspace{-1.4cm}
\includegraphics[trim=2cm 2cm 6.5cm 2cm, scale=0.35]{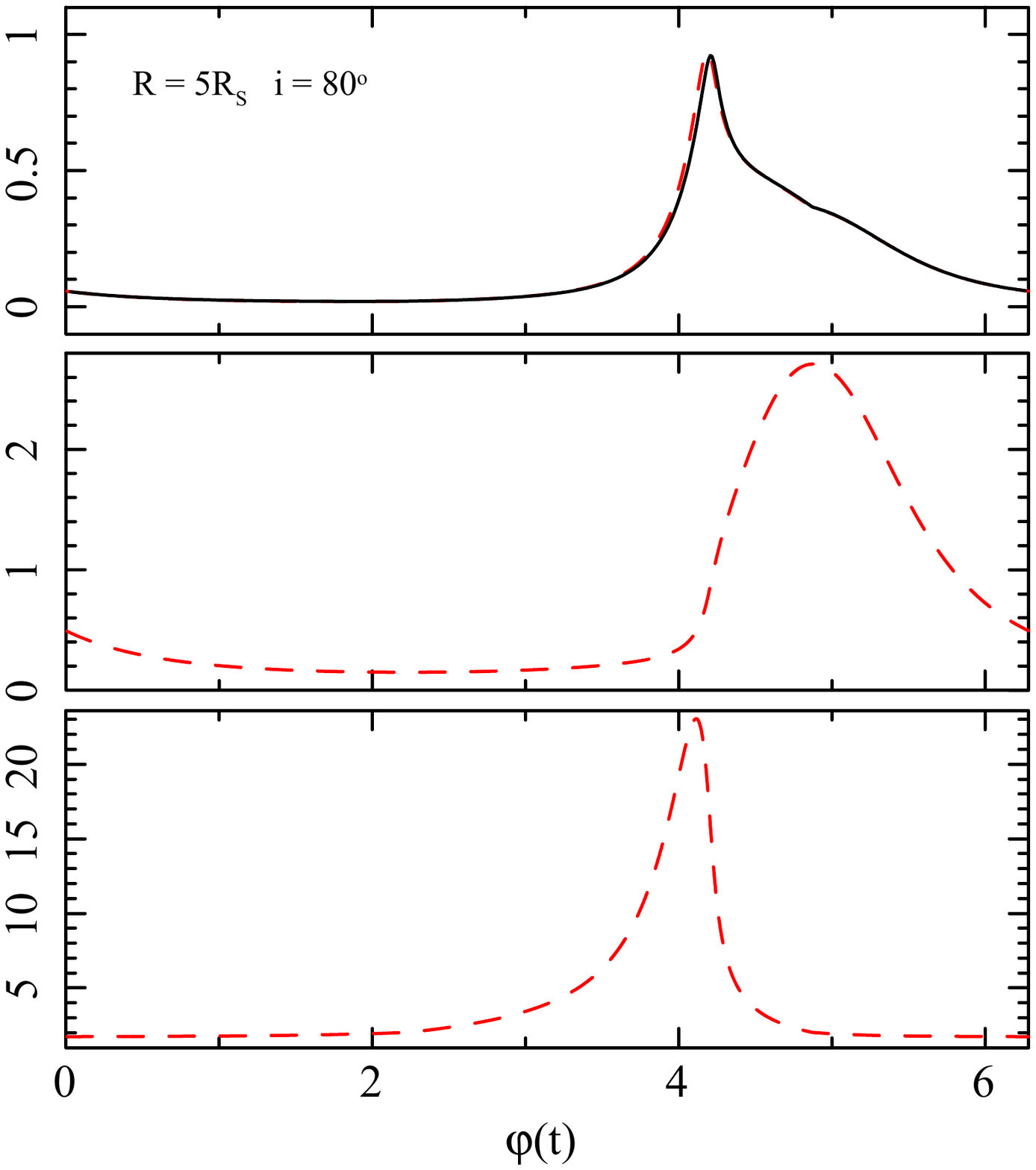}
}
\vspace{-0.1cm}
\caption{Modulated flux (normalized to the maximum), Doppler factor $(1+z)^{-4}$ and solid angle (arbitrary units) in the rest coordinate frame of an emitting clump in a circular orbit around a Schwarzschild BH for different radii and inclinations angles. The continuous black lines are calculated with the exact equations, while the dashed red lines are calculated with the approximate equations. All quantities are plotted as a function of the arrival phase at the observer. In the left panels a self-eclipse of the spot is apparent.
} 
\label{fig:Fig1_8}
\end{figure}

I first consider a clump defined as a small sphere radiating isotropically in its own rest frame, orbiting a Schwarzschild BH in a circular orbit with angular velocity $\omega_{k} = (M/R^{3})^{1/2}$. The geometry is shown in Fig. \ref{fig:Fig1_7}. For simplicity I assume $\frac{\epsilon_0 \xi^{q}}{4\pi}=1$. 
Figure \ref{fig:Fig1_8} shows the modulation of the Doppler factor $(1+z)^{-4}$, solid angle $d\Omega,$ and flux from the orbiting clump as a function of phase, $\varphi(t)$, including light travel time delays. When the clump is behind the BH, gravitational lensing magnifies the solid angle from which the clump is seen by observer; the Doppler factor is greatest when the projected velocity along the photon trajectory reaching the observer is highest. The gravitational effects are stronger for larger inclination angles, and the observed peak flux is not at $\varphi=\pi$, but is significantly shifted especially for large inclination angles due to the travel time delays. The errors between the approximated and the original flux depend only on the emission radius, since the inclination angle figures as a constant. However, it is evident that the main errors derive from the approximated time delay equation (as shown in Sec. \ref{sec:apptide}). 

\subsection{Emission line profile from an accretion disk around a black hole}
\label{sec:Ironline}
\begin{figure}[h]
\centering
\includegraphics[trim=2cm 2cm 6.5cm 2.6cm,scale=0.6]{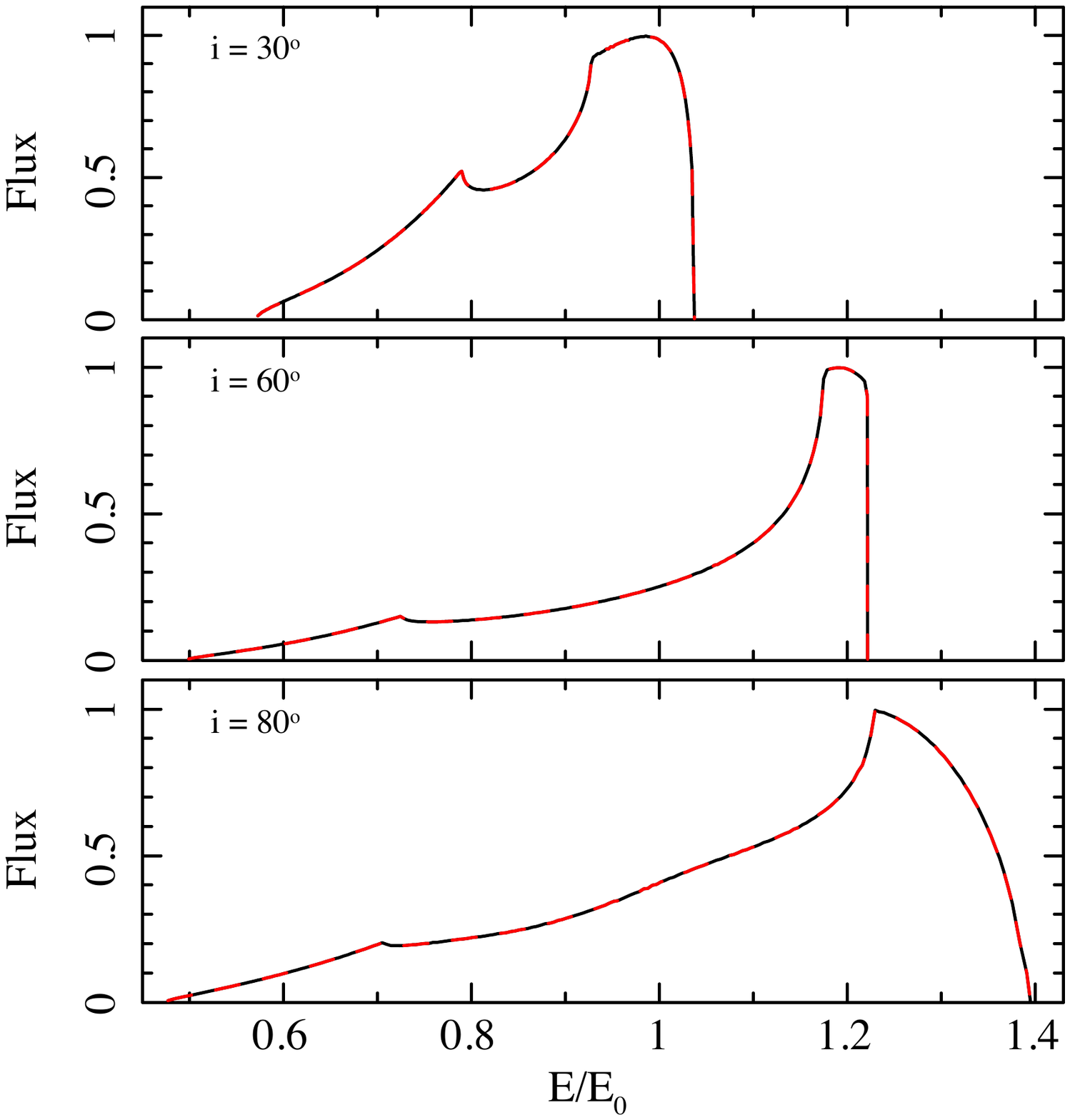} 
\caption{Line profile for isotropic radiation from $R_{\rm in}=3\,r_{\rm s}$ to $R_{\rm out}=50\,r_{\rm s}$ assuming surface emissivity $q=-3$. The continuous black lines represent the original equations and the dashed red lines are the polynomial approximate equations.} 
\label{fig:Fig1_9}
\end{figure}

In Fig. (\ref{fig:Fig1_9}) I calculate the steady relativistically broadened emission line profile from an accretion disk around a Schwarzschild BH (e.g., Refs. \cite{Fabian89,beckwith04} and references therein). Fe $K{\rm \alpha}$ lines at $\sim 6- 7$~keV from a number of accreting stellar mass BHs and NSs in X-ray binaries, as well as supermassive BHs in the nuclei of active galaxies are interpreted based on this model (e.g., Ref. \cite{tomsick14}). I integrate over the disk surface from an inner to an outer disk radius and ignore light propagation delays, as I consider a steady disk. The approximate equations reproduce very accurately the profiles obtained with the exact equations.  A high accuracy is also retained for large inclination angles, even if larger inclination angles enhance the relativistic effects (see Sec. \ref{sec:spotsBH}). 
  
\subsection{Light curve from a hot spot on the surface of a rotating neutron star}
\label{sec:spotsNS}
\begin{figure}[ht]
\hbox{
\includegraphics[trim=4cm 2cm 4.6cm 1cm, scale=0.35]{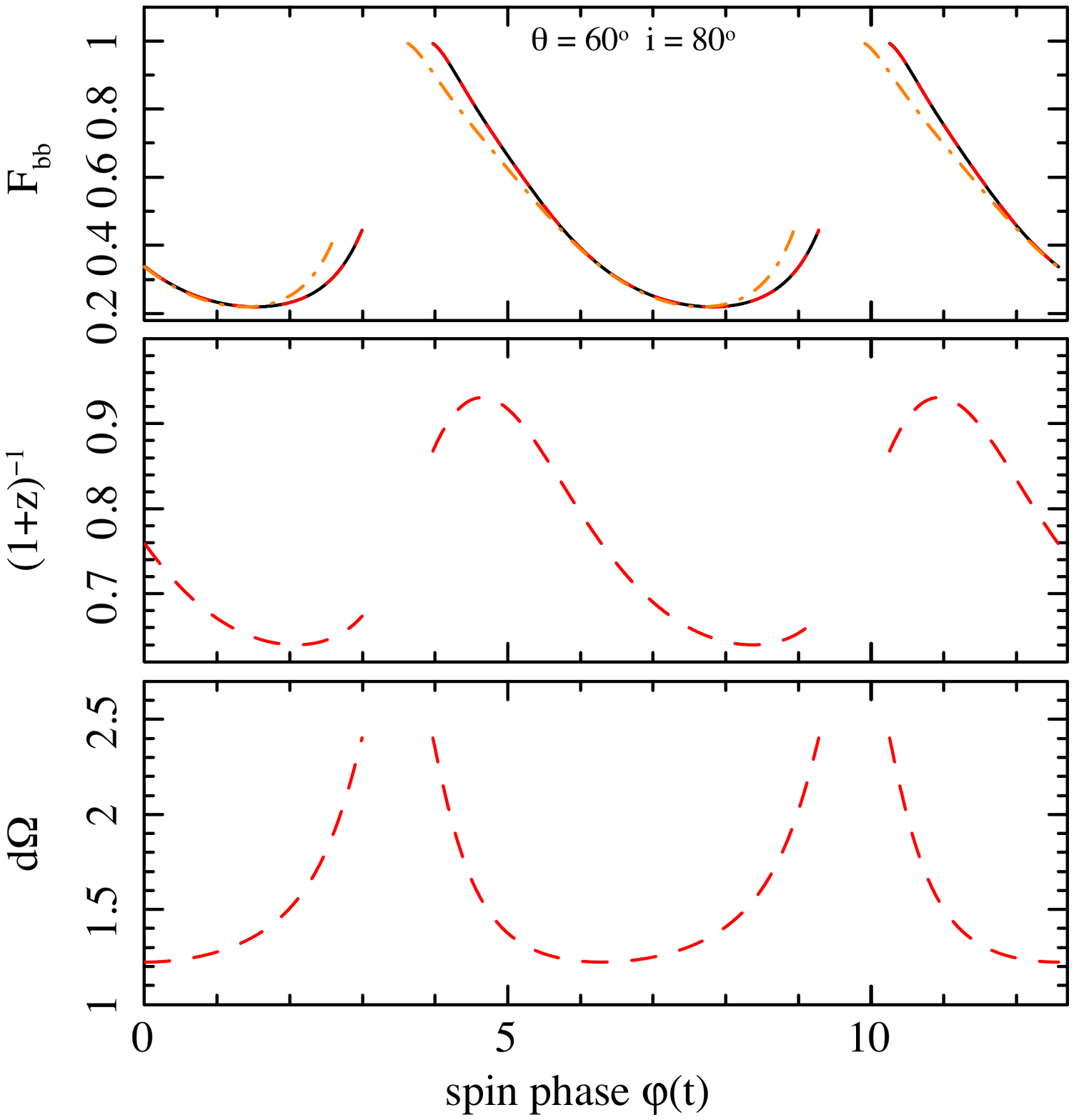}
\hspace{-1.4 cm}
\includegraphics[trim=2cm 2cm 6.5cm 1cm, scale=0.35]{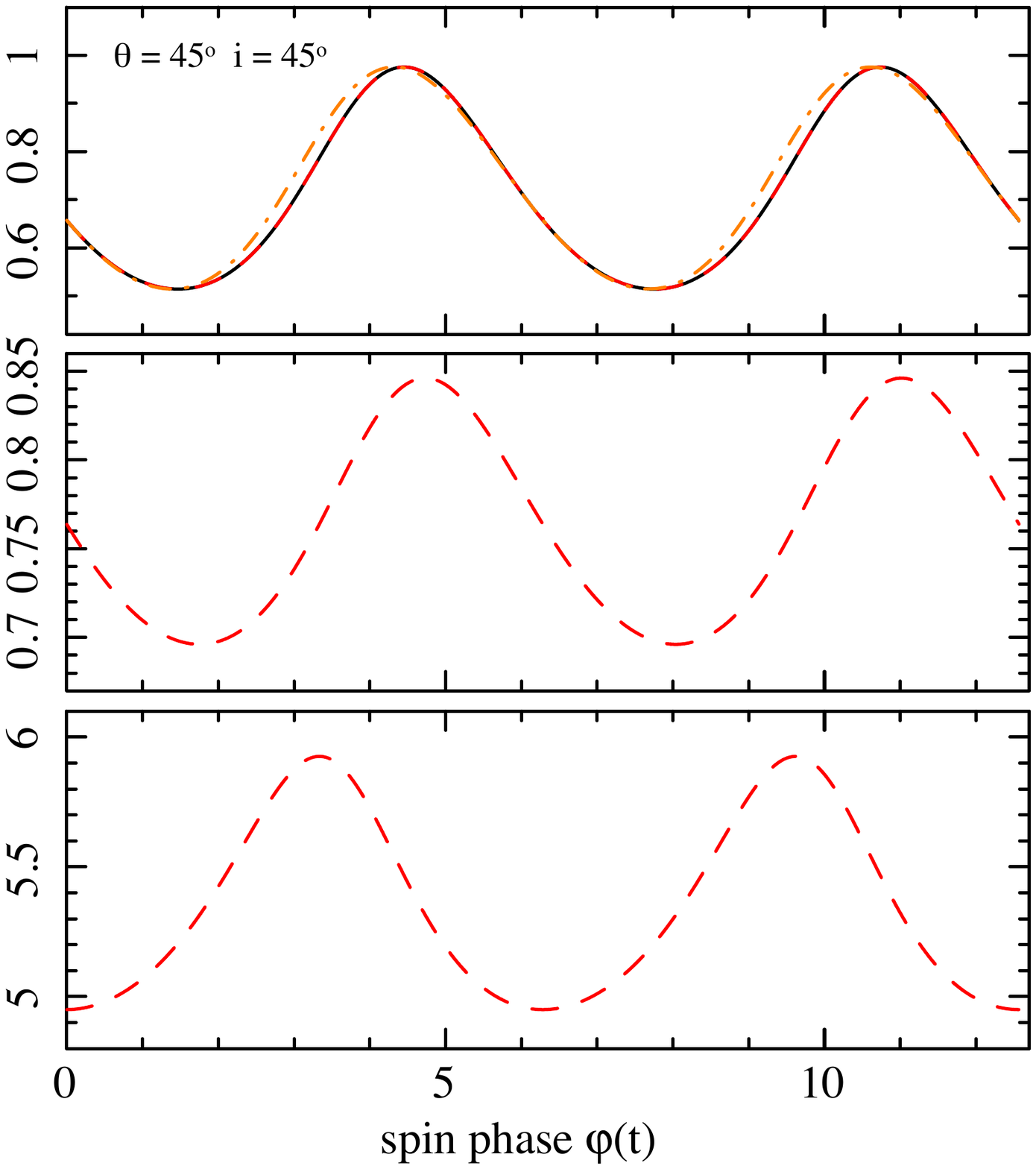}
}
\caption{Modulation from a hot spot on an NS as a function of rotational phase for different inclination angles and hot spot colatitude. Light travel time delays are included.  The continuous black lines represent the results from a numerical integration of the original equations; the dashed red lines are obtained from the polynomial approximate equations. The dashed-dotted orange line does not include light travel time delays.} 
\label{fig:Fig1_10}
\end{figure}

I calculate here the pulse profile generated by a point-like hot spot located on the surface of a NS, which emits like an isotropic blackbody.  Calculations of this type have been carried out extensively to 
model the periodic signals of accreting millisecond pulsars (see, e.g., Refs. \cite{Pechenick83,Poutanen06,leahy11,baubock15}) as well as the so-called burst oscillations during Type-I thermonuclear bursts in NS low-mass X-ray binaries (e.g., Refs. \cite{nath02,miller15}); some of these  calculations also include the angular size of the hot spot, the star oblateness, and the spacetime modifications induced by fast rotation. I use here a canonical NS mass of $1.4M_{\odot}$ and radius $R_{\rm NS} = 12$ km, together different inclination angles, $i$, and colatitudes, $\theta$ of the spot. The NS spin frequency is chosen to be $\nu_s = 600$ Hz. In Fig. \ref{fig:Fig1_10} I report the corresponding pulse profiles; as expected, the case with higher values
of $i$ and $\theta$ displays larger departures from a sinusoidal shape. In this type of applications the value of  $\alpha$ is always limited to $\leq \pi/ 2$, as no turning points are involved. Therefore my approximate equations retain very high accuracy as long as the NS radius is $ \ge 2.5r_s$, a range that encompasses a number of NS 
models for different equations of state, excluding only the upper end of the mass-radius branches. I conclude that my approximate equations can be usefully employed in calculations of the pulse profile of fast spinning NSs over a range of (but not all) models to be tested against the observation that the Neutron Star Interior Composition ExploreR (NICER), and other large-area X-ray missions of the future, such as Athena or LOFT (see Ref. \cite{watts16} and references therein).

\subsection{Applicability regions}
\label{sec:appreg}
\begin{figure}[ht]
\centering
\includegraphics[trim=0cm 2cm 3cm 8cm,scale=0.65]{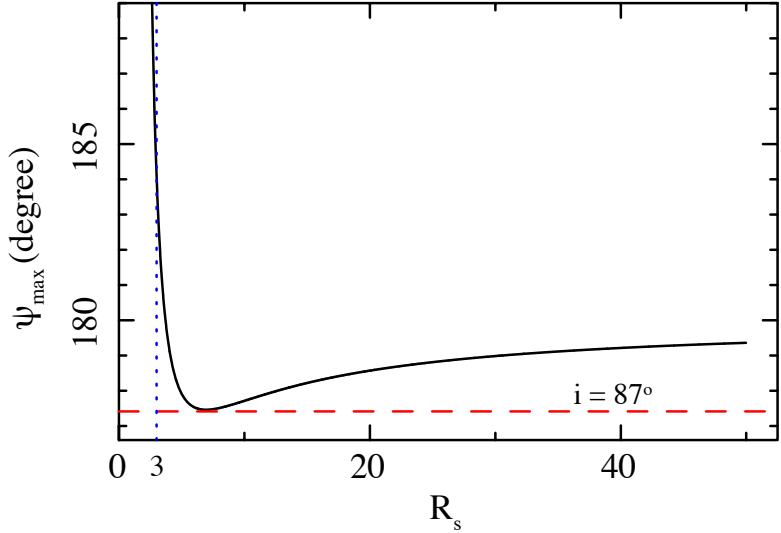} 
\caption{Largest bending angle $\psi_{max}$, vs. the emission radius (continuous black line). For inclination angles below $i=87^\circ$ (dashed red line) the approximate equations provide a high accuracy, since they are below the $\psi_{max}$-value. $R=3r_s$ (dotted blue line) separates the applicability region from the strong-field regime ($R<3r_s$).} 
\label{fig:Fig1_11}
\end{figure}

In Fig. \ref{fig:Fig1_11} I plot $\psi_{\rm max}$ as a function of the emission radius to investigate the applicability regions of the approximate equations. If I consider trajectories with turning points for radii $R<3r_s$, that is,{\it } smaller than the ISCO, then $\psi_{\rm max} \ge 180^\circ$ and a polynomial treatment is no longer accurate because of strong field effects (see also Fig. \ref{fig:Fig1_4}). I note that for $R\longrightarrow1.5r_s$, $\psi_{\rm max}$ my solution approaches asymptotically $270^\circ$. Instead, for $R\ge3r_s$, when the observer is located edge on (i.e., $i=90^\circ$), $\psi_{\rm max}=180^\circ$ is attained; otherwise, for slightly smaller but still extreme inclination angles, for example,{\it } $87^\circ$, photon trajectories always remain below the critical bending angle, which guarantees a high accuracy of my polynomial approximations. This argument is valid for all the emission radii $R\ge3r_s$, since for $R \longrightarrow  \infty$, $\psi_{\rm max}$ approaches $180^\circ$. 

\subsection{Concluding remarks}
\label{sec:conclusions}
I developed an analytical method to approximate the elliptic integrals that describe gravitational light bending and light travel time delays of photon geodesics in the Schwarzschild metric. Based on this, I derived for the first time an approximate polynomial equation also for the solid angle. I discussed the accuracy and range of applicability of the approximate Eqs. (\ref{AFLB}), (\ref{FATIDE}), and (\ref{AFSA}); adopting them can considerably speed up calculations related to a variety astrophysical problems, which normally require time-consuming numerical integrations. I also presented a few simple applications as examples. 

\chapter{Poynting-Robertson effect}
\epigraph{If I have seen further than others, it is by standing upon the shoulders of giants.}{Isaac Newton}

\lettrine{T}{his chapter} deals with the PR effect, that is an efficient mechanism to remove angular momentum and energy from small-sized test particles invested by a radiation field. The explanation of this effect was not very clear at the beginning because the concepts of GR theory was at dawn of full understanding. The description of the PR effect in Newtonian frame was given for the first time by Poynting in 1904 \cite{Poynting03}. Then in 1937 Robertson proposed the special relativistic treatment \cite{Robertson37} and only around 2009--2011 Bini \emph{et al.} \cite{Bini09,Bini11} extended this effect to general relativistic formalism. Therefore, from the initial classical formalism until its general relativistic description, it elapsed nearly one century. I show and compare the orbits in a flat spacetime with the curved spacetimes of Schwarzschild and Kerr. This permits to understand what is the influence of the PR effect when coupled with general relativistic effects. Such phenomenon can be seen as a dissipative system and normally it is relatively difficult to prove that they admits a Lagrangian formulation, especially in GR where matters complicate considerably due to the nonlinearity of the geometrical background. Nevertheless, I will show how it will be possible to describe the PR effect in terms of a Lagrangian and a dissipative Rayleigh potential through the introduction of an integrating factor \cite{Defalco2018}. In another work, I will then show how to extend the general relativistic model of Bini \emph{et al.} framed in a two-dimensional space in a three-dimensional space \cite{Defalco20183D}.

\section{Newtonian framework}
\label{sec:newframe}
Radiation pressure is exerted upon any surface exposed to electromagnetic radiation, interacting via absorption, reflection, or both. Bodies emitting radiation also experience this pressure. Generally, the forces generated by radiation pressure are too small to be detected under everyday circumstances, but they assume a primary role in several astrophysical contexts, e.g., cometary science. 

In 1619 Kepler, following the Newton's corpuscular theory of light, introduced the concept of radiation pressure to explain the form of comet's tails. In 1746 Euler, in the framework of wave theory of light, ascribed theoretically this pressure to solar radiation. In 1756 De Mairan attempted to estimate this pressure in laboratory experiments, but the disturbing action of the gases employed in the measurements, led him to confusing and contradictory results. Only in 1873 Maxwell predicted rigorously this phenomenon based on his electromagnetic theory and, independently of him, in 1876 Bartoli found the same based on thermodynamics arguments. In 1901 Lebedew proposed how to improve the experimental methods in the measurements of radiation pressure \cite{Lebedew02}. Following this line of thought, Nichols and Hull elaborated incredible experiments able to confirm the light pressure in good agreement with Maxwell's theory \cite{Nichols02}. Strong of the outcome of their results, they tried to extend them in the explanation of cometary theory with no great success, due to the lack in discriminating between several contributing influences \cite{Nichols03}.

\subsection{Discovery of the retarding force}
\begin{figure}[h]
\centering
\includegraphics[scale=0.7]{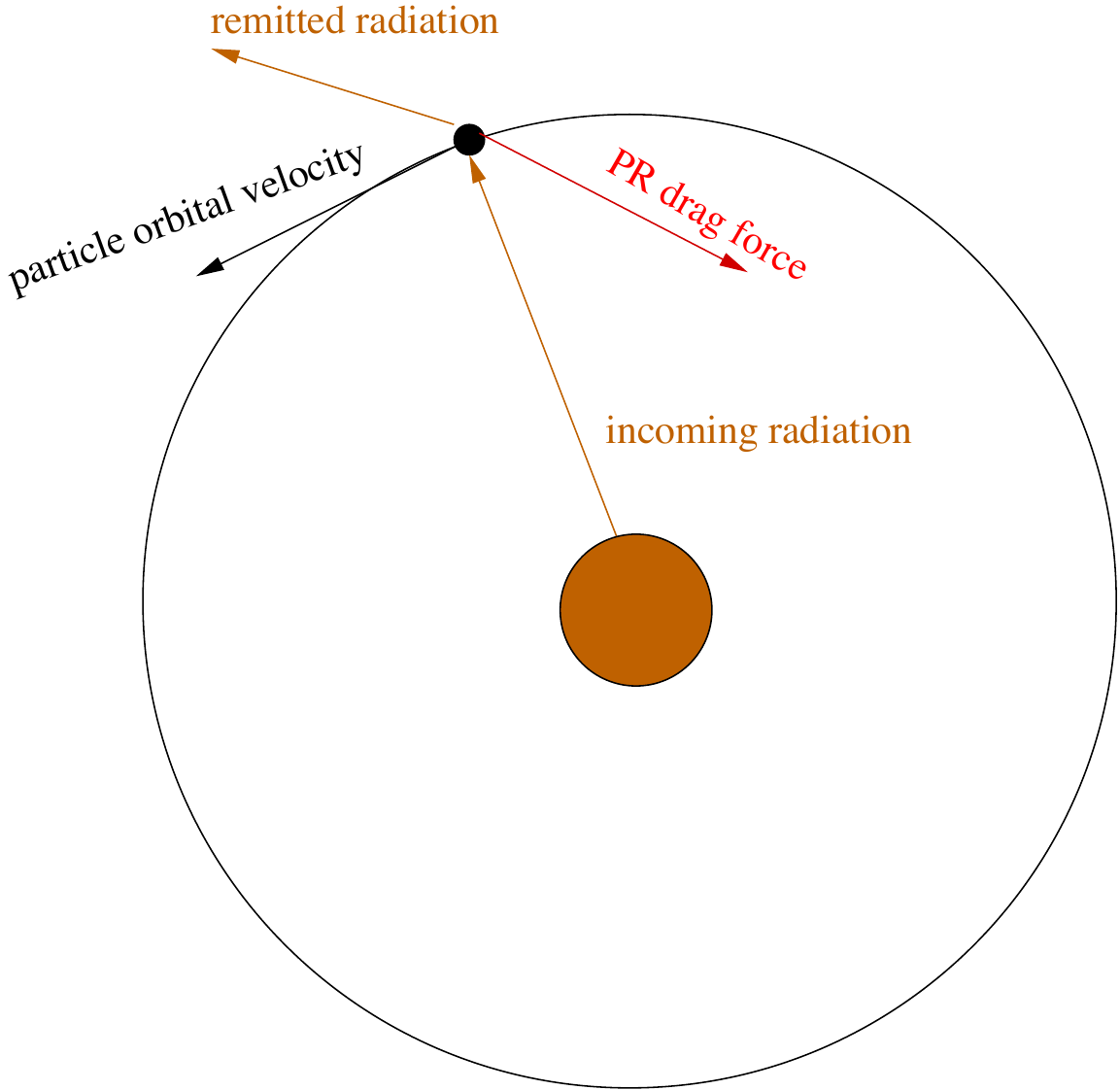}
\caption{Geometry of the PR effect, where a small-sized particle orbits around a radiating source. The particle absorbs and remits the incoming radiation, experiencing so a force opposite to its orbital motion. This radiation drag force removes angular momentum and energy from the particle, forcing it to spiral inward or outward depending on the radiation field intensity.} 
\label{fig:Fig3_0}
\end{figure}

Poynting studied more carefully the radiation pressure effects on the temperature of small bodies \cite{Poynting03}. He analysed the radiation pressure in different situations: radiation absorption and emission, comparison with the gravitational field, and in the action between two mutually radiating bodies. The innovative part of his work relies on the calculation of the radiation pressure against a moving surface with a general velocity in a given direction. He considers the surface invested by the radiation as at rest, meaning that he choices a reference frame co-moving with the surface. In this situation, the radiation crowds up into less space or spreads over more, exerting a tangential pressure opposite to the particle motion. This aberration effect generates a \emph{retarding force}, that removes angular momentum and energy from the body on which it is applied (see Fig. \ref{fig:Fig3_0}). 
The equations describing the orbit of a small absorbing particle of size $\alpha$ and density $\rho$ moving in a stationary medium around the Sun are \cite{Poynting03}
\begin{eqnarray}
\ddot{r}-r\dot{\theta}^2&=&-\frac{\mu}{r^2}-\frac{A}{r^2}\dot{r},\label{EQ1} \\ 
r^2\dot{\theta}&=&C-A\theta, \label{EQ2} 
\end{eqnarray}
where $\mu$ is a constant including the gravitational and radiation force, $\frac{A}{r^2}\dot{r}$ is the retardation force with $A=Sd^2/3c^2\rho \alpha$, $S$ the solar constant at Earth's distance $d$, and $C$ the integration constant. The discovery of this new effect permitted to give a more suitable explanation to some puzzling phenomena, like: small particles drawn into the Sun, motion of meteoric dust, falling dust particles circulating around the Earth, which receives the remitted heat (or radiation) from the planet or directly from the Sun. 

In 1905--1906 Plummer was the first, who applied Poynting's theory for the explanation of comets motion \cite{Plummer05,Plummer06}. A comet is supposed to be spherically symmetric and a perfect absorber of radiation. 
At that time in comet theory there were two compelling issues to understand: the discrepancies between observations and theory in the orbital motions of comets; the constituents of the comet structure were supposed to be uniform and permanent, but the solar radiation pressure might have caused to coalesce them producing a progressive enhancement in size. Exploiting the Poynting's equations, he found a partial agreement with the observed data of the Encke's comet, which at that time received the greatest amount of study for its puzzling behavior. Such a theory implied a great increase of the period, not reputable to the large error in the adopted distance. Therefore although this theory appeared to be attractive, it should have ben improved to explain more precisely such kinds of phenomena.

\section{Special relativistic treatment}
\label{sec:srframe}
This subject was reopened again in 1913 by Larmor \cite{Larmor1917}, who gave an alternative approximate treatment of the retarding force exerted on a body moving with uniform velocity $v$, arising from its own radiation, based on classical electromagnetic theory arguments. A light ray transmitting energy $E$ per unit time pushes backward the body with a force to the first order in $v$ given by $3vA/c^2$, configuring three times Poynting's reported value (see Eqs. (\ref{EQ1}) and (\ref{EQ2})). 

The retardation force experienced by a moving particle due to its own radiation revealed to be contradictory, since it was colliding with the \emph{electromagnetic theory}. Indeed, it would have implied the detection of absolute velocities, in strong disagreement with the \emph{principles of special relativity}. Since special relativity and electromagnetic theory get along, the discordance of the retardation force with the former arose heavy contradictions \cite{Obs20}.

In 1918 Page undertook a detailed examination of such issues in the framework of Maxwell theory, considering the interaction between the radiation pressure and the electrons inside the matter \cite{Page18a,Page18b}. The resulting force $F$ given by the radiation pressure upon the body surface can be decomposed as the sum of two forces: $F_1$, the stress exerted on the body surface, and $F_2$, the decrease of electromagnetic momentum rate within the enclosing envelope. If I assume that the matter is microscopically formed by electron-oscillators, these two forces can be also interpreted as: $F_1$, reaction between incoming radiation and radiation emitted by each oscillator, and $F_2$, force exerted on each oscillator by the radiation from neighboring oscillators. In this frame the result of Larmor is based on the tacit assumption that the periodic motion of the electron is undamped, or $F_2=0$ (leading to a contradiction). It was shown rigorously that classical electrodynamics provided no retardation on a moving and radiating mass, meaning that $F_1=-F_2$, and it is completely in accordance with the principle of relativity. Therefore, according to Page when damping is taken into consideration this inconsistency disappears, and the retarding force vanishes. 

In 1918 Larmor reconsidered his position, claiming that an isolated body cooling in the space would not change its velocity, since the retardation force due to the back thrust of radiation pressure is compensated by increase of velocity due to momentum conservation with diminishing mass \cite{Larmor17}. In addition, a particle orbiting around the Sun is affected by radiation, which restores the energy lost by the particle radiation, establishing so again the retarding force $-Rv/c^2$. However, for Larmor the particle was subject to another drag of the same amount due to the astronomical aberration of light, which doubled the result. Closer examination showed that this was still in contradiction with the theory of relativity and it affected only radial component of the drag force without producing any decrease of the angular momentum.    

\subsection{Solution of the paradox}
In 1937 Robertson clarified all the paradoxical issues and formulated correctly the problem in the special relativity theory \cite{Robertson37}. He considered the motion of a small spherical particle of mass $m$ and 4-velocity $u^\mu$ in a beam of radiation generated by a central object in the Minkowsky metric $ds^2=dt^2-(dx^2+dy^2+dz^2)$. The particle absorbs all the energy falling upon it, and re-emits this energy at the same rate isotropically respect to a reference system $\Sigma$, where it is instantaneously at rest. The radiation field is described by the null vector $l^\mu$, whose components are subjected to the normalization conditions $l^0=1$ and $g_{\mu\nu}l^\mu l^\nu=0$ in $S$, the inertial frame located in the central object. The equations of particle motion into the original inertial system $S$ are \cite{Robertson37}
\begin{equation} \label{EqR} 
\frac{d (mu^\mu)}{ds}=\frac{fw}{c}(l^\mu-wu^\mu),
\end{equation}    
where $w=l_\mu u^\mu$ is a scalar, $f=\sigma d$ is the radiation force acting on the particle in the reference system $S$, $d$ the radiation energy density and $\sigma$ the cross section of the particle invested by the radiation. 

In the case where the relativistic effects can be neglected, Eqs. (\ref{EqR}) reduce to a simpler form, directly comparable with the approximate equations derived by Poynting and Larmor. Defined $v^\alpha=\frac{dx^\alpha}{dt}$, $n^\alpha=\frac{l^\alpha}{c}$, $u^0=1$, $u^\alpha=v^\alpha$, $v_n=v^\alpha n_\alpha$ (the component of the velocity $v^\alpha$ in the direction $n^\alpha$), $w=1-v_n/c$ and considering $v^\alpha/c\ll 1$, the equations at the first order in $(v^\alpha/c)$ are
\begin{equation} \label{NEQ} 
m\frac{dv^\alpha}{dt}=f\left(1-\frac{v_n}{c}\right)n^\alpha-f\frac{v^\alpha}{c}.
\end{equation}
The first term on the right may be interpreted as the radiation pressure in the direction of the incoming beam, but weakened by the Doppler factor $1-v_n/c$; while the second represents the tangential drag because the particle is absorbing energy at the rate $cf$, and reradiating it isotropically at the rate $(cf/c^2)v^\alpha$, but for the conservation of total momentum the particle loses momentum at the rate $fv^\alpha/c$, that is exactly the retarding force. The total drag in the direction of the beam $n^\alpha$ is in magnitude $2fv_n/c$, while that in direction transverse to the beam is $fv'/c$, where $v'$ is the component of $v^\alpha$ in the transverse direction. 

Considering the same case above, but for a particle orbiting around the Sun, I have that $n^\alpha$ is the unit vector radius from the Sun, $d=Sb^2/cr^2$ the energy density, and $A=\frac{\sigma Sb^2}{mc^2}=\frac{3Sb^2}{4c^2\rho\alpha}$ (using the same nomenclature reported for Eqs. (\ref{EQ1}), (\ref{EQ2})). Adding to the Eq. (\ref{NEQ}) the the solar gravitational force $GMm/r^2$ in the direction $n^\alpha$, the equations of motion, in polar coordinates $(r,\theta)$ in the plane of the orbit with pole at the Sun, are
\begin{eqnarray}
\ddot{r}-r\dot{\theta}^2&=&-\frac{\mu}{r^2}-\frac{2A\dot{r}}{r^2},\\ 
\frac{1}{r}\frac{d}{dt}(r^2\dot{\theta})&=&-\frac{A\dot{\theta}}{r}, 
\end{eqnarray}
where the dot indicates the differentiation respect to $t$, $A$ including the effect of the gravitational and radiation field. These equations of motion have a very similar form to those derived by Poynting, showing thus the correctness of Robertson formalism. In this framework, he proposed also a better explanation of Encke's comet motion and possible general relativistic corrections \cite{Robertson37}.

\begin{figure}[h]
\centering
\includegraphics[trim=2cm 1cm 0cm 3cm, scale=0.5]{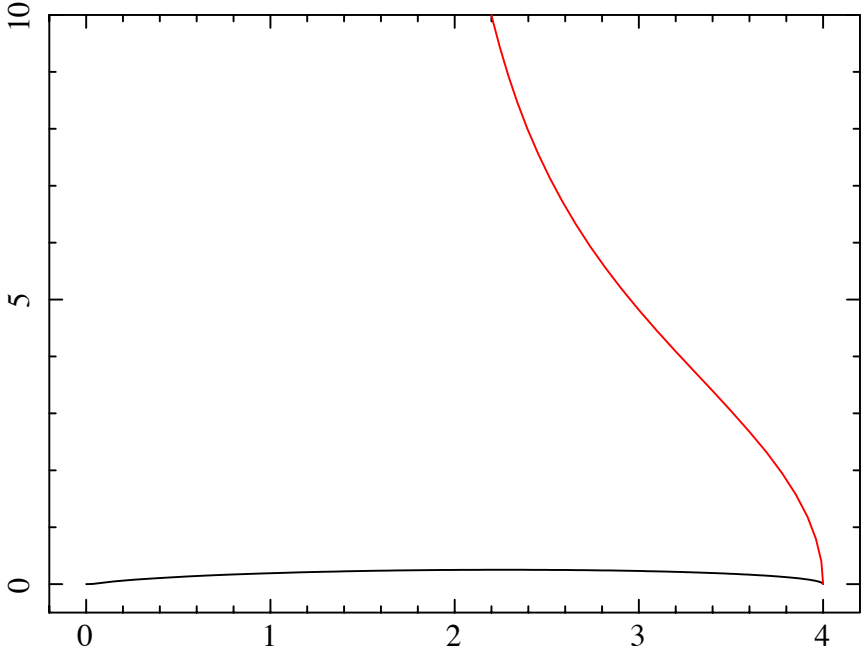}
\caption{Orbits in a flat spacetime expressed in unity of $b$. The black line is given by $(A,T)=(0.6,0.5)$ and $(r_0,\varphi_0,\alpha_0,\nu_0)=(4,0,0,0.01)$, and the red line by $(A,T)=(-0.2,0.6)$ and $(r_0,\varphi_0,\alpha_0,\nu_0)=(4,0,0,1)$.} 
\label{fig:Fig3_1}
\end{figure}

In Fig. \ref{fig:Fig3_1}, the orbits under the PR effect in a flat spacetime are of two kinds: falling toward the central object (gravitational field and PR drag force are stronger than radiation pressure), or going to infinity (radiation field is stronger than the other two forces).    

\section{General relativistic model}
\label{sec:GRmodel}
After the special relativistic formalism has been proposed by Robertson, there were several attempts to apply or slight modify this model for better understanding different astrophysical phenomena. 
\begin{itemize} 
\item In the solar system, small-sized particles (like dust grain) are very rapidly pulled toward the Sun, due to the action of PR effect. To describe such phenomenon, Wyatt and Whipple slightly extended the Robertson's equations for particles moving on slowly elliptical orbits \cite{Wyatt50}. They estimated the collapse times for meteor orbits in terms of initial semi-major axis and eccentricity, and particle radius and density.  
\item Guess considered the radiation source as a spherical body of finite extension, and investigated the influence of PR effect on test particles motion \cite{Guess62}. The equations of motion at the first order in $(v/c)$ give the value of the angular and radial drag force, being $8/3$ and $4/3$ with respect to the point-source case, respectively. Far away from the radiation source, the equations reduce thus to those reported by Robertson.     
\item Burn \emph{et al.} investigated radiation field effects by adding scattering processes to absorption and re-emission of radiation \cite{Burns79}. They derived a new accurate expression for the radiation pressure and the PR drag force. The equations of motion for particles moving around the Sun are expressed in terms of their size, in order to reproduce the proprieties of the interplanetary dust.
\item Abramowicz \emph{et al.} provided a complete study, without any approximation, of the radial motion of a test particle in GR under the influence of a gravitational field, radiation pressure, and PR drag force \cite{Abramowicz90}. The radiation field derives from a spherical star emitting it isotropically from each point of its surface and the related radiation force is assumed to be independent from the radiation frequency. The authors approach the dynamics qualitatively, classifying the possible scenarios without solving the equations analytically or numerically. The problem is completely described by four parameters: star luminosity and radius, initial conditions on position and velocity of the particle. Seven possible types of trajectories were classified according to luminosity ranges (see Fig. 2 in Ref. \cite{Abramowicz90}).
\item Miller and Lamb demonstrated that the radiation forces can be more influent of the general relativistic gravitational forces in altering the motion of accreting matter onto a slowly rotating NS, if the source luminosity is greater than $\sim1\%$ of Eddington limit \cite{Miller93}. In such mechanism, the radiation drag plays a fundamental role, inducing the accreting matter to lose angular momentum and spiral inward. These theoretical results have important observational consequences in X-ray spectra, time variability, and spin evolution of NSs with weak magnetic fields, and testing general relativistic effects. 
\item In a following paper, Lamb and Miller focused their attention on the concept of critical luminosity, which balances the outward radiation force with the inward gravitational pull \cite{Lamb95}. They derived in GR the expressions for the radiation force in the case of particles with arbitrary cross sections and analysed the radiation field produced by radiating matter orbiting around slowly rotating gravitating mass. 
\item Miller and Lamb extended their previous analysis to nonuniform emission, slow rotation of the gravitating mass and radiation source (with their rotation axes co-aligned) \cite{Miller96}. As previously noted, the relativistic effects increase the radiation drag force and enhance also the fraction of angular momentum and energy that can be transferred from the accretion flow to the radiation field. 
\item Srikanth proposed a pedagogical and physical interpretation of the PR effect \cite{Srikanth99}. If the absorbing and re-emission processes are written in terms of two distinct parameters, one discovers that the absorption is determining for producing the PR drag force; instead the re-emission is not important (even in the heliocentric reference frame) and responsible for any kind of PR drag force, as it was initially argued by Robertson. 
\item Kimura \emph{et al.} investigated radiation pressure and PR effect on the motion of fluffy dust \cite{Kimura02}. They rewrote the standard equations takin into account not only a radial radiation field, but also in a general direction. Beside the PR effect, there are other effects to be taken into account to estimate the motion of particles, i.e.: non-sphericity of the particles, rotation rate and axis of particles, interaction between electric charges and solar magnetic field, and material composition of particles.   
\end{itemize}

\subsection{First relativistic geometrical model}
After seventy years of Robertson's model, there was any published article, which extended completely the problem in GR. In 2009--2011, Bini, Jentzen, and Stella proposed for the first time a full general relativistic geometrical model set in the framework of stationary and axisymmetric spacetimes (made explicit for Schwarzschild and Kerr metrics) \cite{Bini09,Bini11}. Such model, albeit is simple in several features, can anyway provide interesting applications in the physics of accretion phenomena around BHs and NSs. 

\begin{figure}[h]
\centering
\includegraphics[trim=0cm 0cm 0cm 0cm, scale=0.5]{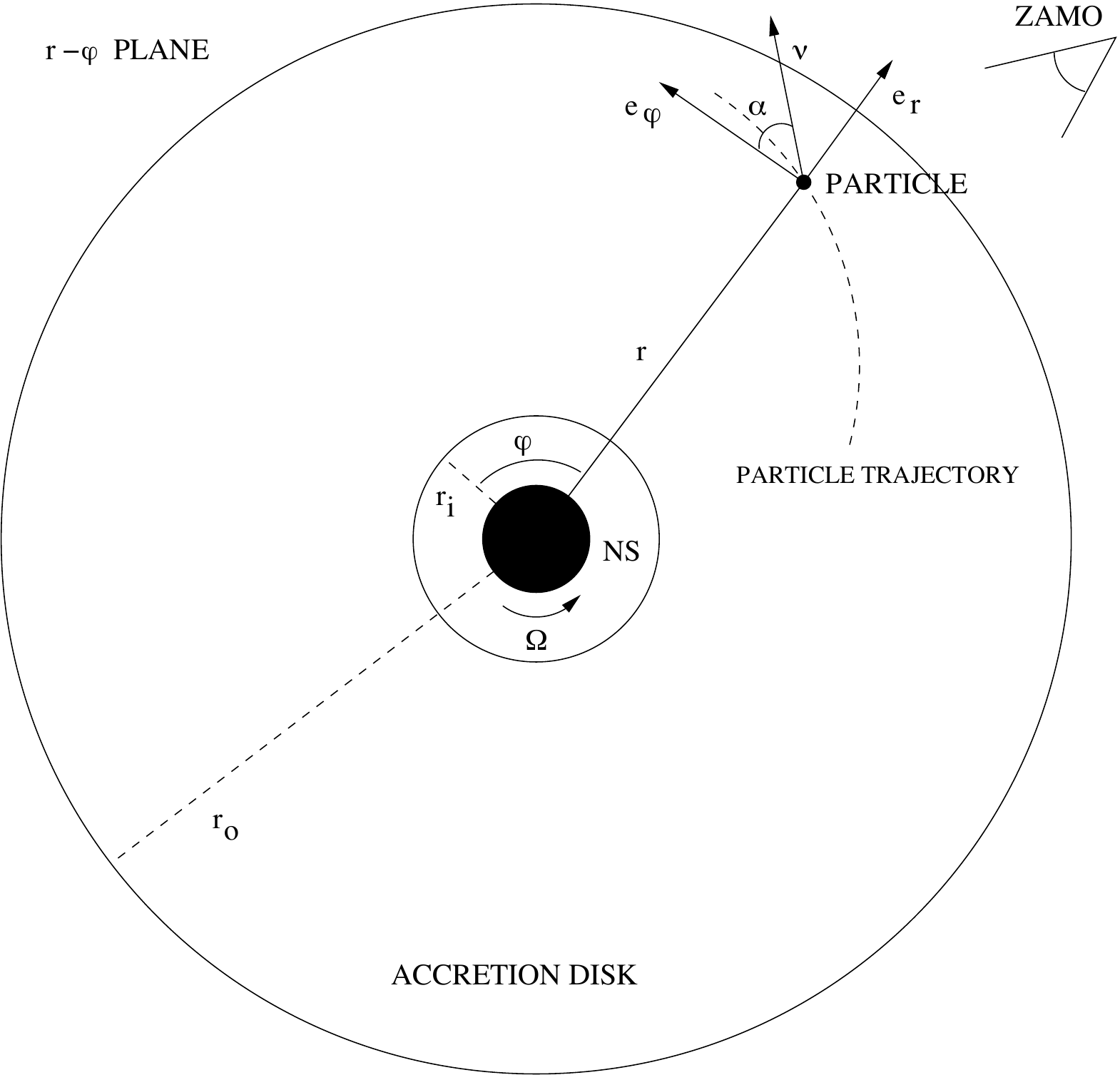}
\caption{Geometry of the system set in the $r-\varphi$ plane, composed by: a compact object rotating with spin $\Omega$, an accretion disk with inner radius, $r_{{\rm i}}$, and outer radius, $r_{{\rm o}}$, and a ZAMO located at infinity and corotating with the central object. An orthonormal basis in the ZAMO coordinates is defined by the vectors $(e_r,e_\varphi)$, where $e_r$ has the direction connecting the central object with a particle inside the disk and $e_\varphi$ is orthogonal to $e_r$ pointing in the rotation direction. A particle in the disk has coordinates $(r,\varphi)$ and velocity $\nu$, that forms an angle $\alpha$ with the unit vector $e_\varphi$.} 
\label{fig:Fig3_2}
\end{figure}

Considering a stationary axisymmetric spacetime around a rotating compact object of mass $M$ and spin $a$, the Kerr metric, written in Boyer-Lindquist coordinates $(t,r,\varphi,\theta)$ and set in the equatorial plane $\theta=\pi/2$, reads as \cite{Kerr1963}
\begin{equation}
ds^2=-\left(1-\frac{2M}{r}\right)dt^2+\frac{r^2}{\Delta}dr^2-\frac{4aM}{r}dt\ d\varphi +\frac{\rho}{r}d\varphi^2,
\end{equation}
where $G=c=1$, $\Delta=r^2+a^2-2Mr$ and $\rho=r^3+a^2r+2a^2M$. A zero angular momentum observer (ZAMO) is located at infinity, falls into the compact object and co-rotate with it due to the frame-dragging effect. The radiation field is considered as a coherent flux of photons traveling along null geodesics of the background metric in some preferred direction given by the photon impact parameter $b$. The equations of motion read as \cite{Bini11}
\begin{eqnarray} 
\frac{d\nu}{dt}&=&-(1-\nu^2)\sin\alpha\sqrt{\frac{r\Delta}{\rho}}P+Q\ [\cos(\alpha-\beta)-\nu],\label{Eqsmot1}\\
\frac{d\alpha}{dt}&=&\frac{\cos\alpha}{\nu}\sqrt{\frac{\rho}{r\Delta}}\left[P-\frac{\nu^2\sqrt{\Delta}(r^3-a^2M)}{r^2\rho}\right]+\frac{Q\ \sin(\beta-\alpha)}{\nu},\label{Eqsmot2}\\
\frac{dr}{dt}&=&\frac{\Delta\ \nu\sin\alpha}{\sqrt{r\ \rho}},\label{Eqsmot3}\\
\frac{d\varphi}{dt}&=&\frac{r\ \sqrt{\Delta}\ \nu \cos\alpha+2aM}{\rho} \label{Eqsmot4},
\end{eqnarray}
where
\begin{eqnarray}
P&=&\frac{M}{r^2\rho\sqrt{\Delta}}\left[(r^2+a^2)^2-4a^2r-2a\ \nu\cos\alpha\sqrt{\Delta}\ (3r^2+a^2)\right],\\
Q&=&\frac{A}{r}\sqrt{\frac{1-\nu^2}{\Delta}}\left(1-\frac{2aMb}{\rho}\right)\left[\frac{1-\nu\cos(\alpha-\beta)}{|\sin\beta|}\right],\\
\beta&=&\arccos\left(\frac{b\ r\sqrt{\Delta}}{\rho-2Mab}\right), 
\end{eqnarray}
$A=(L_{\rm{X}}/L_{\rm{Edd}})\ M$ is the normalized luminosity parameter ranging in the interval $[0,1]$ with $L_{\rm{X}}$ the emitted luminosity at infinity, $\nu$ the magnitude of the particle spatial velocity, $\alpha$ the polar angle of the particle velocity measured clockwise from the positive $\varphi$ direction in the $r-\varphi$ tangent plane, and $\beta$ the photon angle formed with the radial unit vector in the observer frame (see Fig. \ref{fig:Fig3_2}). Equations (\ref{Eqsmot1})--(\ref{Eqsmot4}) are independent from $\varphi$, expressing their rotationally symmetric propriety.  

\section{PR orbits in GR}
In order to understand what is the influence of the PR effect in the motion of particles, it is interesting to explore the orbits in the general relativistic frame. The system of four coupled first order differential equations (\ref{Eqsmot}) for $(\nu,\alpha,r,\varphi)$ admits a critical solution which corresponds to a circular orbit of constant radius $r=r_{\rm (crit)}$, constant speed $\nu=\nu_{\rm (crit)}$, and constant angles $\beta=\beta_{\rm (crit)},\alpha=\alpha_{\rm (crit)}$, where $\beta$ formally is the angle formed by the vector direction of the relative photon velocity with respect to the ZAMOs (i.e., $\nu_{\rm pho}=\sin\beta\ e_{\rm r}+\cos\beta\ e_{\rm \varphi}$). This critical solution is given by the following conditions $d\nu/d\tau=d\alpha/d\tau=0$ and physically corresponds in balancing the gravitational attraction and the radiation force. The equations of the critical region are given by \cite{Bini11}
\begin{eqnarray} 
\cos\beta&=&\frac{b\ r\ \sqrt{\Delta}}{\rho-2aMb}, \label{CR1}\\
\frac{A}{\sqrt{\Delta}}S&=&\frac{M[(r^2+a^2)^2-4a^2Mr]-\nu\cos\alpha\sqrt{\Delta}\ \Sigma}{\sqrt{r\rho^3}}, \label{CR2}
\end{eqnarray}
where
\begin{eqnarray} 
\Sigma&=&2aM(3r^2+a^2)+\nu\cos\alpha\sqrt{\Delta}(r^3-Ma^2), \label{par1} \\
S&=&\sign(\sin\beta)|\sin\beta|^3\left(1-\frac{raMb}{\rho}\right). \label{par2}
\end{eqnarray}
Generally, Eqs. (\ref{CR1}), (\ref{CR2}) can not be solved explicitly for $r$. In the case of radial photon trajectories (i.e., $b=0$), Eq. (\ref{CR1}) becomes $\beta=\pm\pi/2$, while Eq. (\ref{CR2}) reduces to
\begin{equation} 
\frac{A}{M}=\sqrt{\frac{\Delta}{r\rho^3}}[(r^2+a^2)^2-4a^2Mr].
\end{equation}
The critical region is an attractor for some orbits, as it can be seen in the next subsections, because depending on the initial conditions and the value of the photon impact parameters, the orbits can end up at infinity or on this region. The orbits can reach the boarder of the compact object only when $A/M\approx0$, meaning that the radiation field is almost switched off and the orbits are dominated uniquely by the general relativistic effects. Inversely, in the case $A/M\rightarrow1$ the critical radius moves at infinity and of of course for $A/M>1$ there is not critical radius, because the radiation field is so strong that pushes everything at infinity. It will turn out that $A/M<1$ is a necessary condition for the existence of this critical region. In the next subsections I will investigate the orbits in different spacetimes: flat, Schwarzschild, and Kerr.

\subsection{Flat spacetime}
The orbits reduce to those reported in Fig. \ref{fig:Fig3_1}. In the case of purely circular motion (i.e., $\alpha=\pi/2$), it is possible to calculate the intensity of the drag force
\begin{equation} \label{df}
\frac{d\nu}{d\tau}=\frac{A}{r\sqrt{r^2-b^2}}\left(\frac{b}{r}-\nu\right)\left(1-\frac{\nu b}{r}\right)\equiv F_{\rm (drag)}.
\end{equation}
The magnitude of the drag force $F_{\rm (drag)}$ depends strongly on the value of the photon impact parameter $b$, the velocity of the particle $\nu$, and the intensity of the luminosity $A$. This is what it is expected if it is compared with the classical case, the only big difference resides in the new dependence from $b$, that is however connected with the modeling of the radiation field. I note that for $b=0$, Eq. (\ref{df}) describing the strength of the drag force is
\begin{equation} \label{dfb0}
F_{\rm (drag)}\equiv-\frac{A\nu}{r^2}.
\end{equation}
For non relativistic Keplerian speed $\nu=\sqrt{\frac{M}{r}}\ll1$, Eq. (\ref{dfb0}) becomes
\begin{equation} \label{dfb1}
F_{\rm (drag)}\equiv-\frac{A}{r^2}\sqrt{\frac{M}{r}}.
\end{equation}
The Newtonian gravitational force per unit mass is $F_{\rm g}=M/r^2$, that only grows like the inverse square of the distance. Therefore as the particle approaches to the central mass, the drag force is dominant compared to the gravitational free fall behavior of the particle initially in circular motion. Similarly the radial radiation pressure force per unit mass under these conditions is just $A/r^2$ and the ratio of the drag force to the radial pressure force is just $\nu\ll1$, namely very small. This is the approach adopted by Robertson in describing the radiation in Special Relativity.

\subsection{Schwarzschild spacetime}
\label{ss}
In the Schwarzschild spacetime, once it is determined the critical region, I consider two different cases.
\begin{itemize}
\item \emph{Data outside the critical radius:} the orbits can end up on the critical radius or at infinity (see Fig. \ref{fig:Fig3_3}). Especially one can see that orbit 2 is so deflected because it tends to escape at infinity, but the PR effect induces it to come back toward the central object. A particle can escape to infinity if it has enough initial velocity or
the photon angular momentum is high. Regarding the last point, it is important to note that radial photon trajectories act on the particle faster than the others, therefore in non radial case the particle has more time to escape and the drag force effect has no enough time to influence the motion removing angular momentum. 
\begin{figure}[p]
\centering
\includegraphics[trim=1cm 1cm 0cm 3cm, scale=0.5]{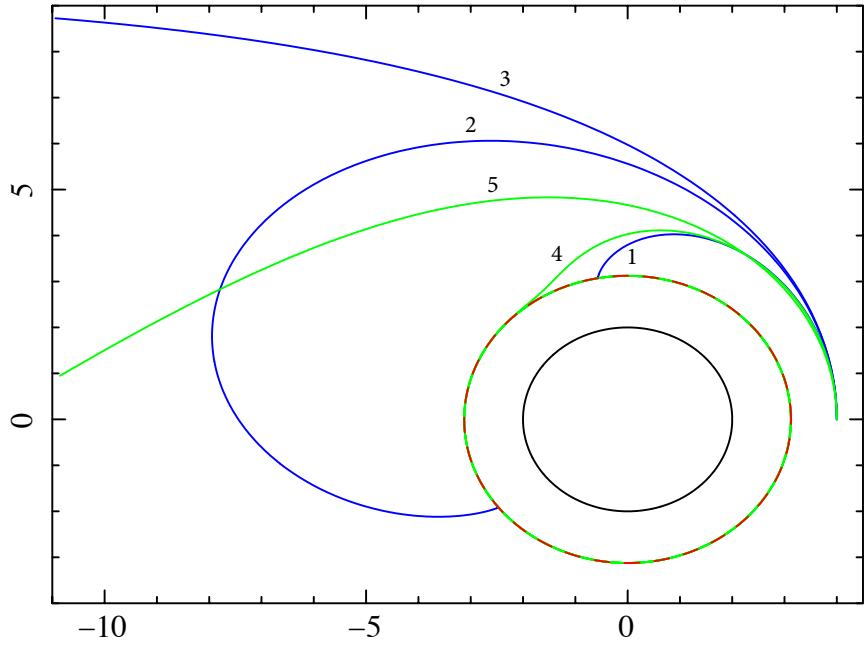}
\caption{The orbits of a particle in the Schwarzschild spacetime with $M=1$ and $A/M=0.6$. The continuous black line is the Schwarzschild radius at $r_{\rm S}=2M$, the dashed red line is the critical orbit with $r_{\rm (crit)}=3.125M$. All the particles have the same initial conditions $(r_0,\varphi_0,\alpha_0)=(4M,0,0)$. The blue lines are for photon angular momentum$b=0$, instead the green lines for photon angular momentum $b\neq0$. Initial conditions have: (1) $\nu_0=0.5$, (2) $\nu_0=0.71$, (3) $\nu_0=0.8$, (4) $(\nu_0,b)=(0.5,0.2)$, and (5) $(\nu_0,b)=(0.71,5.2)$.} 
\label{fig:Fig3_3}
\end{figure}
\item \emph{Data inside the critical radius:} the orbits, even if they start inside the critical radius, cannot end up on the central object, because the radiation field repeal them (see Fig. \ref{fig:Fig3_4}). As in the previous case, they can reach the critical region or escape at infinity. The last case occurs when the initial velocity is high enough or the photon angular momentum is great.
\begin{figure}[p]
\centering
\includegraphics[trim=1cm 1cm 0cm 3cm, scale=0.5]{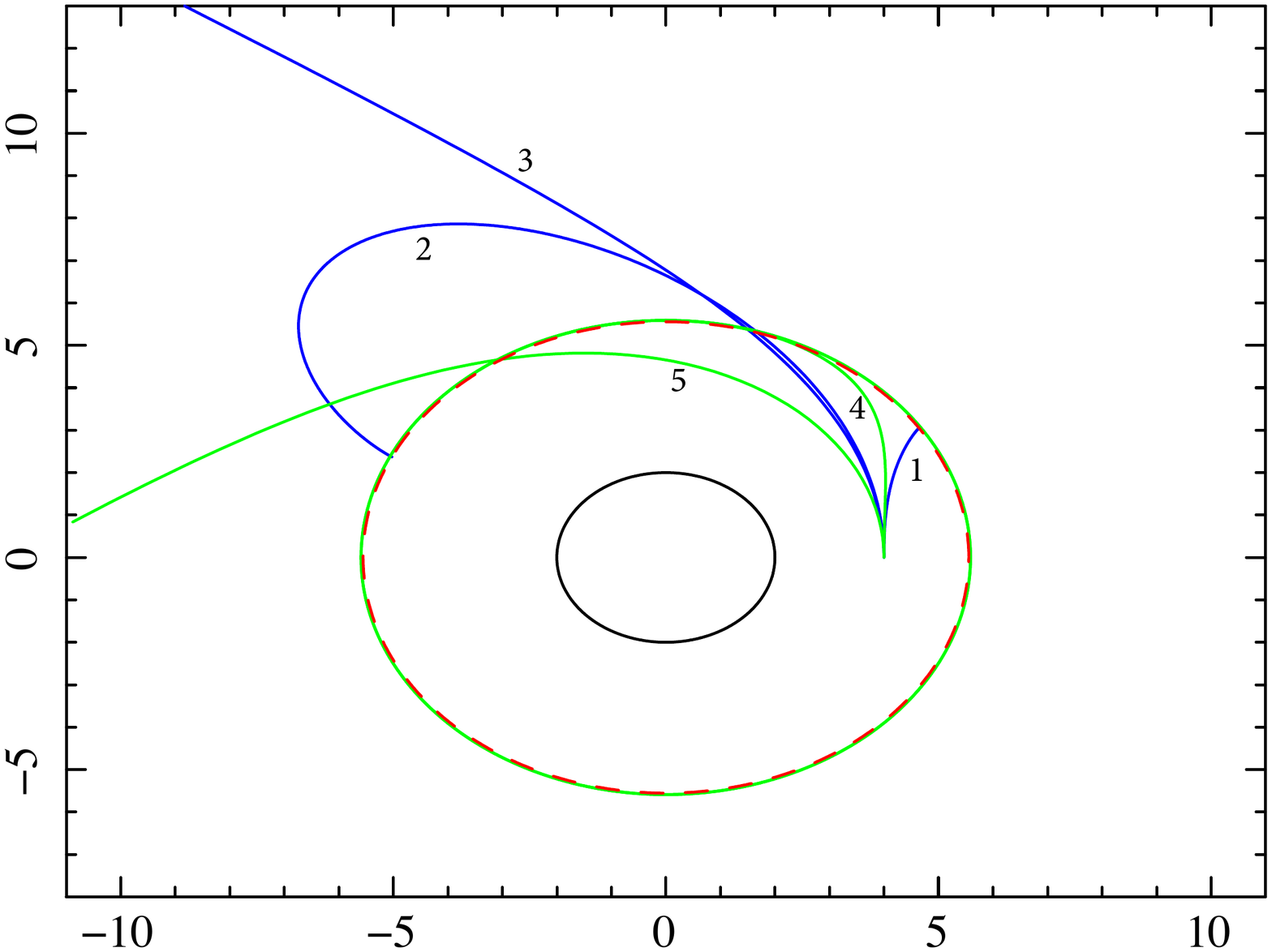}
\caption{The orbits of a particle in the Schwarzschild spacetime with $M=1$ and $A/M=0.8$. The continuous black line is the Schwarzschild radius at $r_{\rm S}=2M$, the dashed red line is the critical orbit with $r_{\rm (crit)}=5.5M$. All the particles have the same initial conditions $(r_0,\varphi_0,\alpha_0)=(4M,0,0)$. The blue lines are for photon angular momentum$b=0$, instead the green lines for photon angular momentum $b\neq0$. Initial conditions have: (1) $\nu_0=0.5$, (2) $\nu_0=0.64$, (3) $\nu_0=0.9$, (4) $(\nu_0,b)=(0.2,0.2)$, and (5) $(\nu_0,b)=(0.64,5.2)$.} 
\label{fig:Fig3_4}
\end{figure}
\end{itemize}
As a special case of data outside of the critical radius, there are the particles spiral slowly towards the compact object, as reported in Fig. \ref{fig:Fig3_5}. In this case $A/M\approx 0$, so the critical radius coincides approximatively well with the Schwarzschild radius. These kinds of orbits are very interesting, because they can represent the trajectories described by matter in an accretion disk. 
\begin{figure}[p]
\centering
\hbox{
\includegraphics[trim=1cm 1cm 0cm 3cm, scale=0.25]{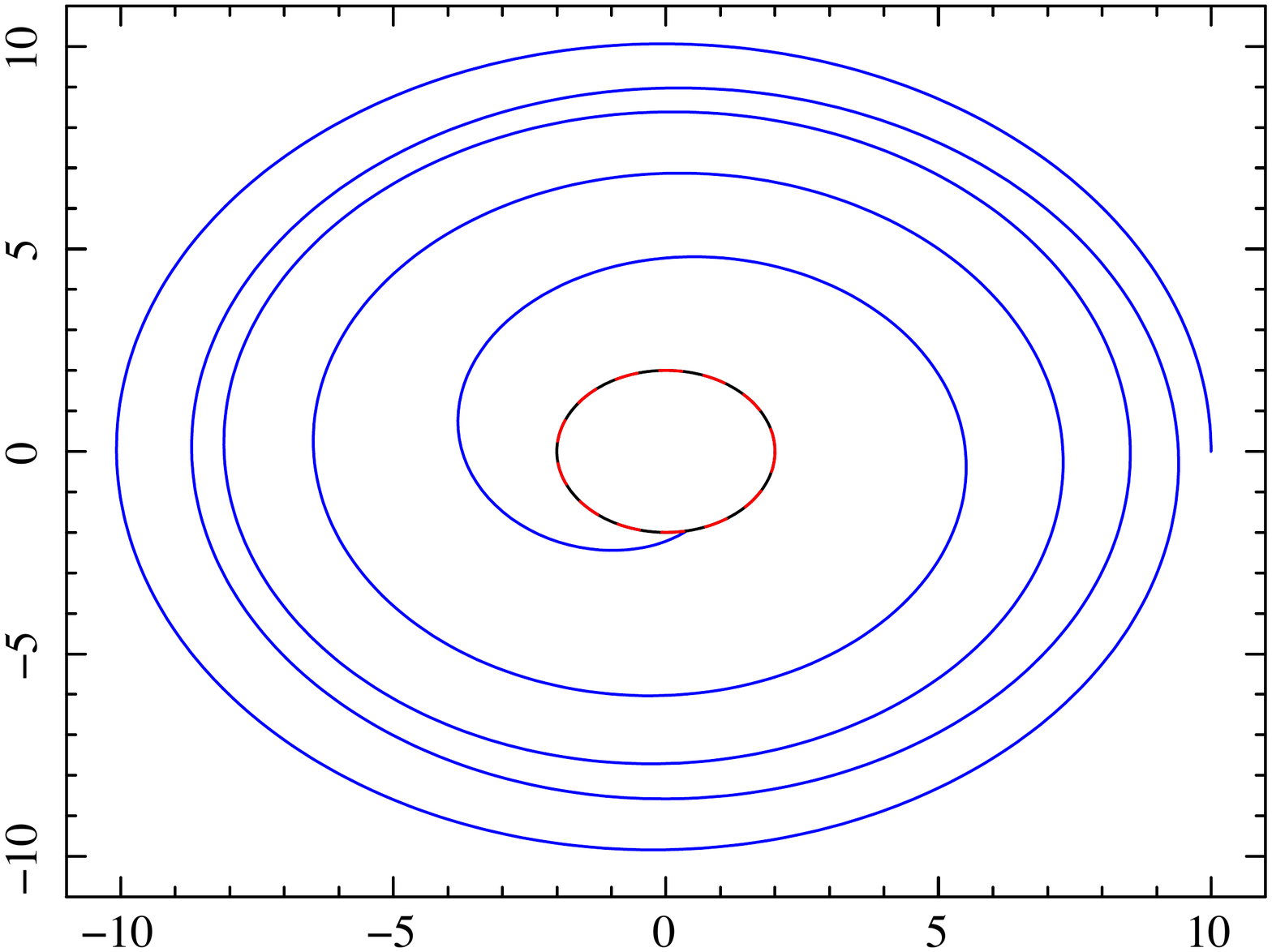}
\hspace{-1cm}
\includegraphics[trim=1cm 1cm 0cm 3cm, scale=0.25]{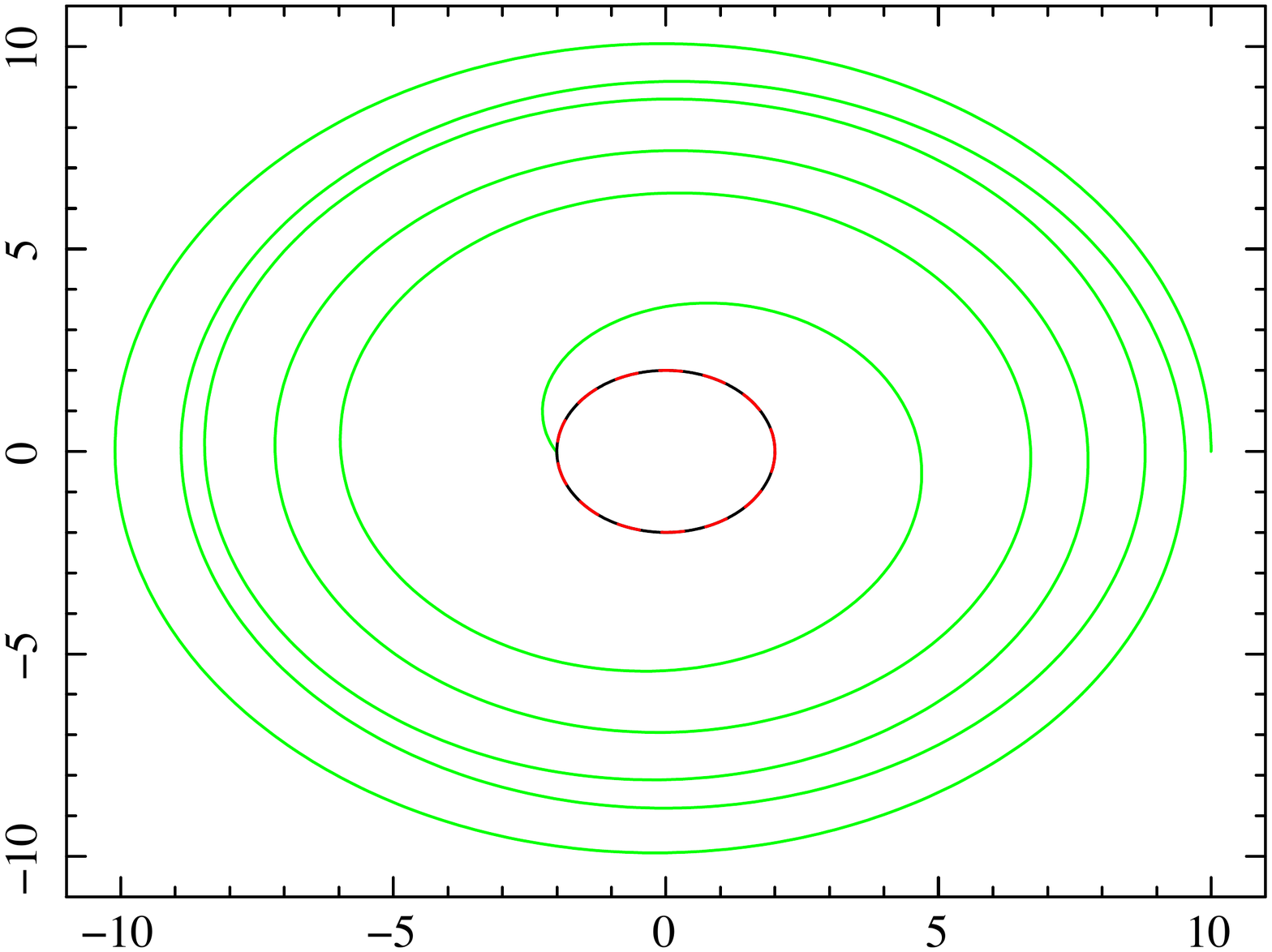}
}
\caption{The orbits of a particle in the Schwarzschild spacetime with $M=1$ and $A/M=0.01$. The continuous black line is the Schwarzschild radius at $r_{\rm S}=2M$, the dashed red line is the critical orbit with $r_{\rm (crit)}\approx r_{\rm S}$. All the particles have the same initial conditions $(r_0,\varphi_0,\alpha_0,\nu_0)=(10M,0,0,0.3536)$, where they have Keplerian velocity. The blue line are for photon angular momentum $b=0$, instead the green line for photon angular momentum $b=0.5$.} 
\label{fig:Fig3_5}
\end{figure}

\subsection{Kerr spacetime}
Kerr spacetimes distinguish from Schwarzschild spacetimes because they are rotating. Therefore, as done previously, I consider four cases depending on the initial conditions and the spin sign $a$.
\begin{itemize}
\item \emph{Data outside the critical radius and $a>0$:} the particles reach the critical radius and after they move in a circular orbit on it, because now the spacetime background is rotating (see Fig. \ref{fig:Fig3_6}). The same arguments of the Schwarzschild cases can be repeated similarly for that spacetime. 
\begin{figure}[t!]
\centering
\includegraphics[trim=1cm 1cm 0cm 3cm, scale=0.5]{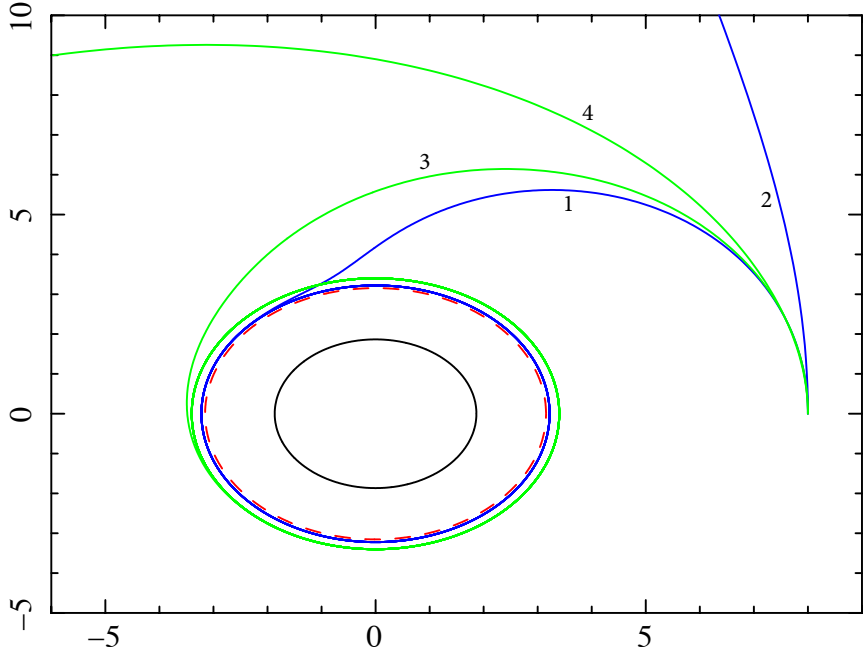}
\caption{The orbits of a particle in the Kerr spacetime with $M=1$, $a=0.5$, and $A/M=0.6$. The continuous black line is the inner radius at $r_{\rm i}=1.866M$, the dashed red line is the critical orbit with $r_{\rm (crit)}=3.154M$. All the particles have the same initial conditions $(r_0,\varphi_0,\alpha_0)=(8M,0,0)$. The blue lines are for photon angular momentum $b=0$, instead the green lines for photon angular momentum $b\neq0$. Initial conditions have: (1) $\nu_0=0.5$, (2) $\nu_0=0.8$, (3) $(\nu_0,b)=(0.2,0.5)$, and (4) $(\nu_0,b)=(0.2,5.2)$.} 
\label{fig:Fig3_6}
\end{figure}
\item \emph{Data inside the critical radius and $a>0$:} the orbits 1 is very interesting because initially the particle crosses the critical radius, but it possesses not enough energy (or velocity) to escape at infinity because the PR effect removes efficiently angular momentum from it (see Fig. \ref{fig:Fig3_7}). In this case it is slightly noticeable the frame dragging because the orbit 1 shows a peculiar hump. The orbit 3 for analogous reasons has a particular shape and it ends up rotating outer than orbit 1.   
\begin{figure}[t!]
\centering
\includegraphics[trim=1cm 1cm 0cm 3cm, scale=0.5]{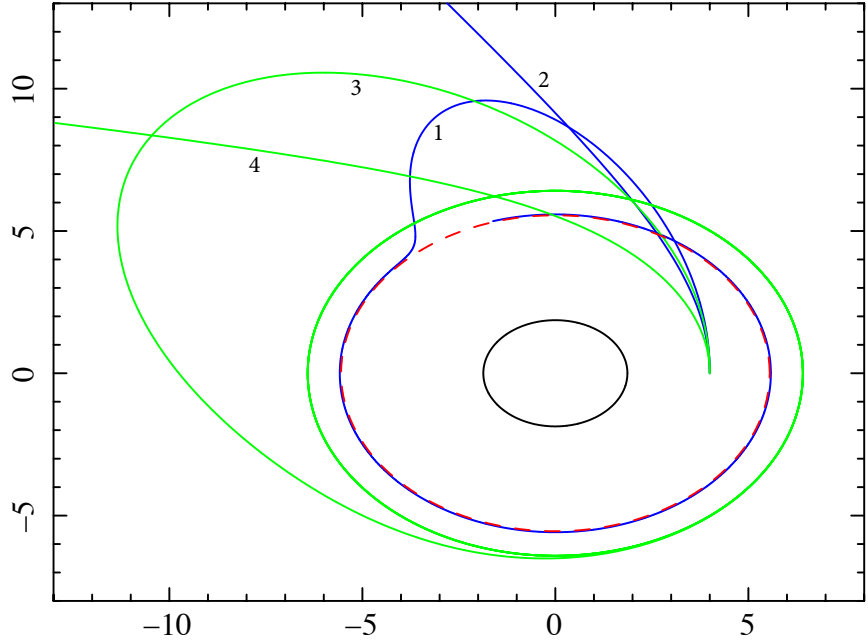}
\caption{The orbits of a particle in the Kerr spacetime with $M=1$, $a=0.5$, and $A/M=0.8$. The continuous black line is the inner radius at $r_{\rm i}=1.866M$, the dashed red line is the critical orbit with $r_{\rm (crit)}=5.551M$. All the particles have the same initial conditions $(r_0,\varphi_0,\alpha_0)=(4M,0,0)$. The blue lines are for photon angular momentum $b=0$, instead the green lines for photon angular momentum $b\neq0$. Initial conditions have: (1) $\nu_0=0.5$, (2) $\nu_0=0.8$, (3) $(\nu_0,b)=(0.5,0.5)$, and (4) $(\nu_0,b)=(0.5,5.2)$.} 
\label{fig:Fig3_7}
\end{figure}
\item \emph{Data outside the critical radius and $a<0$:} in this case and in the next the effects of frame dragging are more relevant since the motion of the particle is against the rotating direction of the central object (see Fig. \ref{fig:Fig3_8}). The orbit 4 is really surprising because the particle makes first a turn around the compact object and after escapes at infinity.
\begin{figure}[t!]
\centering
\includegraphics[trim=1cm 1cm 0cm 3cm, scale=0.5]{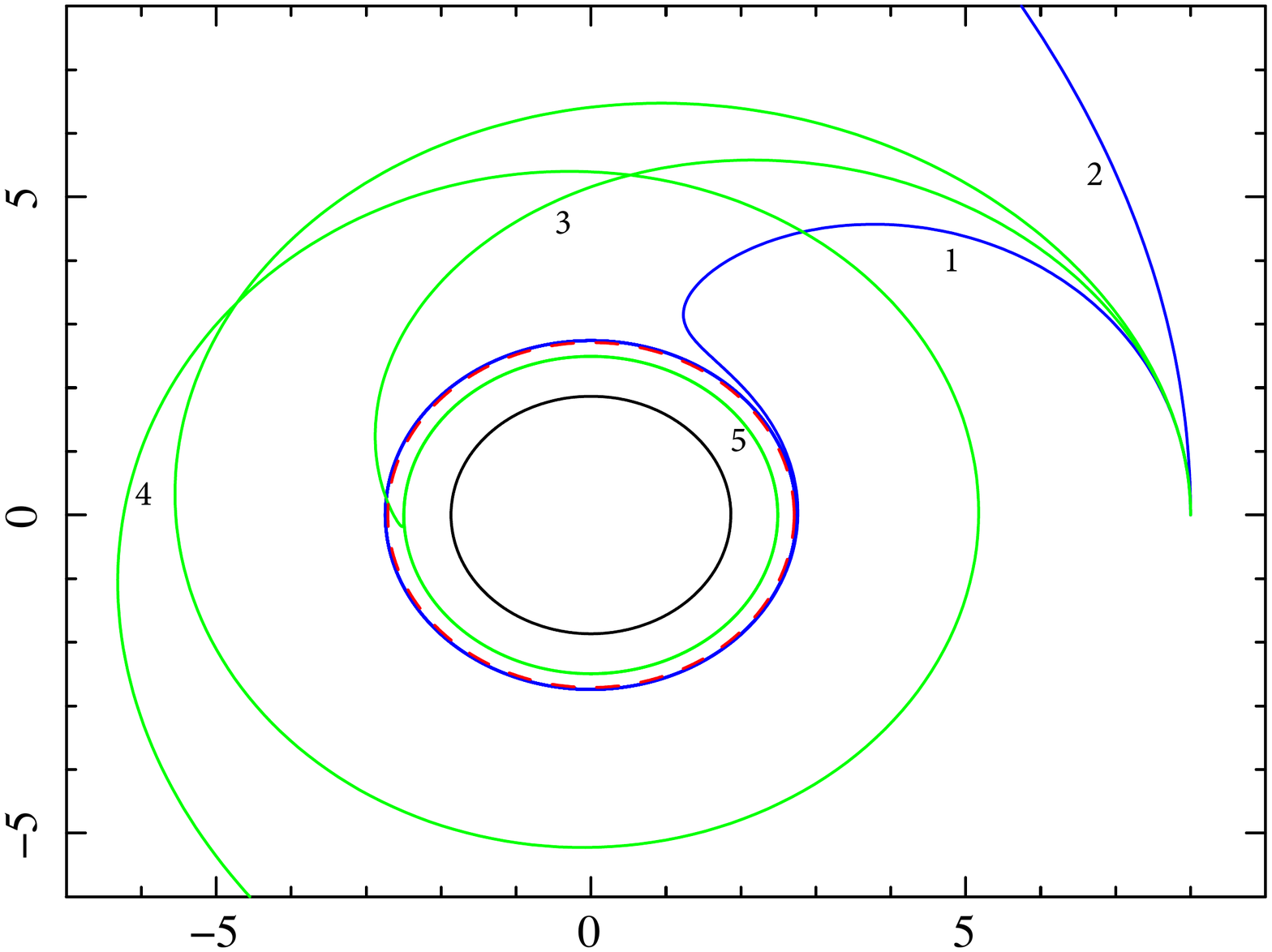}
\caption{The orbits of a particle in the Kerr spacetime with $M=1$, $a=-0.5$, and $A/M=0.6$. The continuous black line is the inner radius at $r_{\rm i}=1.866M$, the dashed red line is the critical orbit with $r_{\rm (crit)}=2.71M$. All the particles have the same initial conditions $(r_0,\varphi_0,\alpha_0)=(8M,0,0)$. The blue lines are for photon angular momentum $b=0$, instead the green lines for photon angular momentum $b\neq0$. Initial conditions have: (1) $\nu_0=0.2$, (2) $\nu_0=0.5$, (3) $(\nu_0,b)=(0.2,1.5)$, and (4) $(\nu_0,b)=(0.2,3.5)$.} 
\label{fig:Fig3_8}
\end{figure}
\item \emph{Data inside the critical radius and $a<0$:} in this case the effects of frame dragging are very strong, since the particles starts inside the critical radius and closer to the compact object (see Fig. \ref{fig:Fig3_9}). The shape of orbits are considerably affected by the background rotation, since they are all distorted backwards the direction of their motion.  
\begin{figure}[t!]
\centering
\includegraphics[trim=1cm 1cm 0cm 3cm, scale=0.5]{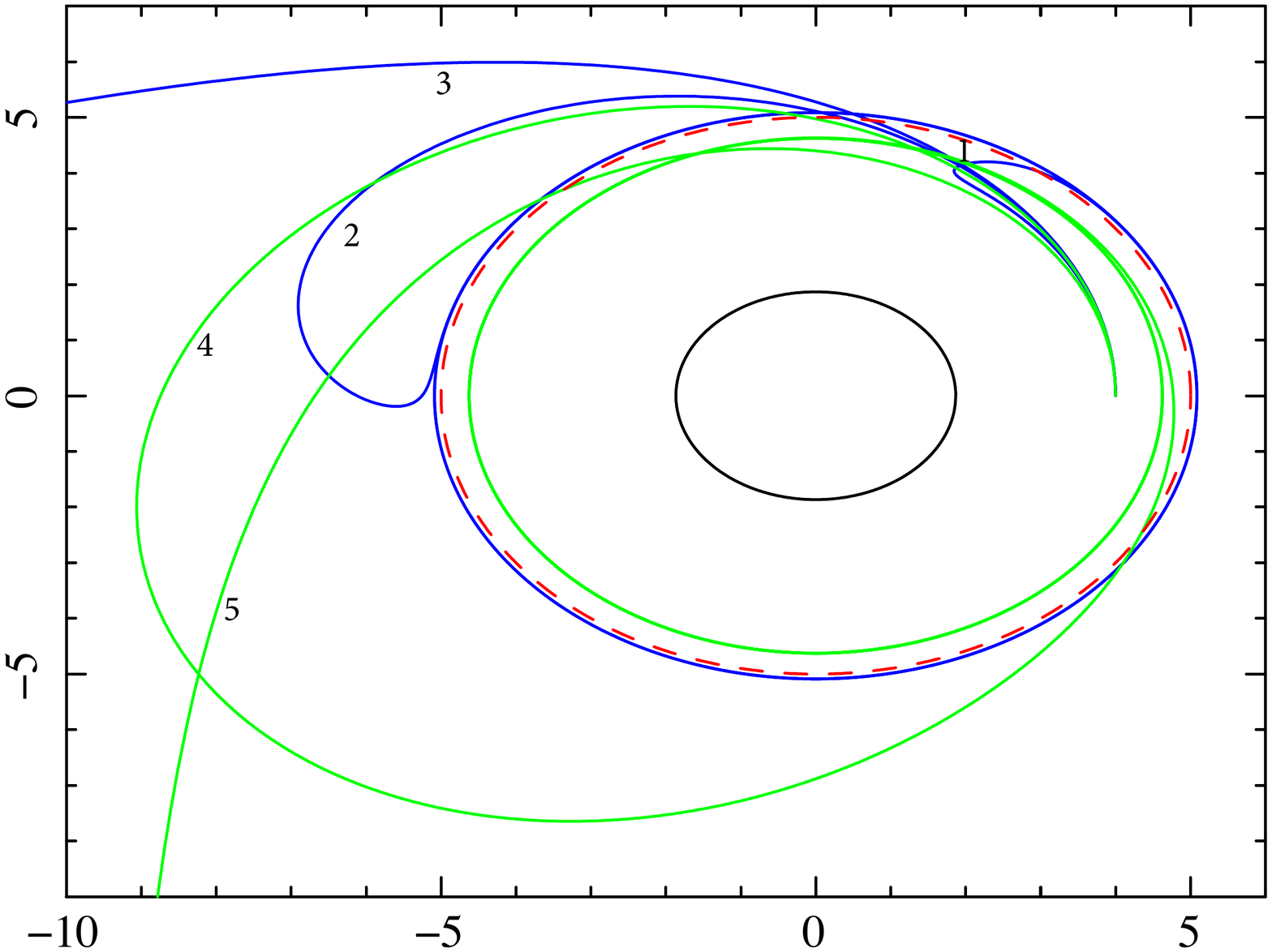}
\caption{The orbits of a particle in the Kerr spacetime with $M=1$, $-a=0.5$, and $A/M=0.8$. The continuous black line is the inner radius at $r_{\rm i}=1.866M$, the dashed red line is the critical orbit with $r_{\rm (crit)}=5M$. All the particles have the same initial conditions $(r_0,\varphi_0,\alpha_0)=(4M,0,0)$. The blue lines are for photon angular momentum $b=0$, instead the green lines for photon angular momentum $b\neq0$. Initial conditions have: (1) $\nu_0=0.5$, (2) $\nu_0=0.8$, (3) $\nu_0=0.9$, (4) $(\nu_0,b)=(0.8,0.5)$, and (5) $(\nu_0,b)=(0.8,3.5)$.} 
\label{fig:Fig3_9}
\end{figure}
\end{itemize}

\section{Lagrangian formulation of the general relativistic PR effect}
In this section, I propose the Lagrangian formulation of the general relativistic PR effect. The innovative part of this work is to prove the existence of the potential linked to the dissipative action caused by the PR effect through the help of an integrating factor, depending on the energy of the system. Generally, such kinds of inverse problems involving dissipative effects might not admit a Lagrangian formulation, especially in GR there are no examples of such attempts in literature so far. I show how the general relativistic Lagrangian formulation reduces to the classical case in the weak field limit. This approach facilitates further studies in improving the treatment of the radiation field and it contains advantageous implications for a deeper comprehension of gravitational waves.

\subsection{Motivations}
\label{sec:intro}
In high-energy astrophysics, it is important to examine the effects of the radiation field on the motion of the matter surrounding a compact object. Such radiation field can be arisen from: a boundary layer around a NS \cite{Inogamov1999}, a thermonuclear flash (type-I X-ray burst) occurring on an accreting NS surface \cite{lewin93}, or a hot corona around a black hole (BH) in X-ray binary systems \cite{Fabian2015}. Many observations confirm that this kind of radiation field, beside to exert an outward radial force, opposite to the gravitational pull from the compact object, can generate also a drag force, produced during the process of absorption and re-emission of the radiation from the affected body \cite{Ballantyne2004,Ballantyne2005,Worpel2013,Keek2014,Ji2014,Worpel2015}. 

The PR effect theory has never been treated from a Lagrangian point of view. This approach is advantageous for the following reasons: ($i$) it is a general, elegant, and effective methodology to attain at the \emph{pure} equations of motion (EoMs), namely once the kinetic, $\mathbb{T}$, and potential, $\mathbb{V}$, energies of the system are given, the constraint forces and the generalized forces, $Q_h$, are identified, and the generalized coordinates, $(q_h,\dot{q}_h)$, are chosen, I can analytically calculate the EoMs through the \emph{Euler-Lagrange equations} (ELEs), i.e.,
\begin{equation} \label{eq:ELeq}
\frac{d}{dt}\frac{\partial\mathcal{L}}{\partial \dot{q}_h}-\frac{\partial\mathcal{L}}{\partial q_h}=Q_h,
\end{equation}
where $\mathcal{L}=\mathbb{T}-\mathbb{V}$ is the \emph{Lagrangian function}; ($ii$) it places emphasis on the geometrical structure of the problem; ($iii$) the formulation of the motion is expressed in integral terms, not anymore in differential form \cite{Chow1995,Goldstein2002}. The nature of the ELEs, Eq. (\ref{eq:ELeq}), admits two points of view, depending whether one consider the \emph{direct or inverse problem}. The direct problem coincides with the ordinary approach, where given the Lagrangian function the EoMs are obtained computing the ELEs, seen as second-order ordinary differential equations (ODEs) \cite{Chow1995,Goldstein2002}; on the contrary, the inverse problem consists in determining a possible Lagrangian, that gives, through the ELEs, the assigned EoMs. Now, the ELEs become a set of second-order partial differential equations (PDEs) \cite{Santilli1978,Santilli1979,Lopuszanski1999}. 

Regarding the forces acting on a system, $Q_h$, they can be classified in two types: \emph{conservative} and \emph{non-conservative}. The conservative forces can be written as the generalized coordinates gradient of a potential, $Q_h=-\partial \mathbb{V}/\partial q_h$, and can be englobed in the potential energy, $\mathbb{V}$, of the Lagrangian function, $\mathcal{L}$, e.g., all the central force fields \cite{Chow1995,Goldstein2002}. The non-conservative forces can be divided ulteriorly in two subclasses: \emph{generalized forces} and \emph{dissipative forces}. The first ones are forces deriving from a generalized potential function, $\mathbb{V}$, such that $\partial\mathbb{V}/\partial q_h-d/dt(\partial\mathbb{V}/\partial \dot{q}_h)=Q_h$, and they can be included in the potential energy, $\mathbb{V}$, of the Lagrangian function, $\mathcal{L}$  \cite{Chow1995,Goldstein2002}. This potential finds its applications in the Lorentz force for an electromagnetic field and in general for all gyroscopic-like forces, or \emph{divergence-free fields}. Instead, the second forces do not admit a potential function, but they can be written as the velocity gradient of a \emph{Reyleigh dissipative function}, $\mathcal{F}$, i.e., $Q_h=-\partial \mathcal{F}/\partial \dot{q}_h$ \cite{Chow1995,Goldstein2002,Minguzzi2015}. Therefore, in this case I have to assign the Lagrangian and the Reyleigh dissipative potential function, expressing how the dissipative forces act on the system. However, there exists forces not collocabile in any of the groups mentioned above, since they might admit a potential function by adding an \emph{integrating factor}. For example, in classical thermodynamics where the increment of heat multiplied by the inverse of the temperature gives an exact differential form represented physically by the entropy \cite{Courant1962,Santilli1979}. 

The aim of this work is to develop the Lagrangian formalism of the general relativistic PR effect \cite{Bini09,Bini11}, which requires the use of different techniques. The work is structured as follows: in Sec. \ref{sec:classicPR}, I present the Lagrangian formulation of the classical PR effect; in Sec. \ref{sec:GR_formalism} I recall the relativity of observer splitting formalism, useful to derive the general relativistic PR EoMs; then in Sec. \ref{sec:GRPR} I derive its Lagrangian formulation; the conclusions are drawn in Sec. \ref{sec:conclusions}. 

\subsection{Classical PR effect}
\label{sec:classicPR}
The classical radiation drag force was described in Secs. \ref{sec:newframe} and \ref{sec:srframe}. I rewrite the equations of motion distinguishing the radiation pressure from the gravitational force, but keeping the same nomenclatures, i.e., 
\begin{eqnarray} 
&&\ddot{r}-r\dot{\varphi}^2+\frac{GM-Ac}{r^2}=-2A\frac{\dot{r}}{r^2}, \label{eqm1}\\
&&r^2\dot{\varphi}=L_0-A\varphi. \label{eqm2} 
\end{eqnarray}
I determine the Lagrangian function $\mathcal{L}_{\rm C}=\mathcal{L}_{\rm C}(q_h,\dot{q}_h)$, depending on the Lagrangian coordinates $q_1=r$ and $q_2=\varphi$ and $\dot{q}_h=d q_h/dt$, in presence of the forces $Q_h$, where $Q_1=-2A\dot{r}/r^2$ and $Q_2=-A\dot{\varphi}$, such that the respective ELEs will give the test particle EoMs, Eqs. (\ref{eqm1})--(\ref{eqm2}). The forces $Q_h$ are dissipative, because mathematically they depend on the velocity field and physically the energy is removed from the particle when the motion takes place. This force can be written in terms of the Rayleigh dissipative function, $\mathcal{F}_{\rm C}$. Therefore, the Lagrangian function is constituted by the kinetic, $\mathbb{T}_{\rm C}$, and potential, $\mathbb{V}_{\rm C}$, components in the following way:
\begin{equation} \label{eq:lagrangian}
\mathcal{L}_{\rm C}\equiv \mathbb{T}_{\rm C}-\mathbb{V}_{\rm C}=\frac{1}{2}\left(\dot{r}^2+r^2\dot{\varphi}^2\right)+\frac{GM-A}{r}.
\end{equation}
Instead, the Reyleigh dissipative function, $\mathcal{F}_{\rm C}$, can be determined by verifying whether the radiation differential form, whose components derived from Eqs. (\ref{eqm1})--(\ref{eqm2}), given by 
\begin{equation} \label{eq:pot_comp1}
\frac{\partial \mathcal{F}_{\rm C}}{\partial \dot{r}}=-\frac{A\dot{r}}{r^2}, \qquad \frac{\partial \mathcal{F}_{\rm C}}{\partial \dot{\varphi}}=-A\dot{\varphi},
\end{equation} 
is close and the domain, where it is defined, is simply connected. Such differential form is defined on all the space $\mathbb{R}^2$. This domain results to be simply connected, because, in polar coordinates, it transforms into a rectangle, defined by $r\in[0,+\infty]$ and $\varphi\in[0,2\pi]$. In addition, it is immediate to verify that it is a close form, i.e., the cross derivatives are equal. Therefore, such potential is obtained by integrating Eqs. (\ref{eq:pot_comp1}), constituting a decoupled system of PDEs.
$\mathcal{F}_{\rm C}$ is easily found as
\begin{equation} \label{eq:pot_comp2}
\mathcal{F}_{\rm C}(\dot{r},\dot{\varphi})=\frac{A}{2}\left(\frac{\dot{r}^2}{r^2}+\dot{\varphi}^2\right)+\mbox{const},
\end{equation} 
as an homogeneous function of order two in $(\dot{r},\dot{\varphi})$, completely determined once the initial conditions are set. 

\subsection{Relativistic Poynting-Robertson effect}
\label{sec:GR_formalism}
In classical mechanics the centrifugal forces are conceived to be fictitious inertial forces, that manifest themselves all the time I am observing the dynamics of an object in a rotating reference frame. Inertial forces have been topics of great interests in GR, because: ($i$) there is a close similarity between the gravitational forces, experienced locally on a massive body, and the fictitious pseudo-forces, felt by an observer in a non-inertial accelerated reference frame (argument based on the \emph{equivalence principle}); ($ii$) there is a strong analogy between the gravitational forces and the electromagnetism description, the so-called \emph{gravitoelectromagnetism} \cite{DeFelice1971}. Therefore, there have been many efforts to generalize the concept of centrifugal force to stationary \cite{Abramowicz1988,Abramowicz90,DeFelice1991a}, axially symmetric \cite{Prasanna1990,DeFelice1991b,Iyer1993,DeFelice1995,Barrabes1995} and finally also to arbitrary spacetimes \cite{Abramowicz1993}, encountering, however, several difficulties. The flaw of such attempts reside in the employment of the \emph{direct spacetime approach}, where the interpretation of the dynamical variables depends on further quantities (see e.g., Refs. \cite{Abramowicz1988,Abramowicz90,Abramowicz1993}). 

The successful approach, in terms of comprehensibility and clearness about the outcoming results, revealed to be the \emph{relativity of observer splitting formalism}, based on the full orthogonal splitting of the test particle motion relative to the observer in: ($i$) \qm{{\it 4=3+1}}: local rest space and local time direction; ($ii$) \qm{{\it3=2+1}}: transverse and longitudinal components of the local rest space (see e.g., Refs. \cite{Bini1997a,Bini1998}). Such formalism entails several advantages: it relies on a logical and unambiguous mathematical structure, it offers a natural link with respect to the classical case and, in the same time, provides an explicit physical interpretation of the involved terms (see e.g., Refs \cite{Bini1997b,Bini1998}). In addition, the general relativistic description, at the contrary of the classical framework, mixes the gravitational field, due to the presence of matter, with those of the accelerated motion of the observers. Therefore, it is significative to choose a Frenet-Serret frame, where it is possible to coherently split the different contributions and make sense to the splitting, reproducing thus the classical case \cite{Bini1997a,Bini1999}. 

In the following sections, I present for completeness the modern approach to derive the general relativistic PR EoMs, focussing the attention on its geometrical aspects. I show how this general relativistic formalism, to describe the non-inertial relative motions, recovers the line of though of the classical formalism. 

\subsubsection{Classical formalism and non-inertial relative motions}
I consider two reference frames $\mathbb{R}\equiv\left\{O,x,y,z\right\}$ and $\mathbb{R'}\equiv\left\{O',x',y',z'\right\}$ in relative motion to each other, observing the dynamical trajectory described by a point $P$. I call $\mathbf{r}(t)=P(t)-O$ and $\mathbf{r'}(t)=P(t)-O'$ the radius vectors in the reference frames $\mathbb{R}$ and $\mathbb{R'}$, respectively. Thus, I have the following relationship between the positions $\mathbf{r}=\mathbf{r'}+OO'$. Using the Poisson's formula on the versors of the reference frame $\mathbb{R'}$, i.e., $d\mathbf{u}/dt=\pmb{\omega}\times\mathbf{u}$, where $\mathbf{u}$ is a versor and $\pmb{\omega}$ is the angular velocity associated to the variation of $\mathbf{u}$, I have the relation between the velocities, 
\begin{equation} \label{eq:velocity}
\mathbf{v}=\mathbf{v_{O'}}+\mathbf{v'}+\pmb{\omega}\times\mathbf{r'},
\end{equation}
where $\mathbf{v_t}=\mathbf{v_{O'}}+\mathbf{v'}$ is the translatory velocity and $\pmb{\omega}\times\mathbf{r'}$ is the rotating velocity. Now passing to the accelerations, I have
\begin{equation} \label{eq:acceleration}
\mathbf{a}=\mathbf{a_{O'}}+\mathbf{a'}+\pmb{\omega}\times\pmb{\omega}\times\mathbf{r'}+2\pmb{\omega}\times\mathbf{v'},
\end{equation}
where $\mathbf{a_t}=\mathbf{a_{O'}}+\mathbf{a'}$ is the translatory acceleration, $\pmb{\omega}\times\pmb{\omega}\times\mathbf{r'}$ is the centrifugal force, and $2\pmb{\omega}\times\mathbf{v'}$ is the Coriolis force. Eqs. (\ref{eq:velocity})--(\ref{eq:acceleration}) are well known in the literature, where all the components have a precise and clear physical meaning \cite{Sommerfeld1964,Arnold2013}. 

\subsubsection{General relativistic formalism}
\label{sec:observer}
I consider a Riemannian manifold endowed with a Lorentzian metric, $g_{\alpha\beta}$, with signature $(-,+,+,+)$, a symmetric Levi-Civita connection, $\Gamma^\alpha_{\beta\gamma}$, and a covariant derivative, $\nabla_\alpha$. I also consider a family of observers with a 4-velocity, defined by a future-pointing unit timelike vector field $u^\alpha$ ($u_\alpha u^\alpha=-1$). Its proper time, $\tau_u$, parametrizes the world lines, that are integral curves of $u^\alpha$, the so-called \emph{observer congruence} (see Chap. 6 of Ref. \cite{Stephani2003}). 
 
\subsubsection{3+1 splitting}
In order to orthogonally decompose each tangent space into local rest space and local time in the direction of the observer $u^\alpha$, I project all the quantities in the hypersurface orthogonal to $u^\alpha$, through the projector operator $P(u)_{\alpha\beta}=g_{\alpha\beta}+u_\alpha u_\beta$. All the tensors having no components along $u^\alpha$, are termed \emph{spatial}. 

\subsubsection{Kinematical decomposition of observer congruence}
I define the acceleration vector related to the observer $u^\alpha$, given by $a(u)^\alpha=u^\beta \nabla_\beta u^\alpha$. Using $u_\alpha u^\alpha=-1$, it can be easily proved, that this acceleration has the propriety to be orthogonal to its velocity field $u^\alpha$, i.e., $u^\alpha a(u)_\alpha=0$ \cite{Misner73}. I note that $(a(u)^\alpha u_\beta+\nabla_\beta u^\alpha)u^\beta=0$, therefore the term in parenthesis is spatial and can be decomposed into its symmetric and antisymmetric parts \cite{Wald1984,Stephani2003} as
\begin{equation}
\nabla_\alpha u_\beta=-a(u)_\alpha u_\beta+\theta_{\alpha\beta}+\omega_{\alpha\beta},
\end{equation}
where $\theta_{\alpha\beta}=\nabla_{(\beta}u_{\alpha)}a_{(\alpha}u_{\beta)}\equiv P^\mu_\alpha P^\nu_\beta \nabla_{(\nu} u_{\mu)}$ is the expansion tensor, describing how an initial spherical cloud of test particles becomes distorted into an ellipsoidal shape; $\omega_{\alpha\beta}=\nabla_{[\beta}u_{\alpha]}a_{[\alpha}u_{\beta]}\equiv P^\mu_\alpha P^\nu_\beta \nabla_{[\nu} u_{\mu]}$ is the vorticity tensor, representing how an initial spherical cloud of test particles tends to rotate. I have used the following notations: $(A,B)=\frac{1}{2}(AB+BA)$ and $[A,B]=\frac{1}{2}(AB-BA)$. This is the so-called \emph{kinematical decomposition of the observer congruence}.

\subsubsection{Spatial derivative operators}
\begin{figure} [h!]
\includegraphics[scale=0.55]{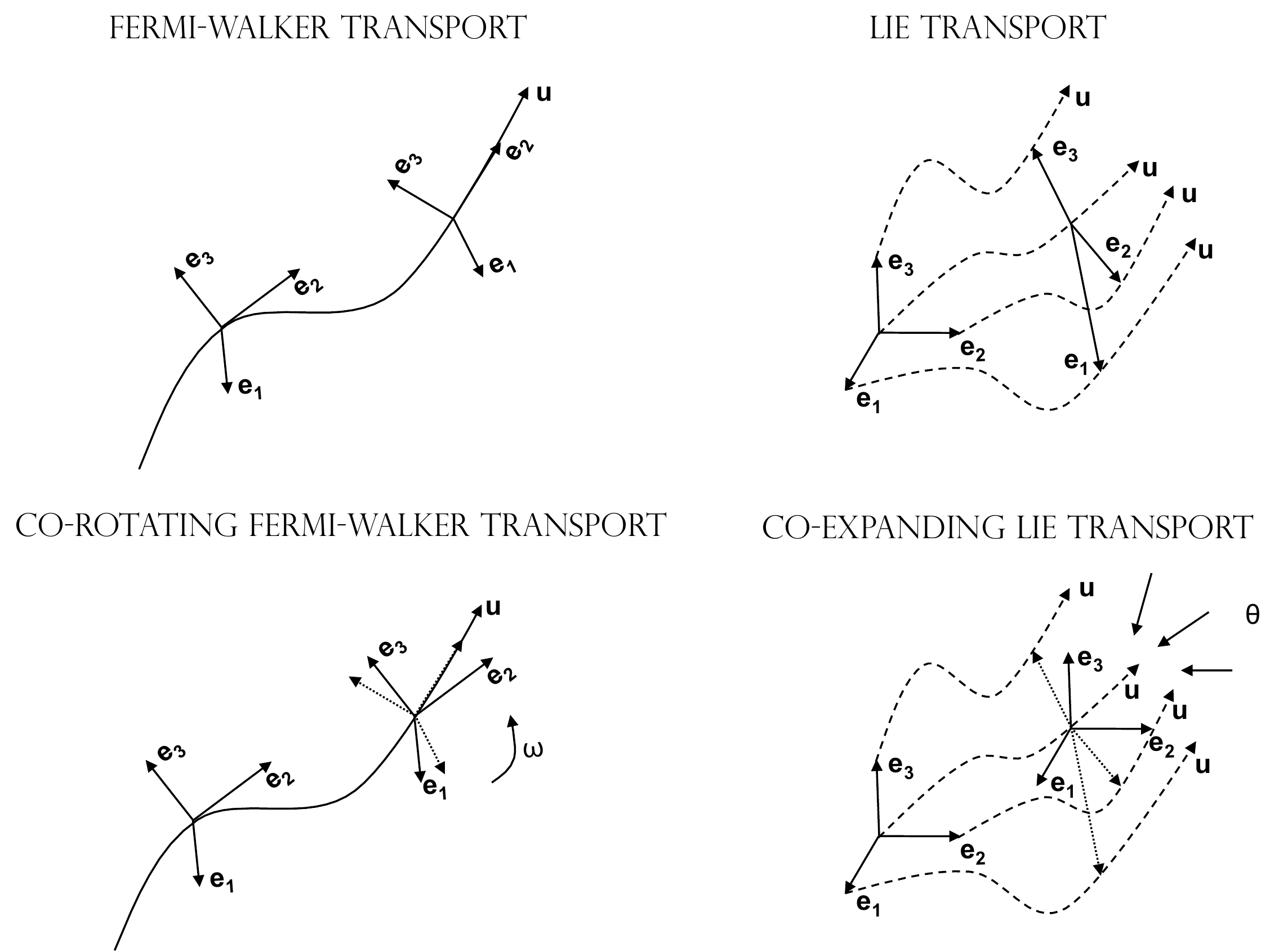}
\centering
\caption{Geometrical representation of the Fermi-Walker, Lie, co-rotating Fermi-Walker, and co-expanding Lie transport.}
\label{fig:SDO}
\end{figure}

It is now appropriate to introduce a set of spatial derivative operators, through the projector $P(u)_{\alpha\beta}$, in order to achieve the 2+1 splitting. Given any spatial vector field, $X^\alpha$, and a generic vector, $v^\alpha$, I define \cite{Bini1997a,Bini1998,Bini1999,Bini2010}:
\begin{itemize}
\item the spatial Lie derivative: 
\begin{equation}
\begin{aligned}
\mathcal{L}(u)_Xv^\alpha&=P(u)\mathcal{L}_Xv^\alpha=P(u)^\alpha_\gamma \left( X^\beta\nabla_\beta v^\gamma+v^\beta\nabla_\beta X^\gamma \right);
\end{aligned}
\end{equation}
\item the spatial covariant derivative: 
\begin{equation}
\nabla(u)_\beta v^\alpha=P(u)^\alpha_\delta P(u)_\beta^\gamma\nabla_\gamma v^\delta;
\end{equation}
\item the temporal Lie derivative: 
\begin{equation}
\nabla_{\rm (Lie)}(u)v^\alpha=P(u)\mathcal{L}_u v^\alpha;
\end{equation}
\item the temporal Fermi-Walker derivative: 
\begin{equation}
\nabla_{\rm (fw)}(u)v^\alpha=P(u)^\alpha_\gamma u^\beta\nabla_\beta v^\gamma.
\end{equation}
\end{itemize}
In the definitions reported above, I have introduced two ways to perform the derivatives, i.e., the \emph{Fermi-Walker} and \emph{Lie transport}. The Fermi-Walker transport with respect to the observer congruence $u^\alpha$ moves rigidly a spatial tetrad, $(\mathbf{e_1},\mathbf{e_2},\mathbf{e_3})$, along the world line described by the vector $u^\alpha$ (see Fig. \ref{fig:SDO} and Ref. \cite{Misner73}); instead the Lie transport with respect to the observer congruence $u^\alpha$ evolves a spatial tetrad, $(\mathbf{e_1},\mathbf{e_2},\mathbf{e_3})$, following the geodesic flux described by $u^\alpha$ (see Fig. \ref{fig:SDO} and Refs \cite{Hawking1973,Stephani2003}). 
There is another transport obtained combining the Fermi-Walker (or Lie) transport with respect to the kinematical decompositions of the observer, i.e., the temporal \emph{co-rotating Fermi-Walker} (or the \emph{co-expanding Lie}) \emph{derivative} (see Fig. \ref{fig:SDO} and Refs. \cite{Bini1997a,Bini1998,Bini1999,Bini2010}): 
\begin{equation}
\begin{aligned}
\nabla_{\rm (cfw)}(u)v^\alpha&=\nabla_{\rm(fw)}(u)v^\alpha+\omega(u)^\alpha_\beta v^\beta=\nabla_{\rm(Lie)}(u)v^\alpha+\theta_\beta^\alpha v^\beta.
\end{aligned}
\end{equation}
This kind of transport with respect to the observer congruence $u^\alpha$ let a spatial tetrad, $(\mathbf{e_1},\mathbf{e_2},\mathbf{e_3})$, unchanged during the evolution, either rigidly along the observer world line and after applying an opportune rotation, or according to the geodesic flow described by $u^\alpha$ and after applying an opportune expansion.

\subsubsection{Nonlinear reference frame}
\label{sec:nrf}
\begin{figure} [h!]
\includegraphics[scale=0.5]{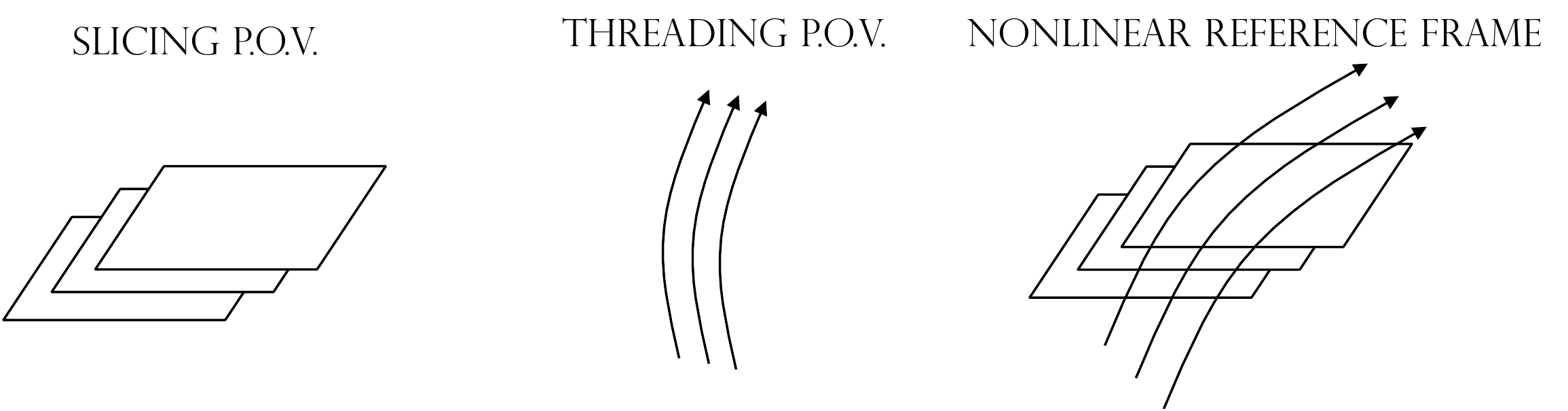}
\centering
\caption{Geometrical representation of the slicing and threading point of view, and nonlinear reference frame.}
\label{fig:pov}
\end{figure}

To further split the local rest space I have to adopt an adequate point of view. A full splitting of spacetime manifold requires both a \emph{slicing} of the spacetime in spatial hypersurfaces and a \emph{threading} of the spacetime along the observer congruence. A slicing together with a transversal threading form a structure dubbed \emph{nonlinear reference frame} (see Fig. \ref{fig:pov} and Refs. \cite{Jantzen1992,Bini1997a,Bini1998}). 
Therefore, I have to introduce a system of coordinates adapted to the observer congruence. Let $\left\{E_a^\alpha\right\}\equiv \left\{E_1^\alpha, E_2^\alpha, E_3^\alpha\right\}$ be a spatial frame, i.e., a basis of each local rest space with respect to the observer congruence $u^\alpha$. The Latin-index refers to the spatial frame components, instead the Greek-index refers to the spacetime components. The frame derivatives of a function, $f$, are denoted by the comma notation, i.e., 
\begin{equation}
u(f)=f_{,0},\qquad E_a^\alpha\partial_\alpha f=f_{,\alpha}.
\end{equation}  
To perform the derivatives of tensor fields, I define temporal and spatial derivatives of the spatial frame vectors, as well as their Lie brackets (see Eqs. (11.2) in Ref. \cite{Bini1997a}):
\begin{eqnarray}
&&\nabla_{\rm (tem)}(u)\,E_a^\alpha=C_{\rm (tem)}(u)^b_a\, E_b^\alpha,\ _{\rm tem=fw,\ Lie,\ cfw,}\label{eq:cc1}\\
&&\nabla_{E_a}\,E_b^\alpha=\Gamma(u)^c_{ab}\, E_c^\alpha,\label{eq:cc2}\\
&&\left(P(u)\,[E_a,E_b]\right)^\alpha=C(u)^c_{ab}\, E_c^\alpha,\label{eq:cc3}
\end{eqnarray}
where the following terms may be called: $C_{\rm (tem)}(u)^b_a$ the \qm{temporal constant structures}, $\Gamma(u)^c_{ab}$ the \qm{spatial connections}, and $C(u)^c_{ab}$ the \qm{spatial constant structures}. 

\subsubsection{Test particle motion}
I consider the motion of a test particle with respect to unit timilike 4-velocity, $U^\alpha$, and I describe such motion relative to the observer $u^\alpha$. The world line of such particle is parametrized by the particle's proper time, $\tau_U$, related to the observer proper time, $\tau_{(U,u)}$, by $d\tau_{(U,u)}/d\tau_U=\gamma(U,u)$. Therefore, the 4-velocity $U^\alpha$ can be decomposed in the component along $u^\alpha$, $U^{(||_u)}$, and in the spatial components with respect to $u^\alpha$, $[P(u)U]^\alpha$, as
\begin{equation} \label{eq:U_split}
\begin{aligned}
U(\tau_U)^\alpha&=U(\tau_U)^{(||_u)}\,u^\alpha+[P(u)\,U(\tau_U)]^\alpha\\
&=\gamma(U,u)\,[u^\alpha+\nu(U,u)^\alpha]=E(U,u)\,u^\alpha+p(U,u)^\alpha,
\end{aligned}
\end{equation}
where $\nu^\alpha\equiv\nu^\alpha(U,u)$ is the relative spatial velocity of $U^\alpha$, $\gamma\equiv\gamma(U,u)=(1-\nu^2)^{-1/2}$ is the Lorentz factor with $\nu\equiv\nu(U,u)=\sqrt{\nu_\alpha \nu^\alpha}$ the module of the relative spatial velocity of $U^\alpha$, $E\equiv E(U,u)=\gamma$ is the energy per unit mass as seen by the observer $u^\alpha$, and $p^\alpha\equiv p(U,u)^\alpha=\gamma(U,u)\nu(U,u)^\alpha$ is the relative spatial momentum of the test particle per unit mass, and $p\equiv p(U,u)=\sqrt{p_\alpha p^\alpha}$ is the module of the relative spatial momentum. 

\subsubsection{Intrinsic derivative along the test particle curve}
In order to operate along the test particle curve, I define the \emph{intrinsic} or \emph{absolute derivative} of a spatial vector field, $X^\alpha$, along the test particle trajectory simply restricting the action of the covariant derivative on the vector field $X^\alpha$ along the test particle curve \cite{Bini1997a,Bini1998,Bini2010}, i.e.,
\begin{equation}
\frac{DX^\alpha(\tau_U)}{d\tau_U}=\frac{dX^\alpha(\tau_U)}{d\tau_U}+\Gamma^\alpha_{\beta\gamma}\,U(\tau_U)^\beta\, X^\gamma(\tau_U).
\end{equation}
Therefore, I can extend the notions of Fermi-Walker, Lie, and co-rotating Fermi-Walker transport along the test particle curve in the following way \cite{Bini1997a,Bini1998,Bini2010}
\begin{equation} \label{eq:Dcurve}
\begin{aligned}
&\frac{D_{\rm(tem)}(U,u)\ X^\alpha(\tau_U)}{d\tau_{(U,u)}}=\left[\nabla_{\rm(tem)}(u)+\nu^\beta\,\nabla(u)_\beta\right]X^\alpha(\tau_U)\\ 
&\hspace{3cm}_{\rm tem=fw,\ Lie,\ cfw}.
\end{aligned}
\end{equation}
A factor $\gamma$ is missing in Eq. (\ref{eq:Dcurve}), because it is included into the reparametrization of the particle world line through the formula $d\tau_{(U,u)}/d\tau_U=\gamma(U,u)$.

\subsubsection{2+1 splitting}
I consider a system of coordinates $\left\{u^\alpha, E^\alpha_a\right\}$ adapted to the observer congruence in a nonlinear reference frame (see Sec. \ref{sec:nrf}). The spatial projection of the test particle's 4-acceleration, $a(U)^\alpha=DU^\alpha/d\tau_U$, measured by the the observer congruence, $A(U,u)^\alpha$, is given by $A(U,u)^\alpha=1/\gamma P(u)^\alpha_\beta a(U)^\beta$. Therefore, I have (see Sec. 9 in Refs. \cite{Bini1997a,Bini1998})
\begin{equation} \label{eq:S+T}
\begin{aligned}
a(U)^\alpha&=\gamma\, P(U)^\alpha_\beta\, A(U,u)^\beta=\gamma\, P(U)^\alpha_\beta\,\left\{\frac{D_{\rm (tem)}(U,u)}{d\tau_{(U,u)}}\left[\gamma\, u^\alpha+p(u,U)^\alpha\right]\right\}\\
&=\gamma^2\frac{D_{\rm (tem)}(U,u)}{d\tau_{(U,u)}}u^\alpha+\gamma\frac{D_{\rm (tem)}(U,u)}{d\tau_{(U,u)}}p(u,U)^\alpha.
\end{aligned}
\end{equation}
It is important to note that the term connected with the derivative of the factor $\gamma$ does not appear in Eq. (\ref{eq:S+T}) because it carries a term $u^\alpha$ that it is vanished by the projector, since I am interested in the spatial part of the test particle acceleration. I have split the test particle acceleration in temporal and spatial part with respect to the observer congruence. The temporal projection along $u^\alpha$ leads to the evolution equation for the observed energy of the test particle along its world line; instead the spatial projection orthogonal to $u^\alpha$ leads to the evolution equation for the three-momentum of the test particle along its world line, where the kinematical quantities linked to the observer motion figure in these equations as \emph{inertial forces}.

The first term in Eq. (\ref{eq:S+T}) is called the \emph{spatial gravitational force}, interpreted as the inertial forces due to the motion of the observers themselves. These inertial forces involve the kinematical quantities of the observer congruence. Indeed, this term can be decomposed in (see Sec. 9 in Refs. \cite{Bini1997a,Bini1998})
\begin{equation} \label{eq:GE+GM}
\gamma^2\frac{D_{\rm (tem)}(U,u)}{d\tau_{(U,u)}}u^\alpha=\gamma^2\left[a(u)^\alpha+H_{\rm (tem)}(u)_\beta^\alpha\,\nu(U,u)^\beta\right],
\end{equation}
where 
\begin{equation}
H_{\rm (tem)}(u)_\beta^\alpha=\begin{cases}
\quad\theta(u)^\alpha_\beta-\omega(u)^\alpha_\beta,& _{\rm tem=fw};\\
\quad2\theta(u)^\alpha_\beta-2\omega(u)^\alpha_\beta,& _{\rm tem=Lie};\\
\quad\theta(u)^\alpha_\beta-2\omega(u)^\alpha_\beta,& _{\rm tem=cfw}.
\end{cases}
\end{equation}
The first term in Eq. (\ref{eq:GE+GM}) leads to the \emph{gravitoelectric gravitational force}, instead the second term in Eq. (\ref{eq:GE+GM}) to the \emph{gravitomagnetic} or \emph{Coriolis gravitational force}. 

The second term in Eq. (\ref{eq:S+T}) can be decomposed into a longitudinal and transverse relative acceleration terms, with respect to the observer congruence, as \cite{Bini1997a,Bini1998,Bini2010}
\begin{equation} \label{eq:CL+SC+TC}
\begin{aligned}
&\gamma\frac{D_{\rm (tem)}(U,u)}{d\tau_{(U,u)}}p(u,U)^\alpha\\
&=\gamma\frac{D_{\rm (tem)}(U,u)\,p(U,u)}{d\tau_{(U,u)}}\hat{\nu}(U,u)^\alpha+\gamma\,p(U,u)\frac{D_{\rm (tem)}(U,u)\,\hat{\nu}(U,u)^\alpha}{d\tau_{(U,u)}}\\
&=\frac{d p(U,u)^\alpha}{d\tau_U}+\gamma^2\, C_{\rm (tem)}(u)^\alpha_\beta\, \nu(U,u)^\beta+\gamma^2\, \Gamma(u)^\alpha_{\beta\gamma}\,\nu(U,u)^\beta\,\nu(U,u)^\gamma,
\end{aligned}
\end{equation}
where I have divided the 4-momentum, $p^\alpha=p\,\hat{\nu}^\alpha$, in the longitudinal part along the versor $\hat{\nu}(U,u)^\alpha$, called the \emph{relative centrifugal force}, and in the transverse part orthogonal to $\hat{\nu}(U,u)^\alpha$, called the \emph{relative centripetal force} \cite{Bini1997a,Bini1998}. The relative centripetal force can be written as:
\begin{equation} \label{eq:curvature}
\frac{D_{\rm (tem)}(U,u)\,\hat{\nu}(U,u)^\alpha}{d\tau_{(U,u)}}=k_{\rm (tem)}(U,u)\, \nu(U,u)^2\, \eta_{\rm (tem)}(U,u)^\alpha,
\end{equation}
where $\eta_{\rm (tem)}(U,u)^\alpha$ is the normal versor relative to the spatial test particle orbit in the osculating plane and $k_{\rm (tem)}(U,u)$ is the relative curvature. This term can be related to the curvature radius of the orbit, $\rho_{\rm (tem)}(U,u)$ through $k_{\rm (tem)}(U,u)=1/\rho_{\rm (tem)}(U,u)$ and also to the spatial connections $\Gamma(u)^\alpha_{\beta\gamma}$ through $\gamma^2\, \Gamma(u)^\alpha_{\beta\gamma}\,\nu(U,u)^\beta\,\nu(U,u)^\gamma$ \cite[][]{Bini1997a}. The evolution of the 4-momentum explicitly reads as in the last row of Eq. (\ref{eq:CL+SC+TC}), where I have removed the contributions coming from the $\theta$-direction, since in my treatment the motion occurs only in the equatorial plane (see Refs. \cite{Bini1997a,Bini1997b,Bini1998,Bini1999}, for the full description). In addition, the term $\Gamma(u)^\alpha_{\beta\gamma}\,\nu(U,u)^\beta\,\nu(U,u)^\gamma$ is called the \emph{space curvature force} in the threading point of view \cite{Bini1997a,Bini1998}.

\subsubsection{General relativistic PR EoMs in stationary and axially symmetric spacetimes}
I consider a stationary and axially symmetric spacetime, parametrized by the nonlinear reference frame associated to the Boyer-Lindquist coordinates $X^\alpha\equiv(t,r,\theta,\varphi)$. In such coordinates, the metric in the equatorial plane, $\theta=\pi/2$, reads as
\begin{equation} \label{eq:metric}
ds^2=g_{tt}dt^2+g_{rr}dr^2+2g_{t\varphi}dt\, d\varphi+g_{\varphi\varphi}d\varphi^2,
\end{equation}
where all metric components depend only on $r$ and $\theta$. In such spacetimes, there are two kinds of observers: $(i)$ the hypersurface normal observers or zero angular momentum observers (ZAMOs) and $(ii)$ the threading observers following the time coordinate line trajectories. Both observers family are accelerated, because the firsts are dragged by the spinning central object, while the seconds are accelerating to contrast the frame dragging effect \cite{Bini1997a,Bini1997b,Bini1998}. In this environment, the ZAMO point of view is the easiest way to describe the motion of a test particle (see Sec. 12.2 in Ref. \cite{Bini1997a}). The ZAMO 4-velocity is $u^\alpha=(N^{-1},0,-N^{-1}N^\varphi,0)$, where $N=(-g^{tt})^{-1/2}$ and $N^\varphi=g_{t\varphi}/g_{\varphi\varphi}$. The frame adapted to the ZAMOs is \cite{Bini09,Bini11}
\begin{equation} \label{eq:ZAMO}
\begin{aligned}
&e^\alpha_t=u^\alpha,\ e^\alpha_r=\left(0,\frac{1}{\sqrt{g_{rr}}},0,0\right),\\
&\quad e^\alpha_\varphi=\left(0,0,\frac{1}{\sqrt{g_{\varphi\varphi}}},0,0\right).
\end{aligned}
\end{equation}
In the ZAMO point of view,  the metric, Eq. (\ref{eq:metric}), becomes \cite{Bini1997a,Bini09,Bini11}
\begin{equation}
ds^2=-N^2dt^2+g_{\varphi\varphi}(d\varphi+N^\varphi dt)^2+g_{rr}dr^2.
\end{equation} 
For a stationary observer congruence, it is useful to exploit the intrinsic spatial Lie derivative, since it is the most appropriate to the spatial geometry without requiring additional kinematic linear transformations of the spatial tangent space, namely $C_{\rm (lie)}(u)^\alpha_\beta=0$, $\omega(u)^\alpha_\beta=0$, $H_{\rm (tem)}(u)^\alpha_\beta=2\theta(u)^\alpha_\beta$, $\Gamma(u)^r_{\varphi\varphi}=k_{\rm (Lie)}(u)^r$, and $\Gamma(u)^\varphi_{\varphi r}=-k_{\rm (Lie)}(u)^r$ (see Sec. 12.1 in Refs. \cite{Bini1997a,Bini1998}). 

The test particle acceleration relative to the observer congruence, given by Eqs. (\ref{eq:S+T}), (\ref{eq:GE+GM}), and (\ref{eq:CL+SC+TC}), reads explicitly as \cite{Bini1997a,Bini1998,Bini2010}
\begin{equation} \label{eq:GRacceleration}
\begin{aligned}
a(U)^\alpha&=\gamma^2\left[a(u)^\alpha+\Gamma(u)^\alpha_{\beta\gamma}\,\nu^\beta(U,u)\, \nu^\gamma(U,u)
+2\theta(u)^\alpha_\beta\, \nu(U,u)^\beta\right]+\frac{d p(U,u)^\alpha}{d\tau(U,u)}\\
&=-F^{\rm (GE)}(U,u)^\alpha-F^{\rm (SC)}(U,u)^\alpha-F^{\rm (GM)}(U,u)^\alpha+\frac{d p(U,u)^\alpha}{d\tau_U},
\end{aligned}
\end{equation}
where the gravitational inertial forces are: gravitoelectric (GE), space curvature (SC), and gravitomagnetic (GM). This splitting, although it is very technical, permits to recognize and give an exact physical meaning to all terms contributing to characterize the acceleration $a(U)^\alpha$, as it happens for the classical case for Eq. (\ref{eq:acceleration}), see Sec. \ref{sec:classicPR} for further details. 

\begin{figure} [h!]
\centering
\includegraphics[trim=0cm 5.5cm 0cm 0cm,scale=0.7]{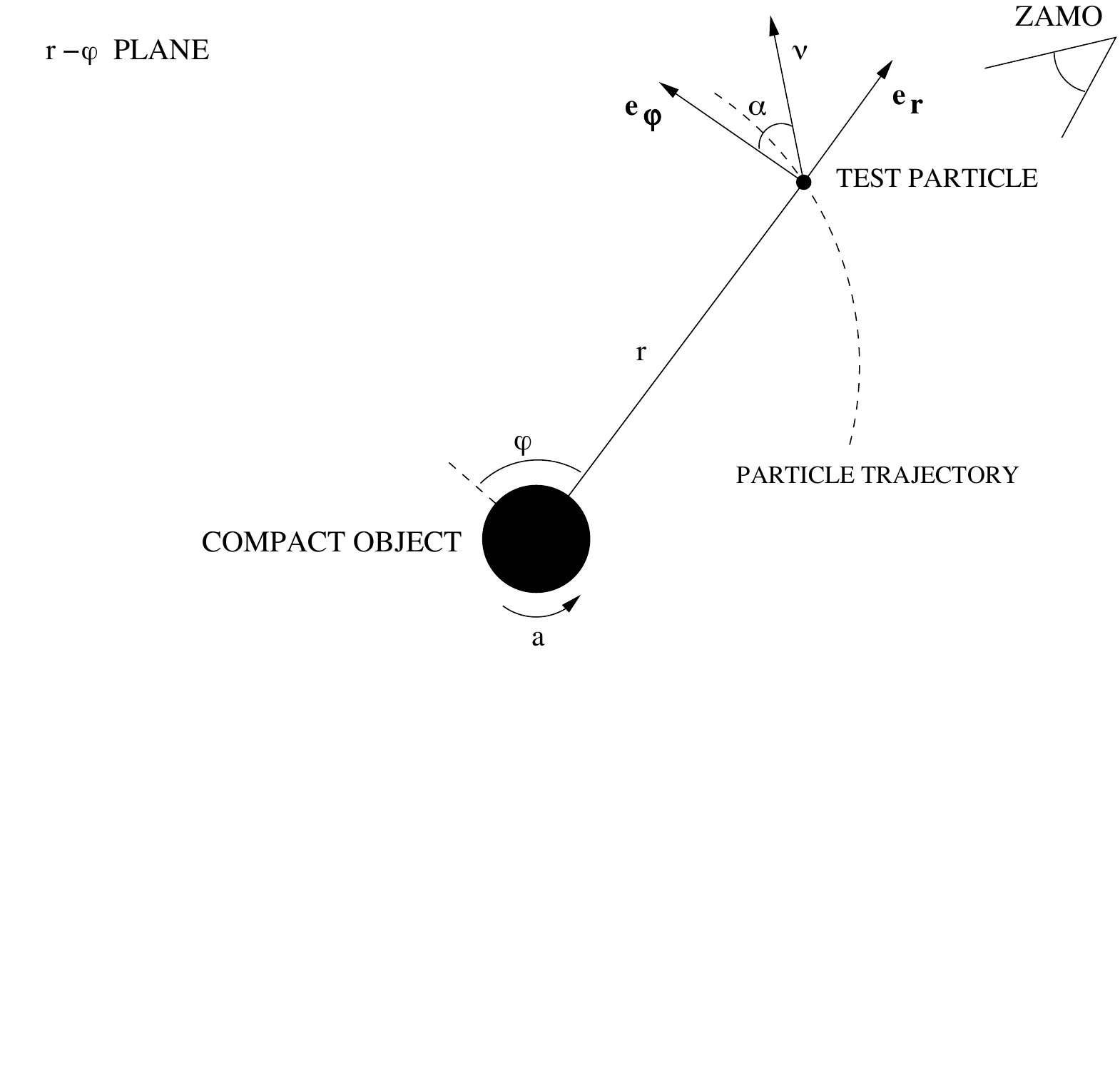}
\caption{The geometry of the problem is given by a test particle orbiting in the equatorial plane around a rotating compact object with spin $a$, at a radius $r$ and azimuthal angle $\varphi$. The test particle motion is described in the ZAMO reference frame $\{\mathbf{e_r},\mathbf{e_\varphi}\}$. The test particle moves with a velocity $\nu$, forming an angle $\alpha$ in the ZAMO reference frame.}
\label{fig:geometry}
\end{figure}

In presence of an external spatial force per unit test particle mass, $f(U)^\alpha$, the test particle EoMs are given by $a(U)^\alpha=f(U)^\alpha$. In my case, the external spatial force is represented by a radiation field, modeled as a coherent flux of photons moving along null geodesics, $k^\alpha$, on the background spacetime. The relative stress-energy tensor is \cite{Bini09,Bini11}
\begin{equation}
T^{\alpha\beta}=\Phi^2 k^\alpha k^\beta, \quad k^\alpha k_\alpha=0,\quad k^\alpha\nabla_\alpha k=\beta=0,
\end{equation}
where $\Phi$ is a parameter connected to the radiation intensity. I consider, that the photons can travel in any direction in the equatorial plane with angular momentum, $b\equiv L/E=-k_\varphi/k_t$. Therefore, it is useful to introduce the parameter $\beta$, defined as the azimuthal angle of the photon 4-momentum measured in the local frame, $\{\mathbf{e_r},\mathbf{e_\varphi}\}$, related to the ZAMO (see Ref. \cite{Bini11}, for more details)
\begin{equation} \label{eq:k_observer}
\begin{aligned}
&k^\alpha=E(u)[u^\alpha+\hat{\nu}(k,u)^\alpha],\quad \hat{\nu}(k,u)^\alpha=e_r^\alpha\,\sin\beta+e_\varphi^\alpha\, \cos\beta.
\end{aligned}
\end{equation}
In addition, the photon 4-momentum can be also decomposed with respect to the test particle velocity, $U^\alpha$, in the following way \cite{Bini09,Bini11}
\begin{equation} \label{eq:k_particle}
k^\alpha=E(U)[U^\alpha+\hat{V}(k,U)^\alpha].
\end{equation}
In this model, the radiation field is given by \cite{Bini09,Bini11}
\begin{equation} \label{eq:radiation}
F_{\rm(rad)}(U)^\alpha=-\frac{\sigma}{m}\, P(U)^\alpha_\beta\, T^\beta_\mu\, U^\mu,
\end{equation}
where $\sigma$ and $m$ are cross section and mass of the test particle, respectively. As done for the test particle velocity, I decompose also the photon 4-momentum relative to the observer congruence in order to get the relative radiation force, $F_{\rm(rad)}(U,u)^\alpha$ (see Refs. \cite{Bini09,Bini11}, for further details). In such decomposition, the intensity of the radiation is associated to the parameter $A$, defined as the emitted luminosity from the central source as seen by an observer at infinity in units of Eddington luminosity \cite{Bini09,Bini11}. The explicit expression of the parameters figuring in Eq. (\ref{eq:GRacceleration}) can be found in the papers of Refs. \cite{Bini09,Bini11}. In such context, the important parameters to determine the motion of the test particle are: $\nu$ and $\alpha$, the velocity and azimuthal angle of the test particle in the ZAMO frame, respectively; $r$ and $\varphi$, the radius and the azimuthal angle in Boyer-Lindquist coordinates (see Fig. \ref{fig:geometry}).
Therefore using Eqs. (\ref{eq:GRacceleration}), (\ref{eq:radiation}), (\ref{eq:U_split}) the test particle EoMs are \cite{Bini11}
\begin{eqnarray} 
&&\frac{d\nu}{d\tau_U}=-\frac{\sin\alpha}{\gamma}[a(u)^{r}+2\nu\cos\alpha\,\theta(u)^{r}_{\varphi}] \label{EoM1}\\
&&\qquad+\frac{A(1+bN^\varphi)}{N^2(g_{\theta\theta}g_{\varphi\varphi})^{1/2}|\sin\beta|}[\cos(\alpha-\beta)-\nu][1-\nu\cos(\alpha-\beta)],\notag\\
&&\frac{d\alpha}{d\tau_U}=-\frac{\gamma\cos\alpha}{\nu}[a(u)^{r}+2\nu\cos\alpha\theta(u)^{r}_{\varphi}+\nu^2k_{\rm (lie)}(u)^{r}]\label{EoM2}\\
&&\qquad+\frac{A}{\nu}\frac{(1+bN^\varphi)[1-\nu\cos(\alpha-\beta)]}{N^2(g_{\theta\theta}g_{\varphi\varphi})^{1/2}|\sin\beta|}\sin(\beta-\alpha); \notag\\
&& U^r\equiv\frac{dr}{d\tau_U}=\frac{\gamma \nu\sin\alpha}{\sqrt{g_{rr}}},\label{EoM3}\\
&& U^\varphi\equiv\frac{d\varphi}{d\tau_U}=\frac{\gamma\nu\cos\alpha}{\sqrt{g_{\varphi\varphi}}}-\frac{\gamma N^\varphi}{N} \label{EoM4},
\end{eqnarray}
where $\Delta=r^2-2Mr+a^2$, and $b=\sqrt{g_{\varphi\varphi}}\cos\beta/[N(1-\nu_{\rm(s)}\cos\beta)]$ is the photon angular momentum with $\nu_{\rm (s)}=-2aM/(r\sqrt{\Delta})$. The test particle velocity $U^\alpha$ is obtained using Eq. (\ref{eq:U_split}), where the spatial velocity, $\nu^\alpha$, is decomposed in the adapted frame $\left\{\boldsymbol{e_t},\boldsymbol{e_r},\boldsymbol{e_\varphi},\boldsymbol{e_\theta}\right\}$ in this way: $\nu^\mu=\nu\sin\alpha\ e_r^\mu+\nu\cos\alpha\ e_\varphi^\mu$ (see Eq. (2.14) in Ref. \cite{Bini09}, for the explicit form).

It is important to note that Eq. (\ref{EoM1}), linked to the time component, is obtained from the orthogonality between $U^\alpha$ and $a(U)^\alpha$, that gives $a(U)^t=\nu[a(U)^r\sin\alpha+a(U)^\varphi\cos\alpha]$ for the acceleration (see Sec. \ref{sec:observer} and Eq. (2.27) in Ref. \cite{Bini09}), and the orthogonality between $U^\alpha$ and $\hat{V}(k,U)^\alpha$, that gives $\hat{V}(k,U)^t=\nu[\hat{V}(k,U)^r\sin\alpha+\hat{V}(k,U)^\varphi\cos\alpha]$ for the force (see Eq. (\ref{eq:k_particle}) and Eq. (2.27) in Ref. \cite{Bini09}. This condition underlines that the motion of the test particle, occurring in the equatorial plane around the central compact object, is determined classically and general relativistically by only two equations. However, $a(U)^t$ is present in the EoMs because it permits to determine the expression of $d\nu/d\tau_U$, that substituted in $a(U)^r$ (or equivalently in $a(U)^\varphi$) permits to get $d\alpha/d\tau_U$ (see Eq. (2.29) in Ref. \cite{Bini09}). The addition of $U^r$ and $U^\varphi$ allows to univocally determine the four parameters $(\nu,\alpha,r,\varphi)$, characterizing the test particle motion (see Sec. \ref{sec:GRtoC}, for a further discussion). Eqs. (\ref{EoM1})--(\ref{EoM2}) describe the dynamics, instead Eqs. (\ref{EoM3})--(\ref{EoM4}) relate the test particle velocity components with respect to the dynamical variables. 

\subsection{Lagrangian formulation of the PR effect}
\label{sec:GRPR} 
I determine the Lagrangian function and the Reyleigh dissipative function, that gives the EoMs (\ref{EoM1})--(\ref{EoM4}). I show, how the general relativistic formulation reduces to the classical case in the \emph{weak field limit} (mass over radius of the compact object tends to zero, $M/r\to0$, and velocities much lower than the speed of light, $v/c\ll1$).

\subsubsection{General relativistic Lagrangian}
\label{sec:GR_lagrangian}
The aim of this section is to find the Lagrangian function, $\mathcal{L}_{\rm GR}$, and the Reyleigh dissipative function, $\mathcal{F}_{\rm GR}$, such that the relative ELEs, Eqs. (\ref{eq:ELeq}) give the general relativistic PR EoMs (\ref{EoM1})--(\ref{EoM4}). It is important to underline that the Lagrangian used in classical mechanics is for discrete particles, each with a finite number of degrees of freedom; instead the one used in field theory is a Lagrangian density applied to continua and fields, which have an infinite number of degrees of freedom. In absence of radiation, i.e., $A=0$, the test particle motion becomes purely geodetic. Therefore, the Lagrangian function coincides with its kinetic part and it is straightforward determined \cite{Misner73,Chandrasekhar83,Stephani2003} as
\begin{equation} \label{eq:GR_lagrangian}
\mathcal{L}_{\rm GR}\equiv \mathbb{T}_{\rm GR}-\mathbb{V}_{\rm GR}=\frac{1}{2}g_{\alpha\beta}\,U^\alpha\, U^\beta+\frac{1}{2},
\end{equation}
where the test particle four velocity, $U^\alpha$, is expressed in the ZAMO reference frame, (see Eq. (2.29) in Ref. \cite{Bini09}). I note that Eq. (\ref{eq:GR_lagrangian}) includes also the contribution from the gravitational field, contained in the metric components $g_{\alpha\beta}$ (see Eq. (\ref{eq:lagrangian}) for comparison). The additive factor $1/2$ permits to obtain the gravitational potential in the weak field limit (see Sec. \ref{sec:GRtoC}). It is remarkable to note that such Lagrangian formalism is very general, because it can also be applied to a test particle moving outside the equatorial plane, since the $\theta$ velocity component, $U^\theta$, would be not null. 

\subsubsection{General relativistic PR Reyleigh dissipative function}
\label{sec:GR_potential}
When the radiation field is present, i.e., $A\neq0$, I need to determine the potential of the radiation force, $F_{\rm(rad)}(U)^\alpha$. Based on the same arguments of Sec. \ref{sec:classicPR}, I have to find the Reyleigh dissipative function, $\mathcal{F}_{\rm GR}$, such that $Q^\alpha=-\partial\mathcal{F}_{\rm GR}/\partial U^\alpha$. In order to calculate such potential I have to verify that the radiation differential form, whose components are the $F_{\rm(rad)}(U)^\alpha$, is exact, namely it admits a primitive. The domain, where the radiation field is defined, is all the equatorial plane minus the region occupied by the compact object including the event horizon, that seems to be not simply connected. However, this domain, transformed in polar coordinates, is a rectangle, defined by $\varphi\in[0,2\pi]$ and $r\in[2M,+\infty]$, see Sec. \ref{sec:classicPR}. Therefore, I have to take care to check if the radiation differential form is closed, i.e., $\partial F_{\rm(rad)}(U)_\alpha/\partial U_\lambda=\partial F_{\rm(rad)}(U)_\lambda/\partial U_\alpha$. Therefore, calculating the cross derivatives, I have
\begin{equation}
\begin{aligned}
\frac{\partial F_{\rm(rad)}(U)_\alpha}{\partial U_\lambda}&=T_{\alpha\lambda}+\delta_{\alpha\lambda} U_\beta T^{\beta\mu}U_\mu+U_\lambda T_\alpha^\mu U_\mu+U_\lambda U_\beta T_\alpha^\beta,\\
\frac{\partial F_{\rm(rad)}(U)_\lambda}{\partial U_\alpha}&=T_{\lambda\alpha}+\delta_{\lambda\alpha} U_\beta T^{\beta\mu}U_\mu+U_\alpha T_\lambda^\mu U_\mu+U_\alpha U_\beta T_\lambda^\beta.
\end{aligned}
\end{equation}
and then equating them, I have
\begin{equation} \label{eq:der_inc1}
U_\lambda\, T_\alpha^\mu\, U_\mu+U_\lambda\, U_\beta\, T_\alpha^\beta=U_\alpha\, T_\lambda^\mu\, U_\mu+U_\alpha\, U_\beta\, T_\lambda^\beta.
\end{equation}
Decomposing $T_{\alpha\beta}$ with respect to $U^\alpha$ as in Eq. (\ref{eq:k_particle}), Eq. (\ref{eq:der_inc1}) becomes:
\begin{equation} \label{eq:der_inc2}
U_\lambda\, k_\alpha=U_\alpha\, k_\lambda\quad \Leftrightarrow\quad U_\lambda\, \hat{V}_\alpha=U_\alpha\, \hat{V}_\lambda.
\end{equation}
If I multiply by scalar product both members of Eq. (\ref{eq:der_inc2}) for $U_\alpha$, I obtain $\hat{V}_\lambda=0$. This is a really strong condition, because it means that the radiation differential form is closed if and only if the radiation field is vanishing. 

An alternative way to find the Reyleigh dissipative function, $\mathcal{F}_{\rm GR}$, is in finding an integrating factor, $\mu=\mu(U)$, such that the new radiation differential form with components $\mu(U)\, F_{\rm(rad)}(U)^\alpha$, results to be exact. Calculating thus the cross derivatives and equating them, I have
\begin{equation} \label{eq:eq:der_inc2}
\begin{aligned}
0&=\left(-k^\alpha\frac{\partial\mu}{\partial U_\lambda}+k^\lambda\frac{\partial\mu}{\partial U_\alpha}\right)
+U^\alpha\left(\frac{\partial \mu}{\partial U_\lambda}k^\beta U_\beta+2\mu k^\lambda\right)\\
&-U^\lambda\left(\frac{\partial \mu}{\partial U_\alpha}k^\beta U_\beta+2\mu k^\alpha\right).
\end{aligned}
\end{equation}
Eq. (\ref{eq:eq:der_inc2}) reduces to solve the PDEs system
\begin{eqnarray} 
&&-k^\alpha\frac{\partial\mu}{\partial U_\lambda}+k^\lambda\frac{\partial\mu}{\partial U_\alpha}=0, \label{eq:difeqformu1}\\
&&\frac{\partial \mu}{\partial U_\lambda}k^\beta U_\beta+2\mu k_\lambda=0, \label{eq:difeqformu2}
\end{eqnarray}
whose general solution, using the separation of variables method, is given by
\begin{equation}
\mu=\frac{E_p^2}{E^2}.
\end{equation}
The moltiplicative factor $E_p^2$ permits to reduce to unity the integrating factor in the weak field limit (see Sec. \ref{sec:GRtoC}). The force field $F_{\rm(rad)}(U)^\alpha$ can be written equivalently as $(\nabla_{U^\alpha}\mathcal{F}_{\rm GR})/\mu$, preserving thus the dynamics described by EoMs (\ref{EoM1})--(\ref{EoM4}). The Reyleigh dissipative function does not depend on the photon geodesic structure, englobed in the stress-energy tensor $T_{\alpha\beta}$, because it is only function of the test particle velocity field $U^\alpha$. Therefore, this procedure can be also applied to more general photon geodesic emission, like for example for photons moving outside of the equatorial plane with variable $\theta$. I do not determine explicitly the functional form of the the general relativistic dissipative function $\mathcal{F}_{\rm GR}$, because it requires further analysis.

\subsubsection{Weak field approximation}
\label{sec:GRtoC}
I show how the general relativistic formalism in the weak field limit reduces to the classical case presented in Sec. \ref{sec:classicPR}. At this aim, I consider the test particle velocity and null geodesics in the Schwarzschild metric \cite{Misner73,Bini11},
\begin{eqnarray}
U^\alpha&=&\left(\frac{\gamma}{\sqrt{1-\frac{2M}{r}}},\gamma\nu\sin\alpha\sqrt{1-\frac{2M}{r}},\frac{\gamma\nu\cos\alpha}{r},0\right),\\
k_\alpha&=&E_p\left\{-1,\frac{\left[1-\frac{b^2}{r^2}\left(1-\frac{2M}{r}\right)\right]^{1/2}}{1-\frac{2M}{r}},b,0\right\},
\end{eqnarray}
where $E_p$ is the photon energy depending on $c$, and $b$ the photon impact parameter. I remind that in the weak field limit $r\to\infty$, $b\to0$, and $\nu/c\to 0$.

The general relativistic Lagrangian, given by Eq. (\ref{eq:GR_lagrangian}), reduces to the classical Lagrangian, given by Eq. (\ref{eq:lagrangian}), i.e., 
\begin{equation}
\mathcal{L}_{\rm GR}\approx -\frac{1}{2}\left(1-\frac{2M}{r}\right)+\frac{\nu^2}{2}+\frac{1}{2}=\frac{\nu^2}{2}+\frac{M}{r}\equiv\mathcal{L}_{\rm C},
\end{equation}
where in polar coordinates I have $\dot{r}=\nu\sin\alpha$ and $r\dot{\varphi}=\nu\cos\alpha$. The time component of the metric plus the additive factor gives the gravitational potential, instead the other components give the kinetic energy.

The integrating factor, $\mu$, does not figure in the classical case, because it reduces to unity. Indeed I have
\begin{equation}
\frac{E_p^2}{E^2}\approx\frac{E_p^2}{E^2_p(1-\dot{r})}\to1
\end{equation}
where $E(U)=-k_\alpha U^\alpha$, see Eq. (\ref{eq:k_particle}).
The integrating factor depends on the energy of the system, that in the newtonian case reduces to one.

The analysis of the equations $a(U)^\alpha=F_{\rm (rad)}(U)^\alpha$ in the weak field limit is very interesting, because it gives a better physical explanation of the involved terms and it is possible to stress the fundamental role played by the general relativistic effects. I have the following approximations in $a(U)^\alpha$ (see Eqs. (2.28) in Ref. \cite{Bini09}): $a(u)^r\approx M/r^2$, namely the gravitoelectric force corresponds to the gravitational force field; $k_{\rm (Lie)}(u)^r\approx -1/r+M/r^2$, where the first term represents the classical curvature radius, instead the second term, that does not figure in Eq. (\ref{eqm1}), is responsabile for the perihelion shift \cite[see Appendix in][for more details]{Bini09}; $\theta(u)^r_\varphi\approx0$, because the spacetime is flat, therefore there is no deformation of the geodesic flow. Approximating $F_{\rm (rad)}(U)^\alpha$ through linear terms in $(\dot{r},\dot{\varphi})$, I have: $F_{\rm (rad)}(U)^r\approx A(1-2\dot{r})/r^2$, $F_{\rm (rad)}(U)^\varphi\approx -A\dot{\varphi}/r$, and $F_{\rm (rad)}(U)^t\approx -A/r^2-A\nu/r^2$. Therefore I have that $a(U)^r=F_{\rm (rad)}(U)^r$ reduces to Eq. (\ref{eqm1}), instead $a(U)^\varphi=F_{\rm (rad)}(U)^\varphi$ reduces to Eq. (\ref{eqm2}), as I would have expected. I underline that in the general relativistic case I have adopted geometrical units, i.e., $c=G=1$, in order not to have missing terms and create confusion with respect to the classical case. It is remarkable to note that $a(U)^t=F_{\rm (rad)}(U)^t$ reduces to the following equation
\begin{equation}
\frac{d}{dt}\left(\frac{\nu^2}{2}+\frac{A-M}{r}\right)=-\frac{A}{r^2}\nu-\frac{A}{r^2}\dot{r}^2,
\end{equation}
describing the balance of the energy, where the right member represents the dissipated energy. In absence of the PR effect, i.e., $-\frac{A}{r^2}\nu-\frac{A}{r^2}\dot{r}^2=0$, or the radiation field, i.e., $A=0$, there is the conservation of the energy.
 
\subsection{Discussions}
\label{sec:conclusions}
In this work, I have developed the Lagrangian formulation of the general relativistic PR effect. The main challenges, that such work addresses and solves, are: the inverse problem, where the EoMs are given by Refs. \cite{Bini09,Bini11}, connected to the general relativistic radiation field including the PR effect, that is a dissipative force. A priori such problem might also not admit a Lagrangian formulation, due to the presence of a dissipative function \cite{Minguzzi2015}. In addition, another critical complication is the geometrical environment, constituted by a general stationary and axially symmetric spacetime, where the general relativistic effects contribute to make issue more thorny. This formulation may constitute an useful approach, among other existing methods, to investigate the general relativistic radiation fields including the PR effect. 

The importance to provide a Lagrangian formulation relies not only on a better understanding of the underlying physics, but also on a deeper analysis of the geometrical aspects and a simpler mathematical derivation of the EoMs, see Sec. \ref{sec:intro} and \ref{sec:GR_lagrangian}. This approach permitted also to have more insight in the radiation force itself and specifically in the PR effect, where the adding of an integrating factor, depending exponentially on the relativistic energy of the system, allow to identify the Reyleigh dissipative function, see Sec.\ref{sec:GR_potential}. The aim of such work is to furnish a complementary point of view in the study of the general relativistic PR effect with respect to the actual relativity of observer splitting formalism, increasing the interest on that subject and on the latter approach. In addition, comparing the classical and general relativistic Lagrangian formulations it is possible to recognize more clearly how GR influences the classical description, implying also an undimmed interpretations of the entailed variables, see Secs. \ref{sec:GR_formalism} and \ref{sec:GRtoC}. 

The results found in this paper pave principally the way at two compelling theoretical projects. The first one is into direction of improving the elementary description of the radiation field, with the inclusion of the PR effect, more adherent to describe the physical world (see e.g., Refs. \cite{Vaidya1951a,Vaidya1951b,Vaidya1973,Lindquist1965,Vaidya1999,Bini11}, for further details). Indeed, the Lagrangian approach permits to more easily derive the relative EoMs and investigate the relative dynamical systems. 

The second proposal is in the actual and highlighted scientific research line of the theoretical study of the gravitational waves. Indeed in the linearized theory of GR a localized source, that is losing energy, emits gravitational waves, because for the energy conservation it must counterbalance the energy carried off by the gravitational radiation (also known as gravitational radiation damping) \cite{Misner73}. This statement has been successfully confirmed by observations of the energy loss from the first discovered binary pulsar system PSR~B1913+16 \cite{Taylor1982} and the most recently observations from two merging BHs (see e.g., Refs. \cite{Abott2016a,Abott2016b}) and a binary NS inspiral \cite{Abbott2017}. There is a strong analogy between gravitational waves and PR effect, because both are dissipative effects in GR. The Lagrangian approach and the results presented in this paper might be a valuable instrument in terms of theoretical understanding and subsequent observational testability of the gravitational waves.

\section{Three-dimensional PR effect in GR}
In this work I investigate the three-dimensional (3D) motion of a test particle in a stationary, axially symmetric spacetime around a central compact object, under the influence of a radiation field. To this aim I extend the two-dimensional (2D) version of the PR effect in GR, see Sec. \ref{sec:GRmodel}. The radiation flux is modeled by photons which travel along null geodesics in the 3D space of a Kerr background and are purely radial  with respect to the ZAMO frames. The 3D general relativistic equations of motion that I derive are consistent with the classical (i.e. non-GR) description of the PR effect in 3D. The resulting dynamical system admits a critical hypersurface, on which radiation force balances gravity. Selected test particle orbits are calculated and displayed, and their properties described. It is found that test particles approaching the critical hypersurface at a finite latitude and with non-zero angular moment are subject to a latitudinal drift and asymptotically reach a circular orbit on the equator of the critical hypersurface, where they remain at rest with respect to the ZAMO. On the contrary, test particles that have lost all their angular momentum by the time they reach the critical hypersurface do not experience this latitudinal drift and stay at rest with respects to the ZAMO at fixed non-zero latitude. 

\subsection{Motivation}
\label{sec:intro}
The range of applications of PR drag to astrophysical problems has grown steadily since 80's coming to encompass also compact objects, especially NSs and BHs that accrete matter down to the very strong gravitational fields in their vicinity. For instance Walker \emph{et al.} \cite{Walker1989,Walker1992} studied the increase in mass accretion rate that is caused by PR drag when a bright thermonuclear flash occurs on the surface of a NS. This has motivated theory developments involving the PR effect in which GR is taken into account. 

Recent theoretical works on the extension of PR drag in GR have included studies of: test particle motion in the Vaidya spacetime \cite{Bini2011v} and around a slowly rotating relativistic star emitting isotropic radiation \cite{Oh2010}; the general relativistic PR effect on a spinning test particle \cite{Bini2010}; finite size effects \cite{Oh2011}, and the Lagrangian formulation of the general relativistic PR effect \cite{Defalco2018}. More astrophysically-oriented studies of PR effect in strong gravitational fields have concentrated on: the development of the {\it Eddington capture sphere concept} around luminous stars, the surface where gravity, radiation, and PR forces balance \cite{Wielgus2012,Stahl2012,Stahl2013,Wielgus2016,Wielgus2016n}; the {\it cosmic battery model} in astrophysical accretion discs \cite{Koutsantoniou2014,Contopoulos2015}; the dynamical evolution of accretion discs suddenly invested by a constant radiation filed (Bakala \emph{et al.}, 2018, A\&A submitted, \cite{Lancova2017}). Research in this area has acquired further momentum from the growing body of observational evidence for PR effect in matter motion around compact objects, especially accreting NSs undergoing thermonuclear flashes \cite{Ballantyne2004,Ballantyne2005,Worpel2013,Ji2014,Keek2014,Worpel2015,Keek2018}. 

Virtually all previous works on the general relativistic properties of the PR effect have been based upon a 2D model of the effect, i.e. planar (and arbitrarily oriented) orbits in spherically symmetric spacetimes (e.g., Schwarzschild's) and equatorial orbits in the (axially symmetric) Kerr metric. A necessary improvement consists in developing the 3D theory of the PR effect in GR. That would allow to investigate the motion of test particles immersed in non-spherically symmetric radiation fields (e.g., latitude-dependent fields) and/or orbiting away from the equatorial plane of the Kerr metric. That is the aim of the present study, which builds on the formalism developed in Refs. \cite{Bini09,Bini11}. My paper is structured as follows: in Sec. \ref{sec:GRPR} I generalise to the 3D case the previous 2D equations for the PR effect in a stationary and axially symmetric general relativistic spacetimes. I adopt a simple prescription for the radiation, namely a field with zero angular momentum. In Sec. \ref{sec:critc_rad} I define the critical hypersurface on which radiation force balances gravity and discuss its salient features.  In Sec. \ref{sec:orbits} I present calculations of selected orbits in the Schwarzschild and Kerr spacetimes; my concluding remarks are in Sec. \ref{sec:end}. 

\subsection{Scenario and spacetime geometry}
\label{sec:GRPR}
\begin{figure*}[t]
\centering
\includegraphics[scale=0.4]{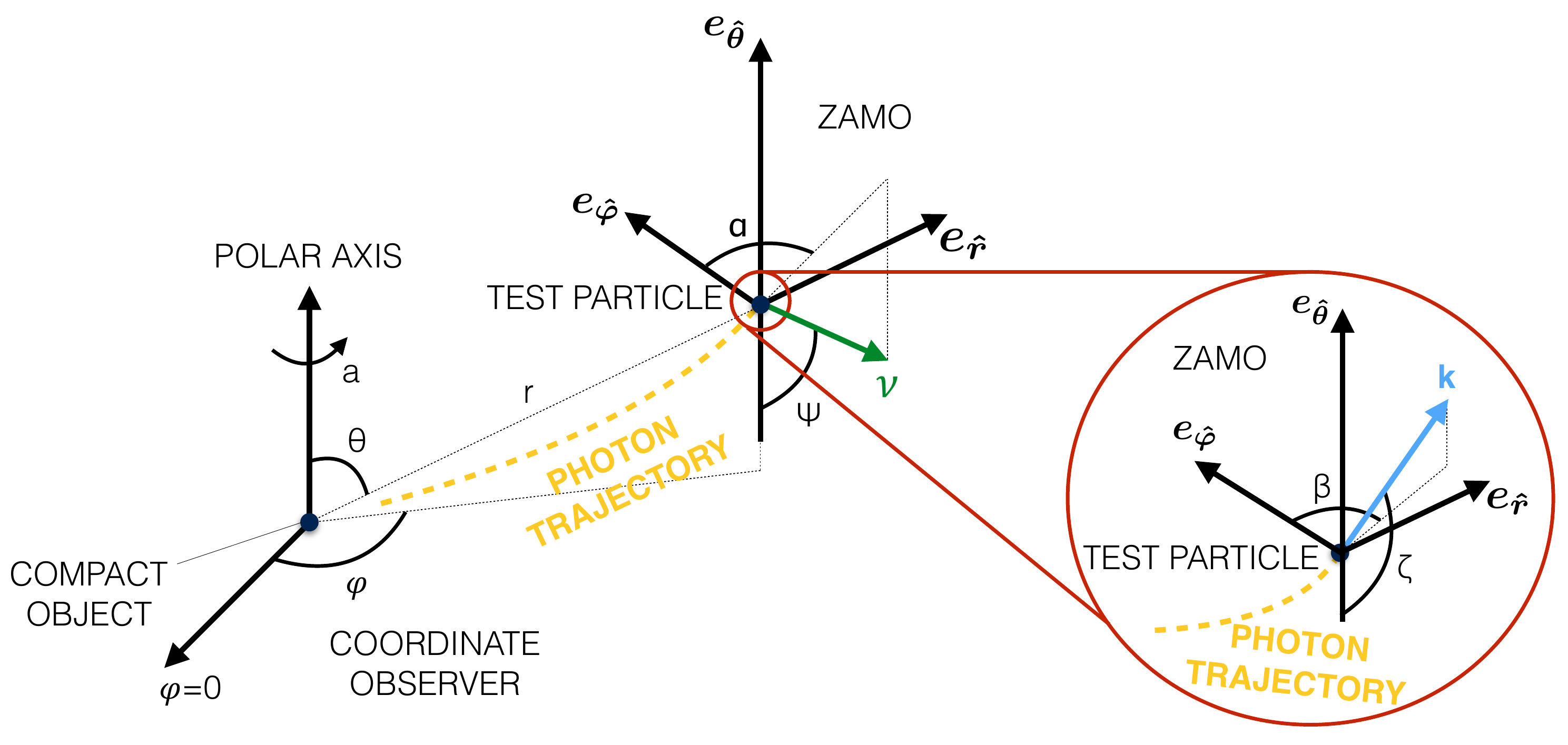}
\caption{Visual representation of the radiation field-test particle interaction geometry in the Kerr metric. The spatial location of the test particle is described by Boyer-Linquist coordinates $\left\{\boldsymbol{r},\boldsymbol{\theta},\boldsymbol{\varphi}\right\}$. The ZAMO local frame is $\left\{\boldsymbol{e_{\hat{t}}},\boldsymbol{e_{\hat{r}}},\boldsymbol{e_{\hat{\theta}}},\boldsymbol{e_{\hat{\varphi}}}\right\}$. The photons of the radiation field travel along null geodesics of the background spacetime with four-momentum $\boldsymbol{k}$.  Two photon impact parameters, $b$ and $q$ are related respectively to the two angles $\beta$ and $\zeta$, formed in the local ZAMO frame. The test particle moves in the 3D space with a velocity $\nu$, forming the azimuthal, $\alpha$, and polar, $\psi$, angles in the local ZAMO frame.}
\label{fig:Fig1}
\end{figure*}

The scenario for the description of the interaction between the radiation field and the motion of a test particle in the extreme gravitational field of a BH, or a NS, is constituted as follows: I consider the radiation field coming from an emitting region, located outside of the event horizon. The test particle motion is determined by its position in spherical coordinates and its velocity field in the ZAMO frame. The photon four-momentum is described by a pair of polar coordinates (see Fig. \ref{fig:Fig1}). In order to derive such set of equations I compute first the quantities in the ZAMO frame and then I transform them in the static observer frame. To deal with the relative motion of two non-inertial observers in GR I use the relativity of observer splitting formalism.

I consider a central compact object (BH or NS), whose outside spacetime is described by the Kerr metric with signature $(-,+,+,+)$ \cite{Kerr1963}. In geometrical units ($c = G = 1$), the line element of the Kerr spacetime in Boyer-Lindquist coordinates, parameterized by mass $M$ and spin $a$, reads as \cite{Boyer1967}
\begin{equation}\label{kerr_metric}
\begin{aligned}
\mathrm{d}s^2 &=-\left(1-\frac{2Mr}{\Sigma}\right) \,\mathrm{d}t^2-\frac{4Mra}{\Sigma} \sin^2\theta\,\mathrm{d}t\, \mathrm{d}\varphi\\
&+\frac{\Sigma}{\Delta}\, \mathrm{d}r^2+\Sigma \,\mathrm{d}\theta^2+\rho\sin^2\theta\, \mathrm{d}\varphi^2, 
\end{aligned}
\end{equation}
where $\Sigma \equiv r^{2} + a^{2}\cos^{2}\theta$, $\Delta \equiv r^{2} - 2Mr + a^{2}$, and $\rho  \equiv r^2+a^2+2Ma^2r\sin^2\theta/\Sigma$. The determinant of the Kerr metric is $g=\sqrt{(\Delta/\rho)g_{rr}g_{\varphi\varphi}g_{\theta\theta}}\equiv-\Sigma^2\sin^{2}\theta$.

\subsubsection{ZAMO frame}
ZAMOs are dragged by the rotation of the spacetime with angular velocity $\Omega_{\mathrm{ZAMO}}=-g_{\phi t}/g_{\phi \phi}$, while their radial coordinate remains constant. The four-velocity $\boldsymbol{n}$ of ZAMOs is the future-pointing unit normal to the spatial hypersurfaces, i.e. \cite{Bini1997a,Bini1997b,Bini09,Bini11},
\begin{equation}
\label{n}
\boldsymbol{n}=\frac{1}{N}(\boldsymbol{\partial_t}-N^{\varphi}\boldsymbol{\partial_\varphi})\,,
\end{equation}
where $N=(-g^{tt})^{-1/2}$ is the time lapse function, $g^{tt}=g_{\varphi\varphi}/(g_{tt}g_{\varphi\varphi}-g_{t\varphi}^2)$, and $N^{\varphi}=g_{t\varphi}/g_{\varphi\varphi}$ the spatial shift vector field. I focus my attention on the region outside the event horizon, where the time coordinate hypersurfaces are spacelike, i.e., $g^{tt}<0$. An orthonormal frame adapted to the ZAMOs is given by \cite{Bardeen1972}
\begin{equation} \label{eq:zamoframes}
\begin{aligned}
&\boldsymbol{e_{\hat t}}=\boldsymbol{n},\quad
\boldsymbol{e_{\hat r}}=\frac1{\sqrt{g_{rr}}}\boldsymbol{\partial_r},\\
&\boldsymbol{e_{\hat \theta}}=\frac1{\sqrt{g_{\theta \theta }}}\boldsymbol{\partial_\theta},\quad
\boldsymbol{e_{\hat \varphi}}=\frac1{\sqrt{g_{\varphi \varphi }}}\boldsymbol{\partial_\varphi}.
\end{aligned}
\end{equation}
The relative dual tetrad of one-forms is given by
\begin{equation} \label{eq:zamoframed}
\begin{aligned}
&\boldsymbol{\omega^{{\hat t}}}= N dt,\quad \boldsymbol{\omega^{{\hat r}}} =\sqrt{g_{rr}} dr,\\
&\boldsymbol{\omega^{{\hat \theta}}}= \sqrt{g_{\theta \theta }} d\theta,\quad
\boldsymbol{\omega^{{\hat \varphi}}}=\sqrt{g_{\varphi \varphi }}(d \varphi+N^{\varphi} dt).
\end{aligned}
\end{equation}
All the indices associated to the ZAMO frame will be labeled by a hat, instead all the quantities measured in the ZAMO frame will be followed by $(n)$. 

\subsubsection{ZAMO kinematical quantities}
Since the accelerated ZAMOs are locally nonrotating, their vorticity vector $\boldsymbol{\omega}(n)$ vanishes, but they have a nonzero expansion tensor $\boldsymbol{\theta}(n)$. For this reason it is more convenient to use the Lie transport (see \cite{Bini1997a,Defalco2018}, for further details). The nonzero ZAMO kinematical quantities (i.e., acceleration $\boldsymbol{a}(n)=\nabla_{\boldsymbol{n}} \boldsymbol{n}$, expansion tensor along the $\hat{\varphi}$-direction $\boldsymbol{\theta_{\hat\varphi}}(n)$, also termed shear vector, and the relative Lie curvature vector $\boldsymbol{k}_{(\rm Lie)}(n)$) have only nonzero components in the $\hat{r}-\hat{\theta}$ plane of the tangent space \cite{Bini09,Bini11,Defalco2018}:
\begin{equation}\label{accexp}
\begin{aligned}
\boldsymbol{a}(n) & = a(n)^{\hat r}\, \boldsymbol{e_{\hat r}} + a(n)^{\hat\theta}\, \boldsymbol{e_{\hat\theta}}\\
&=\partial_{\hat r}(\ln N)\, \boldsymbol{e_{\hat r}} + \partial_{\hat\theta}(\ln N)\, \boldsymbol{e_{\hat\theta}},\\
\boldsymbol{\theta_{\hat\varphi}}(n)& = \theta(n)^{\hat r}{}_{\hat\varphi}\, \boldsymbol{e_{\hat r}} + \theta(n)^{\hat\theta}{} _{\hat\varphi}\, \boldsymbol{e_{\hat \theta}}\\
&= -\frac{\sqrt{g_{\varphi\varphi}}}{2N}\,(\partial_{\hat r} N^\varphi\, \boldsymbol{e_{\hat r}} + \partial_{\hat\theta} N^\varphi\, \boldsymbol{e_{\hat \theta}}),\\
\boldsymbol{k}_{(\rm Lie)}(n)& =  k_{(\rm Lie)}(n)^{\hat r}\, \boldsymbol{e_{\hat r}} + k_{(\rm Lie)}(n)^{\hat\theta}\, \boldsymbol{e_{\hat\theta}}\\
& = -[\partial_{\hat r}(\ln \sqrt{g_{\varphi\varphi}})\, \boldsymbol{e_{\hat r}} + \partial_{\hat\theta}(\ln \sqrt{g_{\varphi\varphi}})\, \boldsymbol{e_{\hat\theta}}].
\end{aligned}
\end{equation}
In Table \ref{tab:ZAMOq} I summarize the expressions of such quantities for the Kerr spacetime. 
\renewcommand{\arraystretch}{1.8}
\begin{table*}
\begin{center}
\caption{\label{tab:ZAMOq} Explicit expressions of metric and ZAMO kinematical quantities for the Kerr metric.}	
\normalsize
\scalebox{0.77}{
\begin{tabular}{l  c} 
\ChangeRT{1pt}
{\bf Metric quantity} & {\bf Explicit expression} \\
\ChangeRT{1pt}
$N=(-g^{tt})^{-1/2}$ & $[\Delta /\rho]^{1/2}$   \\
\hline
$N^\varphi=g_{t\varphi}/g_{\varphi\varphi}$ & $-2Mar/[\Sigma \rho]$ \\
\ChangeRT{1pt}
{\bf ZAMO quantity} & {\bf Explicit expression} \\
\ChangeRT{1pt}
\hline
\hline
\multicolumn{2}{c}{\emph{Radial components}}\\
\hline
\hline
$a(n)^{\hat r}$& $M/[\rho\sqrt{\Sigma^5\Delta}]\left\{\Sigma^2(r^2-a^2)+a^2\sin^2\theta[r^2(3r^2-4Mr+a^2)+a^2\cos^2\theta(r^2-a^2)]\right\}$   \\
\hline
$\theta(n)^{\hat r}{}_{\hat\varphi}$& $aM\sin\theta[(r^2+a^2)(\Sigma-2r^2)-2r^2\Sigma]/[\rho\sqrt{\Sigma^5}]$\\
\hline
$k_{(\rm Lie)}(n)^{\hat r}$ & $-\sqrt{\Delta/\Sigma^5}[r\Sigma^2+a^2M\sin^2\theta(\Sigma-2r^2)]/\rho$\\
\hline
\hline
\multicolumn{2}{c}{\emph{Polar components}}\\
\hline
\hline
$a(n)^{\hat\theta}$ & $-a^2rM\sin(2\theta)[r^2+a^2]/[\rho\sqrt{\Sigma^5}]$\\
\hline
$\theta(n)^{\hat\theta}{} _{\hat\varphi}$ & $a^2rM\sin(2\theta)\sin\theta\sqrt{\Delta}/[\rho\sqrt{\Sigma^5}]$\\
\hline
$k_{(\rm Lie)}(n)^{\hat\theta}$ & $-\sin(2\theta)[(r^2+a^2)(2a^2rM\sin^2\theta+\Sigma^2)+2a^2rM\Sigma\sin^2\theta]/[2\rho\sqrt{\Sigma^5}\sin^2\theta]$\\
\ChangeRT{1pt}
\end{tabular}}
\end{center}
\end{table*}

\subsubsection{Radiation field}
\label{sec:phot}
The stress-energy tensor, describing the radiation field, is modeled as a coherent flux of photons traveling along null geodesics in the Kerr geometry and acting on the test particle in the following manner \cite{Robertson37,Bini09,Bini11}
\begin{equation}
\label{STE}
T^{\mu\nu}=\Phi^2 k^\mu k^\nu\,,\qquad k^\mu k_\mu=0,\qquad k^\mu \nabla_\mu k^\nu=0,
\end{equation}
where parameter $\Phi$ is related to the intensity of the radiation field and $\boldsymbol{k}$ is the four-momentum field describing the null geodesics. The photon four-momentum, $\boldsymbol{k}$, and the photon spatial unit relative velocity with respect to the ZAMOs, $\boldsymbol{\hat{\nu}}(k,n)$, are respectively given by
\begin{equation} \label{photon}
\begin{aligned}
\boldsymbol{k}&=E(n)[\boldsymbol{n}+\boldsymbol{\hat{\nu}}(k,n)],\\
\boldsymbol{\hat{\nu}}(k,n)&=\sin\zeta\sin\beta\ \boldsymbol{e_{\hat r}}+\cos\zeta\ \boldsymbol{e_{\hat\theta}}+\sin\zeta \cos\beta\ \boldsymbol{e_{\hat\varphi}},
\end{aligned}
\end{equation}
where $\beta$ and $\zeta$ are the two angles in the azimuthal and polar direction, respectively (see Fig. \ref{fig:Fig1}). The case of $\sin \beta >0$ corresponds to an outgoing photon beam (increasing $r$) while the case of $\sin \beta <0$ corresponds to an incoming photon beam (decreasing $r$, see Fig. \ref{fig:Fig1}). The photon four-momentum in the background Kerr geometry is identified by two impact parameters $(b,q)$, which are associated with two emission angles $(\beta,\zeta)$, respectively.

Using Eq. (\ref{photon}), the photon energy with respect to the ZAMO, $E(n)$, is expressed in the frame of a distant static observer by
\begin{equation} \label{energyZAMO}
\begin{aligned}
E(n)&=-\boldsymbol{k}(n)\cdot \boldsymbol{n}=-\boldsymbol{k}\cdot\frac{1}{N}\left(\boldsymbol{\partial_t}-N^{\varphi}\boldsymbol{\partial_\varphi}\right)=\frac{E+L_zN^\varphi}{N}\\
&=\frac{E}{N}(1+bN^\varphi),
\end{aligned}
\end{equation}
where $E=-k_t>0$ is the conserved photon energy, $L_z=k_\varphi$ is the conserved angular momentum along the polar $z$ axis orthogonal to the equatorial plane, and $b\equiv -k_{\phi}k_{t}=L_z/E$ is the first (azimuthal) photon impact parameter (constant of motion) \cite{Carter1968}; note that all these quantities are measured by a distant static observer \cite{Bini11}. 

This impact parameter is associated with the relative azimuthal angle $\beta$, measured in the ZAMO frame \cite{Bini11} (see Fig. \ref{fig:Fig1}). The angular momentum along the polar $\hat{\theta}$-axis in the ZAMO frame, $L_z(n)$, is expressed in the distant static observer frame by
\begin{equation} \label{ang1}
\begin{aligned}
E(n)\cos\beta\sin\zeta=L_z(n)=\boldsymbol{k}(n)\cdot\boldsymbol{e_{\hat \varphi}}=\boldsymbol{k}\cdot\frac{\boldsymbol{\partial_\varphi}}{\sqrt{g_{\varphi\varphi}}}=\frac{L_z}{\sqrt{g_{\varphi\varphi}}}
\end{aligned}
\end{equation}
From such equation, I obtain
\begin{equation} \label{ANG1}
\begin{aligned}
\cos\beta&=\frac{b E}{\sin\zeta\sqrt{g_{\varphi\varphi}}E(n)}=\frac{L_zN}{\sin\zeta\sqrt{g_{\varphi\varphi}}(E+L_zN^\varphi)}\\
&=\frac{b N}{\sin\zeta\sqrt{g_{\varphi\varphi}}(1+b N^\varphi)}.
\end{aligned}
\end{equation}
An equation for $\zeta$ is needed to completely determine $\beta$.

The photon specific four-momentum components in the Kerr geometry are given by \cite{Chandrasekhar83}
\begin{align}
     k^{t} &= \Sigma^{-1} \left(a\,b-a^2\sin^{2}\theta+(r^2+a^2)P\,\Delta^{-1}\right)\,,\nonumber\\
     k^{r} &=  s_r\Sigma^{-1} \sqrt{R_{b,q}(r)}\,,\label{CarterEQs}\\
     k^{\theta} &=  s_{\theta}\Sigma^{-1} \sqrt{\Theta_{b,q}(\theta)}\,,\nonumber\\
     k^{\varphi} &=  \Sigma^{-1}\left(b\,\mathrm{cosec}^{2} \theta-a+a\,P\,\Delta^{-1}\right)\,,\nonumber
\end{align} 
where $P\equiv r^{2} + a^{2}-b\,a$, and the pair of signs $s_{r}$, $s_{\theta}$ describes the orientation of the radial and latitudinal evolution, respectively \cite{Carter1968}. The radial and latitudinal effective potentials are respectively \cite{Chandrasekhar83}:
\begin{align}
       R_{b,q} \left( r \right) &=  \left( r^{2} + a^{2}
 - a b \right) ^{2}
       - \Delta \left[ q + \left( b - a \right) ^{2} \right]\,,\label{Rpot} \\
       \Theta_{b,q} \left( \theta \right) &= q + a^{2} \cos^{2}
\theta -b^{2} \mathrm{cot}^{2} \theta\,.\label{Thetapot}
\end{align} 
Here, $q$ is the second (latitudinal) photon impact parameter (constant of motion) related to the covariant components of the photon four-momentum through the relation \cite{Chandrasekhar83}
 \begin{equation} \label{q_def2}
q \equiv \left(\frac{k_{\theta}}{k_{t}}\right)^{2} + \left[b\tan\left(\frac{\pi}{2} - \theta\right)\right]^{2} - a^{2}\cos^{2}\theta\,.    
 \end{equation}\\

\noindent
\textbf{Impact parameters}\\ \\
I consider here a radiation field which consists of photons moving in a purely radial direction at the infinity (this physically admissible because of the asymptotic flatness of the Kerr spacetime). In this case I have
\begin{equation}
\boldsymbol{k}=\boldsymbol{\partial_t}+\boldsymbol{\partial_r}\,. \label{infty_mom}
\end{equation}
From Eq. (\ref{infty_mom}) I note that the azimuthal component of the four-momentum, $k^\varphi$, vanishes at infinity. Moreover, in order to simplify the calculations I assume that the azimuthal impact parameter of the radiation field, $b$, takes null value, i.e., $b = 0$. The latitudinal impact parameter, $q$, can be calculated from the condition
\begin{equation} \label{cond2}
\Theta_{b=0,q}( \theta)=0,
\end{equation}
which results from the absence of latitudinal photon motion ($k^{\theta}=0$). From Eqs. (\ref{Thetapot}) and (\ref{cond2}) I can express  $q$ as a function of the polar angle $\theta$:
\begin{equation} \label{q_r}
q=- a^{2} \cos^{2}\theta\,. 
\end{equation}
This is possible because the latitudinal potential, Eq. (\ref{Thetapot}), is independent of the radial coordinate and therefore the polar angle $\theta$ along a given photon trajectory is conserved.  Photons with a given value of $q$ move only in the radial and azimuthal directions on the surface of the cone with the vertex located in the coordinates origin and with the vertex angle  $\theta$ given by Eq. (\ref{q_r}). Note that the azimuthal motion on finite values of the radial coordinate is caused only by frame dragging.

The above-defined radiation field significantly simplifies the integration of test particle trajectories in that only a single photon beam, described by the constants of motion $b=0,\ q=- a^{2} \cos^{2}\theta$, must be considered at the test particle position. In such case, the radial potential, Eq. ({\ref{Rpot}}), is always positive above the event horizon: this proves that the radiation field reaches every positions (for all $r$ and $\theta$) above the event horizon. The second constant of motion $q$ ranges in the interval $[-a^2, 0]$.  The value $q=0$ corresponds to the motion of photons in the equatorial plane, while the value of $q=  -a^2$ corresponds to the motion of photons along the polar axis on the south or north directions. I note also that since $q \leq 0$ radiation field photons can never cross the equatorial plane.

The  local components of the photon four-momentum in the ZAMO frame are obtained through the following transformation:
\begin{equation}
k^{\hat{\mu}} = \omega^{\hat{\mu}}_{\,\,\,\alpha}\,k^{\alpha},
\end{equation}
where $\omega^{\hat{\mu}}_{\,\,\,\alpha}$ represents the transformation matrix from the holonomic basis $\boldsymbol{\partial_\alpha}$ to the anholomic (tetrad) basis $\boldsymbol{e_{\hat\alpha}}$, see Eq. (\ref{eq:zamoframes}) for determining its components. The local polar direction $\zeta$ of the photon four-momentum is given by (see Fig. \ref{fig:Fig1})
\begin{equation} \label{angle1}
\cos \zeta =-\frac{k^{\hat{\theta}} }{k^{\hat{t} }}.
\end{equation}
For the considered radiation field ($q=- a^{2} \cos^{2}\theta\,,k^{\theta}=0$) I simply obtain
\begin{equation} \label{angle2}
k^{\hat{\theta}} = \omega^{\hat{\theta}}_{\,\,\,\theta}\,k^{\theta}=0.
\end{equation}
Consequently from Eqs. (\ref{angle1}) and (\ref{angle2}), the local polar direction of the radiation field photons in the ZAMO frame is always $\zeta=\pi/2$. From Eq. (\ref{ang1}) and $b=0$, the local azimuthal direction $\beta$ of the photon four-momentum is $\cos\beta=0$ (see Fig. \ref{fig:Fig1}). Therefore, the local azimuthal angle of the test field photons in the  ZAMO frame always take the value of $\beta=\pi/2$. I can conclude that in all ZAMO frames radiation field photons move in a purely radial direction. 
The source of the radiation field can thus be considered as centered in the coordinate origin, (differentially) rotating with a latitude-dependent angular velocity $\Omega_{\mathrm{ZAMO}}$ and emitting photons only along the radial direction in the appropriate, locally co-moving ZAMO frame. \\

\noindent
\textbf{Intensity  parameter}\\ \\
Since the photon four-momentum $\boldsymbol{k}$ is completely determined by  $(b\,,q)$, the coordinate dependence of $\Phi$ then follows from the conservation equations $\nabla_{\beta}T^{\alpha\beta}=0$. Exploiting the absence of photon latitudinal motion ($k^{\theta}=0$) and symmetries of the Kerr spacetime, these can be written as
\begin{equation}
\label{flux_cons}
\begin{aligned}
0&=\nabla_\beta (\Phi^2 k^\beta)=\frac{1}{\sqrt{-g}}\partial_\beta (\sqrt{-g}\,\Phi^2 k^\beta)=\partial_r(\sqrt{-g}\,\Phi^2 k^r).
\end{aligned}
\end{equation}
Therefore, I have 
\begin{equation}
\label{cons_condition}
\begin{aligned}
\sqrt{-g}\,\Phi^2 k^r&=NE(n)\sqrt{g_{\varphi\varphi}g_{\theta\theta}}\sin\zeta\sin\beta=\hbox{const}=E\Phi^2_0,  
\end{aligned}
\end{equation}
where $\Phi_0$ is a new constant related to the intensity of the radiation field at the emitting surface. This equation, however, does not fix the intensity parameter unambiguously. In fact, the conservation equations will be fulfilled, even if I multiply the constant expression $E\Phi_0^2$ by an arbitrary function of the $\theta$ coordinate. Thus this condition determines the class of radiating fields that differ from one another by the latitudinal dependence of the intensity. A radiation field which is independent of latitude (and whole intensity parameter is thus independent of $\theta$) is a natural choice, especially in the Schwarzschild limit because of its spherical symmetry. This can easily achieved by multiplying (\ref{cons_condition}) by a factor $\sin\theta$, such that the intensity parameter becomes 
\begin{equation} \label{eq:int}
\Phi^2=\frac{\Phi_0^2\sin\theta}{\sqrt{g_{\varphi\varphi}g_{\theta\theta}}}\equiv \frac{\Phi_0^2}{\sqrt{(r^2+a^2)^2-a^2\,\Delta\,\sin^{2}\theta}},
\end{equation}
where I have used Eqs. (\ref{ang1}) and (\ref{ANG1}), together with the fact that  $N|b\tan\beta|=\sin\zeta\sqrt{g_{\varphi\varphi}}$ for $b=0$. In a Schwarzschild spacetime limit Eq. (\ref{eq:int}) thus reads
\begin{equation}
\Phi^2=\frac{\Phi_0^2}{r^2}\,,
\end{equation}
which matches the spacetime spherical symmetry. 

\begin{figure*}[th!]
	\centering
	\hbox{
		\includegraphics[scale=0.27]{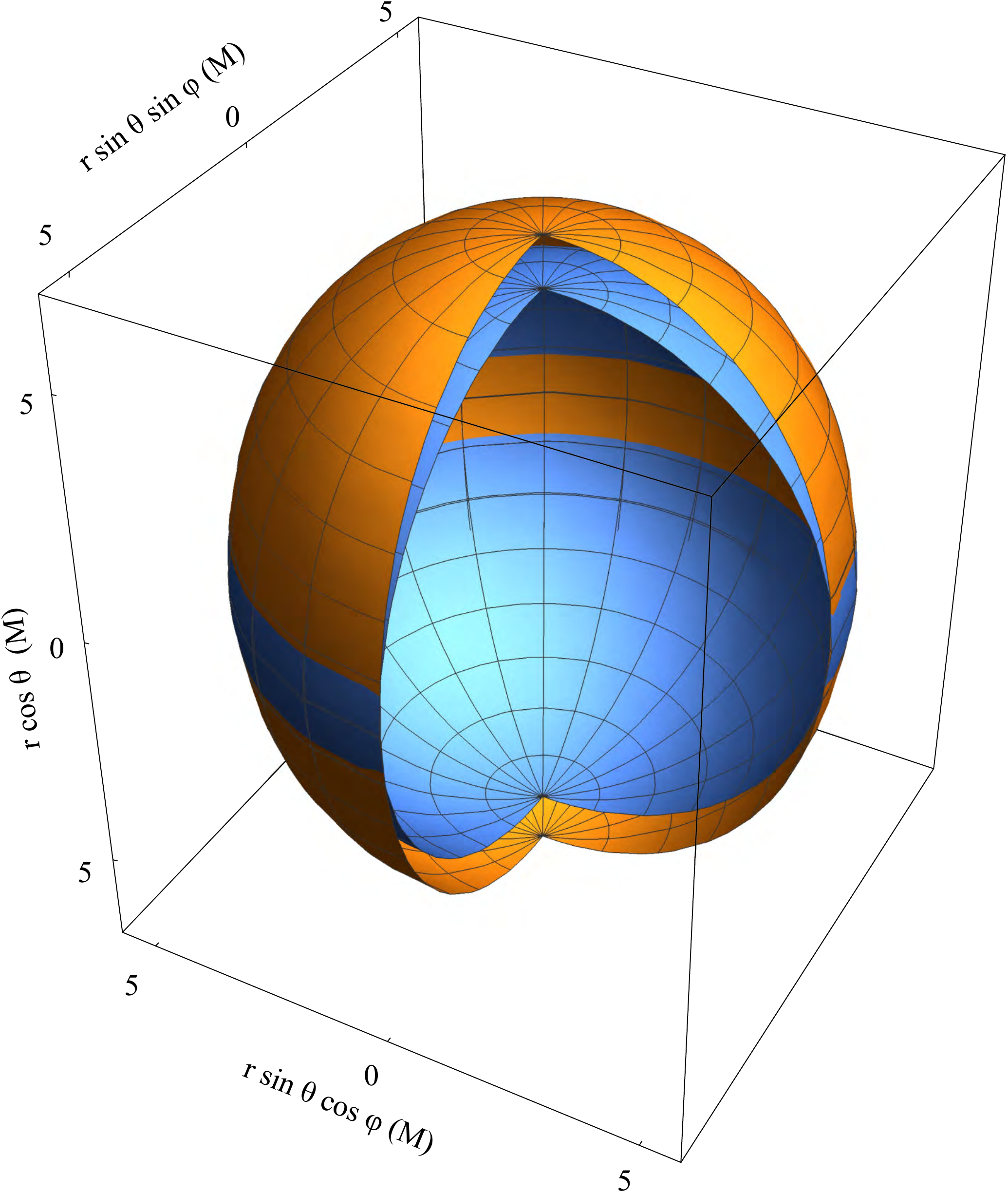}
		\hspace{0.5cm}
		\includegraphics[scale=0.19]{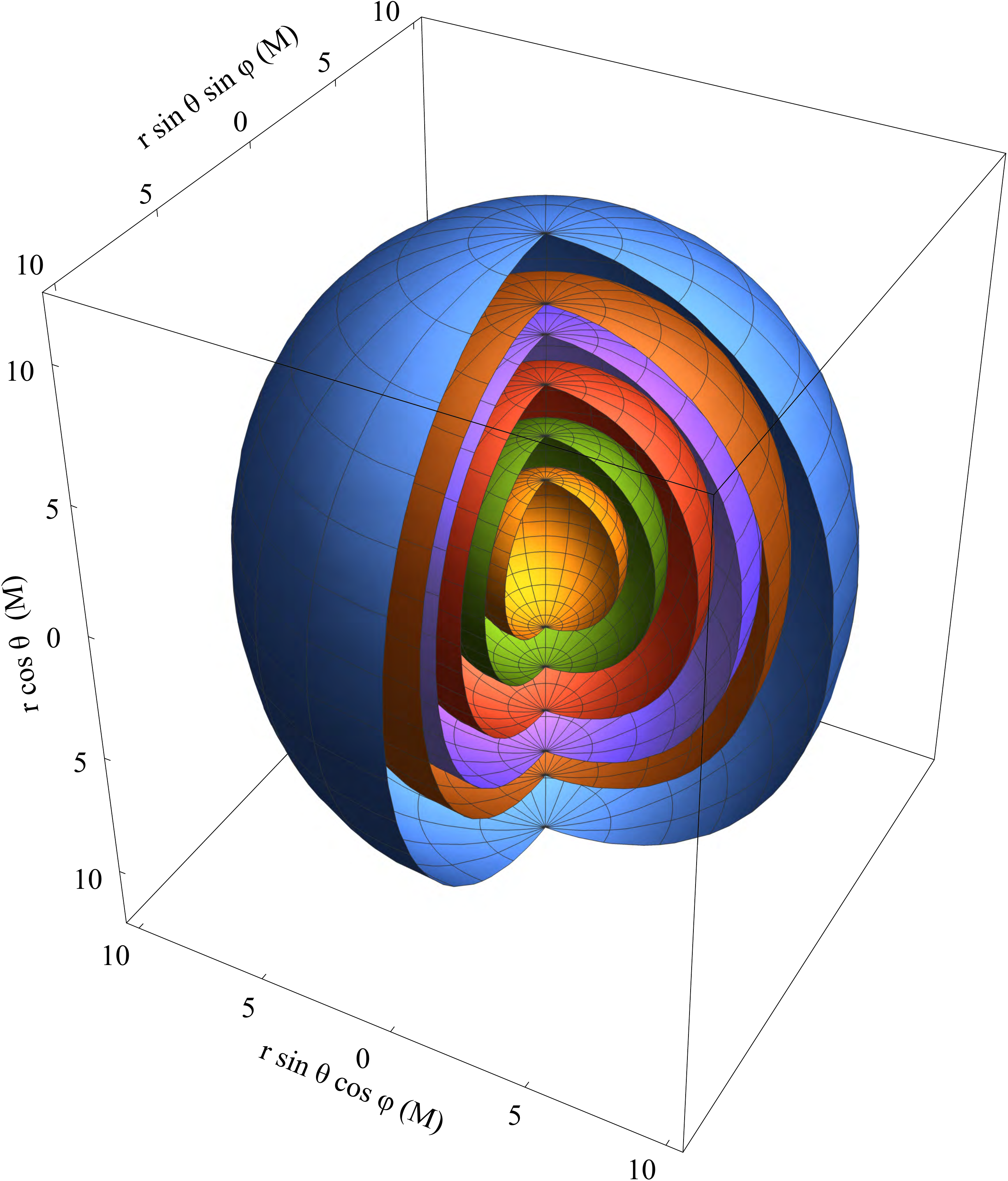}}
	\caption{Left panel: Critical hypersurfaces for the case of high spin $a=0.9995$ (orange) and the case of the Schwarzschild spacetime with $a=0$ (blue). For the Schwarzschild case the critical radius is $r_{\rm (crit)} \sim 5.56M$, while for the Kerr case in the equatorial plane is $r^{\rm eq}_{\rm(crit)} \sim 5.52M$ and $r^{\rm pole}_{\rm(crit)} \sim 6.56M$ at the poles. The relative luminosity of the radiating field takes the value of $A=0.8$. Right panel: Critical hypersurfaces for the values of the relative luminosity $A=0.5, \,0.7, \,0.8, \,0.85, \,0.87, \,0.9$ at a constant spin $a=0.9995$. The respective critical radii in the equatorial plane are $r^{\rm eq}_{\rm(crit)} \sim 2.71M,  4.01M, 5.52M, 7.04M, 7.99M, 10.16M$, while at poles they are $r^{\rm pole}_{\rm(crit)} \sim 2.97M,  4.65M,  6.56M,  8.38M,  9.48M, 11.9M$.}
	\label{fig:Fig2}
\end{figure*}

\subsubsection{Test particle motion}
\label{sec:test_part}
I consider a test particle moving in the 3D space, with four-velocity $\bold{U}$ and spatial three-velocity with respect to the ZAMOs, $\boldsymbol{\nu}(U,n)$:
\begin{eqnarray} 
\boldsymbol{U}&=&\gamma(U,n)[\boldsymbol{n}+\boldsymbol{\nu}(U,n)], \label{testp1}\\
\boldsymbol{\nu}(U,n)&=&\nu^{\hat r}\ \boldsymbol{e_{\hat r}}+\nu^{\hat\varphi}\ \boldsymbol{e_{\hat \varphi}}+\nu^{\hat\theta}\ \boldsymbol{e_{\hat\theta}} \label{testp2}\\ 
&=&\nu\sin\psi\sin\alpha\ \boldsymbol{e_{\hat r}}+\nu\cos\psi\ \boldsymbol{e_{\hat\theta}}+\nu\sin\psi \cos\alpha\ \boldsymbol{e_{\hat\varphi}}\nonumber,\\ \nonumber
\end{eqnarray}
where $\gamma(U,n)=1/\sqrt{1-||\boldsymbol{\nu}(U,n)||^2}$ is the Lorentz factor (see Fig. \ref{fig:Fig1}). I use the following abbreviated notations $\nu^{\hat \alpha}=\nu(U,n)^{\hat \alpha}$, $\nu=||\boldsymbol{\nu}(U,n)||\ge0$, $\gamma(U,n) =\gamma$ throughout this paper. I have that $\nu$ represents the magnitude of the test particle spatial velocity $\boldsymbol{\nu}(U,n)$, $\alpha$ is the azimuthal angle of the vector $\boldsymbol{\nu}(U,n)$ measured clockwise from the positive $\hat \varphi$ direction in the $\hat{r}-\hat{\varphi}$ tangent plane in the ZAMO frame, and $\psi$ is the polar angle of the vector $\boldsymbol{\nu}(U,n)$ measured from the axis orthogonal to the $\hat{r}-\hat{\varphi}$ tangent plane of the ZAMO frame (see Fig. \ref{fig:Fig1}). The explicit expression for the test particle velocity components with respect to the ZAMOs are \cite{Bini09,Bini11}:
\begin{equation} 
\begin{aligned}
&U^t\equiv \frac{dt}{d\tau}=\frac{\gamma}{N},\quad U^r\equiv \frac{dr}{d\tau}=\frac{\gamma\nu^{\hat r}}{\sqrt{g_{rr}}},\\
&U^\theta\equiv \frac{d\theta}{d\tau}=\frac{\gamma\nu^{\hat\theta}}{\sqrt{g_{\theta\theta}}},\quad U^\varphi\equiv \frac{d\varphi}{d\tau}=\frac{\gamma\nu^{\hat\varphi}}{\sqrt{g_{\varphi\varphi}}}-\frac{\gamma N^\varphi}{N},
\end{aligned}
\end{equation}
where $\tau$ is the proper time parameter along $\bold{U}$. \\

\noindent
\textbf{Relativity of observer splitting formalism}\\ \\
\label{sec:rosf} 
The acceleration of the test particle relative to the ZAMO congruence, $\boldsymbol{a}(U)=\nabla_{\bold U} \bold{U}$, is given  by the formula (see Eq. (29) in \cite{Defalco2018} and references therein) \footnote{A complementary approach to the relativity of observer splitting formalism is the general relativistic Lagrangian formulation of the PR effect \cite{Defalco2018}.}:
\begin{equation} \label{tpacc}
\begin{aligned}
a(U)^\alpha&=\gamma^2\left[a(n)^\alpha+\Gamma(n)^\alpha{}_{\beta\gamma}\nu(U,n)^\beta\nu(U,n)^\gamma\right.\\
&\left.+2\theta(n)^\alpha{}_\beta\nu(U,n)^\beta\right]+\frac{d(\gamma\nu(U,n)^\alpha)}{d\tau},
\end{aligned}
\end{equation}
where $\alpha,\beta,\gamma=\hat{r},\hat{\theta},\hat{\varphi}$ run on the spatial indices of the metric coordinates \footnote{
Terms $C_{\rm(Lie)}(n)^\alpha_{\beta\gamma},\ C_{\rm(Lie)}(n)^\alpha_\beta$, representing respectively the temporal and spatial constant structures, are missing in Eq. (\ref{tpacc}), because they vanish in a stationary and axially-symmetric spacetime (see \cite{Bini1997a,Bini1997b,Defalco2018}, for details).}. Calculating the Christoffel symbols $\Gamma(n)^\alpha{}_{\beta\gamma}$, I have \cite{Bini1997a,Bini1997b,Defalco2018}
\begin{equation}
\begin{aligned}
\Gamma(n)^{\hat r}{}_{\hat \varphi \hat\varphi}&=\Gamma(n)^{\hat r}{}_{\hat \theta \hat\theta}=-2\Gamma(n)^{\hat \varphi}{}_{\hat r\hat\varphi}=-2\Gamma(n)^{\hat \theta}{}_{\hat r \hat\theta}=k_{\rm (Lie)}(n)^{\hat r},\\
-2\Gamma(n)^{\hat \varphi}{}_{\hat \varphi\hat\theta}&=\Gamma(n)^{\hat \theta}{}_{\hat \varphi\hat\varphi}=k_{\rm (Lie)}(n)^{\hat \theta}. 
\end{aligned}
\end{equation}
Therefore, Eqs. (\ref{tpacc}) in explicit form become
\begin{eqnarray}
a(U)^{\hat r}&=& \gamma^2 [a(n)^{\hat r}+k_{\rm (Lie)}(n)^{\hat r}\,\nu^2 (\cos^2\alpha\sin^2\psi\label{acc1} \\
&&+\cos^2\psi)+2\nu\cos \alpha\sin\psi\, \theta(n)^{\hat r}{}_{\hat \varphi}]\nonumber\\
&&+\gamma \left(\gamma^2 \sin\alpha\sin\psi \frac{\rm d \nu}{\rm d \tau}+\nu \cos \alpha\sin\psi \frac{\rm d \alpha}{\rm d \tau}\right.\nonumber\\
&&\left.+\nu \cos \psi\sin\alpha \frac{\rm d \psi}{\rm d \tau} \right), \nonumber\\   
a(U)^{\hat \theta}&=&\gamma^2 [a(n)^{\hat \theta}+k_{\rm (Lie)}(n)^{\hat \theta}\,\nu^2 \sin^2\psi\cos^2\alpha\label{acc3}\\
&&-k_{\rm (Lie)}(n)^{\hat r}\, \nu^2\sin\psi\sin\alpha\cos\psi\nonumber\\
&&+2\nu\cos \alpha\sin\psi\, \theta(n)^{\hat \theta}{}_{\hat \varphi}]\nonumber\\
&&+ \gamma\left(\gamma^2 \cos\psi \frac{\rm d \nu}{\rm d \tau}-\nu \sin\psi \frac{\rm d \psi}{\rm d \tau}\right).\nonumber\\
a(U)^{\hat \varphi}&=& -\gamma^2 \nu^2\cos \alpha\sin\psi\left[ \sin \alpha \sin\psi\, k_{\rm (Lie)}(n)^{\hat r}\right.\label{acc2}\\ 
&&\left.+ k_{\rm (Lie)}(n)^{\hat \theta}\cos\psi\right]+ \gamma\left(\gamma^2 \cos \alpha\sin\psi \frac{\rm d \nu}{\rm d \tau}\right.\nonumber\\
&&\left.-\nu\sin \alpha\sin\psi \frac{\rm d \alpha}{\rm d \tau}+\nu\cos\alpha\cos\psi \frac{\rm d \psi}{\rm d \tau}\right),\nonumber
\end{eqnarray}
From the orthogonality between $(\bold{a}(U),\bold{U})$ I have
\begin{equation}\label{acc4}
\begin{aligned}
a(U)^{\hat t}&=\nu[a(U)^{\hat r}\sin\alpha\sin\psi+a(U)^{\hat \theta}\cos\psi+a(U)^{\hat \varphi}\cos\alpha\sin\psi]\\   
&=\gamma^2\nu\left\{\sin \alpha \sin\psi\left[a(n)^{\hat r}\right.\left.+2\nu\cos \alpha\sin\psi\, \theta(n)^{\hat r}{}_{\hat \varphi}\right] \right. \\
&\quad\left.+\cos\psi\left[a(n)^{\hat \theta}\right.\left.+2\nu\cos\alpha\sin\psi \theta(n)^{\hat \theta}{}_{\hat \varphi}\right]\right\}+ \gamma^3\nu \frac{\rm d \nu}{\rm d \tau}. 
\end{aligned}         
\end{equation}

\begin{figure*}[th!]
	\centering
	\hbox{
		\includegraphics[scale=0.19]{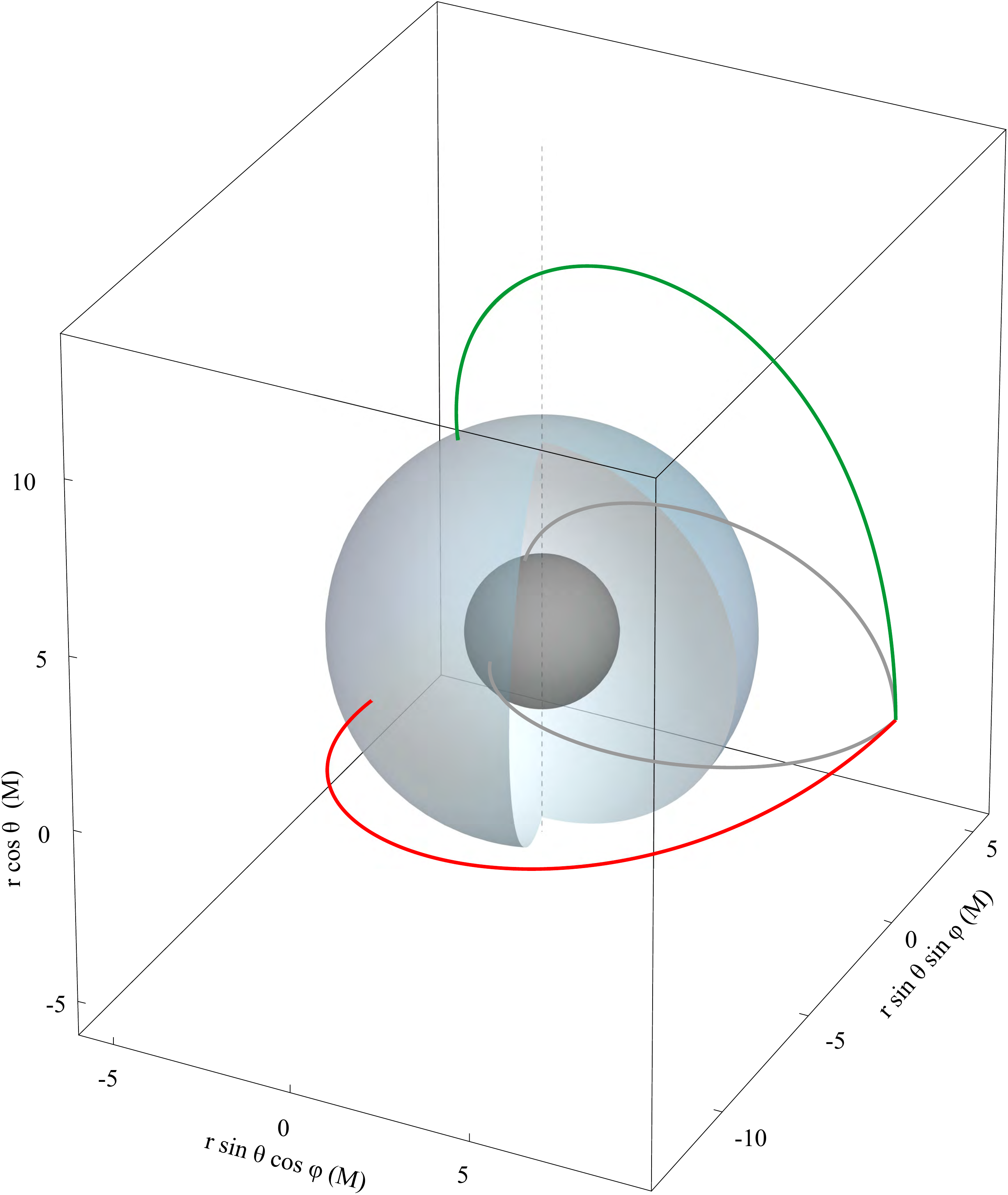}
		\hspace{0.5cm}
		\includegraphics[scale=0.19]{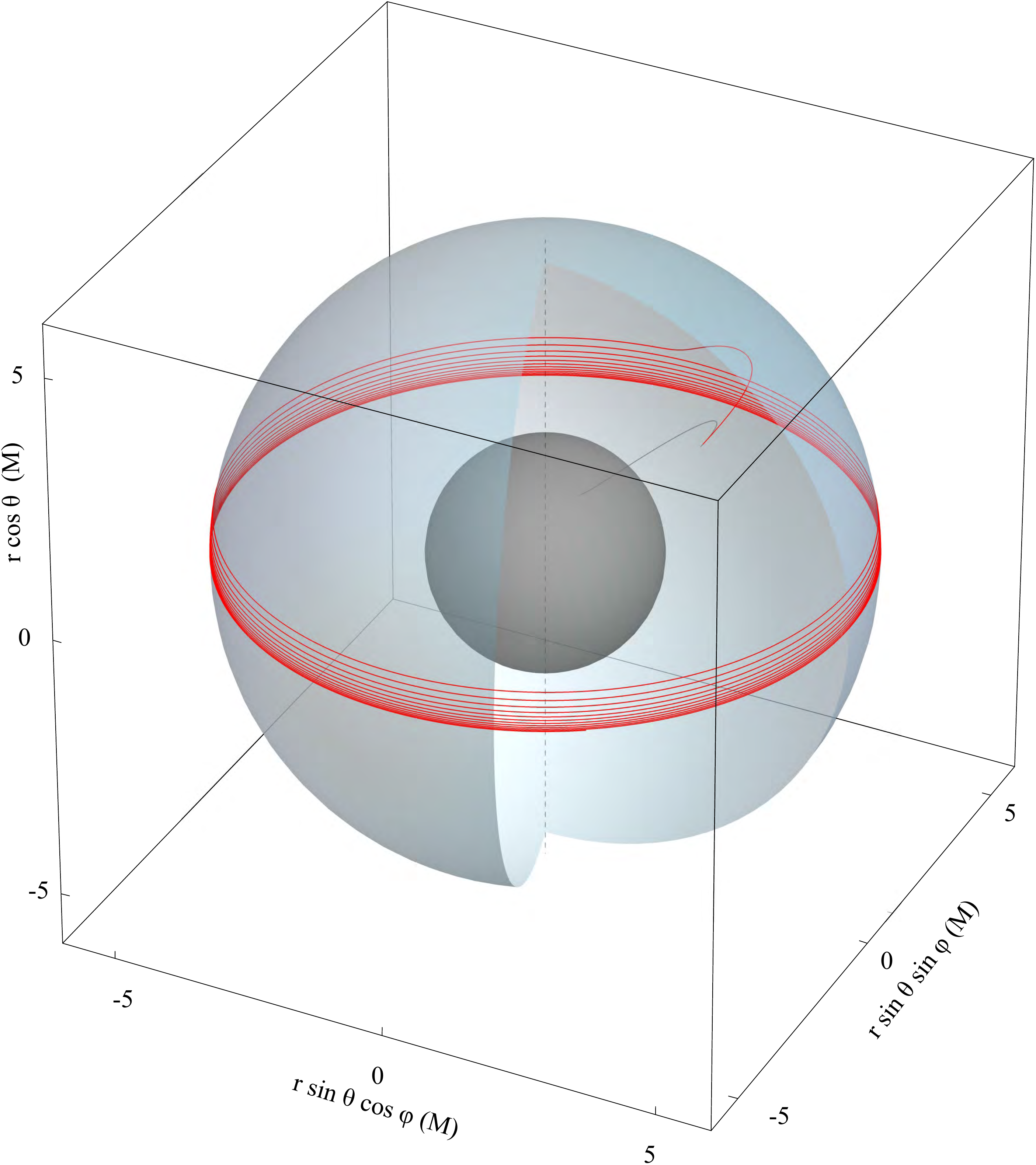}}
	\caption{Left panel: test particle trajectories in a Schwarzschild geometry under the influence of a radiation field  with $A=0.8$. Test particles start  at $r_0=8M$ in the equatorial plane with  initial velocity $\nu_0=0.8$ in azimuthal (red) and polar (green) directions. Right panel: test particle trajectories in a Kerr geometry with small spin ($a=0.05$) under the influence of a radiation field with $A=0.8$. A test particle starts inside the critical hypersurface at $r_0=4M,\, \theta_0=\pi/4$ with initial velocity $\nu_0=0.4$ in the azimuthal direction. In both panels the inner dark surface represents the event horizon and blue-gray, partially open surface represents the critical hypersurface. Gray curves show the geodesic trajectories (i.e. $A=0$) for test particles with initial conditions equal to those described above.}
	\label{fig:Figs_S}
\end{figure*}

\begin{figure*}[th!]
	\centering
	\hbox{
		\includegraphics[scale=0.19]{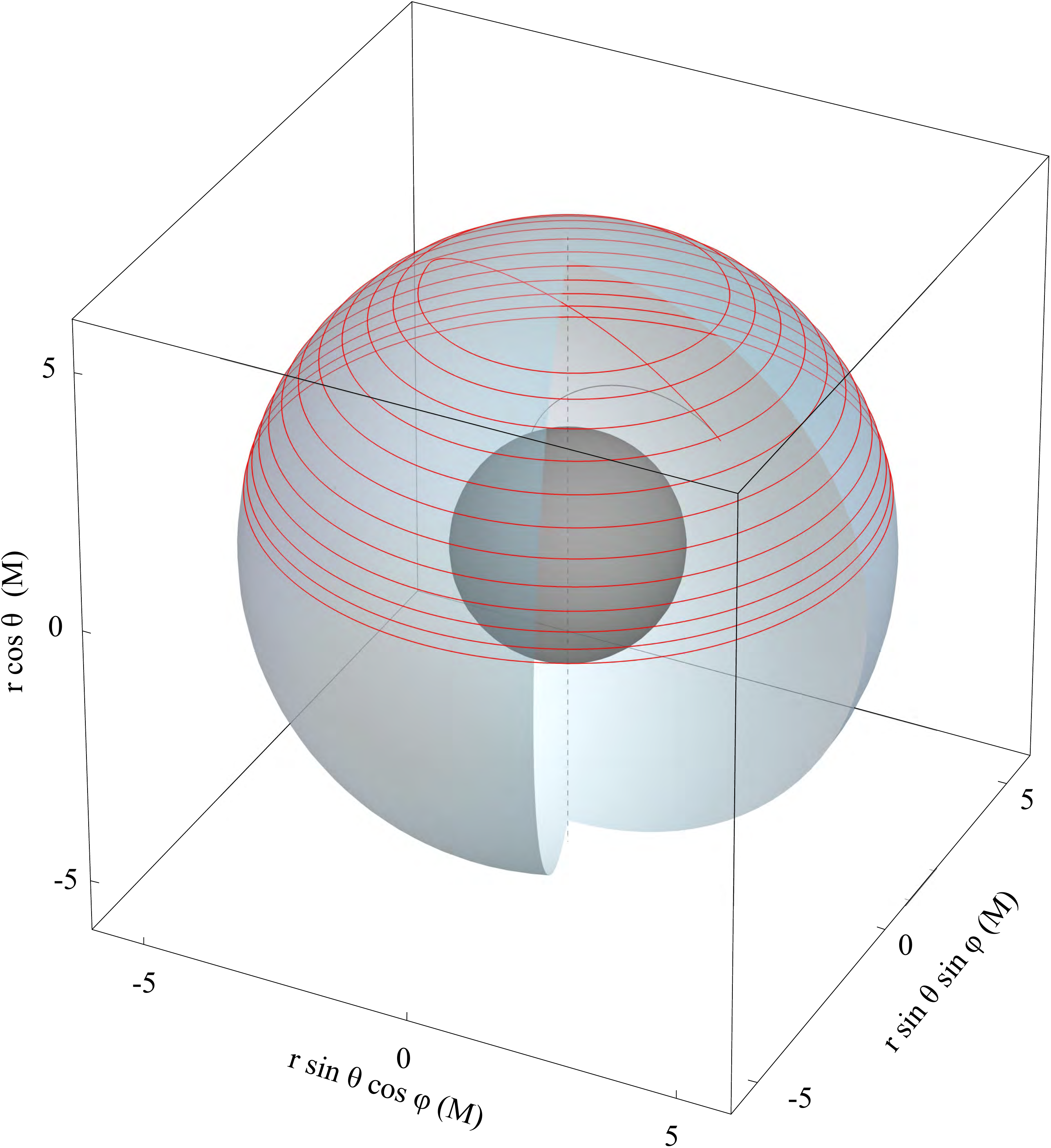}
		\includegraphics[scale=0.21]{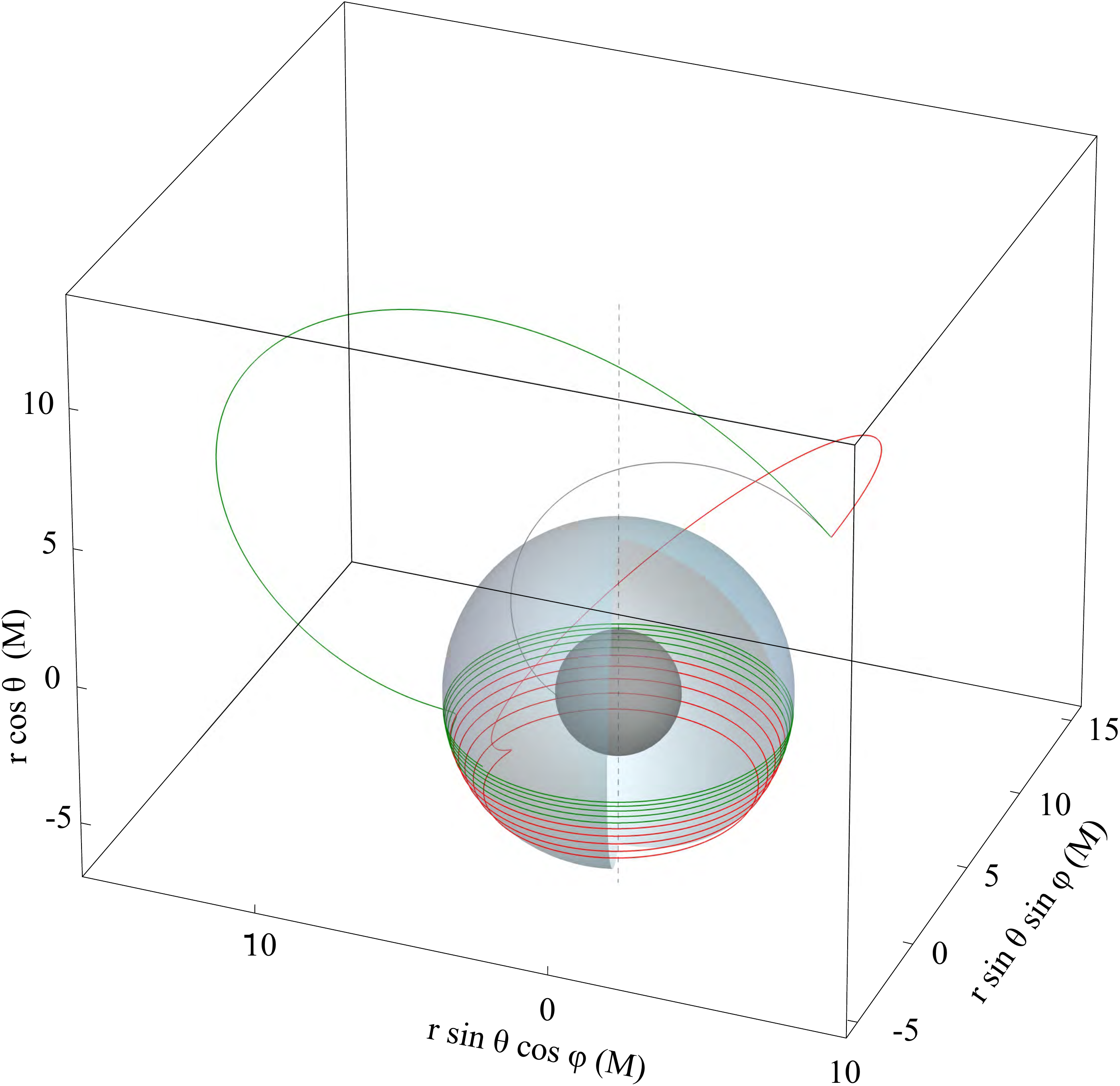}}
	\caption{Test particle trajectories in a Kerr geometry with small spin ($a=0.05$) under the influence of a radiation field with $A=0.8$. Left panel: the test particle starts its motion inside the critical hypersurface at $r_0=4M,\, \theta_0=\pi/4$ with initial velocity $\nu_0=0.4$ in the polar direction. The gray curve denotes the geodesic trajectory (i.e. $A=0$) with $\nu_0=0.4$. Right panel: the test particles start their motion outside the critical hypersurface at $r_0=10M,\, \theta_0=\pi/4$ with initial velocity $\nu_0=0.4$ in the azimuthal (red curve) and polar direction (green curve). In both panels the inner dark surface represents the event horizon and blue-gray, partially open surface represent the critical hypersurface.}
	\label{fig:Figs_KSI1}
\end{figure*}

\subsubsection{Radiation test particle interaction}
\label{sec:eoms}
I assume that the interaction between the test particle and the radiation files takes place through Thomson scattering, characterized by a constant $\sigma$, independent of direction and frequency of the radiation field. The radiation force is \cite{Abramowicz90,Bini09,Bini11}
\begin{equation} \label{radforce}
{\mathcal F}_{\rm (rad)}(U)^\alpha = -\sigma P(U)^\alpha{}_\beta \, T^{\beta}{}_\mu \, U^\mu \,,
\end{equation}
where $P(U)^\alpha{}_\beta=\delta^\alpha_\beta+U^\alpha U_\beta$ projects a vector orthogonally to $\bold{U}$, namely on the spatial hypersurfaces or local rest spaces. The test particle equations of motion then become $m \bold{a}(U) = \boldsymbol{{\mathcal F}_{\rm (rad)}}(U)$, where $m$ is the test particle mass. By definition the radiation force lies in the local rest space of the test particle; to calculated it I decompose the photon four-momentum $\bold{k}$ first with respect to the four-velocity of the test particle, $\bold{U}$, and then to the previous ZAMO decomposition, $\bold{n}$, i.e. \cite{Bini09,Bini11},
\begin{equation} \label{diff_obg}
\bold{k} = E(n)[\bold{n}+\boldsymbol{\hat{\nu}}(k,n)]=E(U)[\bold{U}+\boldsymbol{\hat {\mathcal V}}(k,U)].
\end{equation}
By projecting $\bold{k}$ with respect to the test particle four-velocity, $\bold{U}$ I get
\begin{equation} \label{helpsplit}
\bold{P}(U)\cdot\bold{k}=E(U)\boldsymbol{\hat {\mathcal V}}(k,U)\,,\quad 
\bold{U}\cdot \bold{k}=-E(U).
\end{equation}
Using Eq. (\ref{helpsplit}) in Eq. (\ref{radforce}) I obtain
\begin{equation} \label{Frad0}
\begin{aligned}
{\mathcal F}_{\rm (rad)}(U)^\alpha&=-\sigma \Phi^2 [P(U)^\alpha{}_\beta k^\beta]\, (k_\mu U^\mu)=\sigma \, [\Phi E(U)]^2\, \hat {\mathcal V}
(k,U)^\alpha.
\end{aligned}
\end{equation}
In this way the test particle acceleration is aligned with the photon relative velocity in the test particle local rest space, i.e.,
\begin{equation}\label{geom}
\bold{a}(U)=\tilde \sigma \Phi^2 E(U)^2  \,\boldsymbol{\hat {\mathcal V}}(k,U)\,,
\end{equation}
where $\tilde \sigma=\sigma/m$. Hereafter I use the simplified notation $\boldsymbol{\hat {\mathcal V}}(k,U)=\boldsymbol{\hat {\mathcal V}}$. Multiplying scalarly Eq. (\ref{diff_obg}) by $\bold{U}$ and using Eqs. (\ref{photon}) (i.e. the decomposition of $\bold{k}$ in the ZAMO frame), and (\ref{testp1})--(\ref{testp2}) (i.e. the decomposition of $\bold{U}$ in the ZAMO frame), I find
\begin{equation} \label{enepart}
\begin{aligned}
E(U)&=\gamma E(n)[1-\boldsymbol{\nu}(U,n)\boldsymbol{\hat\nu}(k,n)]\\
&=\gamma E(n)[1-\nu(\sin\zeta\sin\psi\cos(\alpha-\beta)+\cos\zeta\cos\psi)]\\
&=\gamma \frac{E}{N}[1-\nu\sin\psi\sin\alpha],
\end{aligned}
\end{equation}
where I have used Eqs. (\ref{ang1}) and (\ref{ANG1}) and the value of the assumed local angles. Such procedure is very useful for determining the spatial velocity $\boldsymbol{\hat{\mathcal{V}}}$:
\begin{equation}
\boldsymbol{\hat{\mathcal{V}}}=\left[\frac{E(n)}{E(U)}-\gamma\right]\boldsymbol{n}+\frac{E(n)}{E(U)}\boldsymbol{\hat{\nu}}(k,n)-\gamma\boldsymbol{\nu}(U,n).
\end{equation}
The frame components of $\boldsymbol{\hat{\mathcal{V}}}=\hat{\mathcal{V}}^t\boldsymbol{n}+\hat{\mathcal{V}}^r\boldsymbol{e_{\hat r}}+\hat{\mathcal{V}}^\theta \boldsymbol{e_{\hat\theta}}+\hat{\mathcal{V}}^\varphi \boldsymbol{e_{\hat\varphi}}$ are therefore
\begin{eqnarray}
\hat{\mathcal{V}}^{\hat r}&&=\frac{1}{\gamma [1-\nu\sin\psi\sin\alpha]}-\gamma\nu\sin\psi\sin\alpha\nonumber\\
&&=-\gamma\nu^2\left[\frac{1+\sin^2\psi\sin^2\alpha}{1-\nu\sin\psi\sin\alpha}\right],\label{rad1}\\
\hat{\mathcal{V}}^{\hat \theta}&&=-\gamma\nu\cos\psi \label{rad3},\\
\hat{\mathcal{V}}^{\hat\varphi}&&=-\gamma\nu\sin\psi\cos\alpha,\label{rad2}\\
\hat{\mathcal{V}}^{\hat t}&&=\nu(\hat{\mathcal{V}}^{\hat r}\sin\alpha\sin\psi+\hat{\mathcal{V}}^{\hat\theta}\cos\psi+\hat{\mathcal{V}}^{\hat\varphi}\cos\alpha\sin\psi)\nonumber\\
&&=\gamma \nu\left[\frac{\sin\psi\sin\alpha-\nu}{1-\nu\sin\psi\sin\alpha}\right],\label{rad4}
\end{eqnarray}
where the second equality of Eq. (\ref{rad4}) is due to the orthogonality of the $(\boldsymbol{\hat{\mathcal{V}}},\boldsymbol{U})$ pair and I have simplified the components of $\boldsymbol{\hat{\mathcal{V}}}$ of the radiation field.\\ 

\begin{figure*}[th!]
	\centering
	\hbox{
		\includegraphics[scale=0.18]{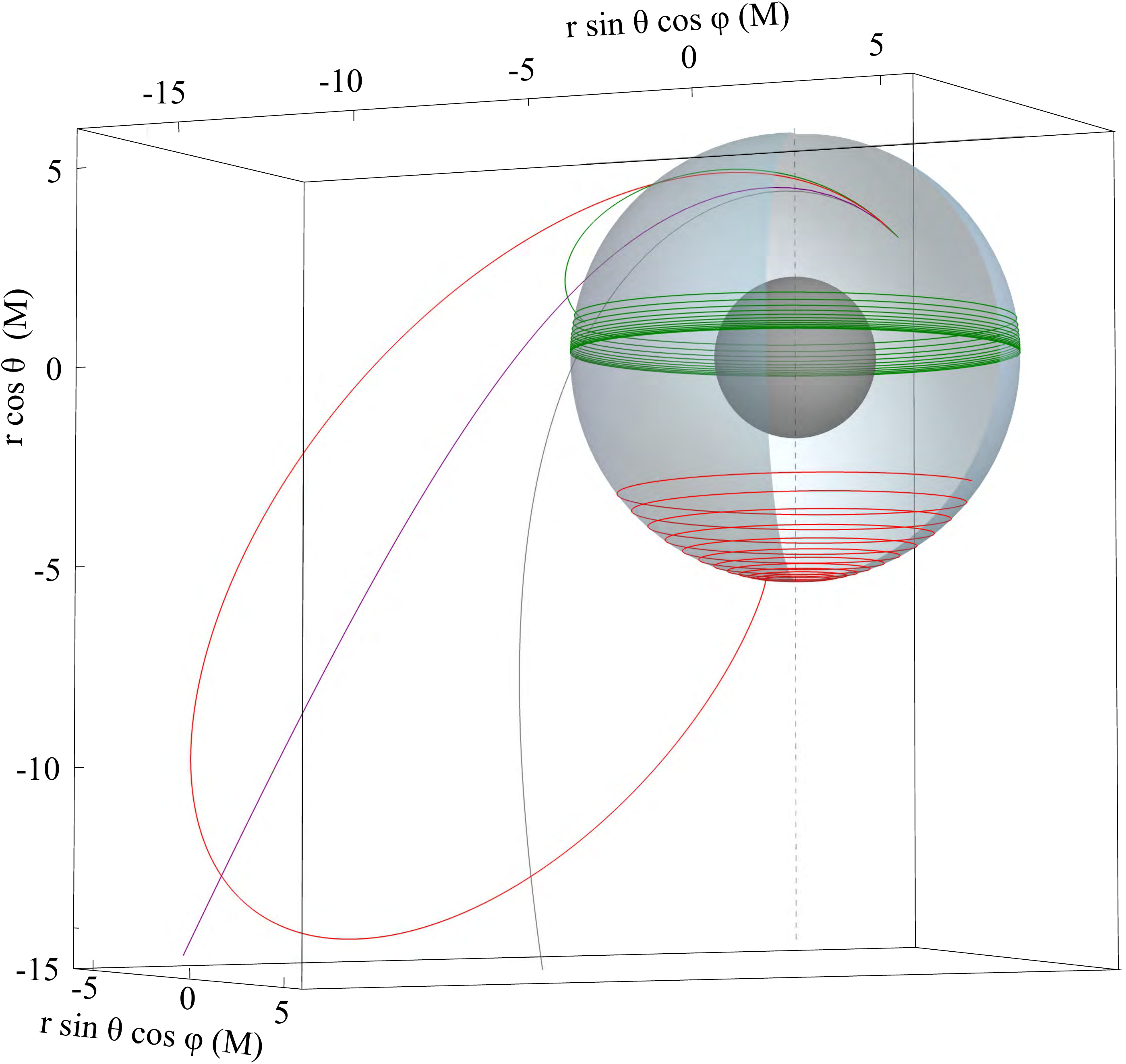}
		\includegraphics[trim=1cm 1cm 0cm 1cm, scale=0.27]{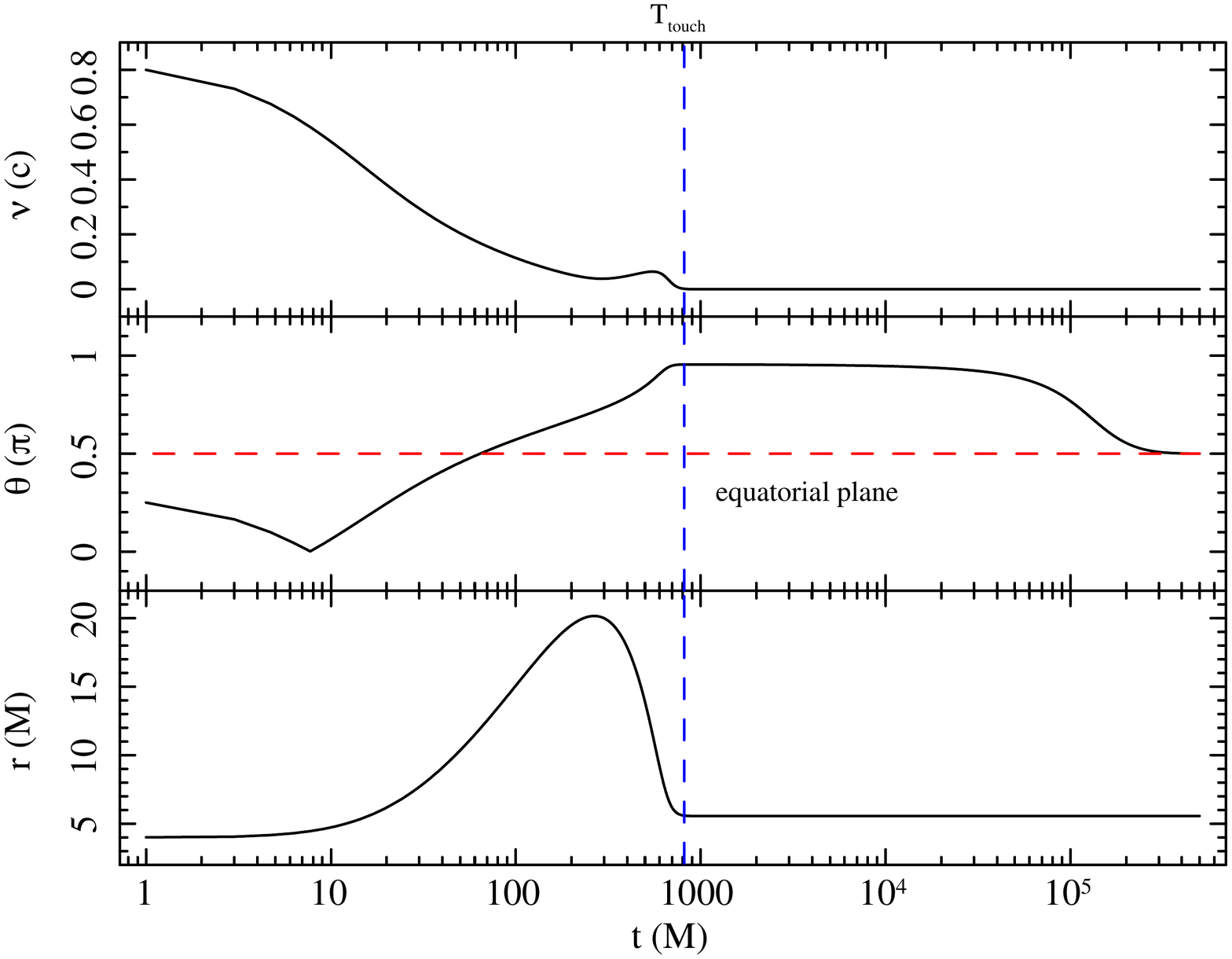}}
	\caption{Test particle trajectories in a Kerr geometry with small spin ($a=0.05$) under the influence of a radiation field with $A=0.8$. Left panel: three test particle starting their motion outside the critical hypersurface at $r_0=4M,\, \theta_0=\pi/4$ with initial velocity along the polar direction and values $\nu_0=0.6$ (green curve), $\nu_0=0.8$ (red curve), $\nu_0=0.87$ (violet curve - escape trajectory). The gray curve shows corresponding the geodesic trajectory (i.e. $A=0$) with initial velocity $\nu_0=0.8$. The inner dark surface represent the event horizon and blue-gray, partially open (spherical or quasi-spherical) surface represents the critical hypersurface. Right panel: velocity profile $\nu$, latitudinal angle $\theta$, and radius $r$ in terms of coordinate time $t$ for the test particle motion with $\nu_0=0.8$ (red curve in left panel). The vertical dashed blue line, $T_{\rm touch}$, represents the time at which the test particle reaches the critical hypersurface; from there on the latitudinal drift on the hypersurface sets in (note the velocity in this stage in much lower  than velocities off the hypersurface). The horizontal dashed red line represents the equatorial plane.}
	\label{fig:Figs_KSI2}
\end{figure*}

\noindent
\textbf{General relativistic equations of motion}\\ \\
In order to make the equations of motion for the test particle moving in a 3D space explicit, Eqs. (\ref{geom}), I consider the ZAMO frame components of the test particle acceleration $\bold{a}(U)$, Eqs. (\ref{acc1})--(\ref{acc4}), and the ZAMO frame components of the radiation force field $\boldsymbol{{\mathcal F}}_{\rm (rad)}(U)$, Eqs. (\ref{enepart}), (\ref{rad1})--(\ref{rad4}). The motion of the test particle is completely defined by the following six parameters $(r,\theta,\varphi,\nu,\psi,\alpha)$, the first three describing the position and the last three the velocity field. The displacement field is simply described by $(U^r,U^\theta,U^\varphi)\equiv(dr/d\tau,d\theta/d\tau,d\varphi/d\tau)$. Instead the velocity field is connected to Eqs. (\ref{geom}) for determining $(d\nu/d\tau,d\psi/d\tau,d\alpha/d\tau)$. I note that using Eq. (\ref{acc4}), it is possible to isolate $d\nu/d\tau$, indeed $a(U)^{\hat t}$ is the energy balance equation (see discussions in \cite{Defalco2018}). Then by using the expression of $d\nu/d\tau$ in $a(U)^{\hat\theta}$, Eq. (\ref{acc3}), it is possible to determine $d\psi/d\tau$. Finally using the expressions of $d\nu/d\tau$ and $d\psi/d\tau$ in $a(U)^{\hat r}$, Eq. (\ref{acc1}) yields $d\alpha/d\tau$. 

Therefore, the general relativistic equations in the Kerr metric for the 3D motion of a test particle immersed in the radiation field defined in Sec. \ref{sec:phot} and \ref{sec:eoms} are the following six coupled ordinary differential equations of the first order
\begin{eqnarray}
&&\frac{d\nu}{d\tau}= -\frac{1}{\gamma}\left\{ \sin\alpha \sin\psi\left[a(n)^{\hat r}+2\nu\cos \alpha\sin\psi\, \theta(n)^{\hat r}{}_{\hat \varphi} \right]\right.\label{EoM1}\\
&&\left.+\cos\psi\left[a(n)^{\hat \theta}+2\nu\cos\alpha\sin\psi\, \theta(n)^{\hat \theta}{}_{\hat \varphi}\right]\right\}+\frac{\tilde{\sigma}[\Phi E(U)]^2}{\gamma^3\nu}\hat{\mathcal{V}}^{\hat t},\nonumber\\
&&\frac{d\psi}{d\tau}= \frac{\gamma}{\nu} \left\{\sin\psi\left[a(n)^{\hat \theta}+k_{\rm (Lie)}(n)^{\hat \theta}\,\nu^2 \cos^2\alpha\right.\right.\label{EoM2}\\
&&\left.\left.\ \quad+2\nu\cos \alpha\sin^2\psi\ \theta(n)^{\hat \theta}{}_{\hat \varphi}\right]-\sin \alpha\cos\psi \left[a(n)^{\hat r}\right.\right.\nonumber\\
&&\left.\left.\ \quad+k_{\rm (Lie)}(n)^{\hat r}\,\nu^2+2\nu\cos \alpha\sin\psi\, \theta(n)^{\hat r}{}_{\hat \varphi}\right]\right\}\nonumber\\
&&\ \quad+\frac{\tilde{\sigma}[\Phi E(U)]^2}{\gamma\nu^2\sin\psi}\left[\hat{\mathcal{V}}^{\hat t}\cos\psi-\hat{\mathcal{V}}^{\hat \theta}\nu\right],\nonumber
\end{eqnarray}
\begin{eqnarray}
&&\frac{d\alpha}{d\tau}=-\frac{\gamma\cos\alpha}{\nu\sin\psi}\left[a(n)^{\hat r}+2\theta(n)^{\hat r}{}_{\hat \varphi}\ \nu\cos\alpha\sin\psi\right.\label{EoM3}\\
&&\left.\ \quad+k_{\rm (Lie)}(n)^{\hat r}\,\nu^2+k_{\rm (Lie)}(n)^{\hat \theta}\,\nu^2\cos^2\psi \sin\alpha\right]\nonumber\\
&&\ \quad+\frac{\tilde{\sigma}[\Phi E(U)]^2\cos\alpha}{\gamma\nu\sin\psi}\left[\hat{\mathcal{V}}^{\hat r}-\hat{\mathcal{V}}^{\hat \varphi}\tan\alpha\right],\nonumber\\
&&U^r\equiv\frac{dr}{d\tau}=\frac{\gamma\nu\sin\alpha\sin\psi}{\sqrt{g_{rr}}}, \label{EoM4}\\
&&U^\theta\equiv\frac{d\theta}{d\tau}=\frac{\gamma\nu\cos\psi}{\sqrt{g_{\theta\theta}}} \label{EoM5},\\
&&U^\varphi\equiv\frac{d\varphi}{d\tau}=\frac{\gamma\nu\cos\alpha\sin\psi}{\sqrt{g_{\varphi\varphi}}}-\frac{\gamma N^\varphi}{N},\label{EoM6}
\end{eqnarray}
where  $\tilde{\sigma}=\sigma/m$ and the two angles $\beta$ and $\zeta$ are calculated in terms of the two impact parameters $b$ and $q$. For $\psi=\zeta=\pi/2$ the  equations of motion reduce to the 2D case \cite{Bini09}. Such set of equations reduce also to the classical 3D case in the weak field limit (see Sec. \ref{sec:classicPR2}).

Following \cite{Abramowicz90,Bini09,Bini11} I define the relative luminosity of the radiation field as 
\begin{equation}\label{eq:A}
A=\tilde{\sigma}\Phi_0^2E^2\,.
\end{equation}
Eq. (\ref{eq:A}) can be recast in the terms of the relative luminosity $A=L_\infty/L_{\rm EDD}$, taking thus the values in $[0,1]$, where $L_\infty$ is the luminosity of the central source as seen by an observer at infinity and $L_{\rm EDD}=4\pi Mm/\sigma$ is the Eddington luminosity at infinity. Then for the investigated radiation field with zero angular momenta ($b=0$, $\beta=\pi/2$) and without latitudinal photon motion ($q=- a^{2} \cos^{2}\theta$, $\zeta=\pi/2$), the term $\tilde{\sigma}[\Phi E(U)]^2$ becomes 
\begin{equation}
\tilde{\sigma}[\Phi E(U)]^2=\frac{ A\,\gamma^2\, [1-\nu\sin\psi\sin\alpha]^2}{N^2 \sqrt{(r^2+a^2)^2-a^2\,\Delta\,\sin^{2}\theta}} \,.
\end{equation}

\begin{figure*}[th!]
	\centering
	\hbox{
		\includegraphics[width=3cm, height=6.7cm]{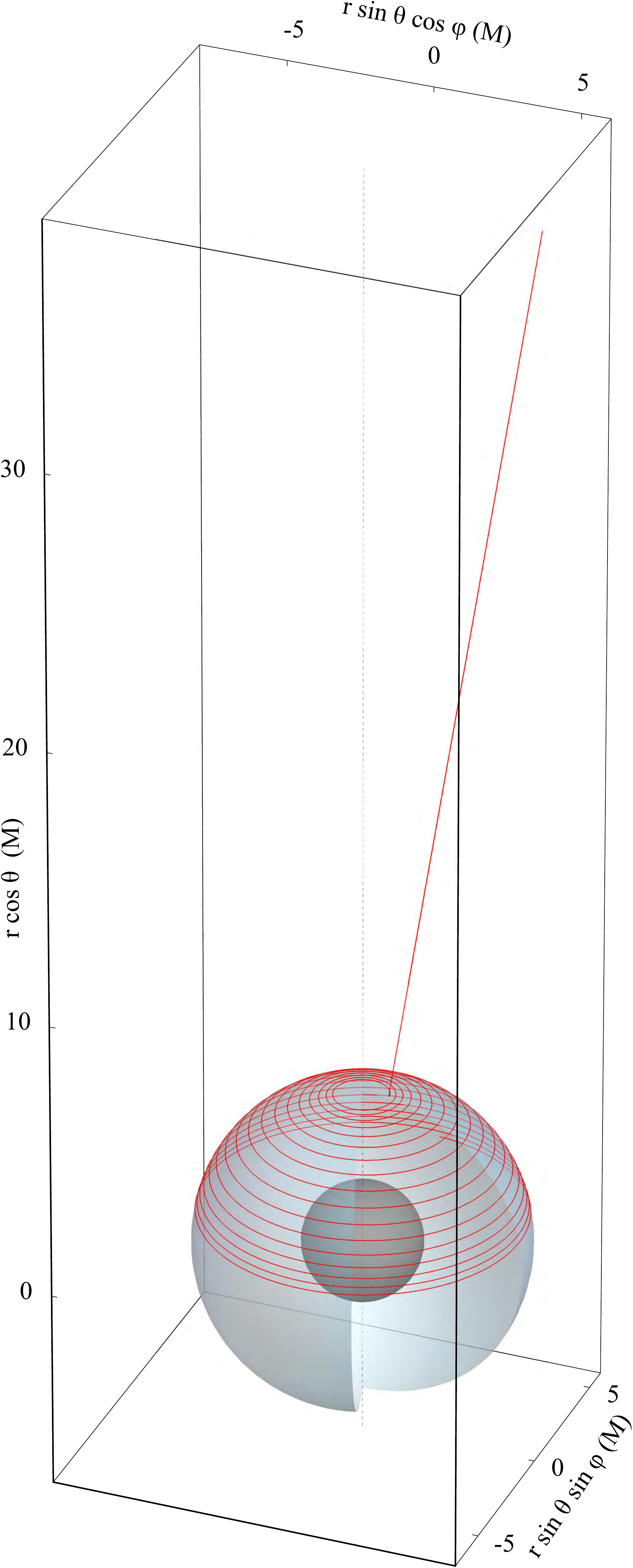}
		\includegraphics[trim=1cm 1cm 0cm 1cm, scale=0.37]{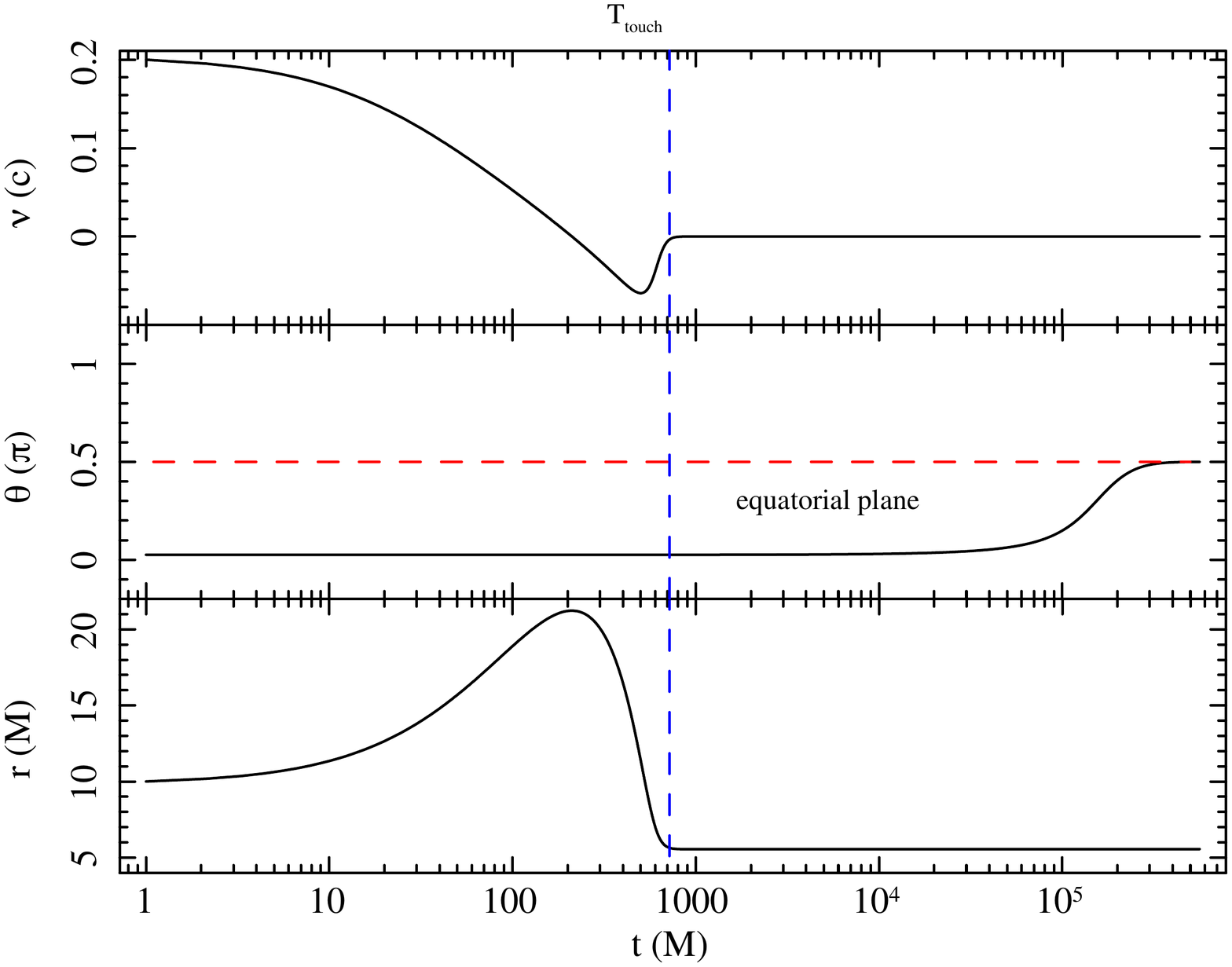}}
	\caption{Test particle trajectories in a Kerr geometry with small spin ($a=0.05$) under the influence of a radiation field with $A=0.8$. Left panel: the test particle starts its motion outside the critical hypersurface at $r_0=10M,\, \theta_0=\pi/20$ with initial velocity $\nu_0=0.4$. The inner dark surface represent the event horizon and blue-gray, partially open (spherical or quasi-spherical) surface represents the critical hypersurface. Right panel: velocity profile $\nu$, latitudinal angle $\theta$, and radius $r$ in terms of coordinate time $t$ for the test particle motion with $\nu_0=0.4$ (red curve in left panel). The vertical dashed blue line, $T_{\rm touch}$, represents the time at which the test particle reaches the critical hypersurface; from there on the latitudinal drift on the hypersurface sets in (note the velocity in this stage in much lower  than velocities off the hypersurface). The horizontal dashed red line represents the equatorial plane.}
	\label{fig:Figs_KSO2}
\end{figure*}

\subsection{Classical 3D PR effect}
\label{sec:classicPR2}
The classical radiation drag force was described and introduced by Poynting (1903) \cite{Poynting03} and Robertson (1937) \cite{Robertson37} in the 2D case. I extend the planar motion to the 3D case, written in spherical coordinates, $(r,\theta,\varphi)$. Noting that the classical drag force can be seen as a viscous effect depending linearly on the test particle velocity \cite{Poynting03,Robertson37,Defalco2018}, and assuming that the radiation 
propagates radially in the whole 3D space, the test particle equations of motion read
\begin{eqnarray}
\ddot{r}-r\dot{\varphi}^2\sin^2\theta-r\dot{\theta}^2+\frac{GM-Ac}{r^2}&=&-2A\frac{\dot{r}}{r^2},\label{eqm1}\\
r\ddot{\theta}+2\dot{r}\dot{\theta}-r\dot{\varphi}^2\sin\theta\cos\theta&=&-A\frac{\dot{\theta}}{r}\label{eqm2},\\
r\ddot{\varphi}\sin\theta+2\dot{r}\dot{\varphi}\sin\theta+2r\dot{\theta}\dot{\varphi}\cos\theta&=&-A\frac{\dot{\varphi}\sin\theta}{r}.\label{eqm3}
\end{eqnarray}

\subsubsection{Weak field approximation of the general relativistic equations}
I show here the way in which the 3D general relativistic equations of motion $ma(U)^\alpha=F_{\rm(rad)}(U)^\alpha$, Eqs. (\ref{EoM1})--(\ref{EoM6}), reduce to the classical 3D case, Eqs. (\ref{eqm1})--(\ref{eqm3}), in the weak field limit ($a\to0,\ r\to\infty,\ \nu/c\to0$). Eqs. (\ref{EoM4})--(\ref{EoM6}) are by definition
\begin{equation}
\begin{aligned}
U^r\equiv\dot{r}\approx\nu\sin\psi\sin\alpha,\ 
U^\theta\equiv \dot{\theta}\approx\frac{\nu\cos\psi}{r},\ 
U^\varphi\equiv \dot{\varphi}\approx\frac{\nu\sin\psi\cos\alpha}{r\sin\theta}. 
\end{aligned}
\end{equation}
The radial components of the ZAMO kinematical quantities reduce to
\begin{equation}
\begin{aligned}
a(n)^{\hat r}\approx \frac{M}{r^2},\ \theta(n)^{\hat r}{}_{\hat\varphi}\approx0,\ k_{\rm (Lie)}(n)^{\hat r}\approx -\frac{1}{r}, 
\end{aligned}
\end{equation}
expressed in geometrical units $G=c=1$, and where the relative Lie radial curvature reduces to the curvature of the osculating sphere (see \cite{Bini09,Bini11,Defalco2018}, for comparisons). Instead for the polar components of the ZAMO kinematical quantities I have 
\begin{equation}
\begin{aligned}
a(n)^{\hat \theta}\approx 0,\ \theta(n)^{\hat r}{}_{\hat\varphi}\approx0,\
k_{\rm (Lie)}(n)^{\hat \theta}\approx -\frac{1}{r\tan\theta}, 
\end{aligned}
\end{equation}
where the relative Lie polar curvature describes the longitudinal Euler acceleration \cite{Bini1997a,Bini1997b,Defalco2018}. Now it is easy to see how the test particle acceleration, $a(U)^\alpha$, reduces to the left members of Eqs. (\ref{eqm1})--(\ref{eqm3}). Approximating the radiation force, $F_{\rm(rad)}(U)^\alpha$, through linear terms in the velocity field, I have (see \cite{Defalco2018}, for comparisons)
\begin{equation}
\begin{aligned}
&F_{\rm(rad)}(U)^{\hat r}\approx \frac{A}{r^2}(1-2\dot{r}),\\  
&F_{\rm(rad)}(U)^{\hat \theta}\approx-\frac{A}{r}\dot{\theta},\\
&F_{\rm(rad)}(U)^{\hat \varphi}\approx-\frac{A}{r}\dot{\varphi}\sin^2\theta,
\end{aligned}
\end{equation}
which reduce to the right members of Eqs. (\ref{eqm1})--(\ref{eqm3}). I note that the time component of the equations of motion, $ma(U)^t=F_{\rm(rad)}(U)^t$, reduces to \cite{Defalco2018}
\begin{equation} 
\frac{d}{dt}\left(\frac{\nu^2}{2}+\frac{A-M}{r}\right)=-A\frac{\nu}{r^2}-A\frac{\dot{r}^2}{r^2},
\end{equation}
which represents the energy conservation equation. Indeed, the left term represents the total mechanical energy, while the right term corresponds to the dissipated energy.

\subsection{Critical hypersurface}
\label{sec:critc_rad}
The system of six differential equations (\ref{EoM1})--(\ref{EoM6}) admits a critical solution of radial equilibrium, which corresponds to the axially-symmetric hypersurface where radiation pressure balances the attraction of the gravitational field. 
Let us consider a test particle moving purely radially with respect to the ZAMO frame ($\alpha=\psi=\pm\pi/2$). Then, at the critical radius $r_{\rm (crit)}$, where the test particle is in rest with respect to the ZAMO frame ($\nu=0, \gamma=1$), the first equation of motion, Eq. (\ref{EoM1}), takes the form
\begin{equation} \label{r-equilibrium}
a(n)^{\hat r}=\frac{A}{N^2 \sqrt{(r_{\rm (crit)}^2+a^2)^2-a^2\,\Delta_{\rm (crit)}\,\sin^{2}\theta}}. 
\end{equation}
In the case of pure radial motion ($\cos\alpha=0, \frac{d\alpha}{d\tau}=0 $), the third equation of motion, Eq. (\ref{EoM3}), is automatically fulfilled. If I multiply the second equation of motion, Eq. (\ref{EoM2}), by the term $\nu^2$, thus removing its divergence, one can easily see that it is fulfilled in the radial equilibrium case ($\cos\psi=0, \frac{d\psi}{d\tau}=0,\nu=0 $). 
For $\theta=\pi/2$ (i.e. in the equatorial plane) relation (\ref{r-equilibrium}) corresponds to  the equilibrium condition Eq. (2.33) derived in \cite{Bini09}, which gives the values $r_{\rm(crit)}$ of the radial coordinate where the test particle co-moves with the ZAMOs in the equatorial circular orbit. However, relation (\ref{r-equilibrium}) generalizes this condition also for the case of test particles with arbitrary polar angle $\theta$ and therefore describes a critical hypersurface which envelops the central compact object and where the test particles co-moves with the local ZAMOs in a bound quasi-circular orbits. \footnote{A different mechanism that leads to the formation of similar off-equatorial circular orbits is the interaction of charged test particles with the magnetic field of a neutron star \cite{Kovar2008}.}
 
 In the case of a non-zero spin, the critical radius given by Eq. (\ref{r-equilibrium}) is function of the polar angle of $r_{\rm(crit)}=r_{\rm (crit)}(A,\theta)$ (in addition to the relative luminosity $A$). The radial equilibrium therefore occurs at the axially symmetric hypersurface, whose shorter axis lies in the equatorial plane and longer axis in the polar direction. This is due to the properties of frame-dragging, as photons (and test particles) are dragged maximally in the azimuthal direction in the equatorial plane $\theta=\pi/2$. Therefore the radial component of the photon four-velocity reaches a maximum (and thus the radial momentum transfer is largest) along the polar axis and decreases for increasing polar angles; that is the reason why the critical hypersurface is elongated along the polar axis. In the case of zero spin (Schwarzschild spacetime), the critical hypersurface turns into a sphere with radius corresponding to the value given by Eq. (2.33) in \cite{Bini09}. The left panel of Fig. \ref{fig:Fig2} compares the shape of the critical hypersurfaces for a high-spin Kerr spacetime $a=0.9995$ and for a Schwarzschild spacetime with $a=0$, where the relative luminosity of the radiating field is in both cases set to the value of $A=0.8$. In the high-spin case the critical radius is $r^{\rm eq}_{\rm(crit)} \sim 5.52M$ in the equatorial plane and $r^{\rm pole}_{\rm(crit)} \sim 6.56M$ at the poles. In the case of a Schwarzschild spacetime, the radius of the critical sphere is $r_{\rm (crit)} \sim 5.56M$. The right panel of Fig. \ref{fig:Fig2} illustrates the shape of the critical hypersurfaces for the values of the relative luminosity in the interval $0.5-0.9$ and for a constant value of the spin $a=0.9995$. 

\subsection{Test particle orbits}
\label{sec:orbits}
I have developed the  \emph{3D PRtrajectories} code to integrate the test particles trajectories described by equations (\ref{EoM1})--(\ref{EoM6}). The integration of the equations of motion in three spatial dimensions turns out to be substantially more sensitive to integration errors than the the 2D case. Therefore I adapted the highly-accurate core for the integration of photon trajectories used in  \emph{LSDCode+}  \cite{Bakala2015} to the case of massive particles. The code
implements the Runge-Kutta method of the eighth order (the Dorman--Prince method) \cite{Press2002} with an adaptive step. Successful integration of the 3D trajectory of test particles influenced by the radiation field (especially in the latitudinal direction) requires advanced monitoring of integration errors. In the  \emph{3D PRtrajectories} code, the  \emph{PI stepsize control} algorithm (see \cite{Press2002} for details) is implemented, which easily attains  an average relative accuracy of $\sim 10^{-14}$. Such a value allow precise and consistent integration of 3D trajectories even in the most sensitive parts, the vicinity of turning points.

I integrated equations (\ref{EoM1})--(\ref{EoM6}) for a set of different boundary conditions and model parameters.  
My results show that the main qualitative features of the 2D case examined in \cite{Bini09,Bini11} remain the same for the trajectories in three spatial dimensions. Similarly to the 2D case, I can divide the orbits into two distinct classes depending on the initial radial position $r(0)=r_0$: inside and outside the critical hypersurface. Also in the 3D case, a test particle trajectory can have only two possible ends, either $(i)$ it goes to infinity or $(ii)$ it reaches the critical hypersurface. Moreover in the presence of (an outgoing) radiation field, the test particle cannot cross the event horizon. I compared representative trajectories of test particles with a polarly and azimuthally-oriented initial velocity for the case of a Schwarzschild spacetime, for the case of the Kerr metrics with a  small spin (a=0.05) that approximates the spacetime in the vicinity of NSs and, finally, for the case of the Kerr metrics with very high spin ($a=0.9995$) that corresponds to the spacetime in the vicinity of almost extreme BHs.
 
In the case of the Schwarzschild metric my results fully agree with those from earlier analyses of the 2D case in which motions are confined to the equatorial plane \cite{Bini09,Bini11} (note however, that this can be chosen arbitrarily for spherically symmetric metrics and radiation fields). The left panel of Fig. \ref{fig:Figs_S} shows examples of 3D trajectories which reflect the spherically-symmetric limit of equations (\ref{EoM1} - \ref{EoM6}) in the case of zero spin. The trajectories of test particles starting from the same location but with initial velocity oriented polarly and azimuthally are identical except for the different orientation of the plane on which they lie. In such a case, the 3D trajectories are easily transformed to the corresponding 2D trajectories through coordinate rotation. The left panel of Fig.\ref{fig:Figs_S} then shows that in a Schwarzschild spacetime, once the trajectory of a test particle  reaches the spherical critical hypersurface it stops precisely there.

Note that the presence of even a very small spin ($a=0.05$) breaks the spherical symmetry of the spacetime geometry and radiation field and introduces qualitatively-new features in test particle trajectories, owing to frame dragging effects. In the Kerr case, the test particle, once captured on the critical hypersurface, gets dragged azimuthally at the angular velocity $\Omega_{\mathrm{ZAMO}}$ and undergoes a \emph{latitudinal drift towards the equatorial plane} (see Sec. \ref{sec:latdrift}, for a detailed explanation). Hence the test particle spirals on the critical hypersurface as it shifts to lower and lower latitudes. The results of my numerical integrations show that, besides the angular velocity $\Omega_{\mathrm{ZAMO}}$, the velocity of the latitudinal drift increases for increasing spins. In fact in for small spin values, test particles caught on the critical hypersurface encircle multiple spirals before attaining the final purely circular equatorial trajectory. Such behavior is exhibited by test particles whose motion starts from the inside (see right panel of Fig. \ref{fig:Figs_S} and left panel of Fig. \ref{fig:Figs_KSI1}), as well as the outside of the critical hypersurface (see Fig. \ref{fig:Figs_KSI2} and right panel of Fig. \ref{fig:Figs_KSI1}). The behaviour of a test particles beginning its motion near the polar axis in a outgoing, purely radial direction is illustrated in Fig. \ref{fig:Figs_KSO2}: the test particle initially travels outward, reaches the turning point and then falls back along a nearly identical trajectory; after being captured near the pole of the critical hypersurface, it drifts toward the equator in tight spiralling trajectory that spans over most of northern hemisphere of the critical hypersurface.

\begin{figure*}[th!]
\centering
\hbox{
\includegraphics[width=5.5cm, height=5.5cm]{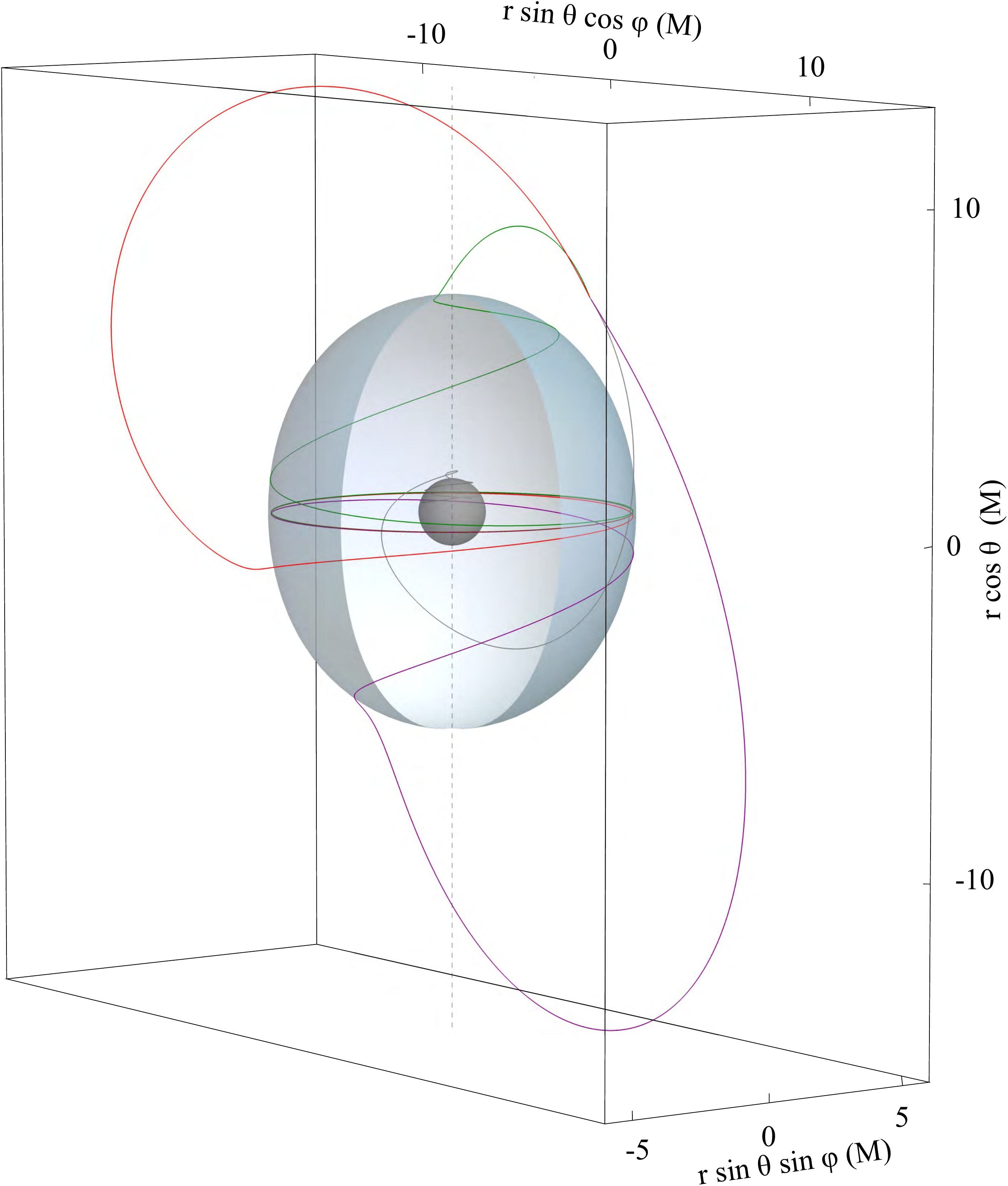}
\includegraphics[trim=1cm 1cm 0cm 1cm, scale=0.27]{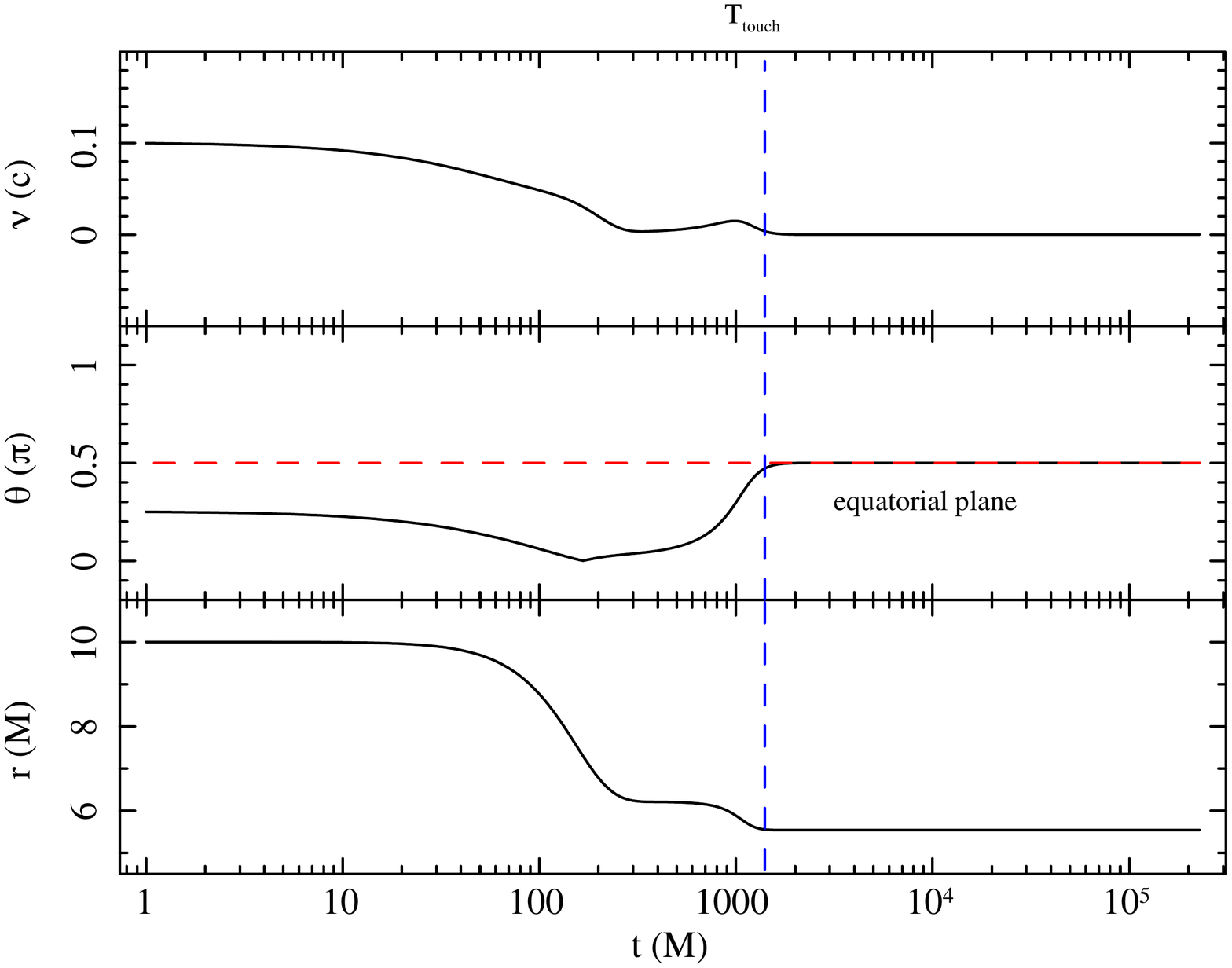}}
\caption{Test particle trajectories in a Kerr geometry with almost extreme spin ($a=0.9995$) under the influence of a radiation field with $A=0.8$. Left panel: three test particle starting their motion outside the critical hypersurface at $r_0=10M,\, \theta_0=\pi/4$ with initial velocity along the polar direction and values oriented towards the north pole $\nu_0=0.1$ (green curve), $\nu_0=0.25$ (red curve), and oriented towards the south pole $\nu_0=0.25$ (violet curve). The inner dark surface represent the event horizon and blue-gray, partially open (spherical or quasi-spherical) surface represents the critical hypersurface. Right panel: velocity profile $\nu$, latitudinal angle $\theta$, and radius $r$ in terms of coordinate time $t$ for the test particle motion with $\nu_0=0.1$ (green curve in left panel). The vertical dashed blue line, $T_{\rm touch}$, represents the time at which the test particle reaches the critical hypersurface; from there on the latitudinal drift on the hypersurface sets in (note the velocity in this stage in much lower  than velocities off the hypersurface). The horizontal dashed red line represents the equatorial plane.}
\label{fig:Figs_KH1}
\end{figure*}

\begin{figure*}[th!]
\centering
\hbox{
\includegraphics[width=0.52\textwidth]{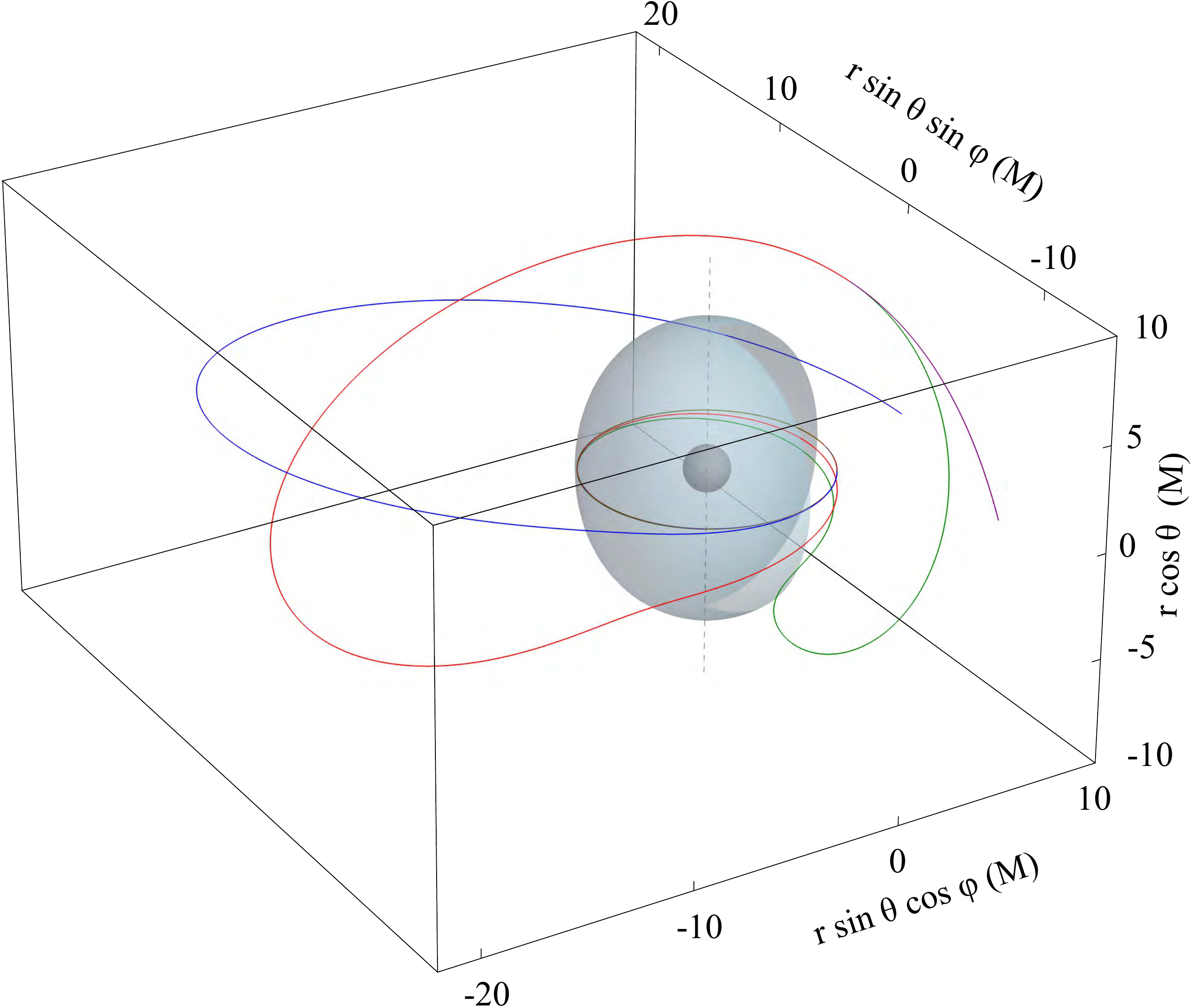}
\hspace{0.1cm}
\includegraphics[width=0.43\textwidth]{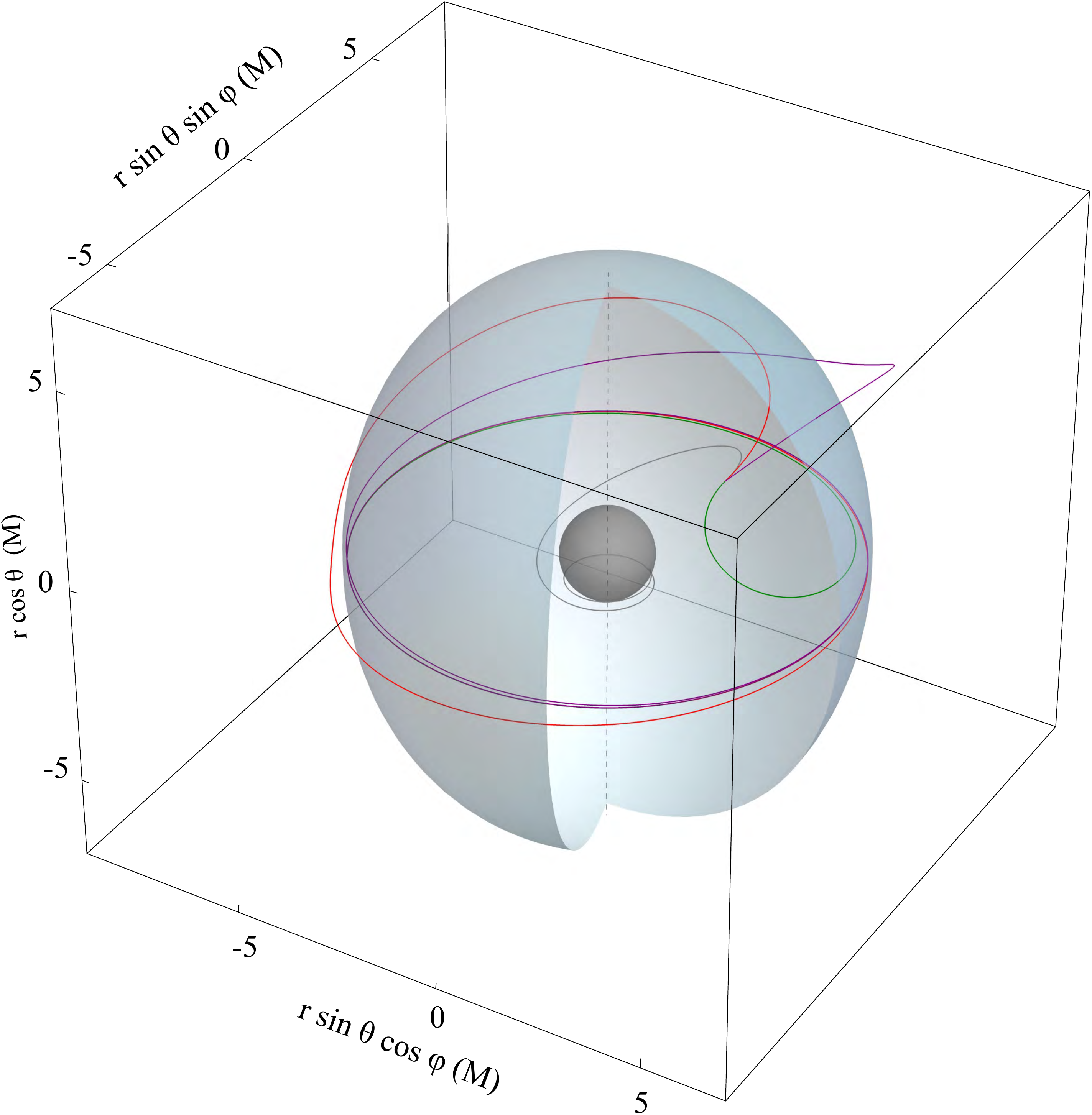}}
\caption{Test particle trajectories in a Kerr geometry with almost extreme spin ($a=0.9995$) under the influence of a radiation field with $A=0.8$. Left panel: three test particle starting their motion outside the critical hypersurface at $r_0=10M$ with initial velocity $\nu_0=0.25$, off-equatorial initial position at $\theta_0=\pi/4$, and in the azimuthal direction co-rotating (red curve) and counter-rotating (green curve) with respect to the compact object and on the equatorial ($\theta_0=\pi/2$) in the azimuthal direction co-rotating with respect to the compact object (blue curve). Right panel: three test particle starting their motion inside the critical hypersurface at $r_0=4M,\,\theta_0=\pi/4$ with initial velocity $\nu_0=0.4$ in the azimuthal direction co-rotating (red curve) and counter-rotating (green curve) with respect to the compact object and in the outgoing radial azimuthal direction (violet curve). The gray curve shows corresponding the geodesic trajectory (i.e. $A=0$) with respect to the red curve. In both panels, the inner dark surface represent the event horizon and blue-gray, partially open (spherical or quasi-spherical) surface represents the critical hypersurface. 
}
\label{fig:Figs_KH2}
\end{figure*}

For a value of the spin ($a=0.9995$) close to that of an extreme Kerr BH, frame dragging is faster and  leads to a faster latitudinal drift, besides a higher $\Omega_{\mathrm{ZAMO}}$. Therefore test particles captured on the critical hypersurface at any value of the polar coordinate $\theta$ are dragged quickly to the equatorial plane where they attain a purely circular trajectory (see  Figs \ref{fig:Figs_KH1} and \ref{fig:Figs_KH2}).


\subsubsection{Orbits bound to the critical hypersurface}
\label{sec:latdrift}
\begin{figure*}[th!]
\centering
\hbox{
\includegraphics[scale=0.19]{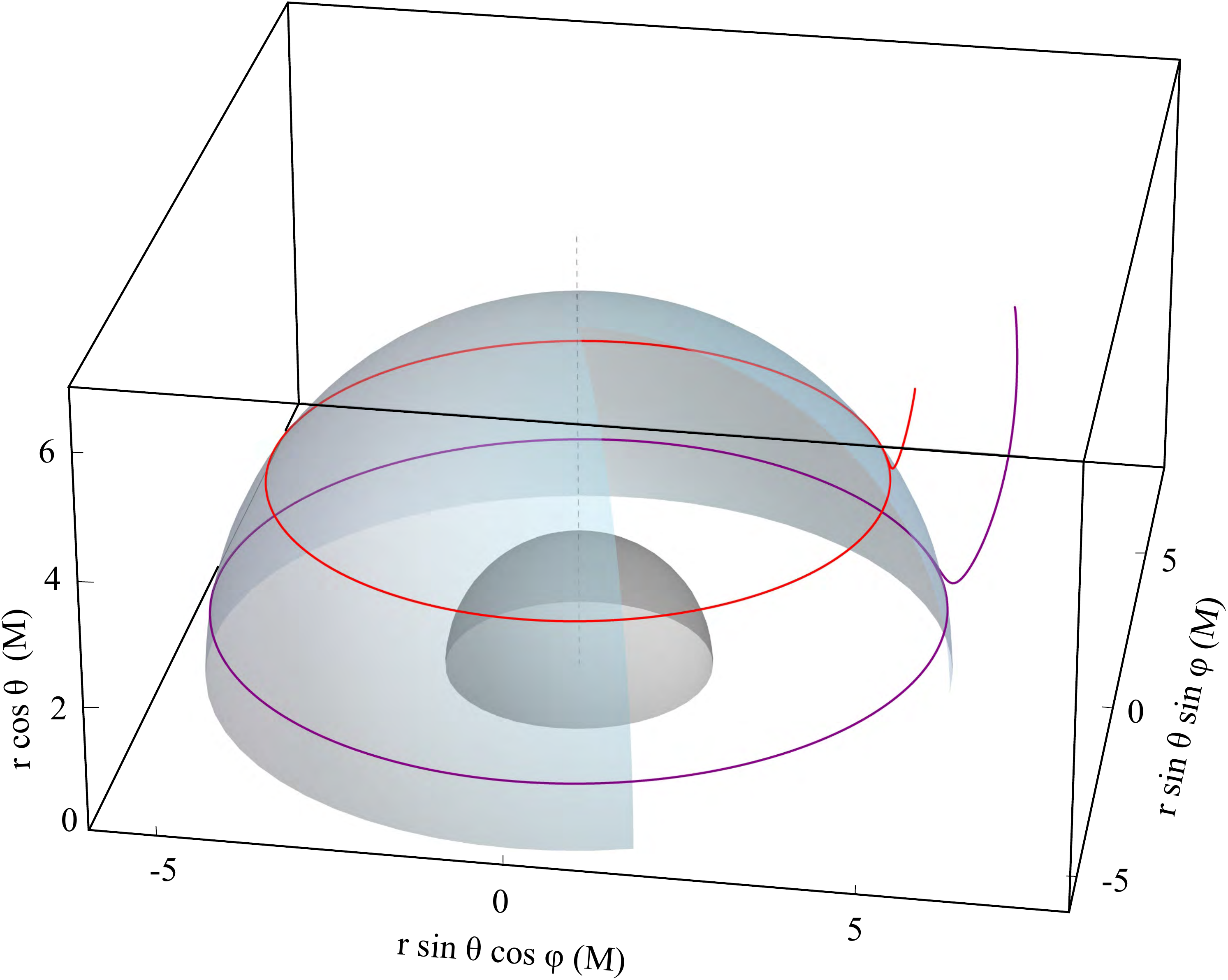}
\hspace{0.2cm}
\includegraphics[scale=0.18]{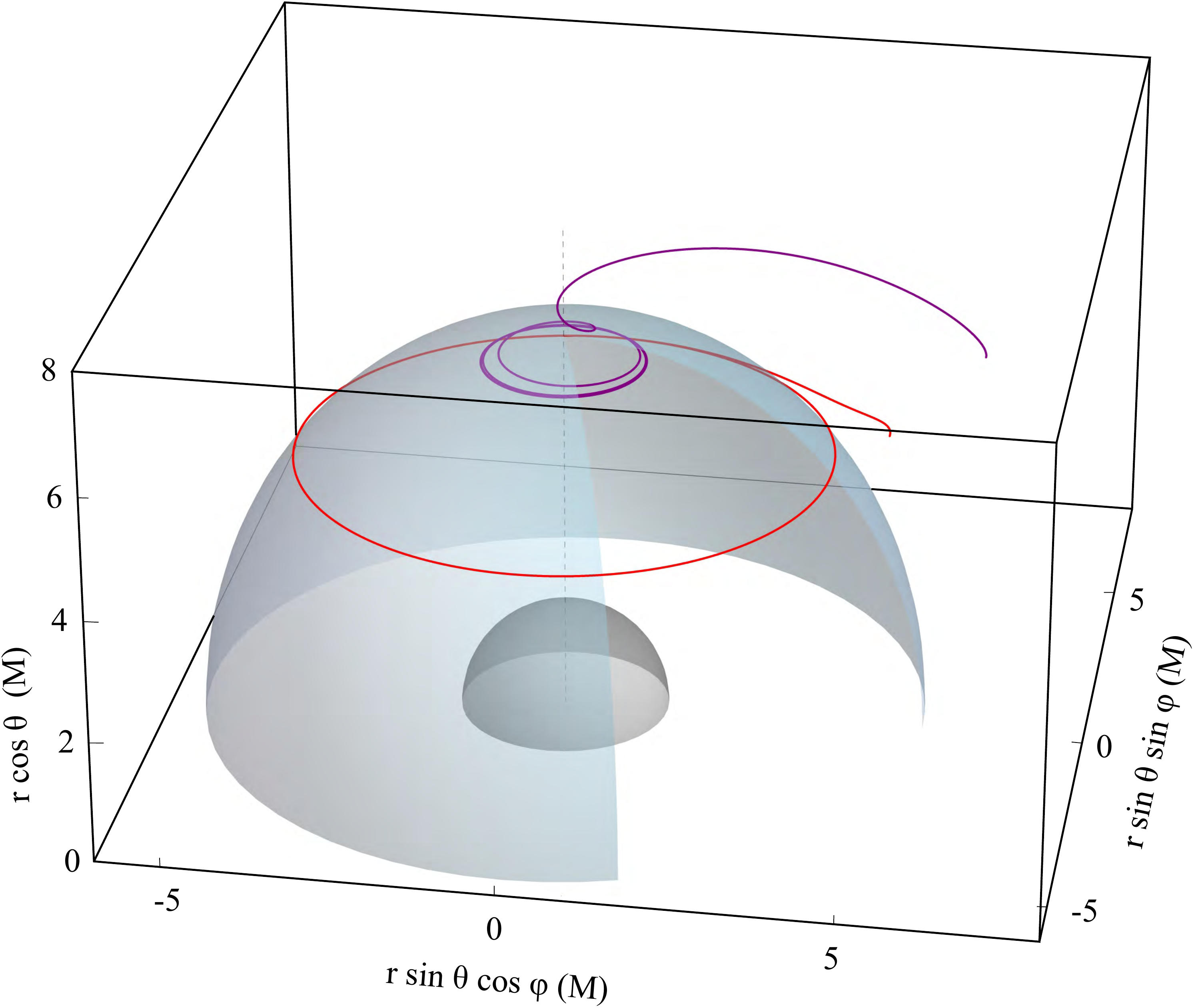}}
\caption{Off-equatorial circular orbits of test particles with zero angular momentum bound to the critical hypersurface. The red curves correspond to the case of test particle trajectory starting from the rest ($\nu=0, \gamma=1$) at $r_0=7M,\,\theta_0=\pi/4$ while the violet ones correspond to the case of test particle trajectory starting from the rest at $r_0=10M,\,\theta_0=\pi/4$. In the right panel the trajectories are plotted for the case of very small spin $a=0.05$ while in the left panel corresponds to the case of almost extreme spin $a=0.9995$. The relative luminosity of the radiating field takes the value of $A=0.8$. The inner black surface denotes the location of the north hemisphere of the event horizon. The blue-gray, partially open surface denotes the location of the north hemisphere of the critical hypersurface.}
\label{fig:Fig_traj_bounded}
\end{figure*}

\begin{figure*}[th!]
\centering
\hbox{
\hspace{-0.5cm}
\includegraphics[scale=0.25]{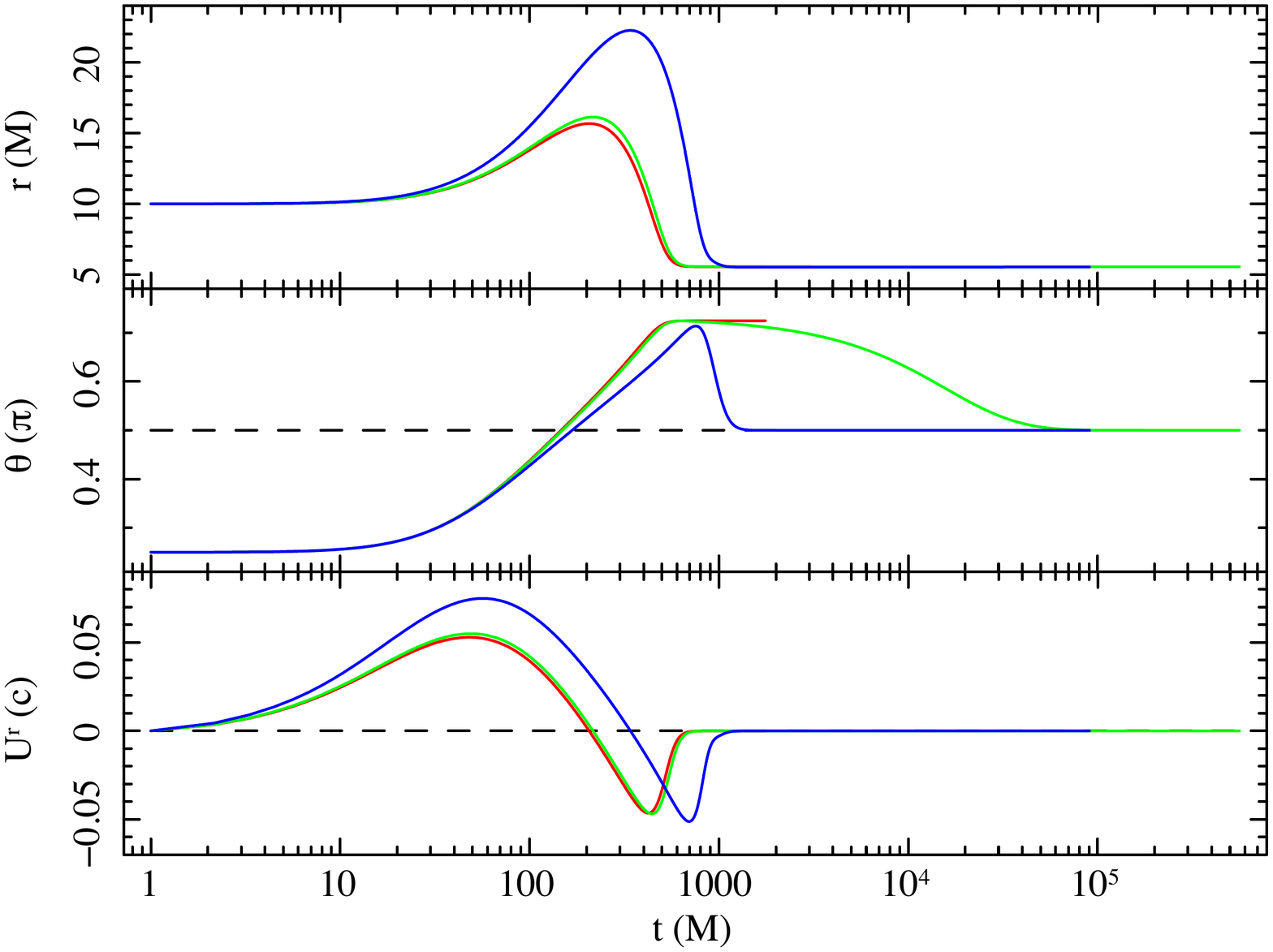}
\hspace{-1cm}
\includegraphics[scale=0.25]{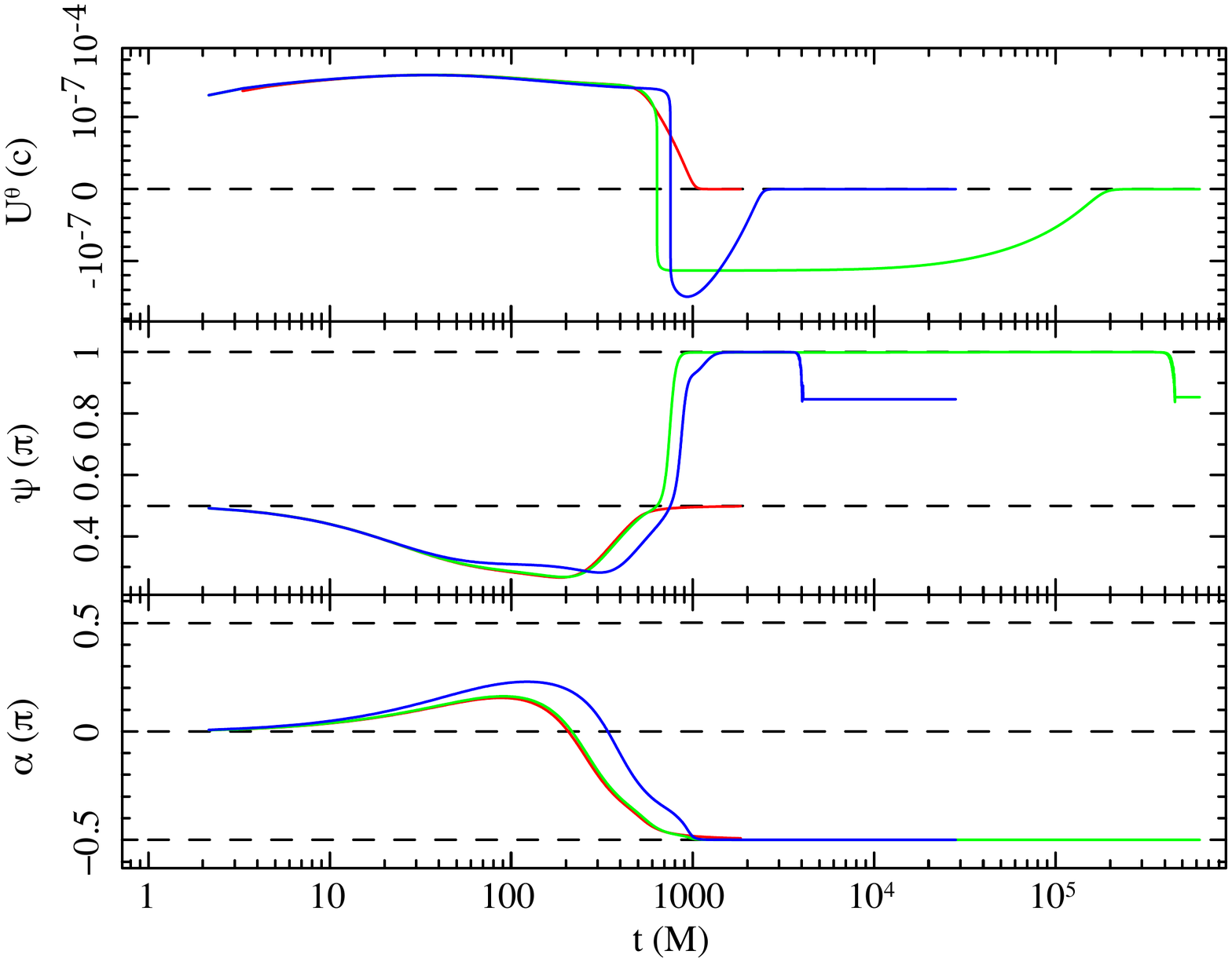}}
\caption{Profiles of $r$ and $\theta$ coordinates, $r$ and $\theta$ components of four-velocity ($U^r,\ U^{\theta}$), and $\psi$, $\alpha$ ZAMO local angles as functions of coordinate time $t$ for the test particles reaching the critical hypersurface with non-zero angular momentum ($\alpha \not=\pm\pi/2$). The function $u^\theta$ has been plotted in symmetric loagrithmic scale. The test particles are emitted outside of the critical hypersurface at $r_0=10M\,\theta_0=\pi/4$ in the azimuthal direction with the initial velocity $\nu_0=0.25$. The plots are constructed for the Schwarzschild case with zero spin case (red curves; compare to the left panel of Fig \ref{fig:Figs_S}), for the Kerr case with very small spin $a=0.05$ (green curves; compare to the right panel of Fig. \ref{fig:Figs_KSI1}) and for the Kerr case with almost extreme spin case $a=0.9995$ (blue curves; compare to the left panel of Fig \ref{fig:Figs_KH2}). The plots clearly illustrate the behavior in the touching point, where the test particles reaches the critical hypersurface. In the Kerr cases, $U^{\theta}$ is zero and ZAMO local polar angle $\psi$ takes the value of is $\pi/2$ at the touching point. Then during the latitudinal drift the angle $\psi$ increases while ZAMO local polar angle $\alpha$ decreases as the angular momentum of the test particle is removed. The local angle $\psi$ reaches the maximum value $\pi$. However, when the orbit is stabilized in equatorial plane and angular momentum is fully removed the angle $\psi$ is going back to the value of $\pi/2$ (The numerical integration of the trajectory is stopped earlier when spatial velocity is less than $10^{-20}$). In the Schwarzschild case, where latitudinal drift does not occur, the angular momentum is removed during the approaching to the critical hypersurface.}
\label{fig:Fig_plot}
\end{figure*}

In this section, I investigate in greater detail test particle trajectories bound to the critical hypersurface and their latitudinal drift towards the equatorial plane. I first emphasise that the condition for the radial balance of (outward) radiation force and gravitational attraction given by the equation (\ref{r-equilibrium}) is satisfied also when the test particle reaches the critical hypersurface with non-zero angular momentum (with its space velocity vector thus forming an arbitrary angle $\alpha$\ in the azimuthal direction; $\alpha \neq\pm\pi/2,\ \psi=\pm\pi/2,\ \nu=0,\ \gamma=1 $). The orbits of the test particles, reaching the critical hypersurface, can be divided into two classes with qualitatively different behavior.
\begin{itemize}
\item[(I)] Test particles with zero angular momentum achieve a complete balance of all forces acting at the critical hypersurface. Such case corresponds to  test particle trajectories which satisfy the condition $\nu=0$ at any $r(0)=r_0$ and $\theta(0)=\theta_0$ and are thus carried around by frame dragging in the azimuthal direction, along with  photons of the radiation field. At the critical hypersurface, such test particles then move along the purely off-equatorial circular orbits at constant latitude with angular velocity  $\Omega_{\mathrm{ZAMO}}$ (see Fig. \ref{fig:Fig_traj_bounded} and Fig. \ref{fig:Fig_plot}) remaining at rest relative to the appropriate ZAMO frame.
\item[(II)] Test particles that reach the critical hypersurface while still endowed with residual (non-zero) angular momentum (not co-aligned with the spin axis, $\alpha \neq\pm\pi/2,\ \psi=\pm\pi/2, \theta\neq\pi/2$) attain radial balance, but  the PR effect still operates on them because the radiation field is  not yet directed in the radial direction in the test particle frame \cite{Bini11}. Such particles exhibit a latitudinal drift on the critical hypersurface under the influence of the polar components of acceleration and consequently experience a polarly-oriented dissipative force originating from the interaction with the radiation field (see Eq. \ref{EoM2}). In the latitudinal drift the residual angular momentum of the test particle is progressively removed. Then in accordance with the reflection symmetry of the Kerr spacetime, full equilibrium ($\alpha=\psi=\pi/2, \nu=0, \gamma=1$) is attained in the equatorial plane, where latitudinal drift stops, the motion stabilises in a circular orbit and the angular momentum of the test particle is completely removed (see right panel of Fig. \ref{fig:Figs_S}, Figs \ref{fig:Figs_KSI1} - \ref{fig:Figs_KH2} and Fig. \ref{fig:Fig_plot}).
\end{itemize}
I note that in the Schwarzschild case, the spin and polar acceleration are absent, and thus latitudinal drift does not occur (see left panel of Fig. \ref{fig:Figs_S}).

\subsection{Concluding remarks}
\label{sec:end}
I developed a fully general relativistic treatment of the 3D PR effect in the Kerr geometry, therefore extending previous works describing 2D PR motion in the equatorial plane of relativistic compact objects. The outgoing radiation field I adopted assumes that photons propagate radially  with respect to the ZAMO frames. Such a boundary condition implies a purely radial propagation of the photons in any local ZAMO frame and may be considered as a simple approximation of the radiation field from a static emitting source very close to the horizon of a Kerr BH.
 
The resulting equations of motion for a test particle moving in the 3D space consist of a system of six coupled ordinary, highly nonlinear differential equations of first order. The non-linearity arises because of the general relativistic environment, further complicated by the PR effect which is a dissipative process and thus always entails nonlinearity. This set of equations is consistent with the previous 2D case for both test particles and photons moving in the equatorial plane \cite{Bini09}.

My analytical and numerical calculations in both the Schwarzschild and Kerr metric, show that 3D PR orbits are strongly affected by general relativistic effects, including frame-dragging. I have demonstrate the existence of a critical hypersurface, where the attraction of gravity is balanced by the outgoing radiation forces. In the case of the Schwarzschild geometry, the critical hypersurface is a sphere, on which the test particles are captured and remain at rest. In the case of the Kerr spacetime (with non-zero spin), the critical hypersurface is elongated in the polar direction. Test particles that are captured by it are dragged at an azimuthal angular velocity $\Omega_{\mathrm{ZAMO}}$, and, if still endowed with a residual and offset angular momentum, they exhibit a latitudinal drift that lead to spiraling towards the equatorial plane. Analysis of the $\nu$ profile shows that the test test particle spatial velocity attain that of the local ZAMO in an infinite time (see Figs. \ref{fig:Figs_KSI2}--\ref{fig:Figs_KSO2}--\ref{fig:Figs_KH1}). The test particle approaches the equatorial plane and radius of the hypersurface asymptotically. In future works I plan to relax some of the simple assumptions of the present study (e.g., by adopting more realistic radiation fields) and to investigate some possible astrophysical applications.  

\chapter{Accreting milliseconds X-ray pulsars}
\epigraph{A man who dares to waste one hour of time has not discovered the value of life.}{The Life \& Letters of Charles Darwin, Charles Darwin}

\lettrine{I}{n this chapter}, I present the accreting millisecond X-ray pulsars (AMXPs), known to be old ($\sim$Gyr) NSs endowed with relatively weak magnetic fields, $B\approx10^{8-9}$ G. These NSs are hosted in transient low mass X-ray binaries (LMXBs) that spend most of their time in quiescence and occasionally undergo weeks to months long outbursts. Coherent X-ray pulsations are observed from these systems with frequencies comprised between 180--600 Hz and their measured orbital periods range from 40 min to 5 hr. The attention then is focused on the recycling scenario theory, that confirms the evolutionary link between the accretion powered LMXBs and the rotation-powered millisecond pulsars. I list some general proprieties of AMXPs. The last sections are dedicated to the analysis of three peculiar AMXPs: \igr, \igrj, and \sax. The analysis of the last source is not reported here, but can be found in Ref. \cite{Li2018}. The contents of these topics are contained in the following papers both published in Astronomy \& Astrophysics Journal \cite{Defalco2017a,Defalco2017b}.

\section{Pulsars}
Pulsars are a highly magnetized, rotating NSs, that are remnants of supernova explosions. They emit electromagnetic radiation, that can be observed only when the beam of emission is pointing toward the observer (lighthouse effect), responsible of the pulsed profile \cite{Longair11}. Normally, stars have radii of $\sim10^6$ km, but under the gravitational collapse they shrink to a size of $\sim10$ km,  leading to densities between $\sim(3.7--5.9)\times10^{14}$ g cm$^{-3}$ comparable to nuclear densities \cite{Shapiro86}. A common propriety of all NSs is they are very dense objects. This happens because their interior consists of zones of increasing density, where the degenerate electron gas becomes relativistic \cite{Bahcall65,Shapiro86}, the total energy of electrons, $E^{\rm e}=\gamma m_{\rm e}c^2$, exceeds the mass difference between the neutron and proton, $E_{\rm max}^{\rm e}=(m_{\rm n}-m_{\rm p})c^2=1.3$ MeV. The condition $E^{\rm e}\geq E_{\rm max}^{\rm e}$ initiates the inverse $\beta$ decay process, $p+e^- \rightarrow  n+\nu_{\rm e}$, where protons are converted into neutrons. This explains from which these compact objects derive the name and why they are so dense. For stable and uncharged NSs, I will prove why the number of neutrons is higher than the number of electrons and protons. 

A \emph{stable} NS means $\epsilon_{\rm F}=E_{\rm max}^{\rm e}$, where \cite{Shapiro86} 
\begin{equation} \label{fe} 
\epsilon_{\rm F}=\frac{\hbar^2}{2m_{\rm e}}\left( \frac{3\pi^2 N}{V} \right)^{\frac{2}{3}}=\frac{\hbar^2}{2m_{\rm e}}\left( \frac{9\pi}{8}\frac{M_{\rm NS}}{m_{\rm n}} \right)^{\frac{2}{3}}\frac{1}{R^2_{\rm NS}},
\end{equation}
is the \emph{Fermi energy} for a typical NS of radius $R_{\rm NS}=10$ km and mass $M_{\rm NS}=1.4\ M_\odot$. Through the relativistic formula of energy, it is possible to calculate the Fermi energy momentum $p_{\rm F}$ as
\begin{equation} \label{pf} 
p_{\rm F}=\sqrt{\frac{\epsilon_{\rm F}^2-(m_{\rm e}c^2)^2}{c^2}}=1.2\ {\rm MeV/c}.
\end{equation} 
To estimate the number of electron particles, $N_{\rm e}$, inside a NS, I need to compute the number density of electron particles, $n_{\rm e}$ \cite{Shapiro86}, i.e.,
\begin{equation} \label{ne} 
n_{\rm e}=\frac{N_{\rm e}}{V}=\int_0^{p_{\rm F}}\omega(p)f(p)dp =\frac{8\pi}{3h^3}p_{\rm F}^3,
\end{equation}
where $\omega(P)=4\pi gp^2/h^3$ is the state density, $f(p)$ is the occupation probability (for energies $E\leq \epsilon_{\rm F}$ is equal to 1), $g=2s+1$ with $s$ being the electron spin (for fermions $s=1/2$). Therefore, $N_{\rm e}=n_{\rm e}V=3.2\times 10^{49}$, where $V=4/3\pi R_{\rm NS}^3$ is the entire volume of the NS. 

A NS is \emph{uncharged}, when the number of protons is equal to the number of electrons, $N_{\rm p}=N_{\rm e}$. The total mass of protons and electrons with respect to the NS mass is
\begin{equation} \label{totnumpe} 
\begin{aligned}
& N_p\cdot m_p\approx 5.34\times10^22\ kg\approx 1.9\times10^{-6}\%,\\
& N_e\cdot m_e\approx 2.92\times10^19\ kg\approx 1\times10^{-9}\%.
\end{aligned}
\end{equation} 
Applying the formula $\epsilon_{\rm F}(n)=\epsilon_{\rm F}(p)+\epsilon_{\rm F}(e)$, I find the following results.
\begin{itemize}
\item \textbf{Classical case.} Using Eq. (\ref{fe}), I obtain
\begin{equation} \label{ratepn} 
\frac{N_n^{\frac{2}{3}}}{m_n}=\frac{N_p^{\frac{2}{3}}}{m_p}+\frac{N_e^{\frac{2}{3}}}{m_e}\quad\Rightarrow
\quad\frac{N_n^{\frac{2}{3}}}{m_n}=\frac{N_p^{\frac{2}{3}}}{m_e}\quad\Rightarrow\quad \frac{N_p}{N_n}\approx\frac{1}{8000}.
\end{equation} 
\item \textbf{Relativistic case.} The following variables are defined \cite{Shapiro86} as
\begin{equation} \label{def} 
x_e=\frac{p_F^e}{m_e c}=2.4,\quad x_p=\frac{p_F^p}{m_p c}=\left(\frac{m_e}{m_p}\right)x_e=13\times 10^{-4}.
\end{equation} 
Since $p_{\rm F}\sim N^{1/3}$, it is $p_{\rm F}^{\rm e}=p_{\rm F}^{\rm p}$. Therefore, I obtain
\begin{equation} \label{rfenfer} 
\begin{aligned}
&\sqrt{m_n^2c^4+(p_F^n)^2 c^2}=\sqrt{m_p^2c^4+(p_F^p)^2 c^2}+\sqrt{m_e^2c^4+(p_F^e)^2 c^2}\quad \Rightarrow\\
&\\
&\Rightarrow\quad\sqrt{(1+x_n^2)}=\frac{m_p\sqrt{(1+x_p^2)}}{m_n}+\frac{m_e\sqrt{(1+x_e^2)}}{m_n}\quad \Rightarrow\\
&\\
&\Rightarrow\quad x_n\approx 5.3\times 10^{-2}.
\end{aligned}
\end{equation} 
This result yields to:
\begin{equation} \label{newratepn} 
\frac{x_p}{x_n}=\left(\frac{N_p}{N_n}\right)^{1/3}\quad \Rightarrow\quad \frac{N_p}{N_n}\approx\frac{1}{7000}.
\end{equation} 
\end{itemize}
Both cases demonstrate clearly that the number of neutrons is higher than the number of protons (electrons).
 
During the gravitational collapse stars conserve the angular momentum, $L_{\rm star}=L_{\rm pulsar}$, therefore for their small size the pulsars posses very short rotational periods. If the mass loss during contraction can be neglected, it obtains
\begin{equation}
\Theta_{\rm star}\omega_{\rm star}=\Theta_{\rm pulsar}\omega_{\rm pulsar} \quad \Rightarrow\quad \omega_{\rm pulsar}=\frac{R^2_{\rm pulsar}}{R^2{\rm star}}\omega_{\rm star},
\end{equation}
where $\Theta$ is the moment of inertia and $\omega$ the frequency. This formula corresponds to pulsar periods of:
\begin{equation}
T_{\rm pulsar}=\frac{R^2_{\rm star}}{R^2_{\rm pulsar}}T_{\rm star}.
\end{equation}  
Therefore for a stellar size $R_{\rm star}=10^5$ km, a pulsar radius $R_{\rm pulsar}=10$ km, and a star rotation period of $T_{\rm star}=12$ days, one obtains $T_{\rm pulsar}=1$ s \cite{Shapiro86}. During the gravitational collapse the original magnetic field is highly amplified. Assuming the conservation of the magnetic flux through the upper hemisphere of a star during the contraction, one obtains
\begin{equation}
\int_{\rm star}B_{\rm star}\cdot dA_{\rm star}=\int_{\rm star}B_{\rm pulsar}\cdot dA_{\rm pulsar}\quad \Rightarrow\quad B_{\rm pulsar}=\frac{R^2_{\rm star}}{R^2_{\rm pulsar}}B_{\rm star}.
\end{equation}   
For $B_{\rm star}=10^4$ G, the magnetic pulsar field is $B_{\rm pulsar}\sim10^{12}$ G. These theoretical results have been experimentally confirmed by measuring quantized energy levels of free electrons in strong magnetic fields. 

\section{Low mass X-ray binaries}
Low-mass X-ray binary systems (LMXBs) host a compact object (NS or BH) and a donar star, whose mass $<1M_{\odot}$. The estimation of the donor star mass could be derived in two ways: through globular cluster membership and/or the measurement of the orbital binary period \cite{lewin93}. The donor's average density, $\bar{\rho}$, is uniquely determined by the orbital period. If it is a low-mass main sequence star with an approximately linear relation between the mass and the radius, one finds that the mass of donor is proportional to the orbital period, through the following formula \cite{frank02}
\begin{equation}
\bar{\rho}=\frac{3M_2}{4\pi R_2^3}=\frac{3}{4\pi a^3}\left(\frac{1+q}{q}\right)\frac{3^4}{8}\approx\frac{3^5\pi M}{8a^3 4\pi^2}=\frac{3^5\pi}{8GP^2}\approx 110P_{\rm hr}^{-2}\ \mbox{g cm}^{-3},
\end{equation}
where $M_2$ is the mass of the donor star, $R_2$ is the Roche lobe radius of donor star, $a$ is the binary separation, $q=M_2/M_1$ with $M_1$ the mass of the compact object, $P_{\rm hr}$ is the period expressed in hours. I have used the approximation $(1+q)/q=M/M_2\approx M$, where $M=M_1+M_2$ is the total mass of the system and to eliminate $a$ I have considered Kepler's third law $4\pi^2 a^2=GMP^2$.   

Normally it is assumed the \emph{mass transfer conservation} from the low-mass to the more massive star, i.e. no mass escapes the system and the total angular momentum of the binary system remains constant. This process can continue if either the donor star expands filling its Roche lobe (occurring when the donor has evolved off the main-sequence branch) or the orbit shrinks requiring a decrease in orbital angular momentum (magnetic braking) \cite{lewin93,frank02}. The formation of the disk is due to the conservation of angular momentum of the infalling matter and viscosity (interactions among the particles of the gas) that permit not to accrete directly on the central object, but induce the matter to spiral around the central object. The gravitational potential energy of the matter is converted for the viscosity effects in thermal X-ray emission, permitting so to be detected, where the intensity emission depends on the mass of the compact object \cite{shakura73, frank02}. 

LMXBs are mostly concentred towards the galactic center and some of them are also found in globular clusters. They are located outside the regions of active star formation, identifying thus them as members of an old population \cite{lewin93,seward10}. LMXBs show some common proprieties \cite{lewin93}: they are bright X-ray sources ($>10^{34}$ erg s$^{-1}$); their star-like optical counterparts are faint; their spectra are normally soft without any normal stellar absorption features; they show no periodic X-ray pulsations such as those are often observed from highly magnetized, rotating NSs; the majority of them exhibits type-I X-ray bursts (see Sec. \ref{burst}, for further details).

Another important propriety characterizing LMXBs is the weak magnetic field. The pulsations are associated to strong magnetic fields, that funnel only a small fraction of the accreted matter. Since the effective accretion rate per unit area is very high, this may suppress the thermonuclear flashes, preventing the starting of type-I X-ray bursts \cite{lewin93}. The magnetic field of old NSs in LMXBs decayed on time scales of $\sim10^7$ yr, but it could be due not to the age, but probably as a result of accretion processes, where the magnetic field lines are "buried" below the surface of the NS by the accreted matter \cite{lewin93}. 

\subsection{Type-I X-ray bursts}
\label{burst}
In LMXBs the ionized accreting gas, coming from the companion star, is funneled by the NS magnetic field onto the polar cap of the NS surface, emitting in the X-ray energy band \cite{lewin93,frank02}. The accumulation of continuos flow creates accretion columns, where the matter is compressed and heated, on a time scale of hours to days. This process continues until the densities and the temperatures become adequate for thermonuclear unstable ignition of hydrogen and/or helium, arising and triggering thermonuclear type-I X-ray bursts \cite{lewin93,strohmayer06}. They are among the most evident signatures of the presence of a NS in LMXBs, and several thousand bursts, from around 110 NSs, have been observed to date \cite{galloway08a}. The Wide Field Cameras (WFC) on BeppoSAX in combination with the All Sky Monitor (ASM) onboard RXTE have provided an unprecedented insight on the burst sources and a long term view of the X-ray sky, encouraging new space discovery for rare events. 

The burst spectra are usually well described by a blackbody spectrum
\begin{equation}
F_{\rm bb}(E)\ dE=\frac{K\ E^3}{\exp^{\frac{E}{kT_{\rm bb}}}-1},
\end{equation}
where $E$ is the energy of the incident photon, $kT_{\rm bb}\approx1-3$ keV is the blackbody temperature, and $K$ is the normalization constant. All bursts have similar shape for all NS sources and are determined only by temperature and normalization constant. The exponential decay time, $\tau$, is attributed to the cooling of the NS photosphere, resulting thus in a gradual softening of the burst spectra \cite{lewin93,strohmayer06}. In most spectral studies of X-ray bursts it is customary to subtract the pre-burst persistent emission from the total signal. Since the burst emission may affect the accretion flow, it is not obvious whether this procedure is correct. When bursts reach their peak luminosities, one might expect a temporary suppression of the accretion. However, it has been argued instead that the accretion rate is enhanced \cite{walker89}, due to the action of radiation drag, that remove angular momentum from the inflowing matter (see Chapter 3 for further details).

\begin{figure}[ht]
\centering
\hbox{\hspace{0.5cm}
\includegraphics[trim=4cm 2cm 4.6cm 6cm, scale=0.27]{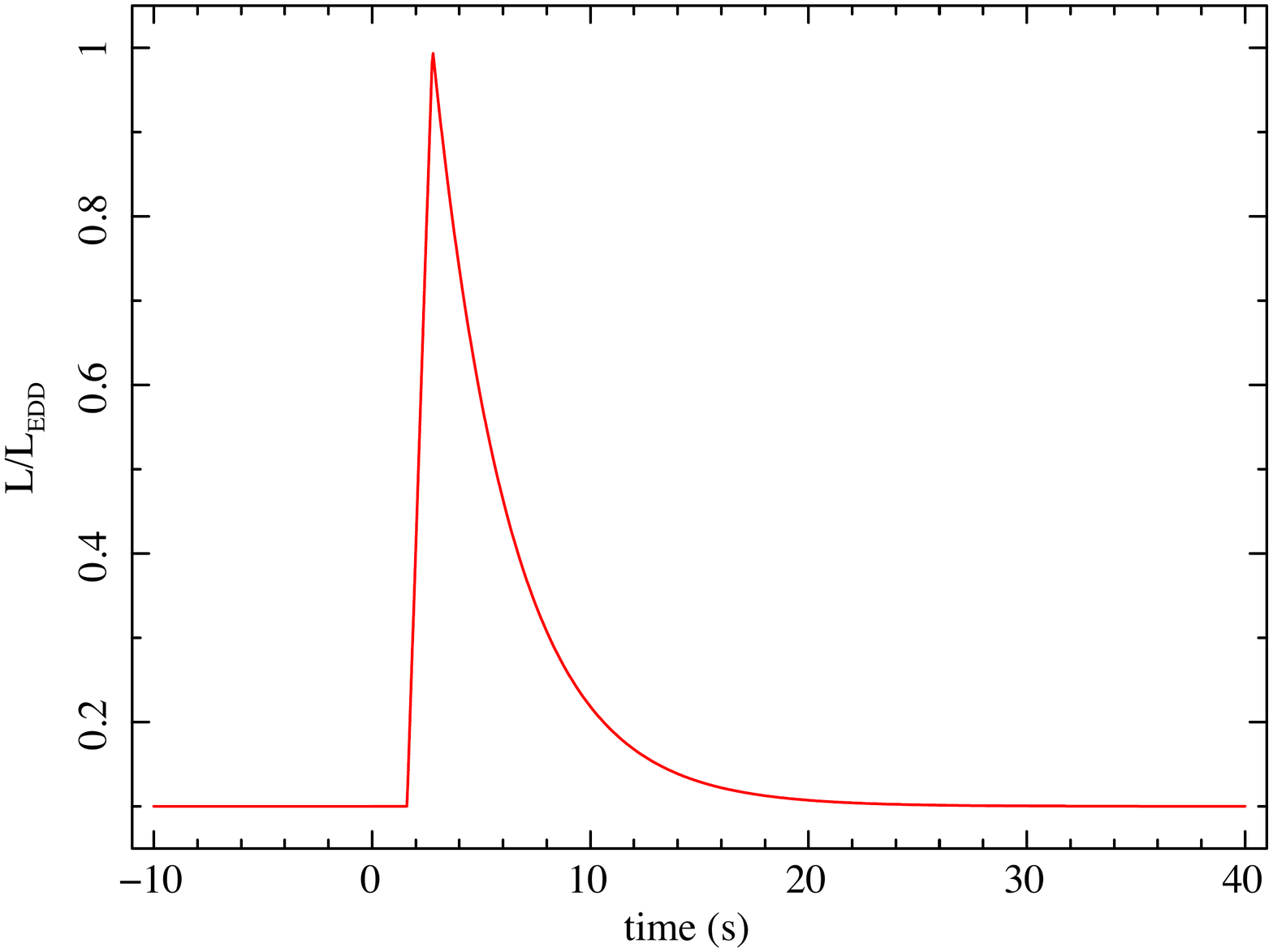}
\hspace{0.cm}
\includegraphics[trim=2cm 2cm 6.5cm 1cm, scale=0.27]{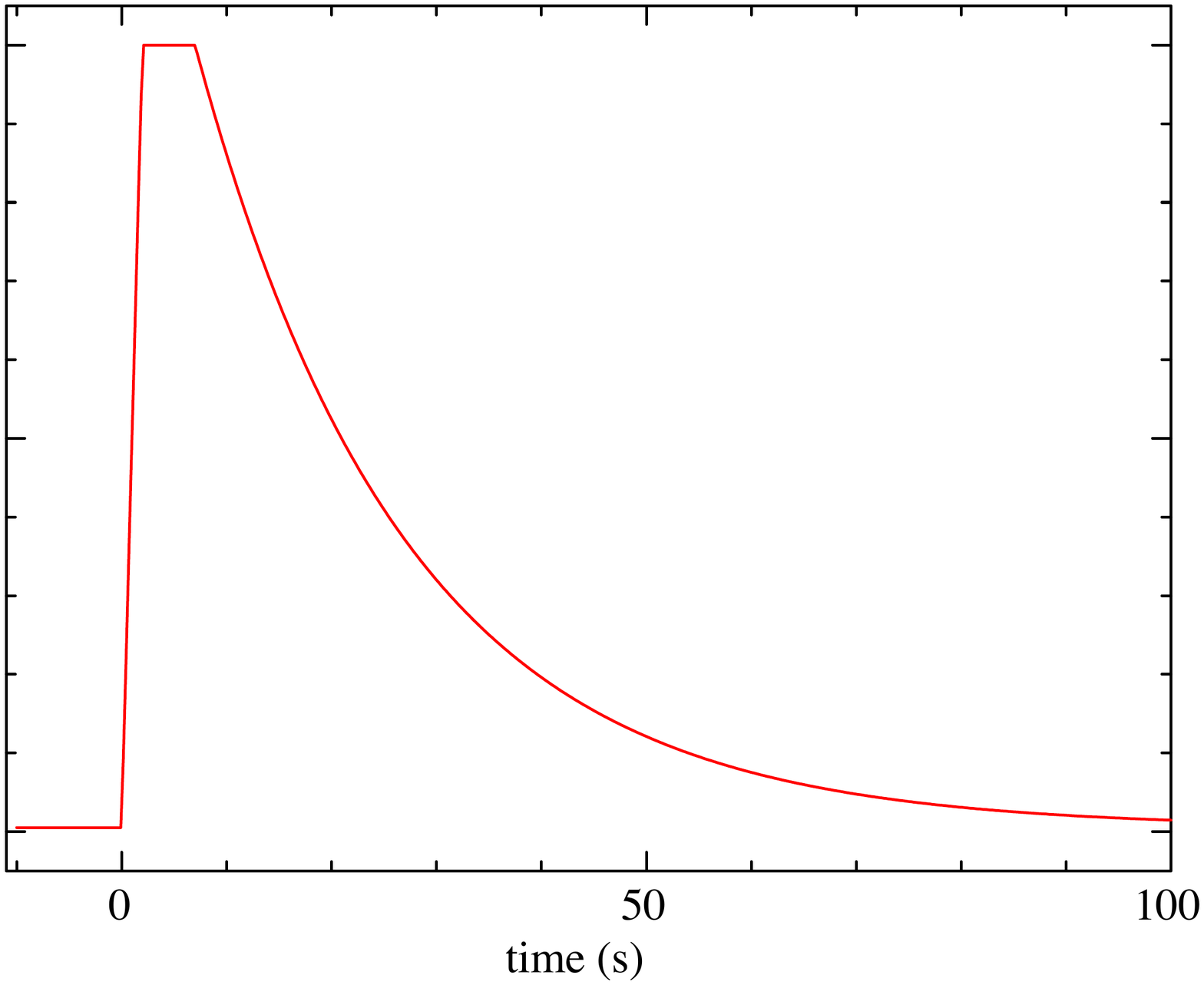}
}
\vspace{0.8cm}
\caption{Example of bursts with and without a PRE.} 
\label{fig:Fig2_1}
\end{figure}

During some type I X-ray bursts it is possible that the emitted luminosity on the NS surface becomes high enough to reach the Eddington limit. At that point the radiation pressure lifts the surface layers from the NS in a photospheric radius expansion (PRE) episode \cite{kuulkers10} and the luminosity remains almost constant near the Eddington limit for few seconds or minutes. Examples of bursts with and without a PRE are reported in Fig. \ref{fig:Fig2_1}.

For a blackbody emission the luminosity is given by $L\sim R^2T^4$, i.e. the blackbody temperature decreases, while the inferred blackbody radius simultaneously increases. The moment when the photosphere falls back to the NS surface (when the temperature is highest) is called "touch down", permitting thus to estimate the NS radius \cite{lewin93}. Radii inferred in this manner are typically in the range of $\sim10$ km, consistent with cooling of an object having the theoretical size of a NS \cite{strohmayer06}. This kind of bursts are considered as standard candle and are used as distance indicators. During the PRE phase, the Eddington luminosity as measured by a distant observer is \cite{lewin93, kuulkers10}
\begin{equation} \label{dist} 
L_{\rm Edd,\infty}=\left(\frac{4\pi cGM}{k}\right)(1+z)^{-1}=4\pi d^2 F_{\rm Edd,\infty},
\end{equation}
where $(1+z)^{-1}=\sqrt{1-2GM/(Rc^2)}=1.31$ is the redshift factor for a common NS ($R=10$ km and $M=1.4M_\odot$), $F_{\rm Edd,\infty}$ is the burst peak flux measured at infinity, $k$ is the opacity, and $d$ is the distance to the source. Regarding $k$, during the PRE phase it is usually considered the electron scattering opacity, that in the low temperature limit is given by $k_0=0.2(1+X)$ cm$^2$ g$^{-1}$, where $X$ is the hydrogen fraction mass of the photospheric matter (for hydrogen-poor/helium is $X=0$ and for cosmic composition/hydrogen-rich is $X=0.73$) \cite{fujimoto81}. In some cases, it may be necessary to take into account the temperature dependence of the electron scattering opacity, that for high temperatures becomes relativistic. This dependence can be described by the following approximation, valid for the low density limit appropriate to NS atmospheres, $k=k_0[1+(kT_{\rm bb}/39.2\ \mbox{keV})^{0.86}]^{-1}$, where $\alpha=2.2\times10^{-9}$ K$^{-1}$ \cite{lewin93}. Since from the observations it is possible to measure the burst temperature and flux, Eq. (\ref{dist}) permits to estimate the distance $d$.  

\begin{figure}[ht]
\centering
\includegraphics[scale=0.65,trim=3.6cm 2cm 4cm 2cm]{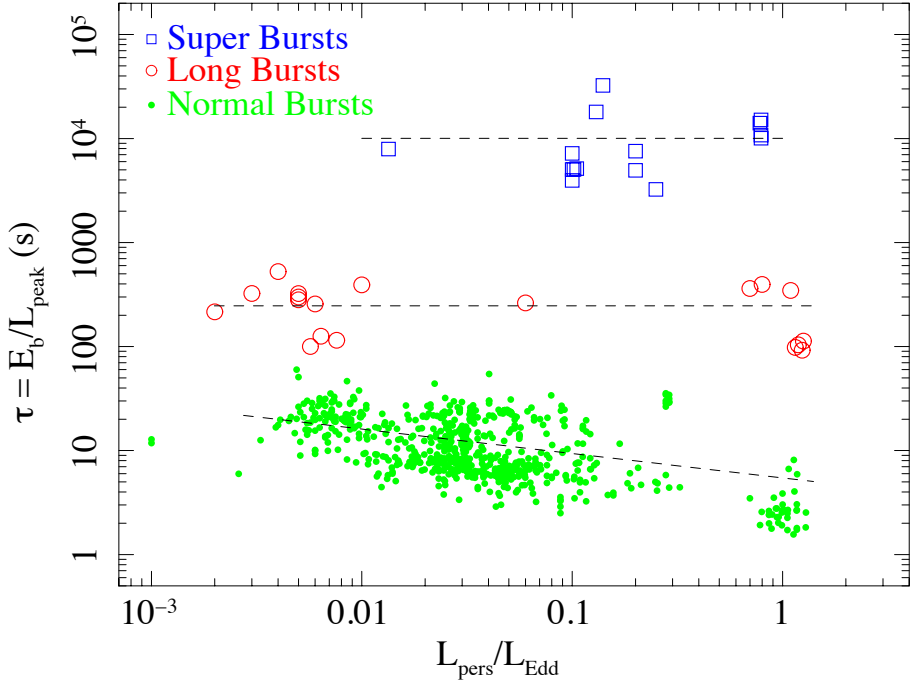}
\caption{Figure 7, adapted from \cite{falanga08}, shows burst effective duration vs. persistent luminosity for normal shortbursts \cite{galloway06}, intermediate long bursts \cite{galloway06,falanga08}, and superbursts \cite{kuulkers04,intzand04,falanga08}.}
\label{fig:Fig2_2}
\end{figure}

From the burst duration it is possible to distinguish three main branches: {\it (i)} normal short-bursts (seconds), {\it (ii)} intermediate long-bursts (minutes) and {\it (iii)} superbursts (hours) \cite{falanga08, kuulkers04}. All sources showing intermediate long bursts or superbursts exhibit also normal bursts \cite{kuulkers04,strohmayer06}. The rise time is $\approx 1-2$ s for short-bursts, $\approx 1-100$ s for long-bursts and $\lesssim 1$ s for superbursts; whereas the decay is approximately exponential, with a duration of a few seconds for short-bursts, tens of minutes for long-bursts, and up to several hours for superbursts, depending on the nuclear reactions involved. All these bursts can be described in terms of different fuel types and accretion rates \cite{lewin93,strohmayer06}. It is thought that: normal short bursts are powered by unstable burning of hydrogen and helium \cite{schatz03}; intermediate long bursts are due to thermally-unstable hydrogen ignition in a mixed hydrogen and helium environment or also pure helium accretion rates \cite{intzand05,cooper07}; the superbursts are powered by an unstable carbon burning in an ocean of heavy nuclei \cite{cumming01,strohmayer06,schatz03}. Normally the short- and long-bursts can show a PRE of $\approx5$ s and $\approx400$ s, respectively; instead the superbursts have never shown a PRE so far \cite{strohmayer06}. The regimes of unstable burning from NSs accreting at sub-Eddington rates are \cite{strohmayer06}:
\begin{itemize}
\item[(1)] $\dot{m}<900$ g cm$^{-2}$ s$^{-1}$ ($\dot{M}<2\times10^{-10}M_\odot\ yr^{-1}$): mixed hydrogen and helium burning triggered by thermally unstable hydrogen ignition;
\item[(2)] 900 g cm$^{-2}$ s$^{-1}$ $<\dot{m}<2\times10^3$ g cm$^{-2}$ s$^{-1}$: pure helium shell ignition following completion of hydrogen burning;
\item[(3)] $\dot{m}>2\times10^3$ g cm$^{-2}$ s$^{-1}$ ($\dot{M}>4.4\times10^{-10}M_\odot\ yr^{-1}$): mixed hydrogen and helium burning triggered by thermally unstable helium ignition.
\end{itemize}
Figure \ref{fig:Fig2_2} reports all kinds of bursts described above in terms of persistent emission and decay time.

\section{Recycling scenario}
In July 1967 a large telescope in the Mullard Radio Astronomy Observatory in Cambridge detected signals, that were repeatedly observed at a fixed declination and right ascension \cite{Hewish68}. In November 1967 Jocelyn Bell Burnell and Antony Hewish analysed the data and discovered the first radio pulsar, PS B1919+21, having a period of 1.3373 s and a pulse width of 0.04 s. Since this discovery, the radio astronomy started to develop and many other radio pulsars were observed. In 1970, since the launch of the Uhuru X-ray satellite, some X-ray pulsars  (e.g., Cen X-3 and Her X-1) were discovered in X-ray binary systems. Therefore it became spontaneous to question why it has never been observed any radio pulsar in a binary system. There were proposed different explanations to solve this conflictual issue, summarized in the following points \cite{bk74}.        
\begin{itemize}
\item The NSs might be totally absent from close binaries, and therefore this is the reason why the radio pulsars do not occur in binary systems.
\item The stellar evolution of a more massive star in a binary system may end up with a supernova explosion, that could destroy the pair. 
\item The intensive accretion on the NS might raise its mass above the limiting value, implying a possible further evolution into a BH.
\item If a radio pulsar was in a binary system, its rotation period, after that the accretion had ceased, should have been almost similar to the observed X-ray pulsar. Therefore in absence of accretion such NS should be observable as a radio pulsar. However performing some calculations, the magnetic field value was estimated to be $B<10^8$, that rendered the radio pulsar unobservable. This weaker magnetic field (by a factor of 100) was thought to decay so rapidly under conditions of intensive accretion. 
\end{itemize}
The last point did not preclude the possibility to find a radio pulsar in a close binary system, but it required further assumption on the magnetic field decay.

\subsection{Radio pulsars in X-ray binary systems}
\begin{figure}[h] 
\centering
\includegraphics[trim=0cm 0cm 0cm 0cm, scale=0.48]{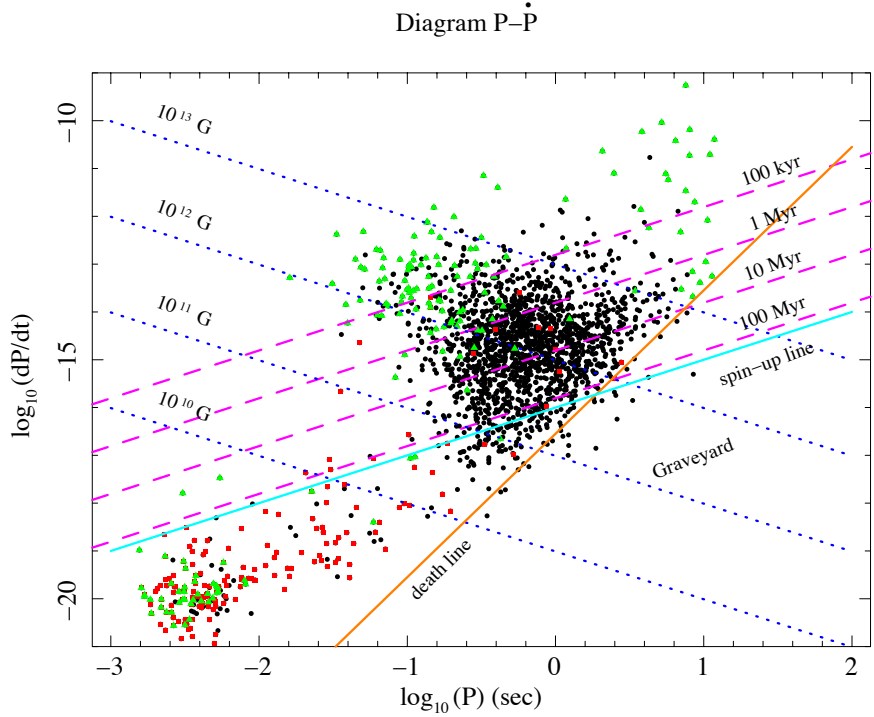}
\caption{$P-\dot{P}$ diagram showing all the actual known pulsars. The circular black points englobe all the pulsars, the green triangular points are pulsars with a companion star in a binary system, the quadratic red points are the pulsars detected at high energies (i.e., X-ray, $\gamma$-ray, infrared, and in general all the wavelength greater than radio). The dotted blu lines represent the magnetic field strengths, the dashed pink lines are the pulsar life times, the continuos orange line is "the death line" where the pulsars are not visible in radio, and the continuos light blue line is the spin-up line setting a maximum limit on the pulsar spin frequencies.}
\label{fig:Fig2_3}
\end{figure}

In 1974 at the Arecibo Observatory in Puerto Rico it was discovered the first radio pulsar, PSR B1913+16, in a binary system \cite{Hulse75}. Since then many other radio pulsars were discovered in X-ray binary systems. The principal results of the timing observations (like pulsar period, pulsar derivative, and celestial coordinates) were remarkably important to have a better understanding about the nature of the pulsar evolution. Particularly interesting is the plot of period derivative versus period for a large sample of pulsars. I report in Fig. \ref{fig:Fig2_3} the $P-\dot{P}$ diagram using all the actual available data. 
I present some formulas and concepts to understand the different features of this plot.
\begin{itemize}
\item {\bf Characteristic age.} A pulsar of spin frequency $\Omega$ and inertial momentum $I$, changes its rotational energy, $E_{\rm rot} = \frac{1}{2} I\Omega^2$, according to
\begin{equation} \label{if} 
\dot{E}_{\rm rot} = I\ \Omega\ \dot{\Omega}, 
\end{equation} 
where I have assumed that $I$ is constant in time. This energy is due to deformation of the magnetic field lines, or outflow of high-energy charged particles, therefore the variation of energy, $\dot{E}_{\rm rot}$, is obtained through the product of the magnetic energy density at the light cylinder \footnote{The light cylinder is a region around a pulsar where the matter rotates at a velocity comparable to the speed of light.}, the effective area of the light cylinder, and the velocity of light, at which the particles move on that region, i.e.,
\begin{equation} \label{newe} 
\dot{E}_{\rm rot}=-\frac{B_{\rm L}^2}{8\pi}\ (4\pi R_{\rm L}^2)\ c,
\end{equation} 
where $B_{\rm L}$ is the magnetic field at the light cylinder, and $R_{\rm L}=c/\Omega$ is the radius of the light cylinder. The magnetic field $B_{\rm L}$ could be considered as a multipole in dependence of the magnetic field at the polar cap, $B_{\rm P}$, i.e.,
\begin{equation} \label{mmfbl} 
B_{\rm L}=B_{\rm P}\cdot \left(\frac{R}{R_{\rm L}}\right)^p,
\end{equation}  
where $p$ is the multipole index, and $R$ is the pulsar radius. So equaling Eqs. (\ref{newe}) and (\ref{if}), I obtain
\begin{equation} \label{tutins} 
\dot{\Omega}=-\frac{1}{8}c^{3-2p}\ I^{-1}\ R^{2p}\ B_P^2\ \Omega^{2p-3}=-k\cdot B_P^2\ \Omega^{n}=-K\cdot \Omega^n,
\end{equation}  
where $k=\frac{1}{8}c^{3-2p}\ I^{-1}\ R^{2p}\ge0$, $K=k\ B_P^2\ge0$, $n=2p-3$ assuming that the magnetic field is independent from the time. The observed regular increase in period is attributed to the loss of rotational energy and angular momentum via ejected particles and/or electromagnetic radiation at the rotation frequency. The characteristic age of a pulsar is given by integrating Eq. (\ref{tutins}), i.e., 
\begin{equation} \label{eta} 
\int_{\Omega_i}^{\Omega}\frac{d\omega}{\omega^n}=-k\int_0^\tau dt\quad \Rightarrow\quad \tau=-\frac{\Omega}{(n-1)\dot{\Omega}}\left[1-\left(\frac{\Omega}{\Omega_i} \right)^{n-1}\right].
\end{equation} 
Eq. (\ref{eta}) can be also expressed in terms of the period $P=2\pi/\Omega$,
\begin{equation} \label{etaper} 
\tau=\frac{P}{(n-1)\dot{P}}\left[1-\left(\frac{P_i}{P} \right)^{n-1}\right].
\end{equation} 
For a dipole magnetic field in the vacuum ($n=3$), Eqs. (\ref{eta}) and (\ref{etaper}) become
\begin{equation} \label{eta3} 
\tau=-\frac{\Omega}{2\dot{\Omega}}\left[1-\left(\frac{\Omega}{\Omega_i} \right)^2\right]=\frac{P}{2\dot{P}}\left[1-\left(\frac{P}{P_i} \right)^2\right].
\end{equation} 
In general, $\Omega_i \ll \Omega$ (or $P_i\gg P$), so Eq. (\ref{eta3}) can be rewritten as
\begin{equation} \label{eta3new} 
\tau=-\frac{\Omega}{2\dot{\Omega}}=\frac{P}{2\dot{P}}.
\end{equation} 
In Fig. \ref{fig:Fig2_3}, Eq. (\ref{eta3new}) is plotted logarithmically, i.e., $\log_{10}(\dot{P})=\log_{10}(2P)-\log_{10}(\tau)$, with dashed pink lines.
\item {\bf Magnetic field.} Using Larmor's formula, the power of the magnetic dipole radiation, $P_{\rm rad}$, from an inclined magnetic dipole, $m$, is
\begin{equation} \label{prad} 
P_{\rm rad}=\frac{2}{3}\frac{\ddot{m}_\bot^2}{c^3},
\end{equation} 
where $\ddot{m}_\bot$ is the perpendicular component of the magnetic dipole moment. For an uniformly magnetized sphere with radius $R$ and a surface magnetic field strength, $B$, the magnetic dipole moment is $m=BR^3$. If the inclined magnetic dipole rotates with angular velocity, $\Omega$, it yields to
\begin{equation} \label{md} 
m=m_0\ e^{-i\Omega t}\quad \Rightarrow \quad \ddot{m}=\Omega^2 m.
\end{equation} 
Substituting Eq. (\ref{md}) in Eq. (\ref{prad}), I have
\begin{equation} \label{prad} 
P_{\rm rad}=\frac{2}{3c^3}(BR^3\sin\alpha)^2\left(\frac{2\pi}{P}\right)^4,
\end{equation}  
where $\alpha>0$ is the inclination angle from the rotational axis. Eq. (\ref{newe}) can be expressed in terms of the period derivative $\dot{P}$ and period $P$ as
\begin{equation} \label{newep} 
\dot{E}_{\rm rot}=-\frac{4\pi^2\ I\ \dot{P}}{P^3}.
\end{equation} 
The pulsar magnetic field $B$ is found equalling Eqs. (\ref{prad}) and (\ref{newep}), i.e., 
\begin{equation} \label{magfieint} 
P_{\rm rad}=\dot{E}_{\rm rot}\ \Rightarrow\ B=\left(\frac{3c^3I}{8\pi^2R^6\sin^2\alpha}\right)^{1/2}(P\dot{P})^{1/2},
\end{equation} 
Since generally the inclination angle $\alpha$ is not known, it is calculated the magnetic field at the pulsar surface, $B_{min}$, as
\begin{equation} \label{magfieint} 
B>B_{min}=\left(\frac{3c^3I}{8\pi^2R^6}\right)^{1/2}(P\dot{P})^{1/2}\approx3.2\times10^{19}\ (P\dot{P})^{\frac{1}{2}}.
\end{equation} 
In Fig. \ref{fig:Fig2_3}, Eq. (\ref{magfieint}) is plotted logarithmically, i.e., $\log_{10}(\dot{P}) = 2 \log_{10}(B)- \log_{10}(3.2\times10^9)-\log_{10}(P)$, with dotted blue lines.
\item {\bf Death line.} When a pulsar reaches a certain magnetic field strength and a sufficiently low period, no more radiation is produced and the pulsar is defined "dead" (occupying so the \emph{graveyard}), because the radio pulsar mechanism is believed to turn off. The radio emission is connected with the formation of electron-positron pair. These particles are accelerated along the magnetic field lines to extreme relativistic energies comparable to $e\Delta V$, where $e$ is the charge of an electron and $\Delta V$ is the potential difference between the center of the polar cap and the edge of the negative current emission region. In the stationary observer's frame it is valid $\boldsymbol{E}=-(\boldsymbol{\Omega}\times\boldsymbol{r})\times\boldsymbol{B}/c$ and using the formula $\Delta V=|\boldsymbol{E}|/e$, it is possible to estimate the potential difference $\Delta V$ in terms of the total flux penetrating the light cylinder $\Phi$ \cite{ruderman75}, i.e.,
\begin{equation} \label{deltaV} 
\Delta V\approx \frac{\Omega}{2\pi c}\Phi\approx \frac{\Omega^2 R^3}{2c^2}(\boldsymbol{B}\cdot\hat{\boldsymbol{r}})\approx6.6\times \frac{B}{P^2}\ \mbox{V},
\end{equation} 
where $\hat{\boldsymbol{r}}$ is the versor defining the direction of interest. It is possible to calculate the maximum possible potential drop $\Delta V_{max}$ along any magnetic field line. In fact the radio emission continues until the potential difference $\Delta V$ generated by the pulsar is greater than a certain maximum potential difference $\Delta V_{\rm max}$, i.e. $\Delta V \ge \Delta V_{\rm max}$. So the condition to determine the death line equation is
\begin{equation} \label{deathline} 
\frac{B}{P^2}=0.17\times10^{12}.
\end{equation}  
In Fig. \ref{fig:Fig2_3}, Eq. (\ref{deathline}) is plotted logarithmically using Eq. (\ref{magfieint}), i.e. $\log_{10}(\dot{P})=3\log_{10}(P)+2\log_{10}\left(\frac{B}{3.2\times 10^{19}}\right)$, with the orange continuos line.
\item {\bf Spin-up line.} The magnetospheric radius, $R_{\rm m}$, representing the region of equilibrium between the magnetic pressure and the ram pressure of the infalling accreting matter, is
\begin{equation} \label{radeq} 
R_{\rm m}=\frac{1}{(8G)^{\frac{1}{7}}}B^{\frac{4}{7}}R^{\frac{12}{7}}M^{-\frac{1}{7}}\dot{M}^{-\frac{2}{7}},
\end{equation}  
The equilibrium period, $P_{\rm e}=2\pi/\Omega_{\rm K}(R_{\rm m})$ where $\Omega_{\rm K}(R_{\rm m})=\sqrt{\frac{GM}{R_{\rm m}^3}}$, can be expressed in terms of Eq. (\ref{radeq}) as
\begin{equation} \label{pereq} 
P_{\rm e}=2.2\ B^{\frac{6}{7}}R^{\frac{18}{7}}m^{-\frac{8}{7}}\dot{m}^{-\frac{3}{7}}.
\end{equation}  
Eq. (\ref{pereq}) is a family of spin-up lines depending on the accretion process $\dot{m}$. The maximum spin-up line is reached in correspondence of the Eddington accretion rate (i.e., $\dot{m}=1$). Using Eq. (\ref{magfieint}) and considering a canonical NS with a radius of $R=10$ km and a mass of $M=1.4\ M_\odot$, it has $\log_{10}(\dot{P})=\log_{10}(P)-16$, that it is plotted in Fig. (\ref{fig:Fig2_3}) with the continuos light blue line. 
\end{itemize} 

In 1982 Backus \emph{et al.} \cite{Backus82} noticed that some sources where in the lower part of the diagram and some of them where also members of binary systems. It was natural to interpret that their unusually short period evolution was influenced significantly by the mass accretion from their companion stars. 

\subsection{Recycled accreting X-ray millisecond pulsars}
In 1982 it was detected the pulsar 4C21.53 with a period of $\sim1.56$ ms. For its extraordinary rapid rotation never observed before, it encouraged a number of observations regarding its possible origin and proprieties. The explanations proposed can be summarized in the following points \cite{r82}:  
\begin{itemize}
\item it acquired its rapid rotation rate in an accreting binary system;
\item there was a strong explosion, that destroyed the companion star and the remanent has been completely dispersed in the interstellar medium, moving so away in distance that was not possible to be observed;
\item from considerations on the energies detected by the X-ray satellite Uhuru it was estimated a period derivative
$<10^{-19}$ s s$^{-1}$, corresponding to a magnetic field of $<4\times10^8$ G.
\end{itemize} 
It appeared reasonable to assume that both the very short period and the very low magnetic field had a common cause and could have been interpreted as a recycled NS previously in an accreting X-ray binary system. All these events led naturally to formulate the \emph{recycling scenario theory}, claiming \cite{alpar82}:\\

\emph{\qm{ Sufficiently low magnetic field NSs which accrete for long times from a surrounding keplerian disk can be spun up to milliseconds periods. After accretion ceases, such stars could become isolated or binary pulsars, depending on whether their companions have subsequently disrupted the binary by either their tidal break ups or their supernova explosions. Such pulsars constitute a new class, with short period, long apparent ages ($P/\dot{P}\gtrsim 10^8$ yr) and pulsed optical, X-ray and $\gamma$-ray fluxes significantly below those expected for canonical pulsars with smaller periods. Moreover the transfer of angular momentum through accretion is the mechanism responsible for the spin-up of pulsars. Once the mass transfer episode terminates, the NS may eventually switch on again as a recycled millisecond radio pulsar. }}

\subsection{Proofs of the recycling scenario}
Nowadays this theory has been widely accepted thanks to a series of observational confirmations. The main discoveries can be summarized as it follows.
\begin{itemize}
\item In 1998 the transient X-ray burst source SAX J1808.4-3658 was the first accreting millisecond X-ray pulsar to be discovered \cite{Wijnands98}, confirming so the theoretical expectations. It was found that this pulsar was a member of a low-mass X-ray binary system; it had a magnetic field $B$ of $(2-6)\times10^8$ G (using standard magnetospheric disk accretion theory); it was reported a clear evidence for X-ray pulsations with a frequency near 401 Hz ($\approx2.49$ ms); the source was found located above the "death line", so it was interpreted that when the accretion phase of the binary finally had ended and the source had turned off as an accretion powered X-ray pulsar, it would have probably switched on as a rotation-powered radio pulsar. Therefore it was thought that the LMXBs could indeed be one of the progenitors of the millisecond radio pulsars.
\item In 2005 it was detected a significant spin-up value $\dot{\nu}=8.4\times10^{-13}$ Hz s$^{-1}$ of the AMXP IGR J00291+5934 \cite{falanga05}. 
\item In 2013 it was detected accretion-powered millisecond X-ray pulsations from the AMXP IGR J18245--2452 previously seen as a rotation-powered radio pulsar \cite{papitto13c}. Within a few days following a month-long X-ray outburst, radio pulses were again detected. Cross-referencing with the known rotation-powered radio pulsars in M28, it was found that the pulsar PSR J1824--2452I had ephemerides identical to those of the INTEGRAL X-ray source IGR J18245--2452. This not only demonstrated the evolutionary link between accretion and rotation-powered millisecond pulsars, but also that some systems can swing between two states on very short timescales. 
\end{itemize}

In Table \ref{tab1} I report the AMXPs discovered so far.

\begin{table}[h]
  \caption{List of the all AMXP discovered so far.}
\centering
\scalebox{0.84}{\hspace{0.5cm}
\begin{tabular}{cccc}
\hline			
Name & Spin frequency & Orbital period & Reference\\
 & (ms) & (h) & \\
\hline			
SAX J1808.4-3658 & 2.5 & 2.0 & Wijnands \& van der Klis 1998\\
XTE J0929-314 & 5.4 & 0.73 & Galloway et al. 2002\\
XTE J1751-305 & 2.3 & 0.7 & Markwardt et al. 2002\\
XTE J1814-338 & 3.2 & 4.0 & Markwardt et al. 2003\\
XTE J1807-294 & 5.3 & 0.67 & Markwardt et al. 2003\\
IGR J00291+5934 & 1.7 & 2.5 & Galloway et al. 2005\\
HETE J1900.1-2455 & 2.7 & 1.4 & Kaaret et al. 2005\\
SWIFT J1756.9-2508 & 5.5 & 0.9 & Markwardt et al. 2007\\
Aql X-1 & 1.8 & 19 & Casella et al. 2007\\
SAX J1748.9-2021 & 2.3 & 8.8 & Altamirano et al. 2007\\
NGC 6440 X-2 & 4.8 & 0.96 & Altamirano et al. 2010\\
IGR J17511-3057 & 4.1 & 3.5 & Markwardt et al. 2009\\
SWIFT J1749.4-2807 & 1.9 & 8.8 & Altamirano et al. 2010\\
IGR J1749.8-2921 & 2.5 & 3.84 & Papitto et al. 2011\\
IGR J18245-2452 & 3.9 & 11.03 & Papitto et al. 2013\\
XSS J12270 & 1.7 & 6.9 & Bassa et al. 2014\\
PSR J1023+0038 & 1.7 & 4.75 & Archibald et al. 2015\\
MAXI J0911-655 & 2.9 & 0.74 & Sanna et al. 2016\\
  \hline  
\end{tabular}
\label{tab1}}
\end{table}  
 
\section{General proprieties of AMXPs}
In this section, some proprieties of the AMXPs are listed. 
\begin{itemize}
\item {\bf Magnetic field.} In 2002 Burderi, Di Salvo et al. proposed a method to constraint the AMXP magnetic field, based on the measurement of the luminosity in quiescence (knowing of course the distance to the source) and on the knowledge of its spin frequency. There are three possible scenarios that can occur in dependence of the location of the magnetospheric radius $R_{\rm m}$ respect to the corotation radius, $R_{\rm co}$, and light cylinder radius $R_{\rm L}$.
\begin{itemize}     
\item[a)] If $R_{\rm m}<R_{\rm co}$, then there is accretion on the NS surface.
\item[b)] If $R_{\rm co}<R_{\rm m}<R_{\rm co}$, then the accretion onto the NS is centrifugally inhibited, but an accretion disk can still be present and emit X-rays.
\item[c)] If $R_{\rm m}>R_{\rm co}$, then there will be no accretion and no disk. The rotating NS emits radiation, bolometric luminosity $L_{\rm bol}$ according to Larmor's formula $L_{\rm bol}=(2/3c^3)\mu^2 \Omega^4$, where $\mu=BR^3$ with $B$ the magnetic field on the NS surface. In this case, the X-ray emission can be produced by: (c1) the reprocessing of part of the bolometric luminosity of the rotating NS or (c2) the intrinsic X-ray emission of the radio pulsar.
\end{itemize}
In all these scenarios it is possible to calculate the expected X-ray luminosity in quiescence, which depends on the NS spin and magnetic field. This can be compared with the observed luminosity (considered as an upper limit), giving therefore an upper limit on the magnetic field, once the NS spin frequency is known. For each of the above scenarios, the upper limits are \cite{Burderi02,DiSalvo03}:
\begin{itemize}
\item[a)] $\mu_{26} \leq 0.08\ L_{33}^{1/2}\ m^{1/3}\ P^{7/6}_{-3}$;
\item[b)] $\mu_{26} \leq 1.9\ L_{33}^{1/2}\ m^{-1/4}\ P^{9/4}_{-3}$;
\item[c1)] $\mu_{26} \leq 0.05\ L_{33}^{1/2}\ P^{2}_{-3}\ \eta^{-1/2}$;
\item[c2)] $\mu_{26} \leq 2.37\ L_{33}^{0.38}\ P^{2}_{-3}$.
\end{itemize}
$\mu_{26}$ is the NS magnetic moment in units of $10^{26}$ G cm$^3$ , $L_{33}$ is the accretion luminosity in units of $10^{33}$ erg s$^{-1}$, and $\eta\sim 0.01-0.1$ is the efficiency in the conversion of the rotational energy into X-rays. Using this technique the AMXP magnetic fields are estimated to be between $\sim10^{8-9}$ G.

\item {\bf Spin up and spin down.} In the general understanding of the AMXPs it becomes particularly important to measure the long-term spin evolution. In 2010 Patruno introduced a new method called \emph{correlation coherent analysis} justifying definitively that the AMXPs were spinning up during an outburst and spinning down during the quiescence (as predicted by the recycling scenario) \cite{Patruno10}. This new method takes into account the possible influence of the X-ray flux on the pulse phase, through a linear relation, not considered by the classical \emph{standard coherent analysis}. It is relevant to note that both kinds of analysis have the same number of parameters to fit (and therefore also the same degrees of freedom), so there is no risk to over-fit the data when using the new method. It really brought improvements in the fit, because the standard coherent analysis was giving some puzzling results, like spin-down during the outburst phase, while it was detected a positive spin frequency derivative. This behaviour can be ascribed to the presence of timing noises, that were neglected by the standard coherent techniques and therefore the outcome quantities became meaningless. This new method permitted to fully confirm and accept the correctness of the recycling scenario, because it was possible to measure the spin-up during the outburst process due to the accretion torque in agreement with the theory of accreting NSs, and the spin-down during the radio emission due to the magneto-dipole torques in agreement with the theory of radio pulsars. Another strong evidence of the validity of this new method was the possibility to measure the NS magnetic field, that was consistent with that inferred from the accretion torques during the outburst \cite{Patruno10}. 
\item {\bf Pulsations.} Since the discovery of AMXPs, pulsations have been detected only in these systems and not from other NSs in LMXBs. This odd phenomenon might be explained in terms of the time-averaged accretion rate, because when it is high, the accreted matter might bury the magnetic field, which has no time to dissipate through the accreted material. Indeed for the AMXPs the time-averaged accretion rate results to be really low, that the magnetic field can dissipate and in the same time it is still strong to disturb the flow of the accreted matter \cite{patruno12}. Therefore the NSs in LMXBs with low time-averaged accretion rate should be found and closely monitored to find other possible AMXP candidates. In this direction, {\emph RXTE} played a remarkable role for the discovery and the study of AMXPs, because it yielded to a better understanding of kHz QPOs and burst oscillations, clarifying therefore the connection between AMXPs and non-pulsating NSs in LMXBs \cite{wijnands06}.     
\item {\bf Pulsar timing parameters.} This class of sources show nearly coherent oscillations for a few seconds during X-ray bursts at frequencies between 270 -- 619 Hz. They have spin frequencies ranging in 180 -- 600 Hz. Normally they reside in very compact binary systems with orbital periods between 40 min -- 4.3 h \cite{Poutanen06}. 
\item {\bf Companion stars.} They have extremely low mass companion stars consistent with degenerate white (helium or carbon-oxygen) or brown dwarfs. All AMXPs are transients with the outburst repeating every few years and lasting a few weeks. They have a rather low time-averaged accretion rate of $\sim10^{-11}M_\odot/yr$, which could be the main reason for magnetic field still to be strong enough for pulsations to be observed \cite{Poutanen06,patruno12}.
\item {\bf Broad band spectra.} The broad-band spectra of the AMXPs are very similar to each other and can be mainly modeled by three components: two soft components which can be ascribed to thermal emission from a colder accretion disc (ranging below a few keV) and a hotter spot on the NS surface; a power-law tail (showing a cutoff around 100 keV) modeled with thermal Comptonization \cite{Poutanen06}. When fitting spectra with thermal Comptonization models, it can be observed that the product of the electron temperature and optical depth is almost invariant. The constancy of the spectral slope means that the emission region geometry does not vary much with the accretion rate \cite{Poutanen06}. If the energy dissipation takes place in a hot shock, the cooling of the electrons is determined by the reprocessing of the hard X-ray radiation at the NS surface. The spectral slope is determined by the energy balance in the hot phase and, therefore, by the geometry \cite{Poutanen06}. 
\item {\bf Pulse profiles.} The pulse profiles from AMXPs are rather close to sinusoidal with peak-to-peak oscillation amplitude between 4 and 12 per cent \cite{wijnands06,Poutanen06}. Deviations from the sine wave are stronger at higher energies. The harmonic content is also stronger when the oscillation amplitude is larger. Pulse profiles at higher energies reach their peaks at an earlier phase relative to the soft photons resulting in the soft time lags. One can point out that the contribution of the black body decreases exponentially with energy and lags increase significantly at the same time. Above $\sim7$ keV the blackbody's contribution is negligible and the lags saturate \cite{Poutanen06}. One believes that the bulk of the X-ray emission observed from AMXPs originates from the polar caps where the gas stream channeled by the NS magnetic field impacts the stellar surface forming a shock. Pulse profile shape, and variability amplitude carry the information about the compactness of the NS, spot and its position at the star, and the emission pattern. The main effects describing the shape of pulse profiles are \cite{Poutanen06}:
\begin{itemize}
\item \emph{a small blackbody spot}, which would produce sinusoidal variations (with possible eclipses) due to a change of the projected area;
\item \emph{the light bending effect}, which reduces the variability amplitude, while the pulse remains almost sinusoidal;
\item \emph{the relativistic aberration and Doppler boosting}, which modify the observed flux for rapidly spinning star;
\item \emph{the light travel time delays}, which slightly modify the profile. 
\end{itemize}
\end{itemize}

\section{The 2015 outburst of the accretion-powered pulsar \igr: \Integ\ and \swift\ observations}
The pulsar \igr\ is the fastest-known accretion-powered X-ray pulsar, discovered during a transient outburst in 2004. In this paper, I report on \Integ\ and \swift\ observations during the 2015 outburst, which lasts for $\sim25$~d. The source has not been observed in outburst since 2008, suggesting that the long-term accretion rate has decreased by a factor of two since discovery. The averaged broad-band (0.1 -- 250~keV) persistent spectrum in 2015 is well described by a thermal Comptonization model with a column density of $N_{\rm H} \approx4\times10^{21}$ cm$^{-2}$, a plasma temperature of $kT_{\rm e} \approx50$ keV, and a Thomson optical depth of $\tau_{\rm T}\approx1$. Pulsations at the known spin period of the source are detected in the \Integ\ data up to the $\sim150$ keV energy band.  I also report on the discovery of the first thermonuclear burst observed from \igr, which lasts around 7 min and occurs at a persistent emission level corresponding to roughly $1.6\%$ of the Eddington accretion rate. The properties of the burst suggest it is powered primarily by helium ignited at a depth of $y_{\rm ign}\approx1.5\times10^9$ g cm$^{-2}$ following the exhaustion by steady burning of the accreted hydrogen. The Swift/BAT data from the first $\sim20$ s of the burst provide indications of a photospheric radius expansion phase. Assuming this is the case, I infer a source distance of $d = 4.2 \pm 0.5$ kpc.

\subsection{The source IGR~J00291+5934} 
\label{sec:intro}
The AMXP \igr\ was discovered during an outburst in 2004 and it is the fastest spinning AMXP discovered so far. Its spin period is 1.67 ms \cite{markwardt04b,falanga05}. IGR~J00291+5934 is also the first AMXP showing a clear spin-up behaviour during outburst \cite{falanga05}. Together with other AMXPs, such as SAX~J1808.4--3658 (the first discovered in 1998; \cite{Wijnands98}), PSR~J1023+0038 \cite{shahbaz15}, and IGR~J18245--2452 \cite{papitto13c}, it confirms the evolutionary link between the accretion powered LMXBs and the rotation-powered millisecond pulsars \cite{alpar82}. \rxte/ASM data suggest that possible outbursts from the source could have occurred already in 1998 and 2001 \cite{remillard04}, tentatively indicating a recurrence time of $\sim3$ yr. This possibility can not be confirmed further, as the following outbursts from IGR J00291+5934 are recorded only in 2008 and 2015 \cite{lewis10,hartman11,lipunov15,sanna15}. The quiescent emission from the source is studied by Jonker et al. (2005) \cite{jonker05}, who record with \chandra\ a luminosity as low as (5 -- 10)$\times10^{31}$~erg~s$^{-1}$ (0.5 -- 10 keV). This value is comparable with that observed from other AMXPs in quiescence. Patruno (2010) \cite{Patruno10} also reports the possible spin-down between the outbursts that would be expected according to the standard magneto-dipole radiation scenario (see, e.g., Ref. \cite{patruno12} for a review). 

The donor star hosted in \igr\ is identified to be a hot brown dwarf with a mass ranging between 0.039 -- 0.16 M$_\odot$ \cite{galloway05}. This is in agreement with stellar evolutionary expectations, which predict that the observed AMXPs with an orbital period in the 1.4 -- 11 hr range (that in this case is 2.5 hr) should host hydrogen-rich white or brown dwarfs (see Ref. \cite{bildsten01}, and references therein). Therefore, all these AMXPs also exhibit helium type-I X-ray bursts after they burned prior to ignition of the accreted hydrogen fuel (see e.g., Refs. \cite{galloway06,watts06,falanga07,galloway07, falanga11,ferrigno11}). 

In this section, I report on \Integ\ and \swift\ observational campaign carried out to monitor the \igr\ July 2015 outburst. I study the properties of the broad-band spectral energy distribution of the source, as well as its timing properties. I also present a detailed analysis of the first thermonuclear X-ray burst observed from \igr\ (see also Refs.  \cite{kuin15,bozzo15b}).   

\subsection{Observations and data} 
The 2015 outburst is covered by \Integ\ (22 -- 250 keV) and \swift\ (5 -- 22 keV) data (see Secs.~\ref{sec:integral} and ~\ref{sec:swift}). To have the full coverage of the outburst trend I have used all the daily-averaged available \swift/BAT data (15 -- 50 keV).

\subsubsection{INTEGRAL} 
\label{sec:integral}  

I analysed the \Integ\ \cite{w03} target of opportunity (ToO) observation performed on \igr\ during its outburst in July 2015. The observation started on 2015 July 27 at 17:45:57.2 UTC and ended on July 29 at 21:07:17.2 UTC, summing up to a total exposure time of 170 ks (see Table \ref{tab:swift}). 

The reduction of the \Integ\ data was performed using the standard {\sc offline science analysis (OSA)} version 10.2 distributed by the ISDC \cite{c03}. \Integ\ data were divided into science windows (ScW), that is, different pointings each lasting $\sim 2-3$\,ks. The algorithms used for the spatial and spectral analysis are described in Ref. \cite{gold03}. The observation, that is in a hexagonal dithering mode aimed at \igr, consisted of 50 ScWs with a source position offset $\lesssim 2^\circ.5$ from the centre of the field of view. I analysed data from the IBIS/ISGRI coded mask telescope \cite{u03,lebr03}, covering the 20 -- 300~keV energy band, and from the two JEM-X monitors \cite{lund03}, covering the 3 -- 20~keV energy range.

The pulsar \igr\ was detected in the IBIS/ISGRI mosaic at a significance level of $60\sigma$ (20 -- 100 keV energy range). I note that the nearby source V709~Cas (detection significance of $9\sigma$ in the 20 -- 100 keV energy band) can be clearly distinguished thanks to the spatial resolution of the instrument and thus is not contaminating the X-ray emission recorded from \igr\ (see also Ref. \cite{falanga05}). \igr\ is also clearly detected in the JEM-X mosaics with a detection significance of $42\sigma$ in the 3 -- 20~keV energy range. The best determined position of \igr\ is at $\alpha_{\rm J2000} = 00^{\rm h} 29^{\rm m} 01^{\rm s}.97$ and $\delta_{\rm J2000} = 59^\circ34'18.''9$, with an associated uncertainty of $3''.5$ at the 90\% confidential level (c.l.; 20 -- 100 keV; \cite{gros03}). The offset with respect to the position of the associated optical counterpart is $0''.2$ \cite{torres08}. I first extracted a number of energy resolved light curves for IBIS/ISGRI and the two JEM-X, at a time scale of one ScW. The analysis of these light curves does not provide evidence for significant spectral variation during the outburst, and thus I extracted a single ISGRI and JEM-X spectrum averaged over the entire exposure time available. These spectra are described in the next section and fit together with the \swift\ data.  

\subsubsection{Swift} 
\label{sec:swift}  
\swift/XRT \cite{burrows05} started to monitor the source $\sim135$ s after a BAT trigger that occurred on 2015 July 25 at 02:12 UTC, which corresponds to the onset of the first type-I X-ray burst observed from this source \cite{kuin15}. The XRT monitoring campaign covers the source outburst from 57227~MJD to 57250~MJD, comprising a total of 13 pointings and an effective exposure time of 21~ks. One XRT observation is carried out on 2015 July 29 (ID~00031258006) simultaneously with the \Integ\ ToO. The effective exposure time of this XRT pointing is 1.4 ks (see Table \ref{tab:swift}). 

I processed the \swift/XRT data by using standard procedures \cite{burrows05} and the calibration files version 20160113. The \swift/XRT data were taken both in window-timing (WT) and photon-counting (PC) modes (processed with the {\sc xrtpipeline} v.0.13.2). Filtering and screening criteria were applied by using FTOOLS contained in the {\sc heasoft} software package (v6.19)\footnote{http://heasarc.gsfc.nasa.gov/docs/software.html.}. I extract source and background light curves and spectra by selecting event grades of 0 -- 2 and 0 -- 12 for the WT and PC mode, respectively. I used the latest spectral redistribution matrices in the HEASARC calibration database. Ancillary response files, accounting for different extraction regions, vignetting and PSF corrections, were generated using the {\sc xrtmarkf} task. When required, I corrected PC data for pile-up, and used the {\sc xrtlccorr} task to account of this correction in the background-subtracted light curve.

I extracted BAT light curve with {\sc batgrbproduct} tool and standard techniques and I also used the public daily-averaged \swift/BAT data in the 15 -- 50 keV energy band retrieved from the Hard X-ray Transient Monitor webpage\footnote{http://swift.gsfc.nasa.gov/results/transients/} \cite{krimm13}. 

\begin{table}[h] 
\caption{Log of all \swift\ and \Integ\ observations used in this work.}
\centering
\begin{tabular}{ccccc} 
\hline 
\hline 
Sequence & Obs/mode & Start time (UT) & Exposure\\
& & (UTC)  & (ks)\\
\hline 
00031258005 & \swift/WT & 2015-07-24 13:23:42 & 1\\ 
00650221000 & \swift/WT & 2015-07-25 01:56:19 & 0.3\\
00650221001 & \swift/WT & 2015-07-25 03:21:27 & 2\\
00031258006 & \swift/WT & 2015-07-29 00:18:11 & 1.4\\
00031258007 & \swift/WT & 2015-07-31 12:46:58 & 1.9\\
00031258008 & \swift/PC & 2015-08-02 01:27:58 & 2.5\\
00031258009 & \swift/PC & 2015-08-04 02:48:58 & 2\\
00031258010 & \swift/PC & 2015-08-06 18:45:58 & 2\\
00031258012 & \swift/PC & 2015-08-09 08:57:57 & 1.9\\
00031258014 & \swift/PC & 2015-08-14 16:57:57 & 1.8\\
00031258015 & \swift/PC & 2015-08-17 08:38:12 & 0.9\\
00031258016 & \swift/PC & 2015-08-19 11:31:58 & 1.1\\
00031258017 & \swift/PC & 2015-08-23 10:00:58 & 1.7\\
\hline  
\hline  
15690004001 & \Integ & 2015-07-27 17:45:57 & 170\\
\hline  
\end{tabular} 
\label{tab:swift}
\end{table} 

\subsection{Outburst properties} 
\subsubsection{The light curve} 
\label{sec:lc} 
\begin{figure}[h] 
\centering 
\includegraphics[trim=3.5cm 1cm 5cm 2.5cm, scale=0.5]{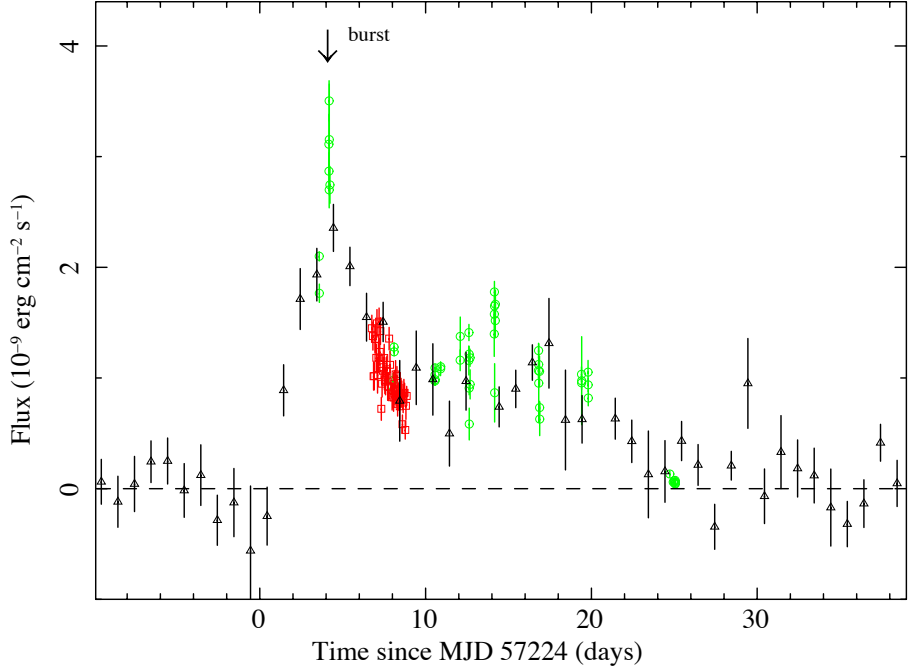} 
\caption{Light curve of the 2015 outburst observed from \igr.\ I show data from \swift/XRT 
(green circles, bin time of 500~s; July 24 -- 15 August), \swift/BAT (black triangles, each measurement is averaged over 1~d; July 11 -- 28 August), and \Integ/ISGRI (red squares, the integration time is one science window of $\sim2$ -- 3 ks; July 27 -- 29). The arrow indicates the time of the only detected type-I X-ray burst. The dashed line shows the quiescent flux level. The source outburst reaches the highest peak flux that has ever been detected in the previous outbursts.}
\label{fig:Fig2_4} 
\end{figure}

In Fig. \ref{fig:Fig2_4} I show the light curve of \igr\ as obtained from all data used in this paper and showing the entire source outburst. The count-rates measured from all instruments were converted to bolometric flux (0.1 -- 250 keV) using the spectral analysis results from Sec.~\ref{sec:specIGRJ}. 

The source displays a clear increase in flux for $\sim3$~d after the onset of the event, reaching a peak value that is about three orders of magnitude larger than the persistent quiescent flux recorded before 57225~MJD. The source then decays back into its quiescent state around 57250~MJD, suggesting that the outburst lasts $\sim25$~d in total. 

The profile of the outburst over time, shown by the source in 2015, is strongly reminiscent of that recorded during the event in 2004, as in both cases the return to quiescence occurs on a time scale of 15 -- 25~d and also the peak flux is roughly comparable (I note that the onset of the outburst in 2004 can not be observed; see Ref. \cite{falanga05}) and references therein). The profile of the source outburst corresponding to the 2008 event is much more peculiar, featuring a double outburst light curve separated by 30~d. On that occasion, the source displayed a clear rise and fast decay in flux during the first 9 d of the event and then remains below the instrument detection level for about a month. A second brightening phase follows, lasting roughly $\sim 15$ d \cite{hartman11}. The origin of this singular outburst profile is still a matter of debate (see also the discussion in Ref. \cite{bozzo15b}). The analysis of the long term BAT light curve shows that no rebrightening episode takes place after the main outburst in 2015. 

A novel feature of the 2015 outburst of \igr\ is the first detection of a thermonuclear burst, occurring 2.8~d after the event onset, close to the time of the peak flux achieved by the source \cite{kuin15,bozzo15b}. I perform a time-resolved spectral analysis of the source X-ray emission during the type-I burst in Sec.~\ref{sec:burst}. No additional thermonuclear bursts are found either in the continuous 170~ks \Integ\ monitoring observation or in the available 13 \swift/XRT pointings lasting 21 ks.

\subsubsection{The recurrence time}
\label{sec:dg} 
Prior to the 2015 outburst, \igr\ has exhibited outbursts roughly every three years, with a gradually increasing interval (Table \ref{tab:outbursts}). The regularity of the outbursts allows the time of the 2008 outburst to be predicted to within 1\% of the actual interval, 3.7~yr \cite{galloway08b}. The same quadratic fit would predict that the next outburst should occur after 4.1~yr, around 56200~MJD (2012 October); instead, it occurs some 2.8~yr later, after an interval of 6.9~yr. 

It is tempting to speculate that the pair of outbursts observed in 2008, separated by one month, may exhaust the disk to an unusual extent, delaying the following outburst. In fact, the total fluence of both 2008 outbursts together is slightly below that of the 2004 outburst. The 2015 outburst is somewhat more energetic again; the integrated flux from the {\it Swift}/BAT and {\it INTEGRAL}/ISGRI measurements indicates a total fluence of $(2.01\pm0.11)\times10^{-3}\ {\rm erg\,cm^{-2}}$. Furthermore, it is usually assumed that such transient outbursts exhaust all the accumulated material in the disk (although admittedly the secondary 2008 outburst is evidence that this is not always the case). I can also rule out any missed outbursts between 2008 and 2015, based on {\it Swift}\/ BAT monitoring. Due to the high declination of the \swift\ satellite, \igr\ is constantly monitored by BAT on a daily timescale, and the typical upper limit outside the outburst is $5\times10^{-10}\ {\rm erg\,cm^{-2}\,s^{-1}}$ (priv. comm., H. Krimm, 2016).

\begin{figure}[h] 
\centering 
\includegraphics[trim=2.8cm 9cm 5cm 0.2cm, scale=0.5]{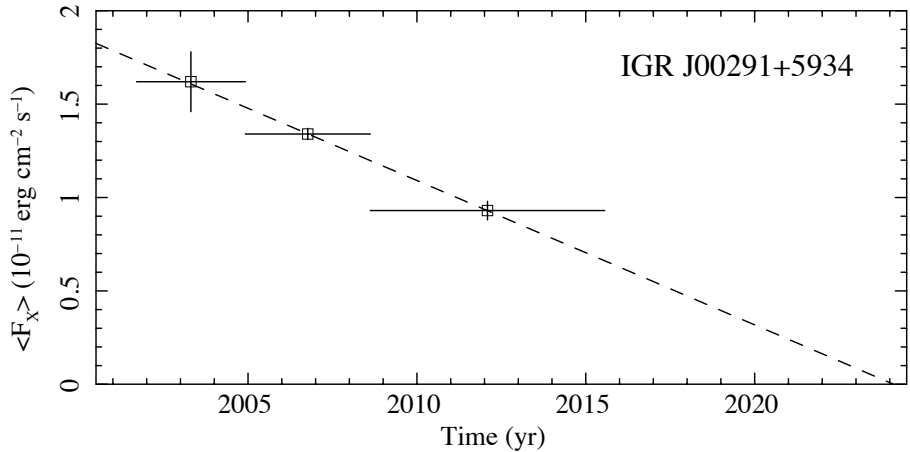} 
\caption{Estimated time-averaged X-ray flux for \igr\ for the outbursts observed between 2001 -- 2015, based on the outburst fluences listed in Table \ref{tab:outbursts}. The 2008 August and September outbursts are taken together to estimate the average rate since 2004. The average flux can be taken as a measure of the accretion rate into the disk between outbursts; the dashed line shows a linear fit, which (extrapolated) decreased to zero by 2024. I predict that the source may not exhibit another outburst as strong as that in July 2015 in the next decade.}
\label{fig:Fig2_5} 
\end{figure}

Instead, I find the long-term outburst history offers strong evidence that the steady mass transfer rate is decreasing, by about 50\% between 2001 and 2015. I calculate in Table \ref{tab:outbursts} the time-averaged X-ray flux as the outburst fluence divided by the outburst interval. This value is $1.6\times10^{-11}\ {\rm erg\,cm^{-2}\,s^{-1}}$ between 2001 -- 2004, but only $9.3\times10^{-12}\ {\rm erg\,cm^{-2}\,s^{-1}}$ between 2008 -- 2015 (see Fig.~\ref{fig:Fig2_5}). 
For a distance of 4~kpc, this corresponds to a range of accretion rates of (0.2 -- 0.01)\%~$\dot{M}_{\rm Edd}$. Such a decrease in the long-term accretion rate is also inferred from the outbursts in SAX~J1808.4$-$3658, on a similar timescale \cite{galloway08b}. Following the same line of thought of \cite{galloway08b}, I also perform a linear fit to the average fluxes over the last three outburst intervals. It may indicate that the time-averaged X-ray flux has dropped steadily with the rate of $\sim0.08\times10^{-11}\ {\rm erg\,cm^{-2}\,s^{-1}\,yr^{-1}}$. While this fit is purely phenomenological, if the trend continues, the wait time for future outbursts (of similar fluence to those in 2004 and 2015) will occur substantially longer than seven years.
\begin{table}[th] 
\caption{Transient outbursts from \igr.}
\centering
\scalebox{0.75}{
\begin{tabular}{ccccccccc} 
\hline 
\hline 
Outburst start & Duration & Interval$^{(a)}$ & $F_{\rm peak,bol}^{(b)}$ & Fluence$^{(c)}$ & $\left<F_X\right>^{(d)}$ & Refs.$^{(e)}$\\
calendar date & d & yr & $10^{-9}\ {\rm erg\,cm^{-2}\,s^{-1}}$ & $10^{-3}\ {\rm erg\,cm^{-2}}$ & $10^{-11}\ {\rm erg\,cm^{-2}\,s^{-1}}$ \\
\hline 
26 Nov 1998 & -- & $>2.9$ & -- & -- & ($<1.8$) & -- \\
11 Sep 2001 & -- & 2.8 & -- & -- & (1.8) & --\\ 
2 Dec 2004  & 14 & 3.2 & $2.9\pm0.2$ & $1.6\pm0.2$ & $1.6\pm0.2$ & [1] \\ 
13 Aug 2008  & 9  & 3.7 & $1.5\pm0.2$ & $0.76\pm0.03$ & $0.65\pm0.03$ & [2]\\
18 Sep 2008  & 15 & 0.1 & $1.1\pm0.1$ & $0.79\pm0.03$ & $25\pm1$ & [2]\\
22 Jul 2015  & 25 & 6.9 & $3.5\pm0.2$ & $2\pm0.1$ & $0.9\pm0.1$ & [3]\\
\hline  
\end{tabular}}
\vspace{0.5cm} 

(a) The epoch for the outburst prior to the first known is assumed to be earlier than the first \rxte/ASM measurements (typically 1996 January 6 or 50088~MJD).
(b) The peak bolometric flux is in the 0.1 -- 250 keV energy range.
(c) Bolometric fluence.
(d) Estimated time-averaged bolometric flux. Values in parentheses are estimated from the \rxte/ASM intensity, since \rxte/PCA measurements are not available at the outburst peak, and are approximate.
(e) References: [1] \cite{galloway05}; [2] \cite{hartman11}; [3] this section.
\label{tab:outbursts}
\end{table} 

\subsection{Spectral analysis} 
\label{sec:specIGRJ}
The spectral analysis is carried out using {\sc xspec} version 12.6 \cite{arnaud96}. All uncertainties in the spectral parameters are given at a $1\sigma$ confidence level for a single parameter. 

I first fitted all the different XRT spectra extracted from the available 13 pointings, excluding the type-I burst. 
This analysis revealed that in all cases the source X-ray emission in the soft energy band can be reasonably well 
described by using a simple absorbed power-law model ($\chi^{2}_{\rm red}{\rm /d.o.f.}= 0.98/97$). In all cases, I measured an absorption column density of $N_{\rm H}=(1.30\pm0.15)\times10^{22}$~cm$^{-2}$, and a photon index of $\Gamma=1.96\pm0.05$, with no evidence of significant spectral variability. I thus performed a more detailed broad-band spectral fit by combining the XRT observations carried out on 2015 July 29 with the \Integ\ data. In order to limit the uncertainties in the calibrations of the different instruments, the fit to the broad-band spectrum is limited to 1.1 -- 7.5 keV for the XRT data in WT mode, 5 -- 22 keV for the two JEM-X and 22 -- 250 keV for ISGRI. A constant factor was included in the fit in order to take into account the inter-calibrations between the different instruments and the possible intrinsic variability of the source (in all cases I assume ISGRI as the reference instrument and fix the corresponding constant to unity). 

\begin{figure}[h] 
\centering 
\includegraphics[trim=3cm 1.5cm 5cm 4cm, scale=0.52]{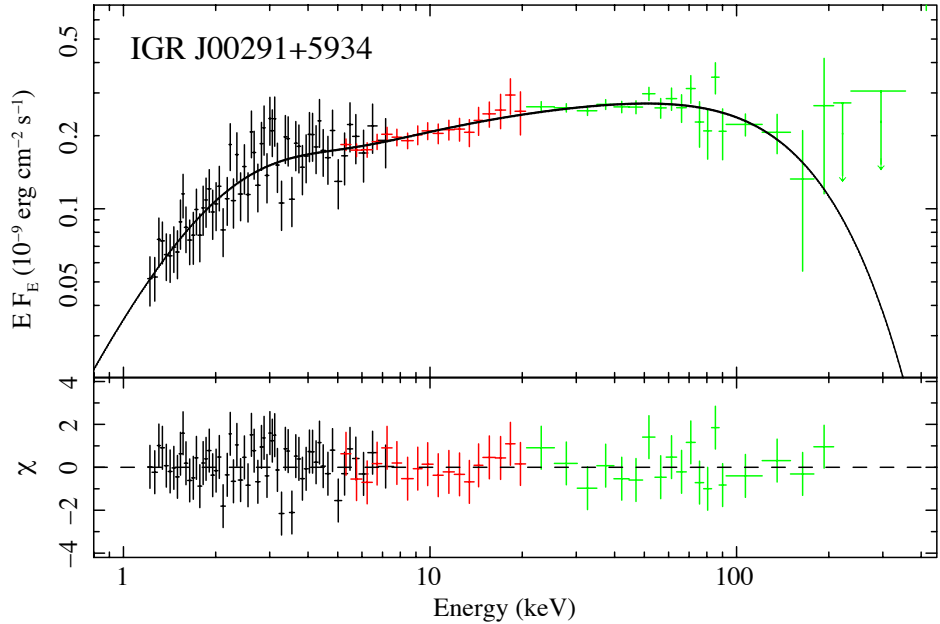}  
\caption{Unfolded measured broad-band spectrum of \igr\ as observed by JEM-X (red points), ISGRI (green points) and \swift/XRT (black points). The best fit is obtained with the \compps\ model with a plasma temperature $kT\approx50$ keV, which is represented in the figure with a solid black line. The residuals from the best fit are shown in the bottom panel.}
\label{fig:Fig2_6} 
\end{figure}

I obtained an acceptable fit to the broad-band spectrum of \igr\  using an absorbed cut-off power-law model ($\chi^{2}_{\rm red}{\rm /d.o.f.}= 0.73/95$). I measured a column density $N_{\rm H}=(1.01\pm0.17)\times10^{22}$~cm$^{-2}$, a photon index of $\Gamma=1.7\pm0.1$ and a cutoff energy of $E_{\rm cut}=147^{+77}_{-39}$ keV. In order to achieve a more physical description of the source X-ray emission, I replaced the cutoff power-law model with the thermal Comptonization model \compps,\ under the assumption of a slab geometry \cite{ps96}. This model has been successfully applied in the past to a number of other AMXPs (see e.g., Refs. \cite{gp05,mfb05,falanga05,mfc07,ip09,bozzo10,falanga11}). The new set of model parameters are the absorption column density $N_{\rm H}$, the plasma temperature of the accretion column $kT_{\rm e}$, the blackbody temperature $kT_{\rm bb}$ of the soft-seed photons assumed to be injected from the bottom of the slab, the Thomson optical depth $\tau_{\rm T}$ across the slab, and the inclination angle $\theta$ between the slab normal and the line of sight. This model provides a fully acceptable result ($\chi^2_{\rm red}/{\rm d.o.f.}=0.72/102$) and I report all values of the best fit parameters in Table \ref{tab:table1}.  
\begin{table}[h] 
\caption{\label{table:spec} Best parameters determined from the fit to the broad-band spectrum of \igr\ with the 
\compps\ model.}
\centering
\scalebox{1}{
\begin{tabular}{ll} 
\hline 
& \compps\ \\
\hline 
$N_{\rm H}\ (10^{21}\ {\rm cm}^{-2})$ & $4_{-3}^{+7}$\\ 
$kT_{\rm e}$ (keV) & $49 \pm 12$\\ 
$kT_{\rm bb}$ (keV) & $0.6^{+0.2}_{-0.1}$\\ 
$\tau_{\rm T}$ & $1.4^{+0.5}_{-0.3}$\\ 
$\cos \theta $ & $0.6\pm0.2$\\
$A_{\rm seed}^{(a)}$ (km$^2$) & $48\pm15$ \\ 
$\chi^{2}_{\rm red}/{\rm d.o.f.}$ & 0.72/102 \\
$L_{\rm bol}^{(a)}$ ($10^{37}$ erg s$^{-1}$) & 0.21$\pm$0.02\\
\hline  
\end{tabular}}
\vspace{0.5cm}
  
(a) Assuming a source distance of 4 kpc (see Sec.~\ref{sec:burst}).
\label{tab:table1} 
\end{table} 
The column density $N_{\rm H}=(4^{+7}_{-3})\times10^{21}$~cm$^{-2}$ turns out to be fully in agreement (to within the uncertainties) with that estimated previously $N_{\rm H}\approx2.8\times10^{21}\  {\rm cm}^{-2}$ from the {\it Chandra} and \swift\ observations \cite{nowak04,torres08,bozzo15b}. The \compps\ model also allows us to determine the apparent area of the hot spot region on the NS surface, $A_{\rm seed}\approx 48\ (d/4 \mbox{kpc})$ km$^2$. At the estimated distance of \igr\ (see Sec.~\ref{sec:burst}), the radius of the hot spot is $\approx 4$ km. I note that the spectral parameters reported in Table \ref{tab:table1} are comparable to those measured by Falanga et al. (2005) \cite{falanga05} during the source outburst in 2004. Fig.~\ref{fig:Fig2_6} shows the unfolded measured broad-band spectrum together with the residuals from the best fit model. In this fit, the normalization constants of the JEM-X and XRT data are 1.02 and 1.8, respectively compared to the ISGRI data fixed at unity. The large variability of the multiplicative normalization factor between XRT and ISGRI data may be due to the different exposure time. In addition, the XRT pointing is coincident with a small flare.

\subsection{Timing analysis}
\label{sec:tmchar}
I study the ephemeris and the pulse profile of \igr\ in the 20 --- 150~keV energy range taking advantage of the relatively good statistics of the ISGRI data. For an AMXP timing analysis I have insufficient time-resolution both in \swift/XRT PC and WT modes. The timing accuracy of the ISGRI time stamps recorded on-board is about $61\mu$s. I convert all ISGRI on-board arrival times from the terrestrial time (TT) reference system to the solar system barycenter (TDB). This barycentering procedure requires an accurate knowledge of the instantaneous spacecraft position and velocity, the best known (optical) pulsar position \cite{torres08}, and the solar system ephemeris (DE200) information. I excluded from this analysis events recorded during time periods where the ISGRI count rate behaves erratically (e.g., near perigeum ingress and egress, or during high solar activity). I further selected only events with rise times between channels 7 -- 90 \cite{lebrun03} from non-noisy pixels having a pixel illumination factor larger than 25\%.

The spin period of \igr\ is known to evolve in time \cite{Patruno10,hartman11,papitto11} and thus the precise rotational frequency of the pulsar, as well as the correspondingly updated binary orbital parameters, can not be accurately predicted a priori for the outburst of 2015. I thus first performed a search for coherently modulated signals in the ISGRI data around the most probable pulsar rotational frequency, assuming binary parameters reasonably similar to previously reported values. In order to maximise the signal to noise ratio of the data I additionally screened out events that are outside the energy range 20 -- 60 keV (see discussion in \cite{falanga05}). The systematic search was carried out by using a code based on a {\sc simplex} optimisation scheme, that finds the global minimum of the $Z_1^2$-test statistics  \cite{buccheri1983} as a function of the spin frequency and the time of the ascending node, $T_{\rm asc}$. The arrival time of each event is first corrected for the binary motion (thus involving the time of ascending node) and subsequently converted into a pulse phase using the assumed spin frequency (the other free parameter). The best parameters determined with this technique are $\nu_{\rm spin}=598.8921299(8)$~Hz at epoch 57231.0 MJD (TDB) and $T_{\rm asc}=57231.847035(25)$~MJD (TDB) \cite{kuiper15} with a detection significance of pulsations of $5.2\sigma$ (uncertainties are given at 1$\sigma$ c. l.). 

The spin frequency measured by ISGRI in the outburst in 2015 is consistent, to within the relatively large uncertainties, with the value that can be predicted from the previous observations in 2008 and taking into account any of the spin down torques reported by Refs. \cite{Patruno10}, \cite{hartman11}, and \cite{papitto11}. The updated value obtained for $T_{\rm asc}$ also allows to refine the knowledge on the system orbital period by fitting together all available measurements of the times of the ascending node. With this technique I obtain $P_{\rm orb}=8844.07672(2)$~s. I find no evidence for a change in the orbital period and the corresponding $2\sigma$ upper limit on the orbital period derivative is $\dot{P}_{\rm orb}=1.1\times 10^{-12}$ s\ s$^{-1}$.
\begin{table}[h] 
\caption{\label{table:burst} Ephemeris of \igr\ obtained from the \Integ\ observations.}
\centering
\begin{tabular}{ll} 
\hline 
Parameter & Value \\
\hline 
Epoch data start/end (MJD) & 57230.7 -- 57232.9\\
Frequency & 598.8921299(8) Hz\\
Solar system ephemeris & DE200\\
Epoch of the period (MJD;TDB) & 57231.0\\
Orbital period & 8844.07672(2) s\\
$a_x \sin i^{(*)}$ & $64.990(1)$ (lt-ms)\\
Time of ascending node (MJD;TDB) & 57231.847035(25)\\
\hline  
\end{tabular}  
\vspace{0.5cm}

(*) From \cite{papitto11}.
\label{tab:table2} 
\end{table}  
\begin{figure}
\centering
\includegraphics[trim=3cm 5.5cm 5cm 5.5cm, scale=0.9]{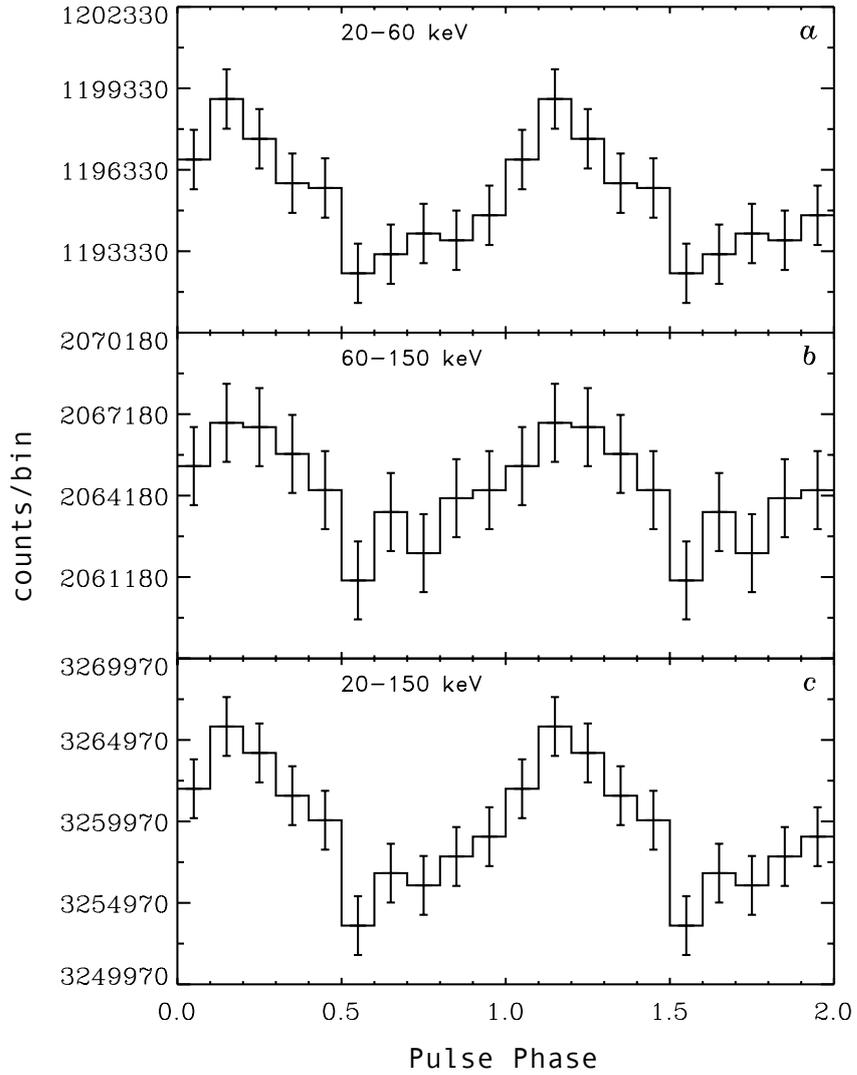} 
     \caption{Pulse profile of \igr\ in different energy bands during its 2015 outburst. 
     The pulsed emission is detected significantly by \Integ\ up to the $\sim150$ keV energy band.}
   \label{fig:Fig2_7} 
\end{figure}

I phase fold the barycentered ISGRI times upon the updated 2015 ephemeris, given in Table \ref{table:burst}. The resulting pulse-phase distributions are shown in Fig.~\ref{fig:Fig2_7} for three different energy bands. The $Z_1^2$-significances are $4.6\sigma, 3\sigma$ and $5.4\sigma$ for the 20 -- 60 keV, 60 -- 150 keV and 20 -- 150 keV bands, respectively. Thus, significant pulsed emission is detected for energies above 60 keV. I inspected the significance of the pulsed signal using the $Z_1^2$ statistics above 60 keV by progressively extending the energy range from 60 -- 90 keV to 60 -- 180 keV in steps of 30 keV. I find that the significance increased as function of the upper integration limit up to 150 keV and then flatten. This is consistent with the source emission being pulsed in the full energy range in which it is significantly detected.

\subsection{Properties of the type-I burst} 
\label{sec:burst} 
\begin{figure}[p] 
\centering 
\includegraphics[trim=1.5cm 1.8cm 3cm 4.35cm, scale=0.8]{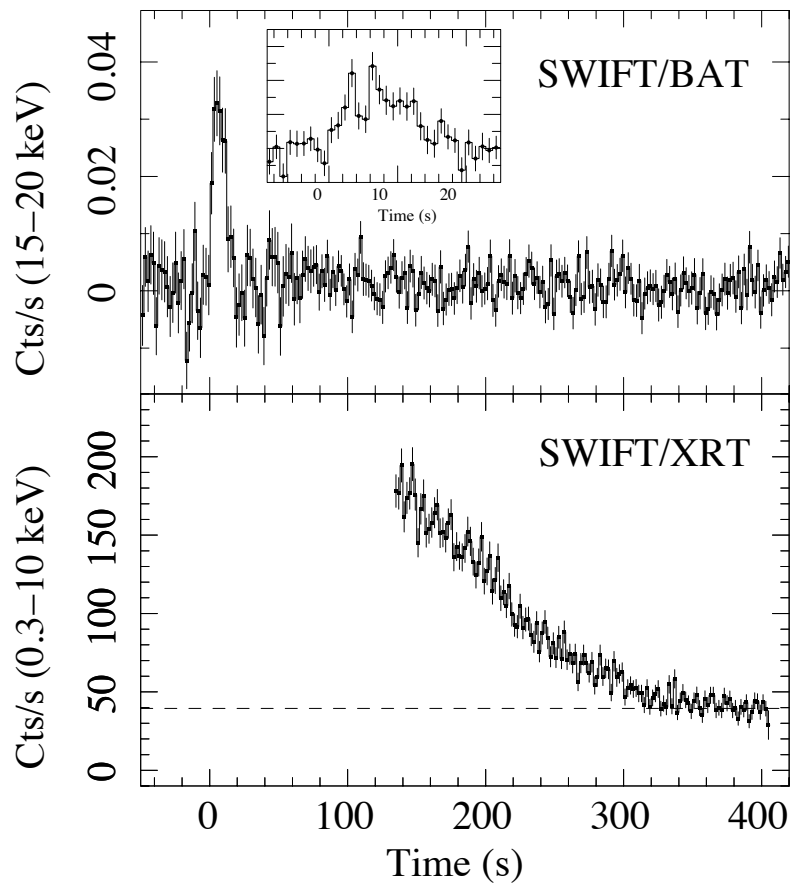}
\caption{Type-I burst detected during its 2015 outburst. The burst start time is 57228.0926~MJD. The upper panel shows the BAT light curve in the 15 -- 25~keV energy range with a time bin of 2~s (1~s for the inset). The XRT light curve in the bottom panel shows that the narrow field instrument on-board \swift\ starts to observe the source $\sim135$ s after the onset of the event. The dashed line in this panel indicates the emission level of the X-ray background as measured $\sim 400$ s after the burst. The burst is very energetic during the first $\sim20$ s with the tail of the blackbody detected above 25 keV.}
\label{fig:Fig2_8} 
\end{figure}

In Fig.~\ref{fig:Fig2_8}, I show the \swift/BAT 15 -- 20 keV (upper panel) and XRT/WT 0.3 -- 10 keV (lower panel) 
light curves of the only type-I burst observed so far from \igr.\ The burst start time of 57228.0926~MJD corresponds to the point at which the X-ray intensity of the source in BAT increases by 15\% with respect to the persistent level. The two plots in the figure suggest that the type-I burst is very energetic during the first $\sim 20$~s, with the tail of the black-body emission becoming detectable above 25 keV. The WT light curve starts with a delay of $\sim135$ s after the onset of the burst caught by BAT. The total duration of the burst, that is, the time to evolve away from and return to the persistent state, is $\sim7$ min. 

\begin{figure}[p] 
\centering 
\includegraphics[trim=2cm 2.6cm 5cm 2.2cm, scale=0.8]{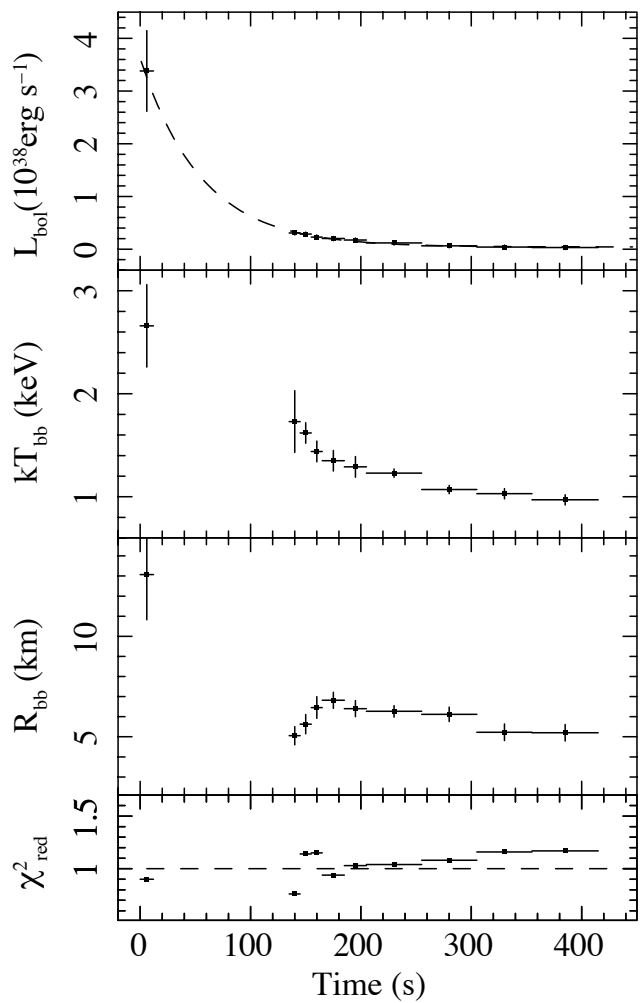}
\caption{Evolution of the spectral parameters during the type-I burst, as measured from BAT (first point) and XRT. 
The bolometric luminosity and the \bbodyrad\ radius are computed by using a distance of 4 kpc. The bottom panel shows the $\chi^2_{\rm red}$ values for all fits. The dashed line in the first upper panel represents the best determined exponential function with which I fit the burst profile. It shows an e-folding time of $\tau_{\rm fit}\approx55$ s.}
\label{fig:Fig2_9} 
\end{figure}

To perform a time-resolved spectral analysis of the burst, I extract 9 XRT spectra during the event. All these spectra can be well fitted by using an absorbed black-body model (\bbodyrad\ in XSPEC) with the absorption column density fixed at the value determined from the broad-band analysis (i.e., $4\times10^{21}$ cm${}^{-2}$; see Table~\ref{tab:table1}). However, if I leave the $N_{\rm H}$ value free to vary, I find a value consistent with $\approx4\times10^{21}$ cm${}^{-2}$.I plot in Fig.~\ref{fig:Fig2_9} the measured values of the apparent radius, $R_{\rm bb}$, the colour temperature, $kT_{\rm bb}$, and the bolometric luminosity, $L_{\rm bol}$, of the thermal emitting region. 

The luminosity at the peak of the burst, $F_{\rm peak}$, is obtained from BAT. The uncertainty derived from the spectral fit is increased by 20\% as the energy range of the instrument is largely outside the peak energy of the black-body (see for more details Fig. 4 in Refs. \cite{falanga09} and \cite{bozzo15a}). Taking into account the errors on the $N_{\rm H}$ value, this does not change the 20\% uncertainty on the burst peak flux. All the other luminosities are obtained from the XRT fluxes extrapolated in the 0.1 -- 35 keV energy range by generating ideal responses.
\begin{table}[h] 
\caption{\label{table:burst} Parameters of the type-I burst observed from \igr\ during its 2015 outburst.}
\centering
\begin{tabular}{ll} 
\hline 
\hline 
$F_{\rm peak}^{(a)}$ ($10^{-8}$ erg cm$^{-2}$ s$^{-1}$) & $18\pm4$\\
$f_{\rm b}^{(b)}$ ($10^{-6}$ erg cm$^{-2}$) & $10.7\pm2.6$\\
$F_{\rm pers,bol}$ ($10^{-9}$ erg cm$^{-2}$ s$^{-1}$) & $3.1\pm1.5$\\
$\tau_{\rm theo}\equiv f_{\rm b}/F_{\rm peak}$ (s) & $60\pm15$\\
$\gamma\equiv F_{\rm pers}/F_{\rm peak}$ & $(17.4\pm1.0)\times10^{-2}$\\
\hline  
\end{tabular}  
\vspace{0.5cm}

(a) Unabsorbed flux (0.1 -- 35 keV). (b) Net burst fluence (0.1 -- 35 keV).
\label{tab:table3} 
\end{table} 

I fitted the burst decay with an exponential function and derive an e-folding time of $\tau_{\rm fit}=55\pm11$ s (see Fig.~\ref{fig:Fig2_8}). This is similar to the value obtained as $\tau_{\rm theo}\equiv f_{\rm b}/F_{\rm peak}=60\pm15$ s. In Table~\ref{tab:table3}, I report the measured burst parameters. The estimated burst fluence is $f_{\rm b}=E_{\rm burst}/(4\pi d^2)=1.1\times 10^{-5}$ erg cm${}^{-2}$, which corresponds to the total energy release of $E_{\rm burst}=2.1\times 10^{40}$ erg (at a source distance of $d = 4$ kpc, see below). 

The BAT burst light curve rebinned at 1s (see the small inset in Fig.~\ref{fig:Fig2_8}) shows a double-peaked profile as evidence for a photospheric radius expansion (PRE; see e.g., Ref. \cite{falanga07}). Assuming that the peak X-ray luminosity of the burst corresponds to the Eddington value $L_{\rm Edd}= 3.8 \times 10^{38}$ erg s${}^{-1}$ (as empirically derived by Ref. \cite{kuulkers03}), I can thus estimate the source distance. I obtain $d=4.2\pm0.5$ kpc in case of a pure helium burst or $d=2.7\pm0.5$ kpc for a thermonuclear burst ignited in material with a solar composition ($X_0=0.7$). For comparison, I note that the theoretical value of the source distance calculated by assuming a helium atmosphere and canonical NS parameters ($M_{\rm NS}=1.4M_\odot$ and $R_{\rm NS}=10$ km) would be $d=3.6\pm0.5$ kpc \cite{lewin93}. The estimated error on the distance is similar to those reported, for example, in Ref. \cite{kuulkers03}.

The amount of fuel liberated during the thermonuclear burning can be estimated as $E_{\rm burst}/\varepsilon_{\rm He}$, where $E_{\rm burst}=2.1\times 10^{40}$ erg is the observed total energy released during the burst and $\varepsilon_{\rm He} \approx 1.6\ \mbox{MeV/nucleon} \approx1.6 \times 10^{18}$ erg g$^{-1}$ is the total available nuclear energy for the transformation of pure helium into iron-peak elements. I thus obtain $E_{\rm burst} /\varepsilon_{\rm He}  \approx 1.3 \times 10^{22}$ g. This value is consistent with the estimated total mass $\Delta M$ accreted from the onset of the outburst to the start time of the thermonuclear burst ($\Delta t= 2.52641$ d). $\Delta M = \dot{M} \Delta t \approx 1.3 \times 10^{22}$ g, where $\dot{M}$ is derived from the X-ray flux at a distance of 4~kpc and the usual accreting equation $L_{\rm X}=GM_{\rm NS}\dot{M}/R_{\rm NS}$ (see, e.g., Ref. \cite{frank02}). For a hydrogen type-I burst, the fuel burned during the thermonuclear burst would be $\approx 2.6 \times 10^{21}$~g, which is one order of magnitude lower than the measured total accreted matter ($\approx 1.2 \times 10^{22}$~g at $d=2.7$~kpc). The energetics of this burst can thus be explained by assuming that it is a helium type-I burst triggered by unstable helium burning, after any accreted hydrogen is exhausted by steady burning prior to the burst. I thus assume $d\approx4$~kpc as the fiducial distance to \igr.\ 

At the distance of 4 kpc, the pre-burst persistent unabsorbed flux $F_{\rm pers,bol}\approx 3.1\times10^{-9}$ erg cm${}^{-2}$ s${}^{-1}$ translates into a bolometric luminosity of $L_{\rm pers,bol}\approx 6\times10^{36}$ erg s${}^{-1}$, or $\approx 1.6\%\ L_{\rm Edd}$. The local accretion rate per unit area is then given by $ \dot{m} = L_{\rm pers} (1+z) (4\pi R^2(GM/R))^{-1}$, that is, $\dot m \approx 3.3\times10^3$ g cm$^{-2}$ s$^{-1}$ (where the gravitational redshift is $1+ z = 1.31$). At this local accretion rate, helium type-I bursts are expected following completion of hydrogen burning, thus providing additional support to the conclusion above (e.g., Ref. \cite{strohmayer06}). 

The observed energy of the type-I burst allows to estimate the ignition depth by using the equation $y_{\rm ign} = E_{\rm burst} (1+z)(4\pi R^2Q_{\rm nuc})^{-1}=1.5\times10^9$ g cm${}^{-2}$, where the nuclear energy generated for helium abundances (assuming a mean hydrogen mass fraction at ignition $\langle X\rangle=0$) is $Q_{\rm nuc}\approx 1.6+4\langle X\rangle\ {\rm MeV/nucleon}\approx1.6\ {\rm MeV/nucleon}$ (see Ref. \cite{galloway04}, and references therein), including losses owing to neutrino emission following Ref. \cite{fujimoto87}. The ignition depth is a particularly interesting parameter, as it regulates the recurrence time between different bursts through the equation $t_{\rm rec} =(y_{\rm ign} /\dot{m})(1 + z)$. At the mass accretion rate corresponding to the peak of the 2015 outburst shown by \igr\ ($1.6\%\ L_{\rm Edd}$ for pure helium), the expected recurrence time is $\sim 7$ d. Since the outburst from \igr\ lasts less than a month and the continuous coverage provided by \Integ\ is only 2.2 d, there is a high probability of missing another type-I burst. For a burst recurrence time of 7 d and for a total BAT exposure time of $\sim2$ ks, the probability of observing a burst is $0.3\%$.

\subsection{Summary and discussion} 
\label{sec:summary} 
I have studied the spectral and the timing behaviours of \igr\ during its outburst in 2015 by using the available \Integ\ and \swift\ observations. I have discussed the outburst recurrence time, that after the last three outburst intervals will drop to zero 8.4 yr after the July 2015 outburst. I have been able to detect the pulsed emission from the source up to the $\sim150$~keV. The standard binary evolution scenario suggests that the orbital period is caused by angular momentum loss through gravitational waves, or by magnetic braking \cite{tauris06}. The AMXP SAX J1808.4--3658 has shown an orbital expansion, but one that is much faster than expected from the standard binary evolution theory (see Refs. \cite{disalvo08,hartman08,patruno12b}, and references therein). For \igr\ $\dot{P}_{\rm orb}$ is consistent with zero, at variance of SAX J1808.4--3658 and orbital evolution prediction. The single \swift\ pointings show no evidence of significant spectral variability and the average broad-band (1.1 -- 250 keV) spectrum is best fitted with a thermal Comptonization model. 

I have also reported on the discovery of the first thermonuclear burst emitted by the source and occurred around the peak of the 2015 outburst, roughly 2.5 d after the onset of the event. I have noted, that the type-I burst lasted $\sim 7$ min, which is most similar to the ``intermediate-duration'' type-I bursts, that last between 15 -- 30 min and are powered by pure helium (see, e.g., Refs. \cite{intzand05,falanga08,falanga09b}). However, in the \igr\ type-I burst the accretion rate is low enough to build up a thick layer of pure helium, and any accreted hydrogen will be exhausted (via the hot CNO cycle) at the base of the layer within $\sim 10$ hr \cite{lampe16}. In this case I consider that the source accretes hydrogen matter with the solar hydrogen fraction $\langle X\rangle=0.7$ and metallicity similar to solar with mass fraction of CNO elements $Z_{\rm CNO} = 0.02$. The burst recurrence times are comparatively longer to reach ignition because there is no contribution from steady hydrogen burning at the base. This is in agreement with the accumulated mass and the inferred ignition column observed by \igr\ for a helium ignition at low mass accretion rates. It is worth mentioning that the calculated ignition depth is a factor of approximately ten larger than the typical short (tens of seconds bursts) for helium burst in high accretion rate sources, which means that the photon diffusion time is also at least a factor of approximately ten larger. The presence of a possible PRE phase is identified by the double-peaked BAT data of the type-I burst, providing the opportunity to estimate the source distance with a relatively good accuracy at $d=4.2\pm0.5$ kpc. This value is  within the previous estimated source distances ranging from 2 to 4 kpc \cite{galloway05,jonker05,torres08}.

\section{The transitional millisecond pulsar IGR J182 45--2452 during its 2013 outburst at X-rays and soft gamma-rays}
\igrj/PSR J1824--2452I is one of the rare transitional accreting millisecond X-ray pulsars, showing direct evidence of switches between states of rotation-powered radio pulsations and accretion-powered X-ray pulsations, dubbed transitional pulsars. \igrj\ with a spin frequency of $\sim254.3$ Hz is the only transitional pulsar so far to have shown a full accretion episode, reaching an X-ray luminosity of $\sim10^{37}$~erg~s$^{-1}$ permitting its discovery with \Integ\ in 2013. In this paper, I report on a detailed analysis of the data collected with the IBIS/ISGRI and the two JEM-X monitors on-board \Integ\ at the time of the 2013 outburst. I make use of some complementary data obtained with the instruments on-board \xmm\ and \swift\ in order to perform the averaged broad-band spectral analysis of the source in the energy range 0.4 -- 250~keV. I have found that this spectrum is the hardest among the accreting millisecond X-ray pulsars. I improved the ephemeris, now valid across its full outburst, and report the detection of pulsed emission up to $\sim60$ keV in both the ISGRI ($10.9 \sigma$) and Fermi/GBM ($5.9 \sigma$) bandpass. The alignment of the ISGRI and Fermi GBM 20 -- 60 keV pulse profiles are consistent at a $\sim25\ \mu$s level. I compared the pulse profiles obtained at soft X-rays with \xmm\ with the soft \gr-ray ones, and derived the pulsed fractions  of the fundamental and first harmonic, as well as the time lag of the fundamental harmonic, up to $150\ \mu$s, as a function of energy. I report on a thermonuclear X-ray burst detected with \Integ, and using the properties of the previously type-I X-ray burst, I show that all these events are powered primarily by helium ignited at a depth of $y_{\rm ign} \approx 2.7\times10^8$ g cm${}^{-2}$. For such a helium burst the estimated recurrence time of $\Delta t_{\rm rec}\approx5.6$ d is in agreement with the observations.

\subsection{The source \igrj}  
\label{sec:intro}
The AMXP \igrj\ was discovered by \Integ\ during observations performed in the direction of the Galactic centre in March 2013 \cite{eckert13}. \swift\ and \chandra\ follow-up observations located the source well within the globular cluster M28 \cite{heinke13, romano13, homan13}, thus providing the first measurement of the source distance at 5.5 kpc \cite{Harris96}. The optical counterpart, confirming the LMXB nature of the system, could be identified by observing large variations in the system magnitude between archival observations during quiescence and follow-up pointings performed shortly after the discovery \cite{Monrad13, Pallanca13a, Cohn13, Pallanca13b}. 

The first thermonuclear burst from the source was caught with \swift/XRT \cite{papitto13a, Linares13} and displayed clear burst oscillations at a frequency of $\sim254.4$ Hz \cite{Patruno13}. A second type-I burst was later reported with \maxi\ \cite{Serino13}. Coherent modulations at a period of $\sim254.33$~Hz were discovered in a dedicated \xmm\ observation campaign, allowing Papitto et al. \cite{papitto13c} to also measure the system orbital period ($\sim11.03$~hr) and its projected semimajor axis ($\sim0.76$ lt-s). These properties firmly associate \igrj\ with the previously known radio pulsar PSR J1824--2452I in M28 \cite{Manchester05}, thus proving that NSs in LMXBs can switch between accretion powered and rotation powered states. LMXBs discovered to undergo such transitions are named ``transitional millisecond pulsars'' \cite{archibald09,demartino10,demartino14,linares14b,patruno14,bassa14,bogdanov14,bogdanov15}. From now on I refer to the source with the name \igrj\, since I focus on its X-ray aspects. Together with other AMXPs like SAX~J1808.4--3658, the first discovered system of this class \cite{Wijnands98}, and IGR~J00291+5934, which displayed the first evidence of a clear spin-up during its outburst \cite{falanga05}, transitional pulsars represent the most convincing proof of the so-called ``pulsar recycling scenario'' \cite{bk74, alpar82,r82}. Among the transitional pulsars, \igrj\ is the only one that has so far displayed a full X-ray outburst, reaching a peak X-ray luminosity comparable to that of other AMXPs in outburst. Its behaviour in X-rays was shown to be particularly puzzling due to a pronounced variability that has been interpreted in terms of intermittent accretion episodes \cite{ferrigno14}. 

In this section, I concentrate on the 2013 outburst from \igrj, carrying out for the first time a detailed spectral and timing analysis of the \Integ\ data. To deepen the study of the accretion event displayed by \igrj, soft X-rays data from \xmm\ and \swift\ are also used to better constrain the hard X-ray results obtained with \Integ. I also report, for the first time at millisecond timescales, on the detection of pulsed emission by Fermi/GBM.

\subsection{Observations and data}
\subsubsection{INTEGRAL}
\label{sec:integral} 
The \Integ\ \cite{w03} dataset comprises all the 196 science windows (ScWs), that is, the different satellite pointings each lasting $\sim 2-3$\,ks, performed in the direction of \igrj\ from 2013, March 26 at 07:12:00 UTC to April 14 at 23:37:49 UTC. The satellite revolutions involved in the analysis were specifically: 1276 -- 1277, and 1279 -- 1280,
and the dedicated Target of Opportunity (ToO) observation covering the entire revolution 1282. The total effective exposure time on the source was of 216.5~ks. I analysed data from the IBIS/ISGRI coded mask telescope \cite{u03,lebrun03}, covering the 20 -- 300~keV energy band, and from the two JEM-X monitors \cite{lund03}, covering the 3 -- 25~keV energy range. The observation in revolution 1282 was the only one performed in the hexagonal dithering mode, which allows the target to be constantly kept within the fields of view of both IBIS/ISGRI and JEM-X. For all other revolutions, I retained for the scientific analysis only those ScWs for which the source was located at a maximum off-set angle with respect to the satellite aim point of $<12^\circ.0$ for IBIS/ISGRI and $< 2^\circ.5$ for JEM-X in order to minimize calibration uncertainties. The reduction of all \Integ\ data was performed using the standard {\sc offline science analysis (OSA)} version 10.2 distributed by the ISDC \cite{c03}. The algorithms for spatial and spectral analysis of the different instruments are described in \cite{gold03}.  

\begin{figure}[h] 
\centering
\includegraphics[trim=4.5cm 6cm 5cm 8cm, scale=0.62]{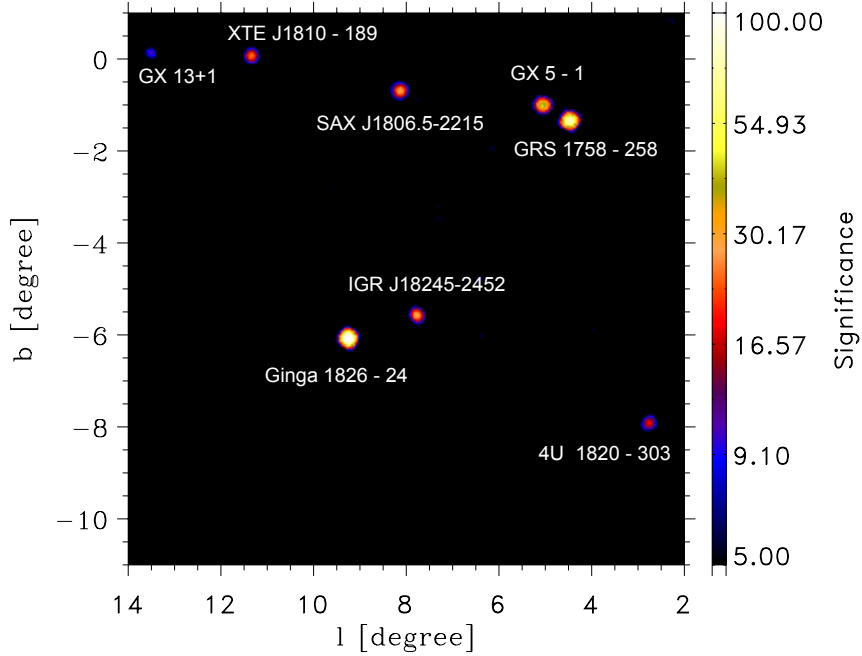} 
\caption{The \Integ\ IBIS/ISGRI mosaic around \igrj\ as obtained in the 20 -- 100 keV energy range 
from the data collected in revolution 1282 (total effective exposure time of $\sim131$~ks). The pixel size in the image corresponds to 3'. \igrj\ is detected in the mosaic with a significance of  $\sim 34.5\sigma$.}
\label{fig:mosa}
\end{figure}

I show in Fig.~\ref{fig:mosa} the ISGRI field of view (significance map) centred on the position of \igrj\ as obtained 
from the data in revolution 1282 (20 -- 100 keV energy range). The source is detected in the mosaic with a significance of $\sim 34.5\sigma$. The best determined source position obtained from the mosaic is at 
$\alpha_{\rm J2000} = 18^{\rm h} 24^{\rm m} 33^{\rm s}6$ and $\delta_{\rm J2000} = -24^{\circ}52' 48.''0$ 
with an associated uncertainty of $0.'9$ at the 90\% confidence level (20 -- 100 keV; \cite{gros03}). I extracted the IBIS/ISGRI light curve of \igrj\ with a resolution of one ScW for the entire observational period covered by \Integ\ (see Sec.~\ref{sec:lc}). The JEM-X and ISGRI spectra were extracted using only the data in revolution 1282, as these occurred simultaneously with one of the two available \xmm\ observations (see Sec.~\ref{sec:xmm}) and permitted the most accurate description of the source-averaged broad-band high-energy emission. Three simultaneous \swift/XRT pointings were also available during the same period (see Sec.~\ref{sec:swift}) to complement the \Integ\ and \xmm\ datasets. The averaged broad-band spectrum of the source, as measured simultaneously by all these instruments, is described in Sec.~\ref{sec:spe}. I describe in Sec.~\ref{sec:tm_IGR} the results of the timing analysis of hard X-ray data as obtained from the ISGRI event files. The resulting pulse profiles are compared to the pulse profiles obtained at the soft X-rays with \xmm\ and reported previously by \cite{ferrigno14}.  

\subsubsection{XMM-Newton}  
\label{sec:xmm}  
\igr\ was observed by \xmm\ \cite{jansen01} twice, on 2013, April 3 -- 4 (obs1, Rev-2439, ID~0701981401) and on 2013, April 13 -- 14 (obs2, Rev-2444, ID~0701981501). The exposure times were 28~ks and 68~ks, respectively. I was interested in the spectra of the source obtained from the RGS (0.4 -- 1.8~keV) and the EPIC-pn (0.5 -- 11~keV) during the observation ID~0701981501 that was carried out during the \Integ\ revolution 1282. I used the average spectra for these instruments (see Sec.~\ref{sec:spe}), as published by \cite{ferrigno14}, and refer the readers to this paper for all the details about the data analysis and spectral extraction techniques. I also made use of the EPIC-pn light curve (see Sec.\ref{sec:lc}). I revisited the EPIC-pn timing analysis and  applied the following event selections, ${\rm FLAG}=0$, ${\rm PATTERN}\leq4$ and ${\rm RAWX}=[33,41]$, to the timing data from both observations.  

\subsubsection{Swift} 
\label{sec:swift}
The \swift/XRT \cite{burrows05} observation campaign carried out to monitor the outburst of \igrj\ started $\sim191$ s after the \swift/BAT trigger caused by the source brightening on 2013, March 30 at 02:22:21 UTC \cite{romano13} and lasted until 2013, April 17 at 07:11:51 UTC. The campaign comprises 43 pointings with a total exposure time of $\sim92.7$~ks. I extracted the source light curve of all observations in the 0.5 -- 10~keV energy band (see Sec.~\ref{sec:lc}) by using the XRT online tool \cite{evans09} and performed a detailed analysis of the three pointings that were carried out simultaneously with the \Integ\ revolution 1282. These were pointings: ID~00032785011 performed on 2013, April 13 from 22:01 to 23:50 UTC, ID~00032785012 performed on 2013, April 14 from 06:26 to 09:49 UTC, and ID~00032785013 performed on 2013, April 15 from 17:25 to 19:06 UTC. I processed the \swift/XRT data using standard procedures \cite{burrows05} and the calibration files v.20160113. The considered XRT data were all taken in window-timing (WT) mode and I analysed them by making use of the {\sc xrtpipeline} (v.0.13.2). Filtering and screening criteria were applied by using the FTOOLS contained in the {\sc heasoft}\footnote{\url{http://heasarc.gsfc.nasa.gov/docs/software.html}.}software package v.6.19. I extracted source and background light curves and spectra by selecting event grades in the range 0 -- 2. I used the latest spectral redistribution matrices in the HEASARC calibration database. Ancillary response files, accounting for different extraction regions, vignetting and PSF corrections, were generated using the {\sc xrtmarkf} task. The considered data were not found to be affected by any significant pile-up.

\subsubsection{Fermi/GBM}
\label{instr_gbm}
The Gamma-ray Burst Monitor (GBM; \cite{meegan09,bissaldi09}) aboard \Fermi\ has as its main goal to increase the science return by observing \gr-ray bursts and other transients below the \Fermi\ LAT \cite{atwood09} passband (20 MeV -- 300 GeV). The GBM comprises a set of 12 sodium iodide (NaI(Tl)) detectors sensitive across the 8 keV to 1 MeV band, and a set of 2 bismuth germanate (BGO) detectors covering the 150 keV to 40 MeV band, and so overlapping with the \Fermi\ LAT passband. The set of non-imaging detectors provides a continuous view on each occulted (by Earth) hemisphere. Since 2012 November 26 (MJD 56257) the GBM in nominal operation mode provides time-tagged events (TTE) with $2\ \mu$s precision, synchronised to GPS every second, in 128 spectral channels, now allowing detailed timing studies at millisecond accuracies. 

\subsection{Outburst light curve} 
\label{sec:lc} 
\begin{figure}[p] 
\centering
\includegraphics[trim=3.5cm 0.5cm 5cm 2.5cm, scale=0.48]{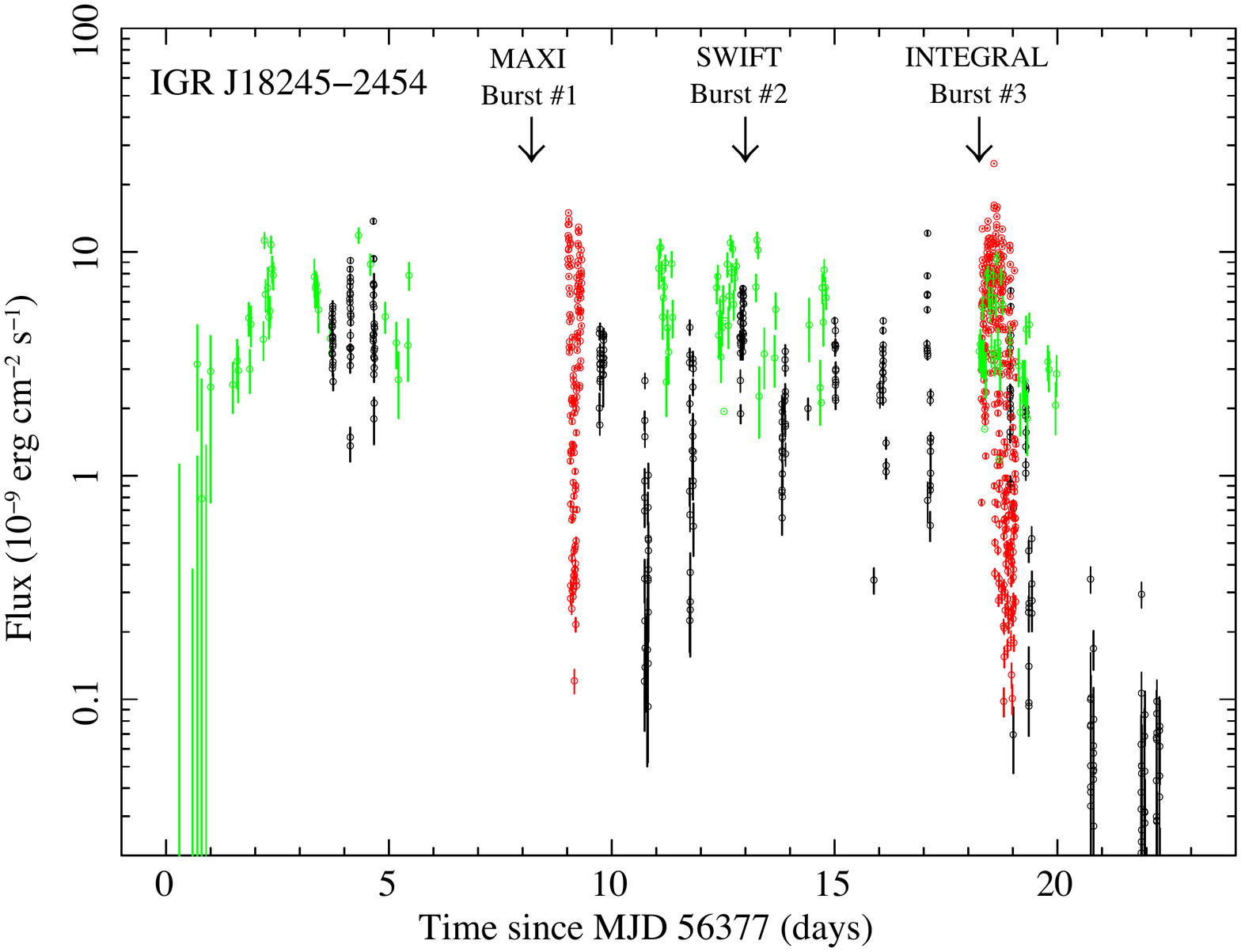}
\caption{Light curve of the 2013 outburst recorded from \igrj\ and observed by \swift/XRT (black circles, bin 
time of 500~s), \xmm/Epic-pn (red circles, with a time bin of 200 s as reported in Ref. \cite{ferrigno14}) and \Integ/ISGRI 
(green circles, the integration time is one science window of $\sim2$ -- 3 ks). The arrows indicate the onset of the 
three type-I X-ray bursts detected from the source (see Sec.~\ref{sec:burst}). \igrj\ shows a rapid flux variability.}
\label{fig:lcr} 
\end{figure}

I report in Fig.~\ref{fig:lcr} the light curve of \igrj\ as obtained from all available X-ray data showing that the entire outburst lasts for about 23~days (from 2013 March 26 to April 17). The count-rates measured from all instruments were converted into bolometric flux values (0.4 -- 250 keV) using the spectral analysis results obtained in Sec.~\ref{sec:spe}. 

The global profile of the outburst observed from \igrj\ is not too dissimilar from that shown by other AMXPs in outburst, typically characterised by a fast rise time ($\sim 1-2$~d) and a slower decay to quiescence ($\sim4-5$~d; see e.g., Ref. \cite{falanga05,bozzo16}). It is the short term variability of the source, observed by all instruments, that is much more peculiar (see Fig.~\ref{fig:lcr} and Ref. \cite{ferrigno14}). A similar variability is also seen in another two transient AMXPs, PSR J1023+0038 and XSS J1227.0--4859 (see e.g., Ref. \cite{linares14b} and reference therein). This rapid flux fluctuation has never been observed in any other AMXP, therefore it constitutes a property remarkably characterising for transient AMXPs. For \igrj, this variability has been interpreted as a transition between accretion state and centrifugal inhibition of accretion, possibly causing the launch of outflows \cite{ferrigno14}. However, other models have been proposed to explain the different kind of variability connected to the other two transient AMXPs (see e.g. \cite{DeMartino13,Papitto14,shahbaz15}). Although \igrj\ undergoes dramatic spectral changes on timescales as short as a few seconds, its average spectral energy distribution was identical between the two \xmm\ observations and, more generally, during the entire flat portion of the outburst (from day 3 to 20 in Fig.~\ref{fig:lcr}; \cite{ferrigno14}). As the fast spectral variations could not be revealed by the reduced sensitivity of the instruments on-board \Integ\ compared to those on-board \xmm, this justified the extraction of a single spectrum for IBIS/ISGRI and the two JEM-X monitors  summing up the data obtained over the entire exposure time available during the revolution 1282. 

During the outburst, three type-I X-ray bursts were recorded from \igrj.\ I analyse and discuss the three bursts in Sec.~\ref{sec:burst}.  

\subsection{Averaged broad-band spectral analysis} 
\label{sec:spe}
In this subsection, I report on the analysis of the averaged broad-band spectral properties of \igrj,\ taking advantage of the hard X-ray energy coverage provided by \Integ.\ I make use of the average JEM-X spectra covering the 5 -- 25 keV energy range and the IBIS/ISGRI spectrum covering the 22 -- 250 keV energy range, as obtained from the data in revolution 1282. I fit these data together with the simultaneous Epic-pn (0.9 -- 11 keV), RGS (0.4 -- 1.8 keV), and XRT spectra (0.4 -- 8 keV). The spectral analysis was carried out using {\sc xspec} version 12.6 \cite{arnaud96}. All uncertainties in the spectral parameters are given at a $1\sigma$ confidential level for a single parameter. All EPIC spectra of the source were optimally rebinned using the prescription in paragraph 5.4 of \cite{Kaastra2016}.

I first fit all the combined spectra by using a simple absorbed power law model, also including a normalisation constant in the fit to take into account uncertainties in the cross-calibrations of the instruments and the source variability (the data are not covering strictly the same time interval, even though they were collected in a compatible portion of the source outburst). This first fit provided an absorption column density of $N_{\rm H}=(0.32\pm0.01) \times 10^{22} {\rm cm}^{-2}$, a power law photon index $\Gamma=1.37\pm 0.01$ and a relatively large $\chi^{2}_{\rm red}{\rm /d.o.f.} = 1.44/2647$, most likely due to the wavy residuals at all energies. Although the fit was not formally acceptable, I noticed that the values of the normalisation constants were all in the range $1.0\pm0.2$, compatible with the finding that the average spectral properties of the source during the considered part of the outburst were stable. A similar range for the normalisation constants was also found in all other fits reported below (I assumed in all cases the EPIC-pn as the reference instrument and fixed its normalisation constant to unity). 

To improve the fit, I modified the power law model by including a cut-off at the higher energies. I measured a cut-off energy of $E_{\rm cut}= 122^{+21}_{-16}$ keV, an absorption column density of $N_{\rm H}=(0.34\pm0.01) \times 10^{22} {\rm cm}^{-2}$ , and a power law photon index of $\Gamma=1.34\pm 0.01$. The improvement in the fit was modest, still resulting in a poorly reduced $\chi^{2}_{\rm red}{\rm /d.o.f.} = 1.38/2647$. 
The addition of a black-body component, possibly related to the thermal emission from (or close to) the NS surface, 
led to a significant reduction of the residuals from the fit and a much more reasonable $\chi^{2}_{\rm red}{\rm /d.o.f.} = 1.16/2640$. The spectral parameters obtained with this model were: $N_{\rm H}=(0.46\pm0.01) \times 10^{22} {\rm cm}^{-2}$, $\Gamma=1.32\pm0.01$, $E_{\rm cut}=94_{-11}^{+14}$ keV, and $kT_{\rm bb}=0.76_{-0.13}^{+0.14}$ keV. Here $kT_{\rm bb}$ represents the temperature of the black-body emission. The black-body radius is $R_{\rm bb}=(6.5\pm0.1)$ m. In all these fits I add a Gaussian line to take into account an iron emission feature around $\sim6.6$~keV, and three Gaussian lines centred at energies of 1.5, 1.8, and 2.2 keV \cite{ferrigno14}.

Following Ref. \cite{ferrigno14}, I also tried to fit the averaged broad-band spectrum of \igrj\ with a thermal Comptonisation model ({\sc nthcomp}, \cite{zdziarski96,zycki99}) to take into account the emission produced by a thermal distribution of electrons which Compton up-scatter the soft seed X-ray photons. This model provided a statistically similar good fit as the phenomenological model described above, once a broad iron line peaking at 6.6~keV is included in the fit (the inclusion of the line leads to an improvement of the fit from $\chi^{2}_{\rm red}{\rm /d.o.f.} = 1.26/2643$ to $\chi^{2}_{\rm red}{\rm /d.o.f.} = 1.19/2640$). Although the results of this fit, summarised in Table~\ref{table:spec}, are quantitatively similar to those previously reported by Ref. \cite{ferrigno14}, I noticed that the measured absorption column density, $N_{\rm H}=(0.24\pm0.01) \times 10^{22} {\rm cm}^{-2}$, is significantly lower than other values reported in the literature (see also Refs. \cite{papitto13c,ferrigno14}). I ascribe this difference to the fact that I are using a much broader energy range and also that a thermal component related to the presence of an accretion disk ({\sc diskbb} in {\sc xspec}) was not needed (see below). 

\begin{figure}[h] 
\centering 
\includegraphics[trim=4.5cm 1cm 6cm 4cm, scale=0.45]{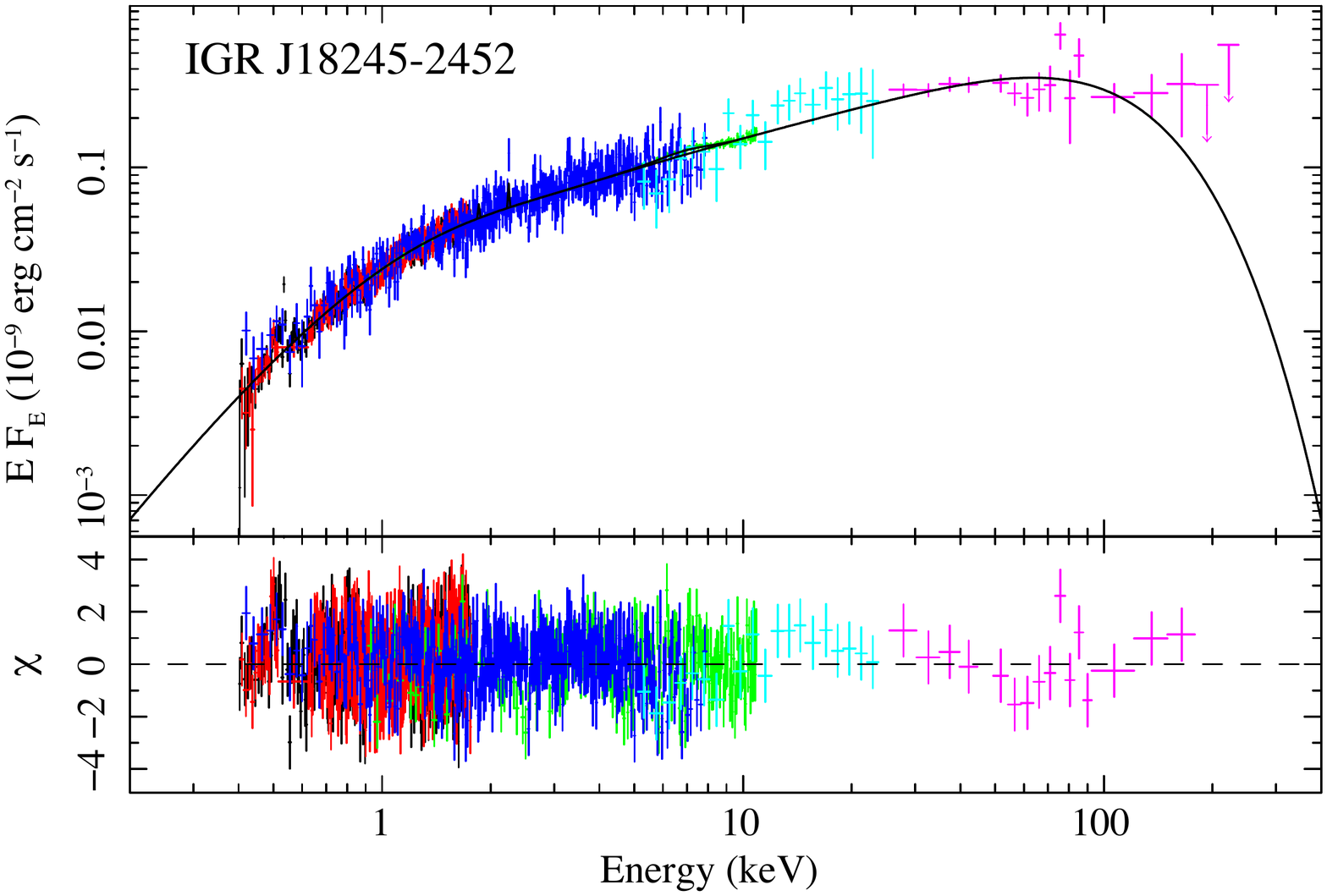} 
\caption{Unfolded absorbed broad-band spectrum of \igrj\ in the 0.4 -- 250 keV energy range. The data points are obtained from the two \xmm/RGSs (red and black points, 0.4 -- 1.8 keV), \xmm/Epic-pn (blue points, 0.9 -- 11 keV), \swift/XRT (green points, 0.4 -- 8 keV), \Integ/JEM-X (light blue points, 5 -- 25 keV), and \Integ/ISGRI (pink points, 22 -- 250 keV). The fit is obtained with the \compps\ model, represented in the top panel with a solid line. The residuals from the best fit are shown in the bottom panel. This source shows a harder spectrum compared to other AMXPs.}
\label{fig:spe} 
\end{figure}

To compare the averaged broad-band spectrum of \igrj\ with those of other AMXPs observed at hard X-rays with \Integ\ (e.g., Refs. \cite{gdb02,gp05,falanga05,mfb05,mfc07,ip09,falanga11,falanga12}), I also performed a spectral fit using a thermal Comptonisation model in the slab geometry (\compps, \cite{ps96}). The main parameters are the absorption column density $N_{\rm H}$, the Thomson optical depth $\tau_{\rm T}$ across the slab, the electron temperature $kT_{\rm e}$, the temperature $kT_{\rm bb}$ of the soft-seed thermal photons (assumed to be injected from the bottom of the slab), and the inclination angle $\theta$ between the slab normal and the line of sight. The results of this fit are also reported in Table~\ref{table:spec} and are similar to those measured from other AMXPs in outburst, but the optical depth was $50\%$ larger compared with, for example, XTE J1751--305, which has nearly identical electron temperature \cite{gp05}. I note that the broad Gaussian iron line at $\sim$6.6~keV was also required in this model. The absorption column density measured from the fit with the \compps\ model is compatible with that obtained before using the {\sc nthcomp} model. The $N_{\rm H}$ value is close to the Galactic value, $0.18\times10^{22}$ cm$^{-2}$, reported in the radio maps of Refs. \cite{dickey90,kalberla05}. I note, that combining the different \xmm\ observations with the other spectra, I could not find any evidence of the {\sc diskbb} component from the residuals of the best fit Comptonisation models (see also Fig.~\ref{fig:spe}). The temperature of the {\sc diskbb} was $kT_{\rm diskbb} = (2\pm1)\times10^{-2}$ keV, two orders of magnitude lower than the value reported by Ref. \cite{ferrigno14}. I noticed that using the {\sc tbnew} photoelectric absorption model with variable abundances of iron and oxygen \cite{wilms00}, the {\sc diskbb} becomes significant in the fit. This may be due to a decrease of the iron abundances and so the {\sc diskbb} component fits  the data well \cite{ferrigno14}.

\begin{table}[h] 
{\small
\caption{\label{table:spec} Optimal spectral parameters determined from the fits to the average broad-band spectrum 
of \igrj\ performed with the {\sc nthcomp} and \compps\ models. The Gaussian lines at 1.5, 1.8, 2.2 keV, and a broad Gaussian iron line were also included in the fit (see Ref. \cite{ferrigno14}, for further details). In both models the {\sc diskbb} component does not improve the fit.}
\centering
\begin{tabular}{lccc} 
\hline 
& \sc{nthcomp} & \compps \\
\hline 
\noalign{\smallskip}  
$N_{\rm H}\ (10^{22} {\rm cm}^{-2})$ & $0.24\pm0.01$ & $0.23\pm0.01$\\ 
$kT_{\rm bb}$ (keV)& $0.34\pm0.01$ & --\\ 
$\Gamma$ & $1.44\pm0.01$ & --\\
$kT_{\rm e}$ (keV)& $23\pm2$ & $30\pm3$\\ 
$kT_{\rm seed}$ (keV)& -- & $0.37\pm0.01$\\ 
$\tau_{\rm T}$ & -- & $2.7^{+1.0}_{-0.1}$\\ 
$\cos \theta $ & -- & $0.76\pm0.02$\\
$A_{\rm seed}\ ({\rm km}^2)$ & -- & $250\pm40$\\
$E_{\rm Fe}$ (keV) & $6.6\pm0.2$ & $6.5\pm0.1$\\
$\sigma_{E_{\rm Fe}}$ (keV) & $1.11\pm0.12$ & $1.09\pm0.13$\\
$\chi^{2}_{\rm red}/{\rm dof}$ & 1.19/2642 & 1.15/2640 \\
$F_{\rm bol}$ ($10^{-10}$ erg cm$^{-2}$ s$^{-1}$)$^{a}$ & $4.25\pm0.02$ & $4.19\pm0.02$\\
\noalign{\smallskip}  
\hline  
\end{tabular}  
\vspace{0.5cm}

(a) Unabsorbed flux in the 0.4 -- 250 keV energy range.
\label{tab:spe} }
\end{table} 

I verified that the relatively large $\chi^2_{\rm red}$ (1.15 -- 1.19) in both fits is mainly due to statistical fluctuations of the data and the high statistics available. I could not find any systematic trends in the residuals that could suggest the presence of additional spectral components or that the selected models are not adequate for the fit. The improvement of the \compps\ compared to the {\sc nthcomp}, as statistically evaluated with the F-test, is highly significant, since I obtained an F-test probability of $2.5\times10^{-20}$. An advantage of the \compps\ model is that it allows to estimate the apparent area of the thermally emitting region on the NS surface, $A_{\rm seed}\approx 250\ (d/5.5\ \mbox{kpc})$ km$^2$. At the distance of \igrj\ ($d=5.5$~kpc), the radius of this region is $\approx 8$~km, assuming a canonical NS radius of 10 km and a spherical geometry. This source shows a harder spectrum, with a larger emitting area, and a smaller seed temperature compared to the other AMXPs (e.g., Refs.  \cite{gdb02,gp05,falanga05,mfb05,mfc07,ip09,falanga11,falanga12}). The spectral slopes for AMXPs have been found in the range $\Gamma\approx(1.8-2.0)$, that is, harder spectra compared to the photon index of $\Gamma\approx1.3$ for \igrj.
I show in Fig.~\ref{fig:spe} the absorbed unfolded averaged broad-band spectrum of \igrj, together with the residuals from the best fit model. 

\subsection{Timing analysis}
\label{sec:tm_IGR}

\subsubsection{The pulsar ephemeris}
\label{subsec:pe}
I studied the pulse profile of \igrj\ at soft X-rays (0.5 -- 10 keV) and hard X-rays/ soft \gr-rays (20 -- 150 keV) using
\xmm/Epic-pn, and \Integ/ISGRI and \Fermi/GBM data, respectively. Irrespective of the instrument involved, the timing analysis starts with the conversion of the arrival times of the (selected) event registered at the satellite to the solar system barycentre. This process uses the instantaneous spacecraft ephemeris (position and velocity) information, the JPL solar system ephemeris information (DE200 or DE405, I used DE200) and an accurate source position to convert the recorded satellite times from Terrestial Time scale (TT or TDT, which differs from Coordinated Universal Time (UTC) by a number of leap seconds plus a fixed offset of 32.184 s) into Barycentric Dynamical Time (TDB) scale, a time standard for Solar system ephemerides. 

I used the \igrj\ source position as listed in Table 1 of Ref. \cite{papitto13c}. This position is consistent at (sub)arcsecond level with the most accurate locations reported at optical wavelengths (see e.g., Ref. \cite{Pallanca13a}) and at radio frequencies (see e.g., Ref. \cite{Pavan13}), and from earlier \chandra\ X-ray observations of M28 \cite{Becker03}. Subsequently, I corrected the TDB arrival times for the acceleration effects along the binary orbit adopting the orbital parameters from Table 1 of Ref. \cite{papitto13c}. I analysed the \xmm/Epic-pn data taken in timing mode (timing accuracy $\sim 30\ \mu$s) from both \xmm\ observations (obs. ids. 0701981401 and 0701981501, which are separated in time by $\sim 9.5$ days; see Sec. \ref{sec:xmm}) performed during the April 2013 outburst. Using the spin frequency of $254.333\,031\,01(62)$ Hz, as derived from the spin period value and its uncertainty as listed in Table 1 of Ref. \cite{papitto13c}, yielded highly significant pulse-phase distributions. However, I noticed a considerable misalignment of $\sim 0.15$ in phase of both pulse-pulse distributions, which is too large when timing data are combined from observations covering periods of weeks, as is the case for \Integ/ISGRI and \Fermi/GBM. Such a shift is indicative of a slightly incorrect spin frequency, or, less likely, is related to the use of the DE200 solar system ephemeris in the barycentering process in this analysis, while the \igrj\ parameters of Table 1 of Ref. \cite{papitto13c} have been derived adopting DE405. 

Irrespective of the cause of the misalignment, I revisited the `best' spin-frequency, now adopting DE200, because in the barycentering process of the \Integ/ISGRI and \Fermi/GBM timing data I will  also adopt the DE200 solar system ephemeris. I derived through $Z_1^2$ test-statistics \cite{buccheri1983} optimisation using the combined \xmm\ datasets a slightly different spin frequency $\nu$ of $254.333\,030\,87(1)$ Hz. The performed timing analysis improves the spin frequency and confirms the other values reported by Ref. \cite{papitto13c} (see Fig \ref{tab:timing}). This procedure also ensures that the pulse phases of the events from both \xmm\ datasets are phase connected automatically across the datagap of $\sim$ 9.5 days. This is demonstrated in Fig. 4 panels a -- c and d -- f showing that the pulse profiles, folded upon one single spin and orbital ephemeris, are nicely aligned. The quoted uncertainty is the statistical error at $3\sigma$ confidential level; the systematical uncertainty in the spin-frequency due to the positional uncertainty (see e.g., Ref. \cite{Sanna17} for the method) in the coordinates of \igrj\ is about $6\times 10^{-8}$ Hz for an assumed uncertainty in source location of $0.''5$, and thus considerably larger than the statistical one. 
\begin{table}[h] 
\caption{\igrj\ orbital and rotational ephemeris, obtained from \xmm\ data during the April 2013 outburst. The performed timing analysys improves the spin frequency and confirms the other values reported by Ref. \cite{papitto13c}.}
\centering
\begin{tabular}{ll} 
\hline 
Parameter & Value \\
\hline 
$\nu$ (Hz) & $254.333\,030\,87(1)$\\
$\dot{\nu}$ (Hz/s) & 0.0\\
Reference Epoch (MJD) & 56386.0\\
Validity range &  56379 -- 56397\\
Solar system ephemeris & DE200\\
Orbital period (hr) & 11.025\,781(2)\\
$a_{\rm x}\sin(i)$ (lt-s) & 0.76591(1)\\
Eccentricity & 0.0\\
$T_{\rm asc}$ (MJD) & $56395.216\,893(1)$\\
\hline  
\end{tabular}  
\label{tab:timing} 
\end{table}  
To assess the effects on phase alignment by using a different solar system ephemeris, DE405, as adopted in \cite{papitto13c}, I repeated the frequency optimisation, and obtained an optimum spin frequency value that differs only $+6 \times 10^{-9}$ Hz from the DE200 value. I, therefore, excluded a different solar system ephemeris, DE405, as the cause of the phase misalignment. 

\begin{figure}[h] 
\centering
\includegraphics[trim=1cm 9cm 1cm 3cm, scale=0.6]{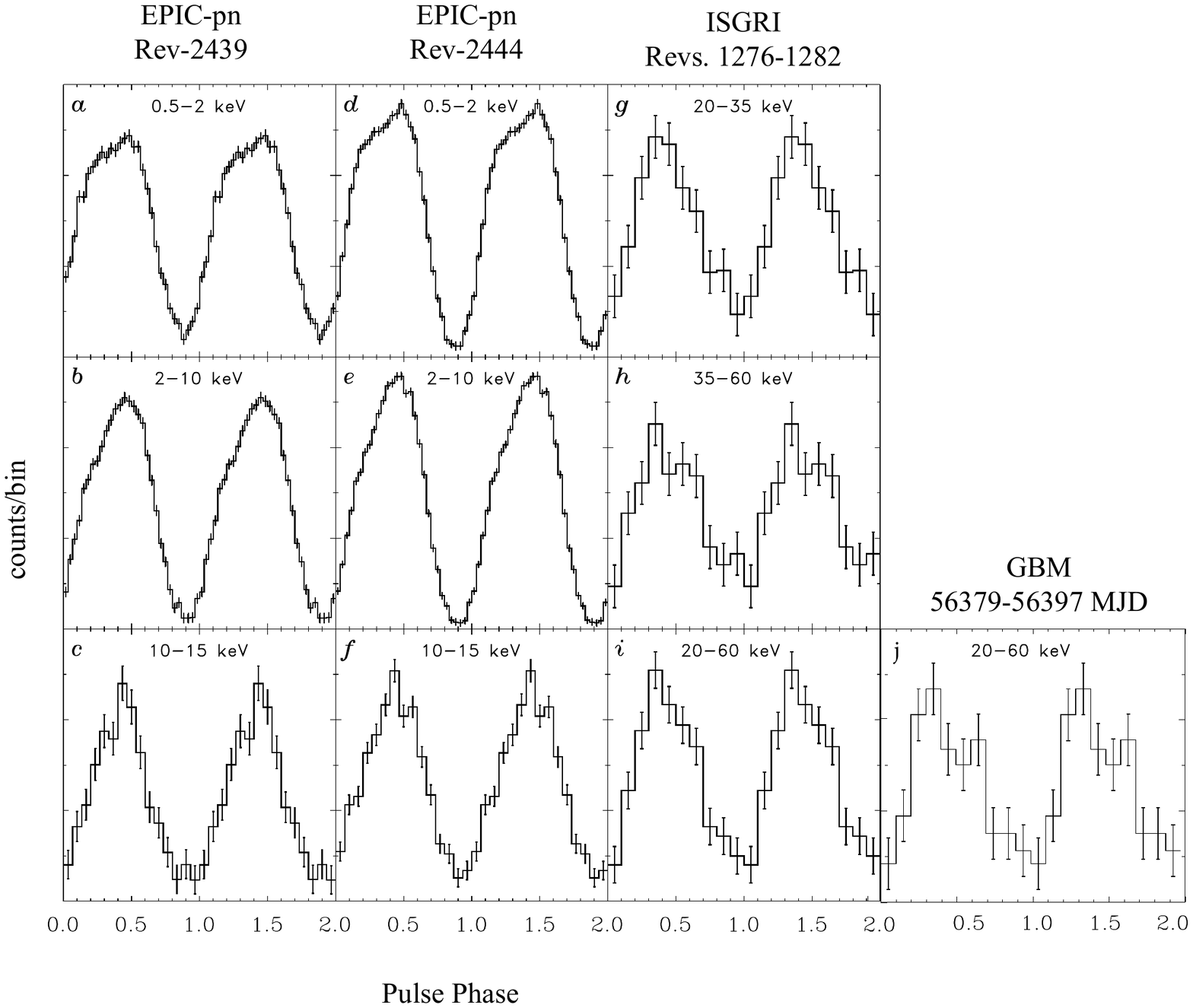}
\caption{Pulse profile compilation of \igrj, showing the X-ray profiles for \xmm/Epic-pn (panel a -- f), \Integ/ISGRI (panel g -- i), and \Fermi/GBM (panel j). The shift between the 20 -- 60 keV ISGRI and GBM profiles is only $23\pm109\ \mu$s indicating a consistent alignment. This consistency is obtained by an improvement of the spin frequency value.}
\label{fig:tim}
\end{figure}

Application of the newly derived spin-frequency value, which is $1.6\times 10^{-7}$ Hz smaller than the value of Ref.  \cite{papitto13c}, in the folding procedure now yielded a consistent alignment between the 2 -- 10 keV \xmm\ pulse-phase distributions of the two different \xmm\ observations (see also Fig.~\ref{fig:tim} panels b and e).

\subsubsection{Pulse profile, pulsed fraction, and time lag}
With an accurate ephemeris, that keeps phase alignment across at least ten days, I proceeded with the timing analysis of ISGRI soft \gr-ray data. I selected \Integ\ observations from the late stage of Rev. 1276 (beyond and including Scw 830010), 1277, 1279 -- 1280 and (ToO) 1282 with \igr\ at $\leq14^\circ.5$ from the pointing
axis. I excluded events recorded during time periods where the ISGRI count-rate behaved erratically (e.g. near perigeum ingress/egress, or during periods corresponding to a high solar activity). Additionally, I selected only those events with rise times between channels 7 and 90 \cite{lebrun03} from non-noisy pixels having a pixel illumination factor at least $25$\%. 

The outburst averaged 20 -- 60 keV pulse-phase distribution deviates from uniformity at a $10.9\sigma$, applying a $Z_1^2$-test. For the 20 -- 35 keV and 35 -- 60 keV bands, separately, I found significances of $8.5\sigma$ and $6.7\sigma$, respectively, while a hint ($\sim 2\sigma$) was seen in the 60 -- 150 keV band. The ISGRI pulse-phase distributions are shown in the right panels of Fig.~\ref{fig:tim} labelled g, h, i. In the same figure, the (time-averaged) phase-distributions are shown from both \xmm\ observations for the 0.5 -- 2 keV (top; a and d), 2 -- 10 keV (middle; b and e) and 10 -- 15 keV (bottom; c and f) energy bands. I would like to point out that the morphology changes as a function of energy for these X-ray profiles.

I have also folded the barycentered time stamps (accuracy $2\ \mu$s; TTE mode) of the NaI detectors of the \Fermi/GBM, collected during 56379 -- 56397 MJD (2013, March 28 -- April 15; continuously monitoring) using the updated \igrj\ DE200 ephemeris. Because of the non-imaging nature of these detectors, I have screened the data only by making selections on observational conditions such as on pointing direction, Earth zenith angle, and spacecraft location with respect to the South Atlantic Anomaly. I also ignored episodes of (intense) bursts. The averaged exposure per NaI detector was 201.3 ks.

For the 20 -- 60 keV band I detected pulsed emission at a $5.9\sigma$ confidential level with a pulse shape fully consistent with the ISGRI 20 -- 60 keV profile, while below 20 keV and above 60 keV I found significances of $3.3\sigma$ and $1.7\sigma$, respectively. The \Fermi/GBM 20 -- 60 keV pulse-phase distribution is shown in Fig.~\ref{fig:tim} (panel j) along with the 20 -- 60 keV ISGRI profile (panel i). Cross-correlation of both profiles (both in 60 bins) shows that the alignment between ISGRI and GBM profiles is fully consistent; I found an insignificant shift of only $23\pm 109\ \mu$s (i.e. $0.006 \pm 0.03$ in phase) between both detectors, validating the updated DE200 ephemeris of \igr.

\begin{figure}[h] 
\centering
\includegraphics[trim=1cm 1cm 1cm 0.7cm, scale=0.5]{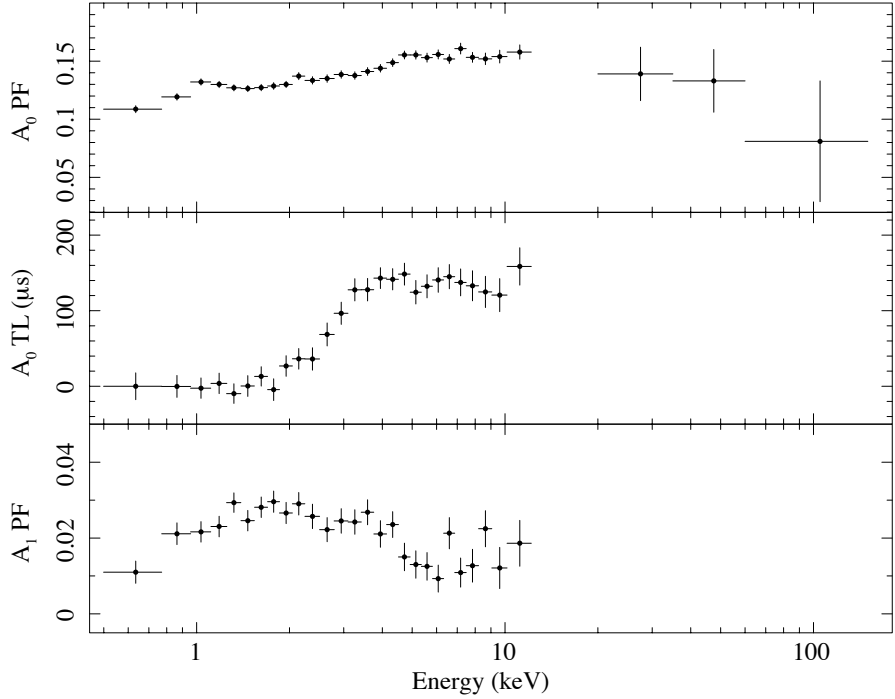}
\caption{Energy-dependent pulsed fraction (PF) for the fundamental, $A_0$, and first, $A_1$, harmonic and phase/time lag (TL) of the pulsed signal for the fundamental, $A_0$, harmonic (see Ref. \cite{ferrigno14}, for more details). The data points between 0.5  and 11 keV are obtained combining the two \xmm\ observations, and the data between 20 and 150 keV are from IBIS/ISGRI. \igrj\ shows a hard lag, normally not observed in other AMXPs (except for IGR J00291+5934).}
\label{fig:pf}
\end{figure}

In Fig. \ref{fig:pf}, I report the pulsed fraction of the fundamental and first harmonics in the 0.5 -- 11 keV and the phase/time lag of the fundamental harmonic obtained combining the two \xmm\ observations. The phase/time lag of the first harmonic is poorly determined, constant to zero, and not reported in this plot. The relative phase/time lags are expressed in microseconds as a function of energy compared to the averaged pulse profile (see Eq. (1) in Ref.  \cite{ferrigno14} and for more details). The zero is arbitrarily taken at the lowest energy band. For ISGRI I derived the time averaged pulsed fraction (across the outburst) in three energy bands of the fundamental component considering the signal-to-noise of the pulse signals above 20 keV. These are derived from fitting a sinusoid to the ISGRI pulse profiles to determine the pulsed counts, converting the pulsed excess counts to photon fluxes and finally dividing this number by the total source flux as derived through ISGRI imaging. The ISGRI pulsed fraction value in the 20 -- 60 keV is $\sim14\%$, and connects well with the $\sim15\%$ near 10 keV in the \xmm\ data. 

For AMXPs, it was found that the low-energy pulses lag behind the high-energy pulses (soft phase/time lags), monotonically increasing with energy, and saturating at about 10 -- 20 keV (see e.g., Refs.  \cite{cui98,ford00,gp05,falanga05,i11}). This soft time lag has been interpreted as the result of photon delay due to down-scattering of hard X-ray photons in the relatively cold plasma of the disk or NS surface \cite{cui98,t02,falanga07}. On the other hand, Refs. \cite{pg03,gp05,ip09} suggested that the lags may be produced by a combination of different angular distributions of the radiation components, and differences in the emissivity pattern. On the contrary, \igrj\ shows a hard lag, that is, low-energy pulsed photons arrive before the hard-energy pulsed photons. A similar trend in pulsed fraction and time lag has been observed for IGR J00291+5934, but only starting at higher energies from $\sim6$ keV \cite{falanga05,Sanna17}. For \igrj\ the thermal seed soft photons, coming from a larger emitting area, may up-scatter off hot electrons in the accretion column and arrive before the hard-energy photons. However, such a Compton-up scattering model is unlikely, since the lags are measured in the pulsed emission and the typical light-crossing time of the emission region is orders of magnitude smaller than the observed lags. It is more probable that the lags reflect variations in the emission pattern as a function of energy. Small deviations of the radiation angular distribution from the Lambert law induce rather large deviations in the pulse profile, leading therefore to lags \cite{pb06,lamb09}. Since the time lags are constant between 0.7 and 2 keV, this might suggest that these photons are coming from the disk. However, an increasing pulsed fraction in the range between $11$ and $14\%$ confirms rather the absence of the accretion disk at this energy range. I note that the maximum observed time delay, $\sim150 \ \mu$s, is comparable in absolute value with other AMXPs, indicating that they share most likely the same geometrical emission size or that the emission pattern has nearly the same energy gradient. The pulsed fraction of the first harmonic is between $1$ and $3\%$, while the time lags are consistent with zero due to large errors. The presence of the first harmonic may indicate that some pulsed emission is coming from the anti-polar cap, not being occulted by a disk, or that the emission pattern is not blackbody-like. 

\subsection{Properties of the type-I X-ray bursts} 
\label{sec:burst} 
During the 2013 outburst of \igrj\ a total of three type-I X-ray bursts were detected (see Table \ref{tab:burst_tot}). The type-I X-ray bursts were separated by similar time intervals of $\Delta t_{\rm rec,1-2}=4.8$ d and $\Delta t_{\rm rec,2-3}=5.2$ d (see Fig. \ref{fig:lcr} and  Table \ref{tab:burst_tot}) and went off during the flat part of the outburst when the source was at the highest X-ray luminosity (i.e. away from the initial rise and final decay phases). 
\begin{table}[h] 
\caption{Observed type-I X-ray bursts during the outburst of \igrj\ in 2013.}
\centering
\begin{tabular}{cccc} 
\hline 
Start  Time & Instrument & $\Delta t_{\rm rec}$ &Reference$^(a)$ \\
(UTC) & & (d) &\\
\hline 
April 3 at 03:10:02 & MAXI & -- &[1]\\
April 7 at 22:18:05 & \swift & 4.8 &[2]\\
April 13 at 04:15:27 & \Integ & 5.2 & this paper\\
\hline  
\end{tabular}  
\vspace{0.5cm}

(a) [1] \cite{Serino13}; [2] \cite{papitto13a,Linares13}.
\label{tab:burst_tot} 
\end{table}  

\begin{figure}[h] 
\centering
\includegraphics[trim=1cm 0cm 3cm 2cm, scale=0.48]{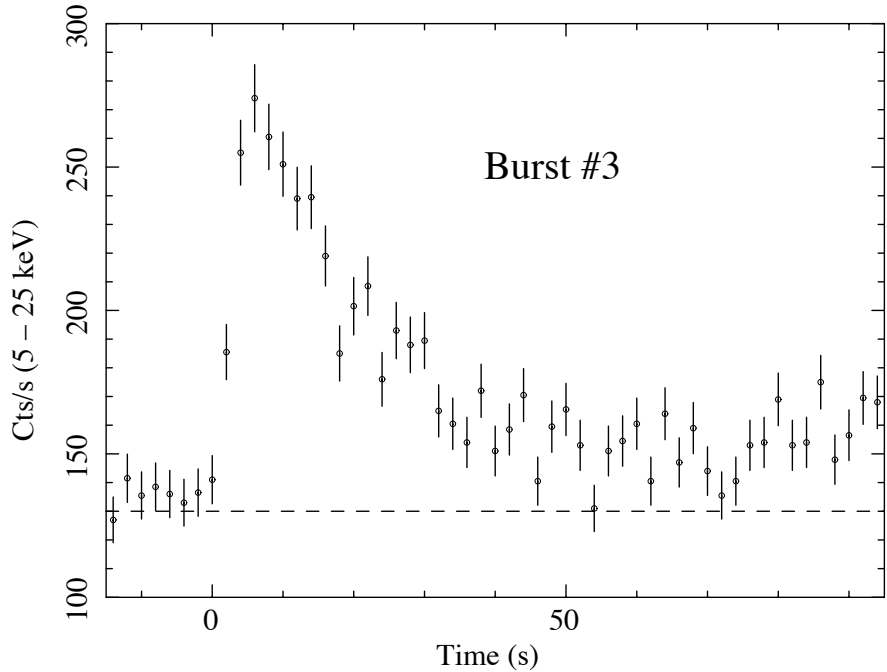}
\caption{The light curve of the type-I X-ray burst detected by JEM-X and reported for the first time in this paper. The burst start time was 56389.9292192~MJD. The JEM-X light curve was extracted in the 5 -- 25 keV energy range with a time bin of 2~s.}
\label{fig:burst}
\end{figure}

The source emission during the MAXI burst could not be analysed in detail because the data were contaminated by the nearby source GS~1826--238. The study of the second burst was reported by Refs. \cite{linares14,papitto13c}. Here I discuss in more detail the third burst that was not yet reported in literature. I discovered this event during a careful analysis of the \Integ\ data in revolution 1282. The ScWs 4, 5, and 6 of this revolution were affected by a high radiation background when the \Integ\ satellite was coming out from the Earth radiation belts and thus I had to specifically force the OSA software to skip the standard GTI selection to obtain the source light curve for this period. The burst was discovered in the ScW 4 and a zoom into the relevant part of the light curve is shown in Fig.~\ref{fig:burst}. 
The burst lasted $\sim90$ s and it showed a rise time of $\sim5$~s. The burst decay profile could be well fitted with an exponential function and the correspondingly derived e-folding time is $\tau_{\rm fit}=20\pm3$ s. The parameters of the JEM-X burst are thus well compatible with those measured by Ref. \cite{linares14} for the XRT burst (see their Table~2). As the spectral analysis of the JEM-X burst could not be carried out due to contamination from the radiation belts, I determined the burst persistent flux after 0.08 d from the JEM-X burst onset. The flux value is $F_{\rm pers}\approx3.6\times10^{-9}$ erg cm$^{-2}$ s$^{-1}$, corresponding to a luminosity of $L_{\rm pers,bol}\approx1.3\times 10^{37}$ erg s${}^{-1}$ (i.e. $3.4\%\ L_{\rm Edd}$, where $L_{\rm Edd}$ is the Eddington luminosity) close to the value reported by Ref. \cite{linares14}. However, since the persistent flux of the source is variable, I discuss later the results considering the flux range $F_{\rm pers}=(3-9)\times10^{-9}$ erg cm$^{-2}$ s$^{-1}$ for 0.11 d after the burst onset. If I assume that the total burst energy released during the JEM-X burst is comparable to that measured during the XRT burst, I obtain a total burst fluence of $f_{\rm b}=1.1\times10^{-6}$~erg~cm$^{-2}$ \cite{linares14}. As the light curve of the JEM-X burst does not show any clear evidence of a plateau at its peak, I conclude that most likely no photospheric radius expansion took place. I summarise in Table~\ref{tab:burst} all measured and extrapolated JEM-X burst parameters. 
\begin{table}[h] 
\caption{Parameters of the type-I X-ray burst observed by \Integ/JEM-X during the outburst of \igrj\ in 2013.}
\centering
\begin{tabular}{ll} 
\hline 
\hline 
$\Delta t_{\rm burst}$ (s)  & $90\pm1$\\
$\Delta t_{\rm rise}$ (s) & $5\pm1$\\
$\tau_{\rm fit}$ (s) & $20\pm3$\\
$f_{\rm b}^*$ ($10^{-6}$ erg cm$^{-2}$) & $1.1\pm0.1$\\
$F_{\rm pers}^*$ ($10^{-9}$ erg cm$^{-2}$ s$^{-1}$) & $3.6\pm0.7$\\
\hline  
\end{tabular}  
\vspace{0.5cm}

$^*$ Extrapolated values.
\label{tab:burst} 
\end{table} 

To constrain the nature of the thermonuclear burning that gave rise to the JEM-X burst, I first determine the local accretion rate per unit area onto the compact object at the time of the event as $\dot{m} = L_{\rm pers} (1+z) (4\pi R^2(GM/R_{\rm NS}))^{-1}$, that is, $\dot m \approx 7.3\times10^3$ g cm$^{-2}$ s$^{-1}$ (where the gravitational redshift is $1+ z = 1.31$ for a canonical NS with a mass $M_{\rm NS}=1.4M_\odot$ and a radius of $R_{\rm NS}=10$ km). I can then estimate the ignition depth at the onset of the burst with the equation $y_{\rm ign} = E_{\rm burst} (1+z)(4\pi R^2Q_{\rm nuc})^{-1}$, where the nuclear energy generated for pure helium (assuming a mean hydrogen mass fraction at ignition $\langle X\rangle=0$) is $Q_{\rm nuc}\approx 1.6+4\langle X\rangle\ {\rm MeV/nucleon}\approx1.6\ {\rm MeV/nucleon}$ and for solar abundances (assuming $\langle X\rangle=0.7$) is $Q_{\rm nuc}\approx4.4\ {\rm MeV/nucleon}$ (see Ref. \cite{galloway04} and references therein). I note that the considered values $Q_{\rm nuc}$ include losses owing to neutrino emission as detailed in Ref. \cite{fujimoto87}. Once the ignition depth is known, the recurrence time between the bursts can be calculated by using the equation $\Delta t_{\rm rec} =(y_{\rm ign} /\dot{m})(1 + z)$. I obtain $y_{\rm ign} =2.7\times10^8$ g cm${}^{-2}$ and $\Delta t_{\rm rec}\sim 5.6$~d in the case of helium burning, while $y_{\rm ign} = 10^8$ g cm${}^{-2}$ and $\Delta t_{\rm rec}\sim 2.1$~d in the hydrogen burning case. 

These results indicate that the JEM-X event could be more likely associated to a helium burst. However, to properly disentangle between the cases of helium and hydrogen bursts, I compare the total accreted matter between two subsequent bursts with the amount of fuel liberated during these events. From Fig.~\ref{fig:lcr} I estimate that the mean persistent fluxes (0.4 -- 250 keV energy band) between the observed bursts are $F_{\rm pers,1-2}=(1.6\pm0.1)\times10^{-9}$ erg cm${}^{-2}$ s${}^{-1}$ and $F_{\rm pers,2-3}=(2.0\pm0.1)\times10^{-9}$ erg cm${}^{-2}$ s${}^{-1}$, respectively. These correspond to a persistent luminosity of $L_{\rm X}\approx(6-7)\times 10^{36}$ erg s${}^{-1}$ (or $(1.5-2)\%\ L_{\rm Edd}$). The mass accretion rate is thus $\dot{M}=(6-8)\times10^{16}$ g s$^{-1}$, as calculated through the standard accretion equation $L_{\rm X}=GM_{\rm NS}\dot{M}/R_{\rm NS}$ (see, e.g., Ref. \cite{frank02}). The estimated total mass accreted between the type-I X-ray bursts is thus $\Delta M = \dot{M} \Delta t \approx (2.6-3.6)\times 10^{21}$ g, using $\Delta t=4.8$~d and $5.2$~d. The amount of fuel liberated during a thermonuclear burst can be estimated as $\Delta M_{\rm He}=E_{\rm burst}/\varepsilon_{\rm He}\approx 2.4 \times 10^{21}$~g and $\Delta M_{\rm H}=E_{\rm burst}/\varepsilon_{\rm H}\approx 4.9 \times 10^{20}$~g. Here $\varepsilon_{\rm He} \approx1.6 \times 10^{18}$ erg g$^{-1}$ is the total available nuclear energy for the transformation of pure helium into iron-peak elements, and $\varepsilon_{\rm H} \approx8 \times 10^{18}$ erg g$^{-1}$ is the total available nuclear energy for the transformation of pure hydrogen into iron-peak elements. As can be immediately noticed, $\Delta M$ and $\Delta M_{\rm He}$ are the closest values. Taking all the results together, I conclude that the JEM-X burst was triggered by unstable helium burning, after all accreted hydrogen was exhausted by the steady burning prior to the burst. If I consider the lower and upper burst persistent flux level of $F_{\rm pers}=(3-9)\times10^{-9}$ erg cm$^{-2}$ s$^{-1}$, this corresponds to a mixed helium/hydrogen burst ignition regime triggered by thermally unstable helium ignition (see Refs. \cite{Strohmayer03,linares14}), with a recurrence time, for solar composition, of between $\Delta t_{\rm rec}=(1$ and $2.5)$ d and for helium between $\Delta t_{\rm rec}=(2.3$ and $6.7)$ d. Such a short recurrence time for solar composition has not been observed during the data set including also the 131 ks \Integ\ continuous data. For helium composition instead, the range is more suitable with the observed value. I note that this conclusion is also in agreement with the light curve profiles of both XRT and JEM-X bursts, as helium bursts typically show a fast rise time of a few seconds at the most \cite{lewin93,Galloway08}. The probabilities of detecting a burst with $\Delta t_{\rm rec}=5.6$ d in the intervals (56377.5 -- 56385.2) MJD, (56385.2 -- 56390) MJD, and (56390 -- 56397) MJD are respectively $P_1=54\%$, $P_2=88\%$, and $P_3=67\%$. Instead for the lower limit recurrence time, $\Delta t_{\rm rec}=2.3$ d, the probabilities decrease as $P_1=39\%$, $P_2=83\%$, and $P_3=56\%$. Therefore, there is higher probability of missing another type-I X-ray burst at the beginning and at the end of the outburst. 

\subsection{Summary and discussion} 
\label{sec:summary} 
In this section, I reported for the first time on a detailed analysis of the \Integ\ data collected during the 2013 outburst of the transitional AMXP \igrj. The source is known for having displayed a peculiarly prominent timing and spectral variability in the X-ray domain during a full accretion episode, on timescales as short as fractions of a second. This variability is not usually found in classical AMXPs, which have never before shown (to the best of the present knowledge) evidence of transitions between phases of rotationally and accretion-powered pulsations. PSR J1023+0038 and XSS J1227.0--4859 are another two transient AMXPs showing a fast flux variability, but they have never been observed in full outburst \cite{linares14b}. In addition, their variability is accompanied by flaring and dip states, that \igrj\ does not show \cite{demartino10,shahbaz15}.

At odds with the peculiar timing and spectral variability displayed by \igrj\ on short timescales, the properties of the source X-ray emission on timescales of a few days seem to be remarkably similar to those of other AMXPs in outburst. The overall outburst profile is also closely reminiscent of that observed from AMXPs, featuring a faster rise ($\sim$ 1 -- 2~days) and a slower decay (several days) (see e.g., Refs. \cite{Gilfanov98,gp05,Powell07,falanga12}). The other two transient AMXPs show highly variable flat X-ray light curves with peculiar low luminosities of $\sim10^{33}$ erg s$^{-1}$ and flares reaching $\sim10^{34}$ erg s$^{-1}$ \cite{demartino10,bogdanov15}. 

The \Integ\ data permitted to carry out an analysis of the averaged broad-band spectral properties of the source, covering the 0.4 -- 250~keV energy range. I showed in Sec.~\ref{sec:spe} that a Comptonisation model can describe the energy distribution of the X-ray photons from \igrj. At the contrary of other AMXPs this source displays a harder spectrum, due to larger emitting area and smaller seed photon temperature (see e.g., Refs. \cite{gdb02,gp05,falanga12}). The X-ray spectra of the other two transient AMXPs are described by an absorbed power law and are softer, showing spectral slopes in the range $\Gamma=(1.6-1.7)$ \cite{demartino10,bogdanov15}.

I improved the \xmm\ ephemeris reported by Ref. \cite{papitto13c}, finding a slightly lower spin frequency. Thanks to this result, I have now a consistent alignment in the folding procedure of the two 2 -- 10 keV \xmm\ pulse-phase distributions and, for the first time, the detection of the millisecond pulsations using the TTE data with \Fermi/GBM. For the 20 -- 60 keV \Fermi/GBM band I detected pulsed emission at a $5.9\sigma$ confidential level with a pulse shape fully consistent with that observed by ISGRI 20 -- 60 keV. The pulse profiles of \igrj\ share many similarities with those of other AMXPs, as their shape is nearly sinusoidal at all energies \cite{patruno12}. The hard time lags of the pulsed emission likely indicate that the emission pattern from the hotspot has a peculiar energy dependence different from that of other AMXPs \cite{falanga05}. Alternatively, the contribution of the secondary cap changes the pulse profile to affect the sign of the lags. It would be interesting to compare the hard lags of \igrj\ with those of the other two transient AMXPs, since they have spin frequencies double that of \igrj\ \cite{Archibald13,demartino14}.

I reported on the discovery of a previously undetected thermonuclear burst from the source, caught by the JEM-X monitors at the beginning of the revolution 1282 when the \Integ\ satellite was coming out from the Earth radiation belts. Even though I could not perform a detailed spectral analysis of the event, the similarity with a previous burst detected by \swift/XRT allowed to demonstrate that type-I X-ray bursts from \igrj\ are most likely triggered by unstable helium burning after the exhaustion of all accreted hydrogen on the NS surface. This conclusion is compatible with both the characteristics of the burst profiles recorded from the source and with their measured recurrence time of roughly 5~days. \igrj\ is the only transient AMXP exhibiting a type-I X-ray burst so far. The donor stars hosted in the other two transient AMXPs were identified to be G-type stars with a mass ranging between (0.2 -- 0.4) M$_\odot$ \cite{Archibald13,demartino14}. This is in agreement with evolutionary expectations, which predict that AMXPs with an orbital period in the hour range (for these two transient AMXPs this is $\sim(5-7)$ hr or half of \igrj\ period \cite{Archibald13,demartino14}) should host a hydrogen-rich companion star \cite{deloye03}. These systems are also expected exhibit pure helium type-I X-ray bursts (see e.g., Refs.  \cite{galloway06,watts06,falanga07,galloway07,ferrigno11,defalco17}). I estimated that the burst recurrence time, for a persistent luminosity of $L=10^{34}$ erg s$^{-1}$ and assuming that the helium burst is similar to that exhibited by \igrj, is $\Delta_{\rm rec}=20$ yr. The transition between radio and X-ray phases may further delay the occurrence of the type-I X-ray bursts.

\chapter{Conclusions and future perspectives}
\epigraph{A thinker sees his own actions as experiments and questions, as attempts to find out something. Success and failure are for him answers above all.}{Friedrich Nietzsche}
 
\lettrine{A}{fter} three years and half of my PhD program (February 2014 -- July 2017), I have completed my project working from theoretical astrophysics to data analysis, and numerical modeling. I have built up a fundamental knowledge about the PR effect and its applications in mass accretion flow physics in GR. My works pave the way for future interesting projects on the general relativistic PR effect and radiation processes in high-energy astrophysics. In the future, I would like to strengthen my skills both in theoretical and numerical modeling fields, very close to my academic background and my natural inclinations. For such reasons, I will list below the possible future projects, which I would prefer to be deeply involved in and work on.\\

\noindent
{\bf Extension of the mathematical method to approximate photon geodesics from Schwarzschild to Kerr spacetimes.} Kerr metric is widely used both in theoretical and observational high-energy astrophysics for its solid predictive power in describing massive rotating compact objects in GR \cite{Misner73}. Schwarzschild metric configures as a first good description of the physical realm for its simplicity in performing calculations in GR. Indeed, the mathematical method introduced to approximate photon geodesics in Schwarzschild spacetime can be reasonably extended to the more challenging geometrical set up of Kerr metric. Such ambitious project would considerably reduce the computational times by developing fast and simple numerical codes helpful for all research groups. However, it entails some difficulties, summarised in the following points: $(i)$ writing the exact photon geodesic formulae in Kerr metric; $(ii)$ associating to the impact parameters $(\lambda,q)$ their respective emission angles $(\alpha,\beta)$; $(iii)$ adapting (or even improving) the mathematical method in two dimensions, implying lengthier calculations; $(iv)$ checking the validity and accuracy of the found approximations in 3D plots. The key part of this work relies on the approximation of \emph{light bending}, since the other effects are expressed in terms of this one. \\

\noindent
{\bf Flux from boundary layers around NSs.} Old NSs in LMXBs endowed with a weak magnetic field accrete matter from their companion stars, and the resulting accretion disk may extend down to the NS surface. The rapidly rotating gas decelerates due to the viscous friction, spreading then all or partially over the NS surface, forming thus the so-called boundary layer. Such structures emits most of their potential energy in the X-ray energy band, which theoretically is comparable to the energy generated in an accretion disk \cite{Syunyaev86,Revnivtsev13}. The emergent X-ray radiation contains potential critical information about the boundary layer, the related NS, and the motion of matter in strong gravitational field regimes. To observationally distinguish the emission of boundary layer from accretion disk, it is important to investigate in which radiation processes and gravitational fields such contributions are enhanced or lowered. The set of approximate equations, which I have derived, can be an useful tool to carry out such study. As underlined in Chap. \ref{chap:1}, the approximate ray-tracing equations do not care about the geometries of emitter and observer, but only on the geometrical background where they are embedded. This propriety permits to adapt our results for different geometries in several astrophysical contexts. \\


\noindent
{\bf General relativistic Rayleigh dissipative potential of PR effect.} The equations of motion proposed by Bini \emph{et al.} constitute the first complete treatment of the PR effect in GR. On the other hand, the relativity of observer splitting formalism, employed to derive them, lay the foundations for modeling other physical phenomena in GR. Such works are both very important for their powerful implications in theoretical physics. The PR effect configures as a dissipative system, therefore I have proposed to analyse it checking whether it admitted a Lagrangian formulation. I have discovered that by adding an integrating factor (physically depending on the energy of the system)  is possibile to prove the existence of the Rayleigh dissipative function. The next step is to derive the analytic expression of the Rayleigh potential. The advantages of such an approach are manifold, because the Rayleigh potential permits to have a direct connection between the theoretical results and the observational data, and in the same time to have a deep insight in the radiation processes in GR from a mathematical and physical perspective. The analytical expression of the Rayleigh potential in GR is something unique, because at best of my knowledge there are no other examples in literature of such attempt so far. In addition, this method is complementary to the relativity of observer splitting formalism, which sometimes configures to be too technical and intricate in the calculation process and the equations of motion are derived once the functional form of the radiation process is given. It would be of fundamental importance to investigate all the consequences of these results and also benchmark the theoretical predictions against the observations. Such project opens up a new research window on the study of dissipative systems in GR and strongly enhance the role of PR effect among astrophysical phenomena in GR.\\


\noindent
{\bf Improvements of the general relativistic PR effect model in 3D.} GR is the best theory of gravity so far available, and is successfully used to describe a wide range of astrophysical systems. However, there are still open questions and it is clear that it is of utmost importance to improve existing assessments and to explore new ways of testing GR. In a future project, my main concern is to improve the radiation field treatment together with the general relativistic PR effect. I would like to begin a systematic study of the PR effect in 3D aimed to improve it in its basic simple assumptions, like: considering a radiation field with photons moving in any direction (not radial at infinity as it is modeled so far); modeling the radiation source as a finite sized emitting object (not anymore seen as a point-source); improving the radiation field-test particle interaction through scattering processes depending on the frequency and angle-dependent cross sections (not anymore through the simple constant Thomson scattering, see \cite{Lamb95}, for further details); extending such effect to a continuous body (not a test particle as it has been done so far). 

The radiation field used in the above outline represents still a toy model and not much realistic, because it is modeled by shooting photons against the test particle. A positive feature of such model relies on the relative simplicity to carry out the calculations to derive the equations of motion or other formulae. Once the above improvements have been completed, the basic radiation field can be substituted with a more physically realistic description (such as Vaidya spacetime \cite{Vaidya1951a,Vaidya1973}), and extend all the previous improvements. 

This model can be then applied in several astrophysical contexts in high-energy astrophysics. It would be interesting to investigate such topics: accretion disc theory, and planetary formation. In accretion disc theory, it will be possible to build up more realistic numerical simulations for following the dynamical evolution of an accretion disc under the influence of a radiation field including the PR effect. Instead in planetary system theory, it can be possible to perform more accurate simulations for having a deeper insight on the fundamental processes and on the initial conditions needed for planetary formation.

\chapter*{Acknowledgments}
\addcontentsline{toc}{chapter}{Acknowledgments} 
\emph{This thesis constitutes another important milestone in my life, because it denotes all my sacrifices, dedication and passion for the scientific research. Before starting my PhD, I was skeptical and diffident about the success of this new life alone abroad due to my total inexperience. I must admit that I had several harsh times, where I felt lost and helpless into breasting various vicissitudes, but then I felt glad to have faced the difficulties counting only on my forces. I experienced the concepts of "distance and time", that from one side brought away from me many dear people or even canceled from my life; from the other side they contributed to strengthen the friendship with some people, teaching me to relish and appreciate all what I have in every occasion. I define my PhD experience as an interesting adventure, that allowed me to bump in disparate situations, to discover new places, and to consolidate the scientific awareness of what I was doing without never losing the sight of my route.} \\

\emph{Thanks to all the constructive people, who supported me blowing in the right direction, and also to the adverse people, who hampered my path giving me more ardor and willfulness to dismantle their criticisms. I hope in the future I will still have the opportunity to be still in Science, that, despite of the usual difficulties as part of the job, still remains a fascinating and marvelous profession made by daily discoveries and inner enrichment.}\\

\emph{Thanks to my family, who has always been the powerful arc launching me toward the future. A special thought goes to my grandmother Anna, who unfortunately passed away, but probably she is carefully following my accomplishments with her usual silent affection. Thanks to all my relatives, who concerned about my developments, being always proud of my choices.}\\

\emph{Thanks to my mentor Maurizio, for having competently educated me as a responsible scientist and for all the beautiful moments spent together; to the Swiss National Science Foundation, who efficiently financed my PhD research project $200021\_149865$; to Professor Friederich Karl Thielemann, who accepted to be my formal supervisor at University of Basel handling smoothly all the procedures during my PhD thesis submission; to Professor Luigi Stella, who, as a second supervisor, gave me wise advises; to Bern, that will be forever my second home, highlighted by my unforgettable "soul places": Rosengarten and the banks of the river Aare, where I got rid of all my stress and I had the necessary concentration to think over some topical moments.}\\

\emph{Thanks to my \qm{swiss family} represented by ISSI, a special small hive where all the bees are queens and the efficiency and productivity are recognised all over the world. Thanks to the executive director Raphael Rodrigo and his wife Concy, who always interested about me and my scientific progresses; to the directors Rudolf von Steiger, Anny Cazenave, and John Zarnecki, who encouraged me tirelessly in my research; to all ISSI staff, composed by Silvia, Saliba, Alexandra, Andrea, Irmela, Jenny, Marco, Greta, Nicholas, Teodolina, and Zhaosheng, with whom I
spent my daily life in Bern. It would be very long to list all what you have done for me, but I can also say that I will always bear in mind the joyful instants, the German and Bernese lessons, and the yoga events. I wish hopefully, that
I will have again the possibility to visit you at ISSI in the future. A special thank goes to Professor Roger Maurice Bonnet, who is for me a great model to admire and take as example. It was a great honor to have met him and it will
always be impressed in my mind the amazing days spent in Alpbach and Beijing and all the useful discussions had together. Thanks to the founder of ISSI, Professor Johannes Geiss, a visionary, charismatic, and polyhedron person, who is another prominent figure to imitate. Thanks to all ISSI visitors, with whom I have interacted and I have learned something not only from a scientific point of view, but also from a social perspective. Thanks to the Professor Antonio Ereditato, who helped me in all needed occasions here in Bern, giving me the feeling to be at home.}\\

\emph{Thanks to all professors of the University of Napoli, who provided me the right preparation to carry out consciously the scientific issues presented during these years. Thanks to Professors Florinda Capone, Salvatore Cuomo, Geatano Festa, and Antonio Romano, who motivated my research interests with illuminating discussions. In particular, thanks to the Professor Salvatore Capozziello, who was my middleman to come to Bern suggesting me always good lines to follow in my scientific career.}\\

\emph{Thanks to the Silesian University in Opava, that hosted me for several times, and to the extraordinary group head by Professor Pavel Bakala, who improved my knowledge about Numerical Analysis and General Relativity, and helped me for developing the main part of my PhD thesis providing sublime results. A special thank goes to Katerina, Debora, Daniela, Bob, and Armand, who demonstrated their friendship since the very beginning.}\\

\emph{Thanks to the ISDC group in Ginevra for having welcomed me with all the appreciated support. In particular, thanks to Enrico and Carlo for having taught me the INTEGRAL data analysis with great patience and high professionalism.}\\

\emph{Thanks to the ISSI Beijing for having given me the opportunity to live a great scientific and cultural experience in China, where I got deeply in touch with a new world, full of secular traditions and spectacular monuments. Thanks to Professor Mario Pinheiro for having given me the possibility to work close to an expert like him and for all the unpredictable Chinese adventures. Thanks to the ISSI Beijing staff, composed by Lijuan, Wang Gang, Chico and Sabrina, for having been a peerless guides and companies in this admirable city.}\\

\emph{Thanks to all people met along my tortuous journey, who enlarged my horizons with their personalities. A great remember goes to Simon, Cristiane, Salvatore, Francesco, Silvia, Daniele, June, Hugo, Vito, Alex, Natascha, Nadia, with whom I spent friendly moments. Thanks to my far friend Alessandro Ridolfi, that albeit our contacts were virtual, we established a very close friendship.}\\

\emph{Thanks to the gushing Italian group, composed by Marta, Valentina, Silvia and Monica, with whom I surely spent one of the best periods of my staying in Bern. A particular thank goes to Monica, also known as "the random variable", with whom I shared magnificent moments, having the miraculous power to remove all the stress, and from whom I learned several things, discovering new surprising sides of my character.}\\

\emph{Thanks to all my incomparable friends in Napoli, who always welcomed me very happily whenever I was back without feeling the distance and the time elapsed among us. Thanks to Floriana, for having always taken me in consideration involving me in her activities, to her family, for the appreciated regard, and in particular to her mother, my ex Professor of Mathematics at high school, Bianca Fagnani, whose teachings dwell limpid inside me, preserving vivid her passion for the mathematics communicated also to me. Thanks to one of my best friend Gianpiero, who always animated me with his happiness and his positive way to approach the life, keeping energetically stronger and stronger our friendship. Thanks to my two colleagues of Mathematics, Concetta and Anna, with whom I spent relaxing moments due to their incalculable sweetness.}\\ 

\emph{Concluding, thanks to my city Napoli, that infuses me that special energy, solarity, adaptation spirit, and force to overcome all the obstacles, which are unique feelings belonging only to this land. The origin of this emotional swirl comes probably from a mysterious and harmonic mixture of the enormous shadow of the high building, the echo of the voices through the narrow alleys, the dynamic chaos of the people crowding the streets, the parfume of the sea, the sensual profile of the Vesuvio during all the seasons of the year, the clothes laid out from the houses coloring the city, the inviting smell of the pizza, the veracious spontaneity of the people, and the the unmistakable music of the mandolin.}

\backmatter
\begin{multicols}{2}
\setstretch{-5.5}
\small{
\bibliography{references}

\begin{thebibliography}{100}

\bibitem{Kippenhahn12}
R.~{Kippenhahn} {\em et~al.}, {\em {Stellar Structure and Evolution: Astronomy
  and Astrophysics Library}}.
\newblock 2012.

\bibitem{Longair11}
M.~S. {Longair}, {\em {High Energy Astrophysics}}.
\newblock 2011.

\bibitem{Shapiro86}
S.~L. {Shapiro} {\em et~al.}, {\em {Black Holes, White Dwarfs and Neutron
  Stars: The Physics of Compact Objects}}.
\newblock 1986.

\bibitem{Misner73}
C.~W. {Misner} {\em et~al.}, {\em {Gravitation}}.
\newblock 1973.

\bibitem{Bini09}
D.~{Bini} {\em et~al.} {\em CQG}, vol.~26, p.~055009, 2009.

\bibitem{Bini11}
D.~{Bini} {\em et~al.} {\em CQG}, vol.~28, p.~035008, 2011.

\bibitem{Beloborodov02}
A.~M. {Beloborodov} {\em ApJL}, vol.~566, pp.~L85--L88, 2002.

\bibitem{Poutanen06}
J.~{Poutanen} {\em et~al.} {\em MNRAS}, vol.~373, pp.~836--844, 2006.

\bibitem{Defalco16}
V.~{De Falco} {\em et~al.} {\em A\&A}, vol.~595, p.~A38, 2016.

\bibitem{Defalco2018}
V.~{De Falco} {\em et~al.} {\em PRD}, vol.~97, p.~084048, 2018.

\bibitem{Defalco20183D}
V.~{De Falco} {\em et~al.} {\em PRD (accepted)}, 2019.

\bibitem{wijnands06}
R.~{Wijnands}, ``{Accretion-Driven Millisecond X-ray Pulsars},'' 2006.

\bibitem{poutanen06b}
J.~{Poutanen}, ``{Accretion-powered millisecond pulsars},'' {\em Advances in
  Space Research}, vol.~38, pp.~2697--2703, 2006.

\bibitem{patruno12}
A.~{Patruno} {\em et~al.} {\em ArXiv e-prints}, 2012.

\bibitem{Li2018}
Z.~{Li} {\em et~al.} {\em A\&A}, p.~A114, Dec. 2018.

\bibitem{Defalco2017a}
V.~{De Falco} {\em et~al.} {\em A\&A}, vol.~599, p.~A88, 2017.

\bibitem{Defalco2017b}
V.~{De Falco} {\em et~al.} {\em A\&A}, vol.~603, p.~A16, 2017.

\bibitem{Luminet79}
J.-P. {Luminet} {\em A\&A}, vol.~75, pp.~228--235, 1979.

\bibitem{Pechenick83}
K.~R. {Pechenick} {\em et~al.} {\em ApJ}, vol.~274, pp.~846--857, 1983.

\bibitem{c18}
R.~D. Carmichael {\em Bull. Amer. Math. Soc.}, 1918.

\bibitem{Fabian89}
A.~C. {Fabian} {\em et~al.} {\em MNRAS}, vol.~238, pp.~729--736, 1989.

\bibitem{Synge64}
J.~L. {Synge}, {\em {Relativity: The general theory}}.
\newblock 1964.

\bibitem{Chandrasekhar92}
S.~{Chandrasekhar}, {\em {The mathematical theory of black holes}}.
\newblock 1992.

\bibitem{Bao1994}
G.~{Bao} {\em et~al.} {\em ApJ}, vol.~435, pp.~55--65, 1994.

\bibitem{beckwith04}
K.~{Beckwith} {\em et~al.} {\em MNRAS}, vol.~352, pp.~353--362, 2004.

\bibitem{tomsick14}
J.~A. {Tomsick} {\em et~al.} {\em ApJ}, vol.~780, p.~78, 2014.

\bibitem{leahy11}
D.~A. {Leahy} {\em et~al.} {\em ApJ}, vol.~742, p.~17, 2011.

\bibitem{baubock15}
M.~{Baub{\"o}ck} {\em et~al.} {\em ApJ}, vol.~811, p.~144, 2015.

\bibitem{nath02}
N.~R. {Nath} {\em et~al.} {\em ApJ}, vol.~564, pp.~353--360, 2002.

\bibitem{miller15}
M.~C. {Miller} {\em et~al.} {\em ApJ}, vol.~808, p.~31, 2015.

\bibitem{watts16}
A.~L. {Watts} {\em et~al.} {\em Reviews of Modern Physics}, vol.~88, p.~021001,
  2016.

\bibitem{Poynting03}
J.~H. {Poynting} {\em MNRAS}, vol.~64, p.~1, 1903.

\bibitem{Robertson37}
H.~P. {Robertson} {\em MNRAS}, vol.~97, p.~423, 1937.

\bibitem{Lebedew02}
P.~{Lebedew} {\em ApJ}, vol.~15, p.~60, 1902.

\bibitem{Nichols02}
E.~F. {Nichols} {\em et~al.} {\em ApJ}, vol.~15, p.~62, Jan. 1902.

\bibitem{Nichols03}
E.~F. {Nichols} {\em et~al.} {\em ApJ}, vol.~17, p.~352, 1903.

\bibitem{Plummer05}
H.~C. {Plummer} {\em MNRAS}, vol.~65, p.~229, 1905.

\bibitem{Plummer06}
H.~C. {Plummer} {\em MNRAS}, vol.~67, p.~63, 1906.

\bibitem{Larmor1917}
J.~{Larmor} {\em Nature}, vol.~99, p.~404, 1917.

\bibitem{Obs20}
 {\em The Observatory}, vol.~43, pp.~291--298, 1920.

\bibitem{Page18a}
L.~Page {\em Phys. Rev.}, vol.~11, pp.~376--400, 1918.

\bibitem{Page18b}
L.~Page {\em Phys. Rev.}, vol.~12, pp.~371--380.

\bibitem{Larmor17}
J.~{Larmor} {\em Nature}, vol.~99, p.~404, July 1917.

\bibitem{Wyatt50}
S.~P. {Wyatt} {\em et~al.} {\em ApJ}, vol.~111, pp.~134--141, 1950.

\bibitem{Guess62}
A.~W. {Guess} {\em ApJ}, vol.~135, pp.~855--866, 1962.

\bibitem{Burns79}
J.~A. {Burns} {\em et~al.} {\em Icarus}, vol.~40, pp.~1--48, 1979.

\bibitem{Abramowicz90}
M.~A. {Abramowicz} {\em et~al.} {\em ApJ}, vol.~361, pp.~470--482, 1990.

\bibitem{Miller93}
M.~C. {Miller} {\em et~al.} {\em AJL}, vol.~413, pp.~L43--L46, 1993.

\bibitem{Lamb95}
F.~K. {Lamb} {\em et~al.} {\em ApJ}, vol.~439, pp.~828--845, 1995.

\bibitem{Miller96}
M.~C. {Miller} {\em et~al.} {\em ApJ}, vol.~470, p.~1033, 1996.

\bibitem{Srikanth99}
R.~{Srikanth} {\em Icarus}, vol.~140, pp.~231--234, 1999.

\bibitem{Kimura02}
H.~{Kimura} {\em et~al.} {\em Icarus}, vol.~157, pp.~349--361, 2002.

\bibitem{Kerr1963}
R.~P. Kerr {\em PRL}, vol.~11, pp.~237--238, 1963.

\bibitem{Inogamov1999}
N.~A. {Inogamov} {\em et~al.} {\em Astronomy Letters}, vol.~25, pp.~269--293,
  1999.

\bibitem{lewin93}
W.~H.~G. {Lewin} {\em et~al.} {\em SSR}, vol.~62, pp.~223--389, 1993.

\bibitem{Fabian2015}
A.~C. {Fabian} {\em et~al.} {\em MNRAS}.

\bibitem{Ballantyne2004}
D.~R. {Ballantyne} {\em et~al.} {\em ApJL}, vol.~602, pp.~L105--L108, 2004.

\bibitem{Ballantyne2005}
D.~R. {Ballantyne} {\em et~al.} {\em ApJ}, vol.~626, pp.~364--372, 2005.

\bibitem{Worpel2013}
H.~{Worpel} {\em et~al.} {\em ApJ}, vol.~772, p.~94, 2013.

\bibitem{Keek2014}
L.~{Keek} {\em et~al.} {\em ApJ}, vol.~789, p.~121, 2014.

\bibitem{Ji2014}
L.~{Ji} {\em et~al.} {\em ApJ}, vol.~782, p.~40, 2014.

\bibitem{Worpel2015}
H.~{Worpel} {\em et~al.} {\em ApJ}, vol.~801, p.~60, 2015.

\bibitem{Chow1995}
T.~Chow, {\em Classical Mechanics}.
\newblock 1995.

\bibitem{Goldstein2002}
H.~Goldstein {\em et~al.}, {\em Classical Mechanics}.
\newblock 2002.

\bibitem{Santilli1978}
R.~Santilli, {\em Foundations of theoretical mechanics}.
\newblock 1978.

\bibitem{Santilli1979}
R.~M. Santilli {\em PRD}, vol.~20, pp.~555--563, 1979.

\bibitem{Lopuszanski1999}
J.~Lopuszanski, {\em The Inverse Variational Problem in Classical Mechanics}.
\newblock 1999.

\bibitem{Minguzzi2015}
E.~Minguzzi {\em European Journal of Physics}, vol.~36, p.~035014, 2015.

\bibitem{Courant1962}
R.~Courant {\em et~al.}, {\em Methods of Mathematical Physics: Partial
  differential equations}.
\newblock 1962.

\bibitem{DeFelice1971}
F.~{de Felice} {\em General Relativity and Gravitation}, vol.~2, pp.~347--357,
  1971.

\bibitem{Abramowicz1988}
M.~A. {Abramowicz} {\em et~al.} {\em General Relativity and Gravitation},
  vol.~20, pp.~1173--1183, 1988.

\bibitem{DeFelice1991a}
F.~{de Felice} {\em MNRAS}, vol.~252, pp.~197--202, 1991.

\bibitem{Prasanna1990}
A.~R. {Prasanna} {\em et~al.} {\em General Relativity and Gravitation},
  vol.~22, pp.~987--991, 1990.

\bibitem{DeFelice1991b}
F.~{de Felice} {\em et~al.} {\em CQG}, vol.~8, pp.~1871--1880, 1991.

\bibitem{Iyer1993}
S.~{Iyer} {\em et~al.} {\em CQG}, vol.~10, pp.~L13--L16, 1993.

\bibitem{DeFelice1995}
F.~{de Felice} {\em CQG}, vol.~12, pp.~1119--1126, 1995.

\bibitem{Barrabes1995}
C.~{Barrabes} {\em et~al.} {\em MNRAS}, vol.~276, pp.~432--438, 1995.

\bibitem{Abramowicz1993}
M.~A. {Abramowicz} {\em et~al.} {\em CQG}, vol.~10, pp.~L183--L186, 1993.

\bibitem{Bini1997a}
D.~{Bini} {\em et~al.} {\em IJMPD}, vol.~6, pp.~1--38, 1997.

\bibitem{Bini1998}
D.~{Bini} {\em et~al.}, ``{The Inertial Forces/Test Particle Motion Game},''
  p.~376, 1999.

\bibitem{Bini1997b}
D.~{Bini} {\em et~al.} {\em IJMPD}, vol.~6, pp.~143--198, 1997.

\bibitem{Bini1999}
D.~{Bini} {\em et~al.} {\em CQG}, vol.~16, pp.~2105--2124, 1999.

\bibitem{Sommerfeld1964}
A.~Sommerfeld, {\em Mechanics}.
\newblock 1964.

\bibitem{Arnold2013}
K.~Vogtmann {\em et~al.}, {\em Mathematical Methods of Classical Mechanics}.
\newblock 2013.

\bibitem{Stephani2003}
H.~{Stephani} {\em et~al.}, {\em {Exact solutions of Einstein's field
  equations}}.
\newblock 2003.

\bibitem{Wald1984}
R.~M. {Wald}, {\em {General relativity}}.
\newblock 1984.

\bibitem{Bini2010}
F.~{de Felice} {\em et~al.}
\newblock 2010.

\bibitem{Hawking1973}
S.~W. Hawking {\em et~al.}, {\em The Large Scale Structure of Space-Time}.
\newblock 1973.

\bibitem{Jantzen1992}
R.~T. {Jantzen} {\em et~al.} 1992.

\bibitem{Chandrasekhar83}
S.~{Chandrasekhar}, {\em {The mathematical theory of black holes}}.
\newblock 1983.

\bibitem{Vaidya1951a}
P.~C. {Vaidya} {\em Physical Review}, vol.~83, pp.~10--17, 1951.

\bibitem{Vaidya1951b}
P.~{Chunilal Vaidya} {\em Indian Academy of Sciences Proceedings Section},
  vol.~33, 1951.

\bibitem{Vaidya1973}
P.~C. {Vaidya} {\em et~al.} {\em PRD}, vol.~7, pp.~3590--3593, 1973.

\bibitem{Lindquist1965}
R.~W. {Lindquist} {\em et~al.} {\em Physical Review}, vol.~137, pp.~1364--1368,
  1965.

\bibitem{Vaidya1999}
P.~C. {Vaidya} {\em General Relativity and Gravitation}, vol.~31, pp.~121--135,
  1999.

\bibitem{Taylor1982}
J.~H. {Taylor} {\em et~al.} {\em ApJ}, vol.~253, pp.~908--920, 1982.

\bibitem{Abott2016a}
B.~P. {Abbott} {\em et~al.} {\em PRL}, vol.~116, p.~061102, 2016.

\bibitem{Abott2016b}
B.~P. {Abbott} {\em et~al.} {\em PRL}, vol.~116, p.~241103, 2016.

\bibitem{Abbott2017}
B.~P. {Abbott} {\em et~al.} {\em PRL}, vol.~119, p.~161101, 2017.

\bibitem{Walker1989}
M.~A. {Walker} {\em et~al.} {\em ApJ}, vol.~346, pp.~844--846, 1989.

\bibitem{Walker1992}
M.~A. {Walker} {\em ApJ}, vol.~385, pp.~642--664, 1992.

\bibitem{Bini2011v}
D.~{Bini} {\em et~al.} {\em CQG}, vol.~28, p.~245019, 2011.

\bibitem{Oh2010}
J.~S. {Oh} {\em et~al.} {\em PRD}, vol.~81, p.~084005, 2010.

\bibitem{Oh2011}
J.~{Sok Oh} {\em et~al.} {\em New Astronomy}, vol.~16, pp.~183--186, 2011.

\bibitem{Wielgus2012}
M.~{Wielgus} {\em et~al.} {\em A\&A}, vol.~545, p.~A123, 2012.

\bibitem{Stahl2012}
A.~{Stahl} {\em et~al.} {\em A\&A}, vol.~546, p.~A54, 2012.

\bibitem{Stahl2013}
A.~{Stahl} {\em et~al.} {\em A\&A}, vol.~555, p.~A114, 2013.

\bibitem{Wielgus2016}
M.~{Wielgus}, ``{Eddington capture sphere around luminous relativistic
  stars},'' vol.~312, pp.~131--134, 2016.

\bibitem{Wielgus2016n}
M.~{Wielgus} {\em et~al.} {\em MNRAS}, vol.~458, pp.~3420--3428, 2016.

\bibitem{Koutsantoniou2014}
L.~E. {Koutsantoniou} {\em et~al.} {\em ApJ}, vol.~794, p.~27, 2014.

\bibitem{Contopoulos2015}
I.~{Contopoulos} {\em et~al.} vol.~805, p.~105, 2015.

\bibitem{Lancova2017}
D.~{Lan{\v c}ov{\'a}} {\em et~al.}, ``{The study on behaviour of thin accretion
  disc affected by Poynting-Robertson effect},'' pp.~127--136, 2017.

\bibitem{Keek2018}
L.~{Keek} {\em et~al.} {\em ApJL}, vol.~855, p.~L4, 2018.

\bibitem{Boyer1967}
R.~H. {Boyer} and other {\em Journal of Mathematical Physics}, vol.~8,
  pp.~265--281, 1967.

\bibitem{Bardeen1972}
J.~M. {Bardeen} {\em et~al.} {\em {ApJ}}, vol.~178, pp.~347--370, 1972.

\bibitem{Carter1968}
B.~Carter {\em Phys. Rev.}, vol.~174, pp.~1559--1571, 1968.

\bibitem{Kovar2008}
J.~{Kov{\'a}{\v r}} {\em et~al.} {\em CQG}, vol.~25, p.~095011, 2008.

\bibitem{Bakala2015}
P.~{Bakala} {\em et~al.} {\em A\&A}, vol.~581, p.~A35, 2015.

\bibitem{Press2002}
W.~H. {Press} {\em et~al.}, {\em {Numerical recipes in C++ : the art of
  scientific computing}}.
\newblock 2002.

\bibitem{Bahcall65}
J.~N. {Bahcall} {\em et~al.} {\em Physical Review}, vol.~140, pp.~1445--1451,
  1965.

\bibitem{frank02}
J.~{Frank} {\em et~al.}, {\em {Accretion Power in Astrophysics: Third
  Edition}}.
\newblock 2002.

\bibitem{shakura73}
N.~I. {Shakura} {\em et~al.} {\em A\&A}, vol.~24, pp.~337--355, 1973.

\bibitem{seward10}
F.~D. {Seward} {\em et~al.}, {\em {Exploring the X-ray Universe}}.
\newblock 2010.

\bibitem{strohmayer06}
T.~{Strohmayer} {\em et~al.}, {\em {New views of thermonuclear bursts}},
  pp.~113--156.
\newblock 2006.

\bibitem{galloway08a}
D.~K. {Galloway} {\em et~al.} {\em ApJ s}, vol.~179, pp.~360--422, 2008.

\bibitem{walker89}
M.~A. {Walker} {\em et~al.} {\em ApJ}, vol.~346, pp.~844--846, 1989.

\bibitem{kuulkers10}
E.~{Kuulkers} {\em et~al.} {\em A\&A}, vol.~514, p.~A65, 2010.

\bibitem{fujimoto81}
M.~Y. {Fujimoto} {\em et~al.} {\em ApJ}, vol.~247, pp.~267--278, 1981.

\bibitem{falanga08}
M.~{Falanga} {\em et~al.} {\em A\&A}, vol.~484, pp.~43--50, 2008.

\bibitem{kuulkers04}
E.~{Kuulkers} {\em Nuclear Physics B Proceedings Supplements}, vol.~132,
  pp.~466--475, 2004.

\bibitem{schatz03}
H.~{Schatz} {\em et~al.} {\em Nuclear Physics A}, vol.~718, pp.~247--254, 2003.

\bibitem{intzand05}
J.~J.~M. {in't Zand} {\em et~al.} {\em A\&A}, vol.~441, pp.~675--684, 2005.

\bibitem{cooper07}
R.~L. {Cooper} {\em et~al.} {\em ApJ}, vol.~661, pp.~468--476, 2007.

\bibitem{cumming01}
A.~{Cumming} {\em et~al.} {\em ApJL}, vol.~559, pp.~L127--L130, 2001.

\bibitem{galloway06}
D.~K. {Galloway} {\em et~al.} {\em ApJ}, vol.~652, pp.~559--568, 2006.

\bibitem{intzand04}
J.~J.~M. {in't Zand} {\em et~al.} {\em A\&A}, vol.~426, pp.~257--265, 2004.

\bibitem{Hewish68}
A.~{Hewish} {\em et~al.} {\em Nature}, vol.~217, pp.~709--713, 1968.

\bibitem{bk74}
G.~S. {Bisnovatyi-Kogan} {\em et~al.} {\em SOVAST}, vol.~18, p.~217, 1974.

\bibitem{Hulse75}
R.~A. {Hulse} {\em et~al.} {\em ApJL}, vol.~195, pp.~L51--L53, 1975.

\bibitem{ruderman75}
M.~A. {Ruderman} {\em et~al.} {\em ApJ}, vol.~196, pp.~51--72, 1975.

\bibitem{Backus82}
P.~R. {Backus} {\em et~al.} {\em ApJL}, vol.~255, pp.~L63--L67, 1982.

\bibitem{r82}
V.~{Radhakrishnan} {\em et~al.} {\em Current Science}, vol.~51, pp.~1096--1099,
  1982.

\bibitem{alpar82}
M.~A. {Alpar} {\em et~al.} {\em Nature}, vol.~300, pp.~728--730, 1982.

\bibitem{Wijnands98}
R.~{Wijnands} {\em et~al.} {\em Nature}, vol.~394, pp.~344--346, 1998.

\bibitem{falanga05}
M.~{Falanga} {\em et~al.} {\em A\&A}, vol.~444, pp.~15--24, 2005.

\bibitem{papitto13c}
A.~{Papitto} {\em et~al.} {\em Nature}, vol.~501, pp.~517--520, 2013.

\bibitem{Burderi02}
L.~{Burderi} {\em et~al.} {\em ApJ}, vol.~574, pp.~930--936, 2002.

\bibitem{DiSalvo03}
T.~{Di Salvo} {\em et~al.} {\em A\&A}, vol.~397, pp.~723--727, 2003.

\bibitem{Patruno10}
A.~{Patruno} {\em ApJ}, vol.~722, pp.~909--918, 2010.

\bibitem{markwardt04b}
C.~B. {Markwardt} {\em et~al.} {\em The Astronomer's Telegram}, vol.~360, 2004.

\bibitem{shahbaz15}
T.~{Shahbaz} {\em et~al.} {\em MNRAS}, vol.~453, pp.~3461--3473, 2015.

\bibitem{remillard04}
R.~{Remillard} {\em The Astronomer's Telegram}, vol.~357, 2004.

\bibitem{lewis10}
F.~{Lewis} {\em et~al.} {\em A\&A}, vol.~517, p.~A72, 2010.

\bibitem{hartman11}
J.~M. {Hartman} {\em et~al.} {\em ApJ}, vol.~726, p.~26, 2011.

\bibitem{lipunov15}
V.~{Lipunov} {\em et~al.} {\em The Astronomer's Telegram}, vol.~7835, 2015.

\bibitem{sanna15}
A.~{Sanna} {\em et~al.} {\em The Astronomer's Telegram}, vol.~7836, 2015.

\bibitem{jonker05}
P.~G. {Jonker} {\em et~al.} {\em MNRAS}, vol.~361, pp.~511--516, 2005.

\bibitem{galloway05}
D.~K. {Galloway} {\em et~al.} {\em ApJL}, vol.~622, pp.~L45--L48, 2005.

\bibitem{bildsten01}
L.~{Bildsten} {\em et~al.} {\em ApJ}, vol.~557, pp.~292--296, 2001.

\bibitem{watts06}
A.~L. {Watts} {\em et~al.} {\em MNRAS}, vol.~373, pp.~769--780, 2006.

\bibitem{falanga07}
M.~{Falanga} {\em et~al.} {\em ApJ}, vol.~661, pp.~1084--1088, 2007.

\bibitem{galloway07}
D.~K. {Galloway} {\em et~al.} {\em ApJL}, vol.~654, pp.~L73--L76, 2007.

\bibitem{falanga11}
M.~{Falanga} {\em et~al.} {\em A\&A}, vol.~529, p.~A68, 2011.

\bibitem{ferrigno11}
C.~{Ferrigno} {\em et~al.} {\em A\&A}, vol.~525, p.~A48, 2011.

\bibitem{kuin15}
P.~{Kuin} {\em et~al.} {\em The Astronomer's Telegram}, vol.~7849, 2015.

\bibitem{bozzo15b}
E.~{Bozzo} {\em et~al.} {\em The Astronomer's Telegram}, vol.~7852, 2015.

\bibitem{w03}
C.~{Winkler} {\em et~al.} {\em A\&A}, vol.~411, pp.~L1--L6, 2003.

\bibitem{c03}
T.~J.-L. {Courvoisier} {\em et~al.} {\em A\&A}, vol.~411, pp.~L53--L57, 2003.

\bibitem{gold03}
A.~{Goldwurm} {\em et~al.} {\em A\&A}, vol.~411, pp.~L223--L229, 2003.

\bibitem{u03}
P.~{Ubertini} {\em et~al.} {\em A\&A}, vol.~411, pp.~L131--L139, 2003.

\bibitem{lebr03}
F.~{Lebrun} {\em et~al.} {\em A\&A}, vol.~411, pp.~L141--L148, 2003.

\bibitem{lund03}
N.~{Lund} {\em et~al.} {\em A\&A}, vol.~411, pp.~L231--L238, 2003.

\bibitem{gros03}
A.~{Gros} {\em et~al.} {\em A\&A}, vol.~411, pp.~L179--L183, 2003.

\bibitem{torres08}
M.~A.~P. {Torres} {\em et~al.} {\em ApJ}, vol.~672, pp.~1079--1090, 2008.

\bibitem{burrows05}
D.~N. {Burrows} {\em et~al.} {\em SSR}, vol.~120, pp.~165--195, 2005.

\bibitem{krimm13}
H.~A. {Krimm} {\em et~al.} {\em ApJ s}, vol.~209, p.~14, 2013.

\bibitem{galloway08b}
D.~{Galloway}, ``{Accreting neutron star spins and the equation of state},''
  vol.~983, pp.~510--518, 2008.

\bibitem{arnaud96}
K.~A. {Arnaud}, ``{XSPEC: The First Ten Years},'' vol.~101, p.~17, 1996.

\bibitem{ps96}
J.~{Poutanen} {\em et~al.} {\em ApJ}, vol.~470, p.~249, 1996.

\bibitem{gp05}
M.~{Gierli{\'n}ski} {\em et~al.} {\em MNRAS}, vol.~359, pp.~1261--1276, 2005.

\bibitem{mfb05}
M.~{Falanga} {\em et~al.} {\em A\&A}, vol.~436, pp.~647--652, 2005.

\bibitem{mfc07}
M.~{Falanga} {\em et~al.} {\em A\&A}, vol.~464, pp.~1069--1074, 2007.

\bibitem{ip09}
A.~{Ibragimov} {\em et~al.} {\em MNRAS}, vol.~400, pp.~492--508, 2009.

\bibitem{bozzo10}
E.~{Bozzo} {\em et~al.} {\em A\&A}, vol.~509, p.~L3, 2010.

\bibitem{nowak04}
M.~A. {Nowak} {\em et~al.} {\em The Astronomer's Telegram}, vol.~369, 2004.

\bibitem{lebrun03}
F.~{Lebrun} {\em et~al.} {\em A\&A}, vol.~411, pp.~L141--L148, 2003.

\bibitem{papitto11}
A.~{Papitto} {\em et~al.} {\em A\&A}, vol.~528, p.~A55, 2011.

\bibitem{buccheri1983}
R.~{Buccheri} {\em et~al.} {\em A\&A}, vol.~128, pp.~245--251, 1983.

\bibitem{kuiper15}
L.~{Kuiper} {\em et~al.} {\em The Astronomer's Telegram}, vol.~7949, 2015.

\bibitem{falanga09}
M.~{Falanga} {\em et~al.} {\em A\&A}, vol.~496, pp.~333--342, 2009.

\bibitem{bozzo15a}
E.~{Bozzo} {\em et~al.} {\em A\&A}, vol.~579, p.~A56, 2015.

\bibitem{kuulkers03}
E.~{Kuulkers} {\em et~al.} {\em A\&A}, vol.~399, pp.~663--680, 2003.

\bibitem{galloway04}
D.~K. {Galloway} {\em et~al.} {\em ApJ}, vol.~601, pp.~466--473, 2004.

\bibitem{fujimoto87}
M.~Y. {Fujimoto} {\em et~al.} {\em ApJ}, vol.~319, pp.~902--915, 1987.

\bibitem{tauris06}
T.~M. {Tauris} {\em et~al.}, {\em {Formation and evolution of compact stellar
  X-ray sources}}, pp.~623--665.
\newblock 2006.

\bibitem{disalvo08}
T.~{di Salvo} {\em et~al.} {\em MNRAS}, vol.~389, pp.~1851--1857, 2008.

\bibitem{hartman08}
J.~M. {Hartman} {\em et~al.} {\em ApJ}, vol.~675, pp.~1468--1486.

\bibitem{patruno12b}
A.~{Patruno} {\em et~al.} {\em ApJL}, vol.~746, p.~L27, 2012.

\bibitem{falanga09b}
M.~{Falanga} {\em et~al.} {\em A\&A}, vol.~496, pp.~333--342, 2009.

\bibitem{lampe16}
N.~{Lampe} {\em et~al.} {\em ApJ}, vol.~819, p.~46, 2016.

\bibitem{eckert13}
D.~{Eckert} {\em et~al.} {\em The Astronomer's Telegram}, vol.~4925, 2013.

\bibitem{heinke13}
C.~O. {Heinke} {\em et~al.} {\em The Astronomer's Telegram}, vol.~4927, 2013.

\bibitem{romano13}
P.~{Romano} {\em et~al.} {\em The Astronomer's Telegram}, vol.~4929, 2013.

\bibitem{homan13}
J.~{Homan} {\em et~al.} {\em The Astronomer's Telegram}, vol.~5045, 2013.

\bibitem{Harris96}
W.~E. {Harris} {\em AJ}, vol.~112, p.~1487, 1996.

\bibitem{Monrad13}
L.~A.~G. {Monard} {\em et~al.} {\em The Astronomer's Telegram}, vol.~4964,
  2013.

\bibitem{Pallanca13a}
C.~{Pallanca} {\em et~al.} {\em The Astronomer's Telegram}, vol.~5003, 2013.

\bibitem{Cohn13}
H.~N. {Cohn} {\em et~al.} {\em The Astronomer's Telegram}, vol.~5031, 2013.

\bibitem{Pallanca13b}
C.~{Pallanca} {\em et~al.} {\em ApJ}, vol.~773, p.~122, 2013.

\bibitem{papitto13a}
A.~{Papitto} {\em et~al.} {\em The Astronomer's Telegram}.

\bibitem{Linares13}
M.~{Linares} {\em The Astronomer's Telegram}, vol.~4960, 2013.

\bibitem{Patruno13}
A.~{Patruno} {\em The Astronomer's Telegram}, vol.~5068, 2013.

\bibitem{Serino13}
M.~{Serino} {\em et~al.} {\em The Astronomer's Telegram}, vol.~4961, 2013.

\bibitem{Manchester05}
R.~N. {Manchester} {\em et~al.} {\em AJ}, vol.~129, pp.~1993--2006, 2005.

\bibitem{archibald09}
A.~M. {Archibald} {\em et~al.} {\em Science}, vol.~324, p.~1411, 2009.

\bibitem{demartino10}
D.~{de Martino} {\em et~al.} {\em A\&A}, vol.~515, p.~A25, 2010.

\bibitem{demartino14}
D.~{de Martino} {\em et~al.} {\em MNRAS}, vol.~444, pp.~3004--3014, 2014.

\bibitem{linares14b}
M.~{Linares} {\em ApJ}, vol.~795, p.~72, 2014.

\bibitem{patruno14}
A.~{Patruno} {\em et~al.} {\em ApJL}, vol.~781, p.~L3, 2014.

\bibitem{bassa14}
C.~G. {Bassa} {\em et~al.} {\em MNRAS}, vol.~441, pp.~1825--1830, 2014.

\bibitem{bogdanov14}
S.~{Bogdanov} {\em et~al.} {\em ArXiv e-prints}, 2014.

\bibitem{bogdanov15}
S.~{Bogdanov} {\em et~al.} {\em ApJ}, vol.~806, p.~148, June 2015.

\bibitem{ferrigno14}
C.~{Ferrigno} {\em et~al.} {\em A\&A}, vol.~567, p.~A77, 2014.

\bibitem{jansen01}
F.~{Jansen} {\em et~al.} {\em A\&A}, vol.~365, pp.~L1--L6, 2001.

\bibitem{evans09}
P.~A. {Evans} {\em et~al.} {\em MNRAS}, vol.~397, pp.~1177--1201, 2009.

\bibitem{meegan09}
C.~{Meegan} {\em et~al.} {\em ApJ}, vol.~702, pp.~791--804, 2009.

\bibitem{bissaldi09}
E.~{Bissaldi} {\em et~al.} {\em Experimental Astronomy}, vol.~24, pp.~47--88,
  2009.

\bibitem{atwood09}
W.~B. {Atwood} {\em et~al.} {\em ApJ}, vol.~697, pp.~1071--1102, 2009.

\bibitem{bozzo16}
E.~{Bozzo} {\em et~al.} {\em A\&A}, vol.~589, p.~A42, 2016.

\bibitem{DeMartino13}
D.~{de Martino} {\em et~al.} {\em A\&A}, vol.~550, p.~A89, 2013.

\bibitem{Papitto14}
A.~{Papitto} {\em et~al.} {\em MNRAS}, vol.~438, pp.~2105--2116, 2014.

\bibitem{Kaastra2016}
J.~S. {Kaastra} {\em et~al.} {\em A\&A}, vol.~587, p.~A151, 2016.

\bibitem{zdziarski96}
A.~A. {Zdziarski} {\em et~al.} {\em MNRAS}, vol.~283, pp.~193--206, 1996.

\bibitem{zycki99}
P.~T. {{\.Z}ycki} {\em et~al.} {\em MNRAS}, vol.~309, pp.~561--575, 1999.

\bibitem{gdb02}
M.~{Gierli{\'n}ski} {\em et~al.} {\em MNRAS}, vol.~331, pp.~141--153, 2002.

\bibitem{falanga12}
M.~{Falanga} {\em et~al.} {\em A\&A}, vol.~545, p.~A26, 2012.

\bibitem{dickey90}
J.~M. {Dickey} {\em et~al.} {\em ARAA}, vol.~28, pp.~215--261, 1990.

\bibitem{kalberla05}
P.~M.~W. {Kalberla} {\em et~al.} {\em A\&A}, vol.~440, pp.~775--782, 2005.

\bibitem{wilms00}
J.~{Wilms} {\em et~al.} {\em ApJ}, vol.~542, pp.~914--924, 2000.

\bibitem{Pavan13}
L.~{Pavan} {\em et~al.} {\em The Astronomer's Telegram}, vol.~4981, 2013.

\bibitem{Becker03}
W.~{Becker} {\em et~al.} {\em ApJ}, vol.~594, pp.~798--811, 2003.

\bibitem{Sanna17}
A.~{Sanna} {\em et~al.} {\em MNRAS}, vol.~466, pp.~2910--2917, 2017.

\bibitem{cui98}
W.~{Cui} {\em et~al.} {\em ApJL}, vol.~504, pp.~L27--L30, 1998.

\bibitem{ford00}
E.~C. {Ford} {\em ApJL}, vol.~535, pp.~L119--L122, 2000.

\bibitem{i11}
A.~{Ibragimov} {\em et~al.} {\em MNRAS}, vol.~415, pp.~1864--1874, 2011.

\bibitem{t02}
L.~{Titarchuk} {\em et~al.} {\em ApJL}, vol.~576, pp.~L49--L52, 2002.

\bibitem{pg03}
J.~{Poutanen} {\em et~al.} {\em MNRAS}, vol.~343, pp.~1301--1311, 2003.

\bibitem{pb06}
J.~{Poutanen} {\em et~al.} {\em MNRAS}, vol.~373, pp.~836--844, 2006.

\bibitem{lamb09}
F.~K. {Lamb} {\em et~al.} {\em ApJ}, vol.~706, pp.~417--435, 2009.

\bibitem{linares14}
M.~{Linares} {\em et~al.} {\em MNRAS}, vol.~438, pp.~251--261, 2014.

\bibitem{Strohmayer03}
T.~{Strohmayer} {\em et~al.} {\em ArXiv Astrophysics e-prints}, 2003.

\bibitem{Galloway08}
D.~{Galloway}, ``{Accreting neutron star spins and the equation of state},''
  vol.~983, pp.~510--518, 2008.

\bibitem{Gilfanov98}
M.~{Gilfanov} {\em et~al.} {\em A\&A}, vol.~338, pp.~L83--L86, 1998.

\bibitem{Powell07}
C.~R. {Powell} {\em et~al.} {\em MNRAS}, vol.~374, pp.~466--476, 2007.

\bibitem{Archibald13}
A.~M. {Archibald} {\em et~al.} {\em ArXiv e-prints}, 2013.

\bibitem{deloye03}
C.~J. {Deloye} {\em et~al.} {\em ApJ}, vol.~598, pp.~1217--1228, 2003.

\bibitem{defalco17}
V.~{De Falco} {\em et~al.} {\em A\&A}, vol.~599, p.~A88, 2017.

\bibitem{Syunyaev86}
R.~A. {Syunyaev} {\em et~al.} {\em Soviet Astronomy Letters}, vol.~12,
  pp.~117--120, 1986.

\bibitem{Revnivtsev13}
M.~G. {Revnivtsev} {\em et~al.} {\em MNRAS}, vol.~434, pp.~2355--2361, 2013.

\end{thebibliography}
}
\end{multicols}
\addcontentsline{toc}{chapter}{Bibliografy} 

\end{document}